\documentclass[12pt,twoside]{article}

\usepackage{epsfig}

\usepackage{axodraw}

\usepackage{graphicx}

\usepackage{rotate}

\title{CP Violation in the B System}

\author{Robert Fleischer\\
Deutsches Elektronen-Synchrotron DESY, Hamburg}

\date{\today}

\def\kpnn{K^+\rightarrow\pi^+\nu\overline{\nu}}
\def\klnn{K_{\rm L}\rightarrow\pi^0\nu\overline{\nu}}

\newcommand{\gsim}{
\mathrel{\hbox{\rlap{\hbox{\lower4pt\hbox{$\sim$}}}\hbox{$>$}}}}

\newcommand{\lsim}{
\mathrel{\hbox{\rlap{\hbox{\lower4pt\hbox{$\sim$}}}\hbox{$<$}}}}

\begin{document}


\begin{titlepage}
\begin{flushright}
\begin{tabular}{r}
DESY--THESIS--2002--022\\
\vspace*{1.0truecm}
July 2001
\end{tabular}
\end{flushright}

\vspace*{0.5truecm}

\begin{center}
\boldmath
{\Large \bf CP Violation in the $B$ System and}\\
\vspace*{0.5truecm} 
{\Large \bf Relations to $K\to\pi\nu\overline{\nu}$ Decays}
\unboldmath

\vspace*{2.0cm}

\renewcommand{\thefootnote}{\fnsymbol{footnote}}
{\sc{\large Robert Fleischer}}\footnote[1]{E-mail: {\tt 
Robert.Fleischer@desy.de}}
\renewcommand{\thefootnote}{\arabic{footnote}}

\vspace*{0.4cm} 

{\it Deutsches Elektronen-Synchrotron DESY, 
Notkestr.\ 85,\\ 
D--22607 Hamburg, Germany}

\vspace*{2.1truecm}

{\large\bf Abstract\\[10pt]} \parbox[t]{\textwidth}{
A review of CP violation in the $B$ system and strategies to determine the 
unitarity triangle of the CKM matrix is given. We begin with an introduction 
to the description of CP violation in the Standard Model of electroweak 
interactions, and discuss the basic features of the theoretical framework to 
deal with non-leptonic $B$ decays, which play the main r\^ole in this review. 
After a brief look at CP violation in the kaon system and a discussion of the 
rare decays $K^+\to\pi^+\nu\overline{\nu}$ and 
$K_{\rm L}\to\pi^0\nu\overline{\nu}$, we turn to the formalism of 
$B^0_{d,s}$--$\overline{B^0_{d,s}}$ mixing, allowing us to explore 
important $B$-factory benchmark modes and the $B_s$-meson system. We then 
focus on charged $B$ decays, $B\to\pi K$ modes and the phenomenology of 
$U$-spin related $B$ decays, including the $B_d\to\pi^+\pi^-$, 
$B_s\to K^+K^-$ system. Finally, we discuss a particularly simple -- but 
very predictive -- scenario for new physics, which is provided by models 
with ``minimal flavour violation''. In this framework, various bounds can 
be derived and interesting connections between the $B$ system and the rare 
kaon decays $K^+\to\pi^+\nu\overline{\nu}$ and 
$K_{\rm L}\to\pi^0\nu\overline{\nu}$ arise.}

\vspace*{0.8truecm}

(To appear in {\it Physics Reports})

\end{center}

\end{titlepage}
 


\newpage                      
\thispagestyle{empty}
\mbox{}
\newpage

\pagenumbering{roman}

\setcounter{page}{1}

\markboth{}{}

\tableofcontents

\newpage

\section{Introduction}

\pagenumbering{arabic}

\setcounter{page}{1}

\setcounter{equation}{0}

The violation of the CP symmetry, where C and P denote the 
charge-conjugation and parity-transformation operators, respectively,
is one of the fundamental and most exciting phenomena in particle 
physics. Although weak interactions are not invariant under P (and C) 
transformations, as discovered in 1957, it was believed for several 
years that the product CP was preserved. Consider, for instance, the 
process
\begin{equation}
\pi^+\to e^+\nu_e~\stackrel{C}{\longrightarrow}~\pi^-\to 
e^-\nu_e^C~\stackrel{P}{\longrightarrow}~\pi^-\to e^-\overline{\nu}_e,
\end{equation}
where the left-handed $\nu_e^C$ state is not observed in nature; only 
after performing an additional parity transformation we obtain the usual 
right-handed electron antineutrino. Consequently, it appears as if
CP was conserved in weak interactions. However, in 1964, it was then 
discovered through the observation of $K_{\rm L}\to \pi^+\pi^-$ decays 
that weak interactions are {\it not} invariant under CP transformations 
\cite{CP-discovery}. 

\subsection{Motivation}
Since its discovery in 1964, CP violation has only been accessible in 
the kaon system, and we still have few experimental insights into this 
exciting phenomenon. However, a new era has just begun through the
observation of CP violation in the $B$ system 
\cite{babar-CP-obs,belle-CP-obs}, which will provide several decisive tests 
of the Standard-Model description of CP violation. In this respect, it is 
obviously crucial to have CP-violating $B$-decay processes available that 
can be analysed reliably within the framework of the Standard Model. As an 
interesting by-product, strategies to explore CP violation with the help of 
$B$ decays provide usually also valuable insights into hadron dynamics, 
including aspects such as ``factorization'' and final-state interaction 
effects. A remarkable bridge to the kaon system is provided by the 
loop-induced decays $K^+\to\pi^+\nu\overline{\nu}$ and 
$K_{\rm L}\to\pi^0\nu\overline{\nu}$, which are very clean and exhibit 
interesting correlations with CP violation in the ``gold-plated''
channel $B_d\to J/\psi K_{\rm S}$. These issues will be the main focus of 
this review. For a collection of basic references on CP violation and 
alternative reviews, the reader is referred to \cite{basics}--\cite{Brev01}.

At present, we are still in the early stage of the $B$-factory era in 
particle physics: in the summer of 2000, the BaBar (SLAC) and Belle (KEK) 
collaborations reported their first results, and the CLEO-III detector 
(Cornell) started taking data. In the summer of 2001, BaBar and Belle 
could establish CP violation in the $B$ system, and run II of the Tevatron 
(Fermilab) is expected to provide data soon, allowing -- in addition to 
studies of $B_u$- and $B_d$-mesons -- a promising exploration of the $B_s$ 
system. Several interesting aspects will also be left for 
``second-generation'' $B$ experiments at hadron colliders, BTeV (Fermilab) 
and LHCb (CERN), which should enter the stage around 2005. Detailed studies 
of the $B$-physics potentials of BaBar, run II of the Tevatron, and the LHC 
can be found in \cite{Studies}. The rare kaon decays 
$K^+\to\pi^+\nu\overline{\nu}$ and $K_{\rm L}\to\pi^0\nu\overline{\nu}$
will be studied in dedicated experiments at Brookhaven, Fermilab and KEK. 
The experimental prospects for these modes were recently reviewed in 
\cite{littenberg}.

The reason why studies of CP-violating effects are so exciting is very 
simple: physics beyond the Standard Model is usually associated with 
new sources for CP violation. Important examples are non-minimal 
supersymmetry, left-right-symmetric models, models with extended Higgs 
sectors, and many other scenarios for ``new'' physics \cite{new-phys}. 
Moreover, the indirect constraints on CP-violating effects in $B$ 
decays that are implied by the CP violation in the kaon system and 
$B^0_{d}$--$\overline{B^0_{d}}$ mixing may also be affected strongly 
by new physics which does not yield additional sources for CP violation,
for instance by models with ``minimal flavour violation'' (MFV) 
\cite{MFV,UUT}. Then we would encounter discrepancies with the directly 
measured CP asymmetries in $B$ decays. Therefore, explorations of CP 
violation may well indicate physics beyond the Standard Model, or may be 
very helpful to distinguish between various realizations of one particular 
kind of new physics after the corresponding new-physics particles have been 
observed directly. In this context, it is worth mentioning that the evidence 
for neutrino masses we got over the recent years points also towards physics 
beyond the Standard Model \cite{akhmedov}, raising -- among other things -- 
the question of CP violation in the neutrino sector \cite{lindner}. In the 
more distant future, these fascinating topics may be studied at dedicated 
$\nu$-factories. 

\boldmath
\subsection{CP Violation in Cosmology and Beyond $K$ and $B$ Decays}
\unboldmath
Interestingly, indirect information on CP violation is also provided 
by cosmology. One of the characteristic features of our Universe is the 
dominance of matter over antimatter. From the observed baryon to photon 
number of the Universe, it can be concluded that there was, a few 
microseconds after the Big Bang, a very small excess of about $10^{-10}$ 
of matter with respect to antimatter. As was pointed out by Sakharov 
\cite{sakharov}, one of the necessary conditions to generate such an 
asymmetry of the Universe is -- in addition to baryon number violation 
and deviations from thermal equilibrium -- that the elementary interactions 
violate CP (and C) \cite{buchmueller}. Model calculations indicate, however, 
that the CP violation present in the Standard Model is too small to generate 
the observed matter--antimatter asymmetry of ${\cal O}(10^{-10})$ 
\cite{shapos}. It is conceivable that the particular kind of new physics 
underlying the baryon asymmetry is associated with very short-distance
scales. In this case, it could not be seen in CP-violating effects in 
weak meson decays. However, as we have noted above, there are also 
plenty of scenarios for physics beyond the Standard Model that would 
affect these processes. Moreover, we do not understand the observed 
patterns of quark and lepton masses, their mixings and the origin of 
flavour dynamics in general. It is likely that the new physics required 
to understand these features is also related to new sources for CP 
violation.

Let us note that not only $K$- and $B$-meson decays allow an exploration 
of CP violation in the laboratory. There are of course also other 
systems which are interesting in this respect, for example the $D$-meson 
system \cite{bigi-D,Ddecay}, where CP-violating and mixing effects are 
very small in the Standard Model, electric dipole moments \cite{dipole}, or 
CP violation in hyperon decays \cite{hyperon}. However, apart from
$D^0$--$\overline{D^0}$ mixing, which may also affect a certain $B$-physics 
approach to explore CP violation, we shall not consider these topics in 
further detail and refer the reader to the corresponding papers and 
references therein.

\subsection{Outline}
This review can be divided into three main parts, where 
Sections~\ref{sec:SM}--\ref{sec:neutral} set the stage for our 
discussion, Sections~\ref{sec:benchmark}--\ref{sec:charged} give 
an up-dated presentation of various ``standard'' approaches, 
and Sections~\ref{sec:BpiK}--\ref{sec:Other} deal with more recent 
developments.

To be more specific, the outline is as follows: in Section~\ref{sec:SM}, 
we discuss the description of CP violation within the framework of the
Standard Model, where the key elements are the quark-mixing matrix 
and the associated unitarity triangles. Since non-leptonic $B$ decays
play the main r\^ole for the exploration of CP violation in the $B$ system,
we discuss the basic features of the theoretical framework to deal with
these transitions in Section~\ref{sec:nonlept}. Before turning to
CP-violating effects in $B$ decays, we have a brief look at the present 
status of CP violation in the kaon system in Section~\ref{sec:kaon},
where we also give an introduction to the rare decays 
$K^+\to\pi^+\nu\overline{\nu}$ and $K_{\rm L}\to\pi^0\nu\overline{\nu}$.
In Section~\ref{sec:neutral}, we then discuss the formalism to describe
the time evolution of neutral $B$ decays, arising from 
$B^0_{d,s}$--$\overline{B^0_{d,s}}$ mixing, and introduce the 
corresponding CP-violating observables. 

The formalism developed in Section~\ref{sec:neutral} is applied to 
important $B$-factory benchmark modes in Section~\ref{sec:benchmark}, 
where we present detailed discussions of the $B\to J/\psi K$, 
$B\to \phi K$ and $B\to\pi\pi$ systems, as well as of 
$B_d\to D^{(\ast)\pm} \pi^\mp$ decays. Because of its outstanding r\^ole for 
$B$-physics experiments at hadron machines, we have devoted a separate 
section to the $B_s$ system: in Section~\ref{sec:Bs}, we discuss contraints 
on the unitarity triangles implied by $B^0_{s}$--$\overline{B^0_{s}}$ mixing, 
the width difference $\Delta\Gamma_s$ between the $B_s$ mass eigenstates, 
``untagged'' $B_s$ decay rates,  and strategies to explore CP violation
with pure tree decays and the ``golden'' mode $B_s\to J/\psi\phi$. 
In Section~\ref{sec:charged}, we turn to CP-violating effects in decays 
of charged $B$-mesons. In the case of $B^\pm_u \to K^\pm D$ and 
$B^\pm_c\to D_s^\pm D$ modes, hadronic uncertainties can be eliminated 
with the help of certain amplitude relations, thereby allowing
studies of CP violation. There we give also a brief discussion 
of $D^0$--$\overline{D^0}$ mixing.

Amplitude relations play also a key r\^ole in the subsequent two
sections: in Section~\ref{sec:BpiK}, we turn to the phenomenology of 
$B\to\pi K$ decays, whereas we focus on certain $U$-spin-related decays 
in Section~\ref{sec:Uspin}, including the $B_{s(d)}\to J/\psi K_{\rm S}$,
$B_{d(s)}\to D^+_{d(s)}D^-_{d(s)}$, $B_{d(s)}\to K^0\overline{K^0}$ and
$B_d\to\pi^+\pi^-$, $B_s\to K^+K^-$ systems, as well as $B_{(s)}\to\pi K$
modes. The physics potential of $B\to\pi K$ decays is particularly 
interesting for the $B$-factories, while the $U$-spin-related modes -- 
in particular the $B_d\to\pi^+\pi^-$, $B_s\to K^+K^-$ system -- is very 
promising for run II of the Tevatron and ideally suited for BTeV and LHCb. 
Although general features of the
impact of physics beyond the Standard Model on the decays listed above
will be emphasized throughout this review in a model-independent way, 
in Section~\ref{sec:MFV}, we discuss a particularly simple -- but very 
predictive -- scenario for new physics in more detail, which is provided 
by models with ``minimal flavour violation''. Within this framework, 
interesting bounds on CP violation in $B_d\to J/\psi K_{\rm S}$ decays 
can be derived, and remarkable connections with the rare kaon decays 
$K^+\to\pi^+\nu\overline{\nu}$ and $K_{\rm L}\to\pi^0\nu\overline{\nu}$ 
arise. Finally, we make a few remarks on further interesting aspects of $B$ 
physics in Section~\ref{sec:Other}, and give the conclusions and a 
brief outlook in Section~\ref{sec:concl}.

\section{CP Violation in the Standard Model}\label{sec:SM}
\setcounter{equation}{0}
\subsection{Charged-Current Interactions}\label{subsec:CC}
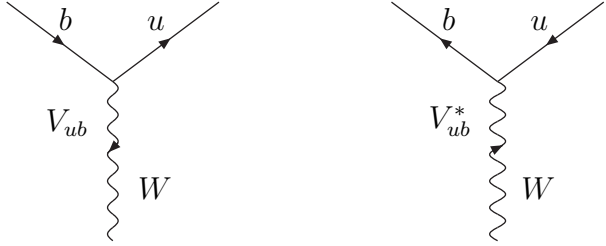
\begin{figure}[t]
\begin{center}
\begin{picture}(360,100)(0,60)
\ArrowLine(60,150)(100,120)
\ArrowLine(100,120)(140,150)
\ArrowLine(245,120)(205,150)
\ArrowLine(285,150)(245,120)
\Photon(100,120)(100,60){2}{6}
\ArrowLine(102,97)(98,93)
\Photon(245,120)(245,60){3}{6}
\ArrowLine(243,94)(247,96)
\Text(80,140)[bl]{$b$}
\Text(120,140)[br]{$u$}
\Text(75,100)[bl]{$V_{ub}$}
\Text(225,140)[bl]{$b$}
\Text(265,140)[br]{$u$}
\Text(220,100)[bl]{$V_{ub}^\ast$}
\Text(110,80)[l]{$W$}
\Text(255,80)[l]{$W$}
\end{picture}
\end{center}
\caption{Charged-current interactions in the Standard Model.}\label{fig:CC}
\end{figure}

In the framework of the Standard Model of electroweak interactions 
\cite{SM}, which is based on the spontaneously broken gauge group
\begin{equation}
SU(2)_{\rm L}\times U(1)_{\rm Y}
\stackrel{{\rm SSB}}
{\longrightarrow}U(1)_{\rm em},
\end{equation}
CP violation is related to the Cabibbo--Kobayashi--Maskawa (CKM) matrix 
\cite{cab,km}, connecting the electroweak eigenstates $(d',s',b')$ of 
the down, strange and bottom quarks with their mass eigenstates 
$(d,s,b)$ through the following unitary transformation:
\begin{equation}\label{ckm}
\left(\begin{array}{c}
d'\\
s'\\
b'
\end{array}\right)=\left(\begin{array}{ccc}
V_{ud}&V_{us}&V_{ub}\\
V_{cd}&V_{cs}&V_{cb}\\
V_{td}&V_{ts}&V_{tb}
\end{array}\right)\cdot
\left(\begin{array}{c}
d\\
s\\
b
\end{array}\right)\equiv\hat V_{\mbox{{\scriptsize CKM}}}\cdot
\left(\begin{array}{c}
d\\
s\\
b
\end{array}\right).
\end{equation}
The elements of the CKM matrix describe charged-current couplings, as
can be seen easily by expressing the non-leptonic charged-current 
interaction Lagrangian in terms of the mass eigenstates appearing in
(\ref{ckm}):
\begin{equation}\label{cc-lag2}
{\cal L}_{\mbox{{\scriptsize int}}}^{\mbox{{\scriptsize CC}}}=
-\frac{g_2}{\sqrt{2}}\left(\begin{array}{ccc}
\overline{u}_{\mbox{{\scriptsize L}}},& \overline{c}_{\mbox{{\scriptsize L}}},
&\overline{t}_{\mbox{{\scriptsize L}}}\end{array}\right)\gamma^\mu\,\hat
V_{\mbox{{\scriptsize CKM}}}
\left(
\begin{array}{c}
d_{\mbox{{\scriptsize L}}}\\
s_{\mbox{{\scriptsize L}}}\\
b_{\mbox{{\scriptsize L}}}
\end{array}\right)W_\mu^\dagger\,\,+\,\,\mbox{h.c.,}
\end{equation}
where the gauge coupling $g_2$ is related to 
$SU(2)_{\mbox{{\scriptsize L}}}$, and the $W_\mu^{(\dagger)}$ field 
corresponds to the charged $W$-bosons. In Fig.~\ref{fig:CC}, we show the 
$b\to u\,W$ vertex and its CP conjugate. The important feature is that 
these processes are related by the replacement 
\begin{equation}\label{CKM-CP}
V_{ub}\stackrel{CP}{\longrightarrow}V_{ub}^\ast.
\end{equation}
Consequently, in the Standard Model, CP violation is due to complex
phases of CKM matrix elements. The origin of these phases, as the
origin of quark mixing and flavour dynamics in general, lies of course
beyond the Standard Model.

\subsection{The Strong CP Problem}
The CP-violating effects discussed in this review are essentially due 
to phases appearing in the CKM matrix, i.e.\ they arise from weak 
interactions. However, there is also another source for CP violation 
in the Standard Model: in Quantum Chromodynamics (QCD), which is described 
by the gauge group $SU(3)_{\rm C}$, topologically non-trivial quantum 
fluctuations, so-called instantons, induce a parity- and 
time-reversal-violating term of the QCD Lagrange density of the following 
form:
\begin{equation}
S_\theta=\theta\frac{g_s^2}{32\pi^2} G^a_{\mu\nu}\tilde G^{a\mu\nu}.
\end{equation}
Here $\theta$ is the ``QCD vacuum angle'', $g_s$ the QCD gauge coupling, 
$G^a_{\mu\nu}$ the QCD field strength tensor, and $\tilde G^{a\mu\nu}$ 
its dual. Observable quantities depend on the parameter
\begin{equation}
\overline{\theta}\equiv\theta-\mbox{arg}(\det M_q),
\end{equation}
where $M_q$ is the non-diagonal quark-mass matrix in the electroweak 
basis. The experimental limits on the electric dipole moment of the 
neutron imply 
$\overline{\theta}
\mathrel{\hbox{\rlap{\hbox{\lower4pt\hbox{$\sim$}}}\hbox{$<$}}}10^{-10}$.
Since QCD gauge invariance would perfectly allow a value of 
$\overline{\theta}={\cal O}(1)$, we arrive at a very puzzling situation, 
suggesting a formidable fine tuning of the $\theta$ parameter to yield
such an extremely tiny value of $\overline{\theta}$. A detailed discussion 
of this ``strong CP problem'', as well as ideas to overcome it, can be 
found, for instance, in \cite{peccei}. 

\subsection{Parametrizations of the CKM Matrix}\label{subsec:CKM-par}
As we have seen in (\ref{CKM-CP}), within the Standard Model, CP violation 
is related to complex phases of CKM matrix elements. However, the phase 
structure of the CKM matrix is not unique, as we may perform the 
phase transformations
\begin{equation}\label{CKM-trafo}
V_{UD}\to\exp(i\xi_U)V_{UD}\exp(-i\xi_D),
\end{equation}
which correspond to the following redefinitions of the up- and down-type 
quark fields:
\begin{equation}\label{field-trafo}
U\to \exp(i\xi_U)U,\quad D\to \exp(i\xi_D)D. 
\end{equation}
Using these transformations, it can be shown that the general $N$-generation 
quark-mixing-matrix is described by $(N-1)^2$ parameters, consisting of
\begin{equation}
\frac{1}{2}N(N-1)
\end{equation}
Euler-type angles, and
\begin{equation}
\frac{1}{2}(N-1)(N-2)
\end{equation}
complex phases. In the two-generation case \cite{cab}, 
we have therefore to deal with a real matrix, involving just one Euler-type
angle, and arrive at the well-known Cabibbo matrix
\begin{equation}\label{Cmatrix}
\hat V_{\rm C}=\left(\begin{array}{cc}
\cos\theta_{\rm C}&\sin\theta_{\rm C}\\
-\sin\theta_{\rm C}&\cos\theta_{\rm C}
\end{array}\right),
\end{equation}
where $\sin\theta_{\rm C}=0.22$ can be determined from 
semi-leptonic $K\to\pi e^+\nu_e$ decays.

\subsubsection{Standard Parametrization}
In the case of three generations, three Euler-type angles and one
{\it complex phase} are needed to parametrize the CKM matrix. This 
complex phase allows us to accommodate CP violation in the Standard 
Model, as was pointed out by Kobayashi and Maskawa in 1973 \cite{km}. 
In the ``standard parametrization'' \cite{SMpar}, the three-generation 
CKM matrix takes the following form:
\begin{equation}\label{standard}
\hat V_{\rm CKM}=\left(\begin{array}{ccc}
c_{12}c_{13}&s_{12}c_{13}&s_{13}e^{-i\delta_{13}}\\ -s_{12}c_{23}
-c_{12}s_{23}s_{13}e^{i\delta_{13}}&c_{12}c_{23}-
s_{12}s_{23}s_{13}e^{i\delta_{13}}&
s_{23}c_{13}\\ s_{12}s_{23}-c_{12}c_{23}s_{13}e^{i\delta_{13}}&-c_{12}s_{23}
-s_{12}c_{23}s_{13}e^{i\delta_{13}}&c_{23}c_{13}
\end{array}\right),
\end{equation}
where $c_{ij}\equiv\cos\theta_{ij}$ and $s_{ij}\equiv\sin\theta_{ij}$. 
Performing appropriate redefinitions of the quark-field phases, the real 
angles $\theta_{12}$, $\theta_{23}$ and $\theta_{13}$ can all be made to 
lie in the first quadrant. The advantage of this parametrization is that
the generation labels $i,j=1,2,3$ are introduced in such a way that
the mixing between two chosen generations vanishes if the corresponding
mixing angle $\theta_{ij}$ is set to zero. In particular, for 
$\theta_{23}=\theta_{13}=0$, the third generation decouples, and we arrive
at a situation characterized by the Cabibbo matrix given in (\ref{Cmatrix}).

\subsubsection{Fritzsch--Xing Parametrization}
Another interesting parametrization of the CKM matrix was proposed by 
Fritzsch and Xing \cite{FX}:
\begin{equation}
\hat V_{\rm CKM}=\left(\begin{array}{ccc}
s_{\rm u} s_{\rm d} c + c_{\rm u} c_{\rm d} e^{-i\varphi} & 
s_{\rm u} c_{\rm d} c - c_{\rm u} s_{\rm d} e^{-i\varphi} &s_{\rm u} s\\
c_{\rm u} s_{\rm d} c - s_{\rm u} c_{\rm d} e^{-i\varphi} & 
c_{\rm u} c_{\rm d} c + s_{\rm u} s_{\rm d} e^{-i\varphi} &c_{\rm u} s\\
-s_{\rm d}s & -c_{\rm d}s & c
\end{array}\right).
\end{equation}
It is inspired by the hierarchical structure of the quark-mass spectrum, 
and is particularly useful in the context of models for fermion masses and
mixings. The characteristic feature of this parametrization is that
the complex phase arises only in the $2\times2$ submatrix involving
the up, down, strange and charm quarks.

\begin{figure}[t]
\vspace{0.10in}
\centerline{
\epsfysize=5.0truecm
\epsffile{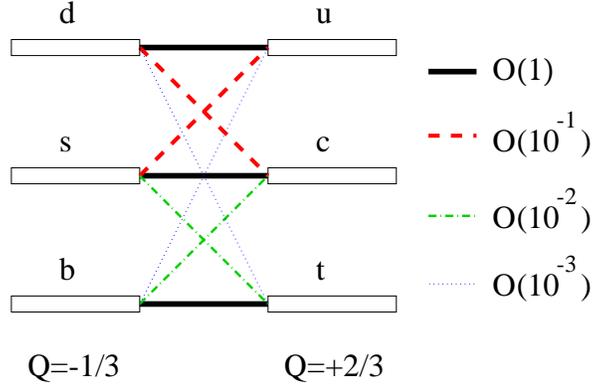}}
\caption[]{Hierarchy of the quark transitions mediated through charged 
currents.}\label{fig:term}
\end{figure}

\subsubsection{Wolfenstein Parametrization}\label{subsubsec:Wolf}
In Fig.~\ref{fig:term}, we have illustrated the hierarchy of the 
strengths of the quark transitions mediated through charged-current 
interactions: transitions within the same generation are governed by CKM 
elements of ${\cal O}(1)$, those between the first and the second 
generation are suppressed by CKM factors of ${\cal O}(10^{-1})$, those 
between the second and the third generation are suppressed by 
${\cal O}(10^{-2})$, and the transitions between the first and the third 
generation are even suppressed by CKM factors of ${\cal O}(10^{-3})$. 
In the standard parametrization (\ref{standard}), this hierarchy 
is reflected by
\begin{equation}
s_{12}=0.222 \,\gg\, s_{23}={\cal O}(10^{-2}) \,\gg\, 
s_{13}={\cal O}(10^{-3}). 
\end{equation}
If we introduce a set of new parameters $\lambda$, $A$, $\rho$ and $\eta$
by imposing the relations \cite{schu,blo}
\begin{equation}\label{set-rel}
s_{12}\equiv\lambda=0.222,\quad s_{23}\equiv A\lambda^2,\quad 
s_{13}e^{-i\delta_{13}}\equiv A\lambda^3(\rho-i\eta),
\end{equation}
and go back to the standard parametrization (\ref{standard}), we obtain
an {\it exact} parametrization of the CKM matrix as a function of $\lambda$ 
(and $A$, $\rho$, $\eta$). Now we can expand straightforwardly each CKM 
element in the small parameter $\lambda$. Neglecting terms of 
${\cal O}(\lambda^4)$, we arrive at the famous ``Wolfenstein 
parametrization'' of the CKM matrix \cite{wolf}:
\begin{equation}\label{W-par}
\hat V_{\mbox{{\scriptsize CKM}}} =\left(\begin{array}{ccc}
1-\frac{1}{2}\lambda^2 & \lambda & A\lambda^3(\rho-i\eta) \\
-\lambda & 1-\frac{1}{2}\lambda^2 & A\lambda^2\\
A\lambda^3(1-\rho-i\eta) & -A\lambda^2 & 1
\end{array}\right)+{\cal O}(\lambda^4).
\end{equation}
Since this parametrization makes the hierarchy of the CKM matrix explicit, 
it is very useful for phenomenological applications. 

In several cases, next-to-leading order corrections in $\lambda$ play an 
important r\^ole. Using the exact parametrization following from 
(\ref{set-rel}), they can be calculated 
in a similar manner by expanding each CKM element to the desired accuracy 
in $\lambda$ \cite{Brev01,blo}:
\begin{displaymath}
V_{ud}=1-\frac{1}{2}\lambda^2-\frac{1}{8}\lambda^4+{\cal O}(\lambda^6),\quad
V_{us}=\lambda+{\cal O}(\lambda^7),\quad
V_{ub}=A\lambda^3(\rho-i\,\eta),
\end{displaymath}
\begin{displaymath}
V_{cd}=-\lambda+\frac{1}{2}A^2\lambda^5\left[1-2(\rho+i\eta)\right]+
{\cal O}(\lambda^7),
\end{displaymath}
\begin{equation}\label{NLO-wolf}
V_{cs}=1-\frac{1}{2}\lambda^2-\frac{1}{8}\lambda^4(1+4A^2)+
{\cal O}(\lambda^6),
\end{equation}
\begin{displaymath}
V_{cb}=A\lambda^2+{\cal O}(\lambda^8),\quad
V_{td}=A\lambda^3\left[1-(\rho+i\eta)\left(1-\frac{1}{2}\lambda^2\right)
\right]+{\cal O}(\lambda^7),
\end{displaymath}
\begin{displaymath}
V_{ts}=-A\lambda^2+\frac{1}{2}A(1-2\rho)\lambda^4-i\eta A\lambda^4
+{\cal O}(\lambda^6),\quad
V_{tb}=1-\frac{1}{2}A^2\lambda^4+{\cal O}(\lambda^6).
\end{displaymath}
It should be noted that here
\begin{equation}
V_{ub}=A\lambda^3(\rho-i\eta)
\end{equation}
receives {\it by definition} no power corrections in $\lambda$. Introducing 
the following modified Wolfenstein parameters \cite{blo}:
\begin{equation}\label{rho-eta-bar}
\overline{\rho}\equiv\rho\left(1-\frac{1}{2}\lambda^2\right),\quad
\overline{\eta}\equiv\eta\left(1-\frac{1}{2}\lambda^2\right),
\end{equation}
we may simply write, up to corrections of ${\cal O}(\lambda^7)$,
\begin{equation}\label{Vtd-expr}
V_{td}=A\lambda^3(1-\overline{\rho}-i\,\overline{\eta}).
\end{equation}
Moreover, we have to an excellent accuracy
\begin{equation}\label{Def-A}
V_{us}=\lambda\quad \mbox{and}\quad 
V_{cb}=A\lambda^2,
\end{equation}
as these CKM elements receive only corrections at the $\lambda^7$ and
$\lambda^8$ levels, respectively. In comparison with other generalizations
of the Wolfenstein parametrization found in the literature, the advantage
of (\ref{NLO-wolf}) is the absence of relevant corrections to $V_{us}$
and $V_{cb}$, and that $V_{ub}$ and $V_{td}$ take similar forms as in
(\ref{W-par}). 

Let us finally note that physical observables, for instance CP-violating
asymmetries, cannot depend on the chosen parametrization of the CKM matrix, 
i.e.\ on the phase transformations given in (\ref{field-trafo}).

\subsection{Requirements for CP Violation}
As we have seen in Subsection~\ref{subsec:CKM-par}, at least three 
generations are required to accommodate CP violation in the Standard 
Model. However, still more conditions have to be satisfied for observable
CP-violating effects. They can be summarized as follows 
\cite{jarlskog,BBG}:
\begin{eqnarray}\label{CP-req}
\lefteqn{(m_t^2-m_c^2)(m_t^2-m_u^2)(m_c^2-m_u^2)}\nonumber\\
&&\times(m_b^2-m_s^2)(m_b^2-m_d^2)(m_s^2-m_d^2)\times
J_{\rm CP}\,\not=\,0,
\end{eqnarray}
where
\begin{equation}
J_{\rm CP}=|\mbox{Im}(V_{i\alpha}V_{j\beta}V_{i\beta}^\ast 
V_{j\alpha}^\ast)|\quad(i\not=j,\,\alpha\not=\beta)\,.
\end{equation}
The factors in (\ref{CP-req}) involving the quark masses are related 
to the fact that the CP-violating phase of the CKM matrix could be 
eliminated through an appropriate unitary transformation of quark 
fields if any two quarks with the same charge had the same mass. 
Consequently, the origin of CP violation is not only closely related 
to the number of fermion generations, but also to the hierarchy of 
quark masses and cannot be understood in a deeper way unless we have
insights into these very fundamental issues, usually referred to as
the ``flavour problem''. 

The second ingredient of (\ref{CP-req}), the ``Jarlskog Parameter'' 
$J_{\rm CP}$ \cite{jarlskog}, can be interpreted as a measure of the 
strength of CP violation in the Standard Model. It does not depend on 
the chosen quark-field parametrization, i.e.\ is invariant under 
(\ref{field-trafo}), and the unitarity of the CKM matrix implies that 
all combinations 
$|\mbox{Im}(V_{i\alpha}V_{j\beta}V_{i\beta}^\ast V_{j\alpha}^\ast)|$ 
are equal. Using the Standard and Wolfenstein parametrizations, we obtain
\begin{equation}
J_{\rm CP}=s_{12}s_{13}s_{23}c_{12}c_{23}c_{13}^2\sin\delta_{13}=
\lambda^6A^2\eta={\cal O}(10^{-5}),
\end{equation}
where we have taken into account the present experimental information
on the Wolfenstein parameters in the quantitative estimate (see 
Subsection~\ref{subsec:CKM-fits}). Consequently, CP violation is a small 
effect in the Standard Model. Typically, new complex couplings are present 
in scenarios for new physics, yielding additional sources for CP violation.

\subsection{The Unitarity Triangles of the CKM Matrix}
Concerning tests of the Kobayashi--Maskawa picture of CP violation, 
the central targets are the ``unitarity triangles'' of the CKM 
matrix. The unitarity of the CKM matrix, which is described by
\begin{equation}
\hat V_{\mbox{{\scriptsize CKM}}}^{\,\,\dagger}\cdot\hat 
V_{\mbox{{\scriptsize CKM}}}=
\hat 1=\hat V_{\mbox{{\scriptsize CKM}}}\cdot\hat V_{\mbox{{\scriptsize 
CKM}}}^{\,\,\dagger},
\end{equation}
implies a set of 12 equations, consisting of 6 normalization relations 
and 6 orthogonality relations:
\begin{itemize}
\item Normalization relations:
\begin{eqnarray}
|V_{ud}|^2+|V_{cd}|^2+|V_{td}|^2 & = & 
1\quad\mbox{[1st column]}\label{UT-norm1}\\
|V_{us}|^2+|V_{cs}|^2+|V_{ts}|^2 & = & 1\quad\mbox{[2nd column]}\\
|V_{ub}|^2+|V_{cb}|^2+|V_{tb}|^2 & = & 1\quad\mbox{[3rd column]}\\
&&\nonumber\\
|V_{ud}|^2+|V_{us}|^2+|V_{ub}|^2 & = & 1\quad\mbox{[1st row]}\\
|V_{cd}|^2+|V_{cs}|^2+|V_{cb}|^2 & = & 1\quad\mbox{[2nd row]}\\
|V_{td}|^2+|V_{ts}|^2+|V_{tb}|^2 & = & 1\quad\mbox{[3rd row].}
\end{eqnarray}
\item Orthogonality relations:
\begin{eqnarray}
V_{ud}V_{us}^\ast+V_{cd}V_{cs}^\ast+V_{td}V_{ts}^\ast & = &
0\quad\mbox{[1st and 2nd column]}\label{UT-ort1}\\
V_{ud}V_{ub}^\ast+V_{cd}V_{cb}^\ast+V_{td}V_{tb}^\ast & = &
0\quad\mbox{[1st and 3nd column]}\label{UT-ort2}\\
V_{us}V_{ub}^\ast+V_{cs}V_{cb}^\ast+V_{ts}V_{tb}^\ast & = &
0\quad\mbox{[2nd and 3rd column]}\label{UT-ort3}\\
&&\nonumber\\
V_{ud}^\ast V_{cd}+V_{us}^\ast V_{cs}+V_{ub}^\ast V_{cb} & = &
0\quad\mbox{[1st and 2nd row]}\label{UT-ort4}\\
V_{ub}^\ast V_{tb}+V_{us}^\ast V_{ts}+V_{ud}^\ast V_{td} & = &
0\quad\mbox{[1st and 3rd row]}\label{UT-ort5}\\
V_{cd}^\ast V_{td}+V_{cs}^\ast V_{ts}+V_{cb}^\ast V_{tb} & = & 
0\quad\mbox{[2nd and 3rd row].}\label{UT-ort6}
\end{eqnarray}
\end{itemize}
The orthogonality relations are of particular interest, since they can
be represented as six ``unitarity triangles'' in the complex plane 
\cite{AKL}. It should be noted that the set of equations 
(\ref{UT-norm1})--(\ref{UT-ort6}) is invariant under the phase 
transformations specified in (\ref{CKM-trafo}). If one performs such 
transformations, the triangles corresponding to 
(\ref{UT-ort1})--(\ref{UT-ort6}) are rotated in the complex plane. 
However, the angles and sides of these triangles remain unchanged and 
are therefore physical observables. It can be shown that all six 
unitarity triangles have the same area \cite{JS}, which is given by the 
Jarlskog parameter as follows:
\begin{equation}
A_{\Delta}=\frac{1}{2}J_{\rm CP}.
\end{equation}

The shape of the unitarity triangles can be analysed with the help of
the Wolfenstein parametrization, implying the following structure for
(\ref{UT-ort1})--(\ref{UT-ort3}) and (\ref{UT-ort4})--(\ref{UT-ort6}):
\begin{eqnarray}
{\cal O}(\lambda)+{\cal O}(\lambda)+{\cal O}(\lambda^5)&=&0\\
{\cal O}(\lambda^3)+{\cal O}(\lambda^3)+{\cal O}(\lambda^3)&=&0\\
{\cal O}(\lambda^4)+{\cal O}(\lambda^2)+{\cal O}(\lambda^2)&=&0.
\end{eqnarray}
Consequently, only in the triangles corresponding to (\ref{UT-ort2}) 
and (\ref{UT-ort5}), all three sides are of comparable magnitude 
${\cal O}(\lambda^3)$, while in the remaining triangles, one side is 
suppressed with respect to the others by ${\cal O}(\lambda^2)$ or 
${\cal O}(\lambda^4)$. The two ``non-squashed'' orthogonality relations 
(\ref{UT-ort2}) and (\ref{UT-ort5}) agree at the $\lambda^3$ level and 
differ only through ${\cal O}(\lambda^5)$ corrections. If we neglect 
these subleading contributions, we arrive at {\it the} unitarity triangle 
of the CKM matrix \cite{JS,ut}, which appears usually in the literature. 
To be more specific, at leading non-vanishing order in $\lambda$, i.e.\
${\cal O}(\lambda^3)$, (\ref{UT-ort2}) and (\ref{UT-ort5}) imply the 
triangle relation
\begin{equation}\label{UTLO}
(\rho+i\eta)A\lambda^3+(-A\lambda^3)+(1-\rho-i\eta)A\lambda^3=0.
\end{equation}
If we now rescale all terms by $A\lambda^3$, the basis of the resulting 
triangle, which coincides with the real axis, is normalized to one, and 
its apex is given by $(\rho,\eta)$. 

\begin{figure}
\begin{tabular}{lr}
   \epsfysize=4.5cm
   \epsffile{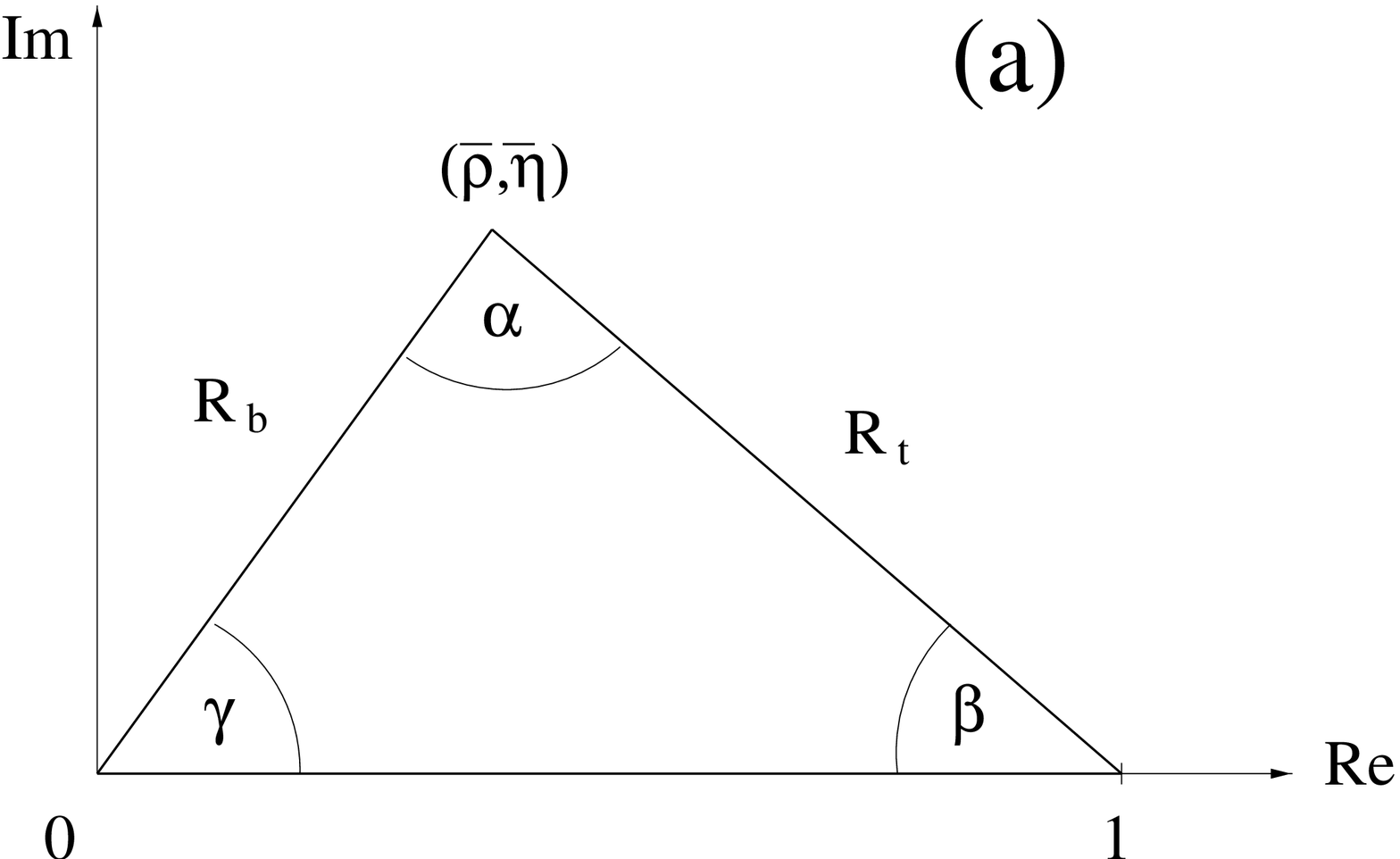}
&
   \epsfysize=4.5cm
   \epsffile{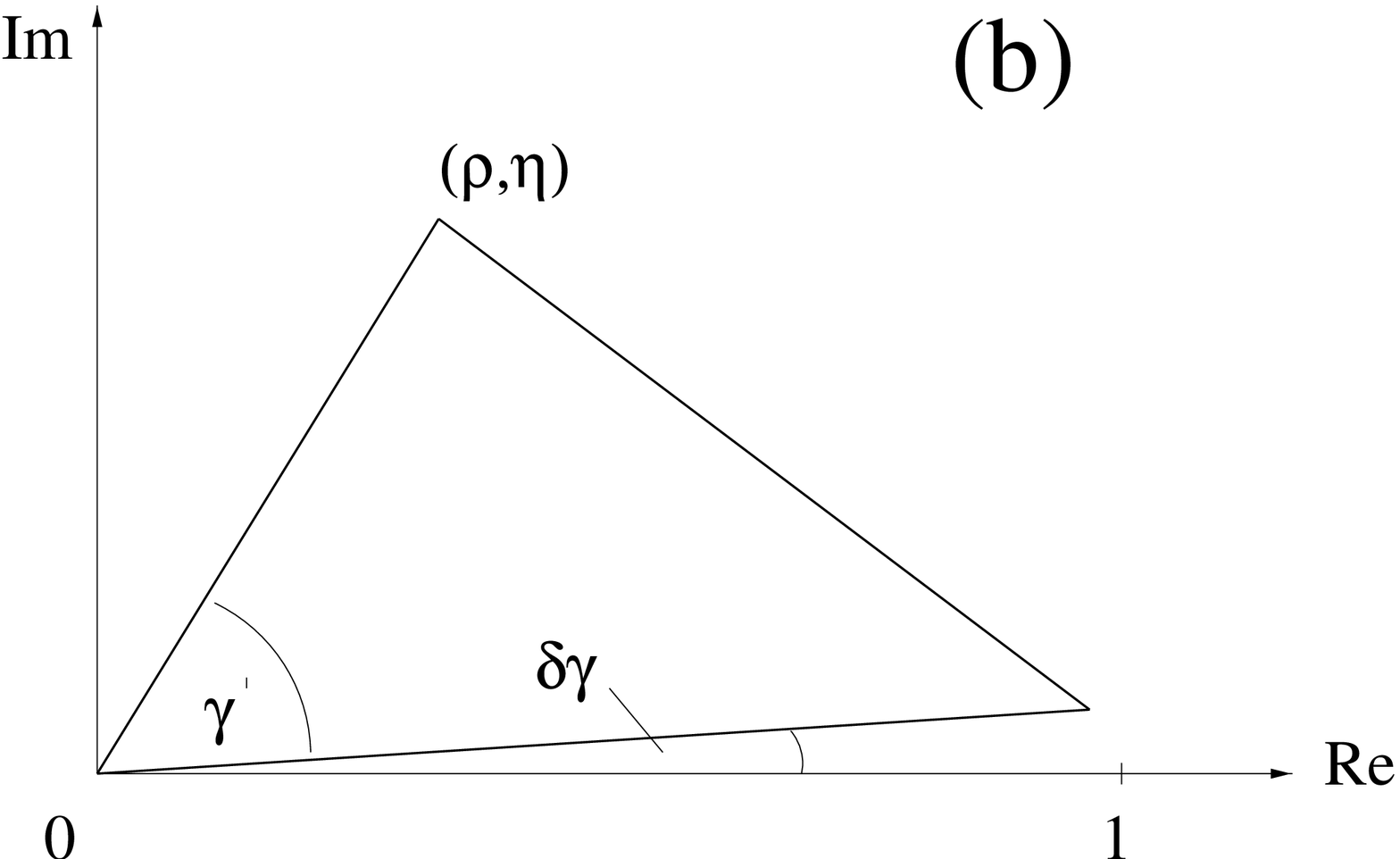}
\end{tabular}
\caption[]{The two non-squashed unitarity triangles of the CKM matrix: 
(a) and (b) correspond to the orthogonality relations (\ref{UT-ort2}) and 
(\ref{UT-ort5}), respectively.}
\label{fig:UT}
\end{figure}

In the era of ``second-generation'' $B$ experiments, the experimental 
accuracy will be so tremendous that we will also have to take into account 
the next-to-leading order terms of the Wolfenstein expansion, and will have 
to distinguish between the unitarity triangles described by (\ref{UT-ort2}) 
and (\ref{UT-ort5}). To this end, the generalized Wolfenstein parametrization 
introduced in Subsection~\ref{subsubsec:Wolf} is particularly convenient. 
If we keep terms up to ${\cal O}(\lambda^5)$ in the orthogonality relations 
(\ref{UT-ort2}) and (\ref{UT-ort5}), and rescale them as above by 
$A\lambda^3$, we arrive at the two unitarity triangles illustrated in 
Fig.~\ref{fig:UT}, which differ at the $\lambda^2$ level. In the unitarity 
triangle corresponding to (\ref{UT-ort2}), we have just to switch to 
the modified Wolfenstein parameters (\ref{rho-eta-bar}), i.e.\ the 
apex is given by $(\overline{\rho},\overline{\eta})$ instead of 
$(\rho,\eta)$. Consequently, here we have a straightforward generalization
of {\it the} unitarity triangle, and whenever we refer to a unitarity
triangle in the remainder of this review, we mean the one shown in
Fig.\ \ref{fig:UT} (a). On the other hand, in the case of 
(\ref{UT-ort5}), the apex is still given by $(\rho,\eta)$, whereas the 
angle $\gamma'$ differs from $\gamma$ through
\begin{equation}
\gamma-\gamma'\equiv\delta\gamma=\lambda^2\eta,
\end{equation}
implying a small angle between the basis of the triangle and the real 
axis. 

The angles $\alpha$, $\beta$, $\gamma$ and $\delta\gamma$ can be probed 
directly through CP-violating effects in $B$ decays. Whereas $\beta$ is 
very accessible and various promising strategies to extract $\gamma$ 
were proposed, an experimentally feasible measurement of $\alpha$ is 
unfortunately difficult. The small angle $\delta\gamma$ plays a 
key r\^ole for CP violation in the $B_s$-meson system.

\begin{figure}
\vspace{0.10in}
\centerline{
\epsfysize=6.7truecm
\epsffile{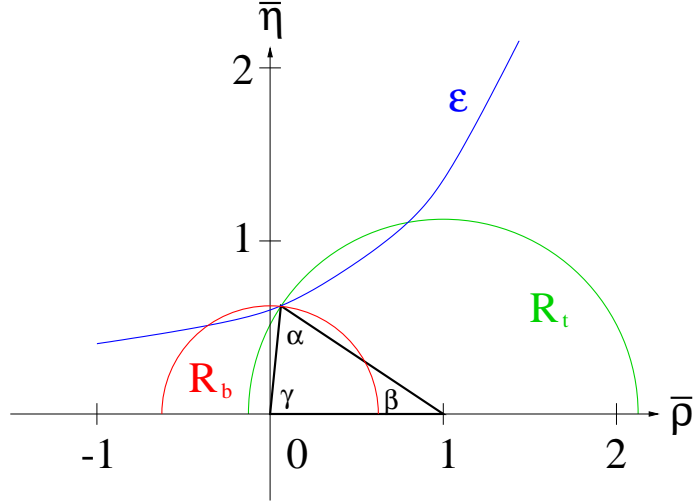}
}
\caption[]{Contours to determine the unitarity triangle in the
$\overline{\rho}$--$\overline{\eta}$ plane.}\label{fig:cont-scheme}
\end{figure}

\boldmath
\subsection{Towards an Allowed Range in the 
$\overline{\rho}$--$\overline{\eta}$ Plane}\label{subsec:CKM-fits}
\unboldmath
Let us now discuss how the apex of the unitarity triangle can be constrained 
in the $\overline{\rho}$--$\overline{\eta}$ plane without using direct 
measurements of the angles $\alpha$, $\beta$ and $\gamma$. To accomplish 
this task, measurements of the sides $R_b$ and $R_t$ have to be performed. 
These quantities can be expressed in terms of CKM matrix elements as follows:
\begin{eqnarray}
R_b&\equiv&\left|\frac{V_{ud}V^*_{ub}}{V_{cd}V^*_{cb}}\right|
=\left(1-\frac{\lambda^2}{2}\right)\frac{1}{\lambda}\left|
\frac{V_{ub}}{V_{cb}}\right|\,=\,\sqrt{\overline{\rho}^2+
\overline{\eta}^2}\label{Rb-intro}\label{Rb-def}\\
R_t&\equiv&\left|\frac{V_{td}V^*_{tb}}{V_{cd}V^*_{cb}}\right| 
=\frac{1}{\lambda}\left|\frac{V_{td}}{V_{cb}}\right|\,=\,
\sqrt{(1-\overline{\rho})^2+\overline{\eta}^2},\label{Rt-def}
\end{eqnarray}
and will show up at several places throughout this review. Other useful
expressions are 
\begin{equation}\label{CKM-UT-ANGLES}
V_{ub}=A\lambda^3\left(\frac{R_b}{1-\lambda^2/2}\right)e^{-i\gamma},\quad 
V_{td}=A\lambda^3 R_t e^{-i\beta},
\end{equation}
since they make the dependence on $\gamma$ and $\beta$ explicit; they
correspond to the phase convention chosen both in the standard 
parametrization (\ref{standard}) and in the generalized Wolfenstein 
parametrization (\ref{NLO-wolf}). Moreover, taking into account 
(\ref{set-rel}), we obtain
\begin{equation}
\delta_{13}=\gamma.
\end{equation}

Looking at (\ref{Rb-def}) and (\ref{Rt-def}), we observe that these 
expressions describe circles in the $\overline{\rho}$--$\overline{\eta}$ 
plane around $(0,0)$ and $(1,0)$ with radii $R_b$ and $R_t$, respectively. 
Employing, moreover, an observable $\varepsilon$, which measures ``indirect'' 
CP violation in the neutral kaon system, as we will discuss in more detail 
in Section~\ref{sec:kaon}, a hyperbola in the 
$\overline{\rho}$--$\overline{\eta}$ plane can be fixed. These contours 
are sketched in Fig.~\ref{fig:cont-scheme}; their intersection 
gives the apex of the unitarity triangle. As can be seen in
Fig.~\ref{fig:cont-scheme}, an upper bound on $R_b$ implies \cite{blo}
\begin{equation}\label{Rb-bounds}
\left(\sin\beta\right)_{\rm max}=R_b^{\rm max},\quad 
\left(\sin2\beta\right)_{\rm max}=2R_b^{\rm max}\sqrt{1-(R_b^{\rm max})^2},
\end{equation}
where the quantity $\sin2\beta$ is of particular interest, since it governs 
CP violation in the ``gold-plated'' decay $B_d\to J/\psi K_{\rm S}$, as we 
will see in Subsection~\ref{subsec:BpsiK}

The CKM matrix element $|V_{cb}|$, which is required -- among other things -- 
for the normalization of (\ref{Rb-def}) and (\ref{Rt-def}), can be extracted 
with the help of the ``Heavy-Quark Effective Theory'' (HQET) from 
semi-leptonic $B$ decays into charmed mesons \cite{Neubert-rev}. 
The key r\^ole is played 
by the decay $\overline{B^0_d}\to D^{\ast+}\ell^-\overline{\nu}_\ell$, which
is particularly clean. Measuring its differential rate at maximum $q^2$, 
which denotes the mass squared of the lepton--antineutrino system, $|V_{cb}|$ 
can be extracted with corrections based on the HQET. An alternative strategy 
is provided by inclusive $b\to c\,\ell^-\overline{\nu}_\ell$ decays. 
In a recent analysis \cite{Rome-rev}, the following average over the
presently available measurements performed by the CLEO 
and LEP collaborations was obtained:
\begin{equation}\label{A-range}
|V_{cb}|=(41.0\pm1.6)\times 10^{-3}\quad\Rightarrow\quad A=0.832\pm0.033,
\end{equation}
where we have used (\ref{Def-A}) with $\lambda=0.222$ to evaluate the 
CKM parameter $A$.

In order to determine $|V_{ub}|$, semi-leptonic $b\to u$ transitions are
used. The CLEO collaboration has employed the exclusive decay 
$\overline{B^0_d}\to \rho^+\ell^-\overline{\nu}_\ell$, where the extraction
of $|V_{ub}|$ requires models for decay form factors \cite{CLEO-Vub}. 
On the other hand, the LEP collaborations have obtained a value for 
$|V_{ub}|$ by developing algorithms to be sensitive to a 
large fraction of the inclusive $b\to u\,\ell^-\overline{\nu}_\ell$ rate 
\cite{LEP-Vub}. The corresponding results are given by
\begin{equation}
|V_{ub}|=
\left\{\begin{array}{ll}
(32.5\pm2.9\pm5.5)\times 10^{-4}&\mbox{(CLEO \cite{CLEO-Vub})}\\
(41.3\pm6.3\pm3.1)\times 10^{-4}&\mbox{(LEP \cite{LEP-Vub})},
\end{array}\right.
\end{equation}
where the second uncertainty has a theoretical origin. In 
\cite{Rome-rev}, the following average value has been obtained: 
$|V_{ub}|=(35.5\pm3.6)\times 10^{-4}$. On the other hand, in the 
review of the Particle Data Group \cite{PDG}, a more conservative 
uncertainty is given for $|V_{ub}|$, yielding 
$|V_{ub}/V_{cb}|=0.090\pm0.025$. In this review, we shall follow 
\cite{Brev01}, and take 
\begin{equation}\label{Rb-range}
|V_{ub}/V_{cb}|=0.085\pm0.018\quad\Rightarrow\quad R_b=0.38\pm0.08,
\end{equation}
which lies between these two ``limiting'' cases. Using (\ref{Rb-bounds}),
we obtain the bounds
\begin{equation}\label{ubound}
\beta\lsim28^\circ, \quad \sin2\beta \lsim 0.82.
\end{equation}

As we have just seen, $R_b$ can be determined through semi-leptonic 
$b\to u\,\ell^-\overline{\nu}_\ell$ and $b\to c\,\ell^-\overline{\nu}_\ell$ 
decays. The second side $R_t$ of the unitarity triangle shown in 
Fig.\ \ref{fig:UT} (a) can be determined through 
$B^0_{d,s}$--$\overline{B^0_{d,s}}$ mixing, which will be discussed 
in Sections~\ref{sec:neutral} and \ref{sec:Bs}. The present data imply 
\begin{equation}
R_t={\cal O}(1). 
\end{equation}
In particular the experimental constraints on $B^0_{s}$--$\overline{B^0_{s}}$ 
mixing play an important r\^ole, excluding values for $\gamma$ larger than 
$90^\circ$ \cite{AL}. 

Because of theoretical und experimental uncertainties, things are not 
as simple as sketched in Fig.~\ref{fig:cont-scheme}. Instead of contours, 
we have actually to deal with bands, whose intersection gives an allowed 
range in the $\overline{\rho}$--$\overline{\eta}$ plane. Moreover, there 
are strong correlations between theoretical and experimental uncertainties.
For instance, the circle and the hyperbola fixed through 
$B^0_{d}$--$\overline{B^0_{d}}$ mixing and $\varepsilon$, respectively, 
depend on $|V_{cb}|$, the top-quark mass, perturbative QCD corrections, 
and certain non-perturbative parameters, as we will see below. It is 
hence rather involved to convert the experimental information into an 
allowed range in the $\overline{\rho}$--$\overline{\eta}$ plane, and 
various analyses can be found in the literature:
\begin{itemize}
\item The simple scanning approach, where experimental and theoretical
parameters are both scanned independently within reasonable ranges (see, 
for example, \cite{Brev01}). 
\item The Gaussian approach, where experimental and theoretical input 
parameters are both treated with Gaussian errors (see, for instance, 
\cite{AL}).
\item The ``BaBar 95\% Scanning Method'' \cite{PS,GNPS}, where one sets 
the theoretical input parameters to some fixed values and determines 
the 95\% C.L. range in the $\overline{\rho}$--$\overline{\eta}$ plane 
by using a Gaussian error analysis for the experimental input parameters.
This procedure is repeated for all possible sets of theoretical input 
parameters lying within their allowed ranges, and an envelope of the
95\% C.L. regions is obtained.
\item The Bayesian approach \cite{Rome-rev}, where the theoretical 
uncertainties are taken into account in a way similar to the experimental 
ones through probability distribution functions (p.d.f.). If the 
uncertainties are dominated by statistical effects or if there are 
many comparable contributions to the systematic error, a Gaussian model 
is chosen. On the other hand, a uniform p.d.f.\ is employed when the 
parameter can be assumed to lie (almost) certainly within a given range, 
with points that can be considered as being equally probable. 
\item The statistical approach developed in \cite{hoecker}, allowing a
non-Bayesian treatment of theoretical parameters and theoretical 
systematics of measurements. Using a likelihood function 
${\cal L}(y_{\rm mod})={\cal L}_{\rm exp}(x_{\rm exp}-x_{\rm theo}
(y_{\rm mod})){\cal L}_{\rm theo}(y_{\rm QCD})$, $\chi^2\equiv-
2\ln{\cal L}(y_{\rm mod})$ is minimized in the fitting procedure. 
The experimental likelihood ${\cal L}_{\rm exp}$ measures the agreement
between measurements $x_{\rm exp}$ and theoretical predictions $x_{\rm theo}$,
which are functions of model parameters $y_{\rm mod}$. On the other hand,
the theoretical likelihood ${\cal L}_{\rm theo}$ expresses the knowledge
of QCD parameters $y_{\rm QCD}\in\{y_{\rm mod}\}$, where the allowed ranges 
are specified through the theoretical uncertainties. The agreement between
theory and experiment is gauged by the global minimum $\chi^2_{{\rm min};
y_{\rm mod}}$, which is determined by varying all model parameters 
$y_{\rm mod}$. Besides a detailed discussion of the technicalities of this
procedure, also comparisons with other approaches can be found in 
\cite{hoecker}. A similar comment applies to \cite{Rome-rev}.
\end{itemize}
\begin{table}[h]
\begin{center}
\begin{tabular}{|l|c|c|c|c|c|}
\hline
Group & $\alpha$ & $\beta$ & $\gamma$ & $\sin2\alpha$ & $\sin2\beta$ \\
\hline
Buras \cite{Brev01} & $78.8^\circ$--$120^\circ$ & $15.1^\circ$--$28.6^\circ$
& $37.9^\circ$--$76.5^\circ$ & $-0.87$--$0.38$ & $0.50$--$0.84$\\
Ali \& London \cite{AL} & $77^\circ$--$127^\circ$ & $14^\circ$--$35^\circ$
& $34^\circ$--$81^\circ$ & $-0.96$--$0.45$ & $0.46$--$0.94$\\
H\"ocker {\it et al.}~\cite{hoecker} & $80^\circ$--$126^\circ$ & 
$14^\circ$--$27^\circ$ & $34^\circ$--$82^\circ$ & $-0.95$--$0.33$ & 
$0.47$--$0.81$\\
\hline
\end{tabular}
\end{center}
\caption{Recent results of fits to the parameters of the unitarity 
triangle.}\label{tab:CKM-fits}
\end{table}
In Table~\ref{tab:CKM-fits}, we have collected recent results, which agree 
rather well with one another. On the other hand, considerably narrower 
ranges are given by Ciuchini {\it et al.}\ in \cite{Rome-rev}:
\begin{equation}\label{Rome-gamma}
\gamma=(54.8 \pm 6.2)^\circ,\quad 
\sin2\alpha=-0.42\pm0.23,\quad
\sin2\beta=0.698\pm0.066,
\end{equation}
corresponding to
\begin{equation}
\overline{\rho}=0.224\pm0.038,\quad \overline{\eta}=0.317\pm0.040.
\end{equation}

The question of combining the theoretical and experimental uncertainties
in the ``standard analysis'' of the unitarity triangle illustrated in
Fig.~\ref{fig:cont-scheme} in an optimal way and to obtain a realistic 
range in the $\overline{\rho}$--$\overline{\eta}$ plane will certainly 
continue to be a hot topic in the future. This is also reflected by 
the vast debate going on at present, dealing with the key issue of how
to deal with the corresponding uncertainties (for instance, Bayesian 
\cite{Rome-rev} vs.\ non-Bayesian \cite{hoecker} approach).
The importance of CP violation in $B$ decays is related to the fact that 
such effects allow us to determine the angles of the unitarity triangle 
{\it directly}. Comparing the thus obtained values with the indirect ranges 
determined as sketched above, we may well encounter discrepancies, which 
may shed light on the physics beyond the Standard Model. In this context, 
non-leptonic $B$-meson decays play the key r\^ole. 

We shall come back to the allowed range in the 
$\overline{\rho}$--$\overline{\eta}$ plane in Section~\ref{sec:Bs}, 
where we discuss the important impact of $B^0_{s}$--$\overline{B^0_{s}}$ 
mixing in more detail (see Fig.~\ref{fig:UT-constr}), and in 
Section~\ref{sec:MFV}, where we consider models with minimal flavour 
violation \cite{MFV,UUT}, a simple class of extensions of 
the Standard Model. In this framework, various interesting bounds on
the unitarity triangle, $\sin2\beta$ and $K\to\pi\nu\overline{\nu}$
decays can be derived \cite{BB-Bound,BF-MFV}.

\boldmath
\section{Non-leptonic $B$ Decays}\label{sec:nonlept}
\unboldmath
\setcounter{equation}{0}
\subsection{Preliminaries}
We may divide weak decays of $B$-mesons into leptonic, semi-leptonic 
and non-leptonic transitions. The first ones, the leptonic 
modes $B^-\to \ell\,\overline{\nu}_\ell$, allow -- at least in principle --
a determination of the $B$-meson decay constant $f_{B}$, which is 
defined by
\begin{equation}
\langle0|\overline{u}\gamma_5\gamma_\mu b|B^-(p)\rangle=if_{B}p_\mu.
\end{equation}
However, the corresponding rates suffer from a helicity suppression 
proportional to $(m_\ell/M_B)^2$, and are hence very small. For $\ell=e$ 
and $\mu$, we have branching ratios at the $10^{-10}$ and $10^{-7}$
levels, respectively. Consequently, these decays are very hard to
measure. In the case of $\ell=\tau$, the helicity suppression is not
effective, but the experimental reconstruction of the $\tau$-leptons 
is unfortunately very challenging \cite{fulvia}. 

Using heavy-quark arguments, exclusive and inclusive semi-leptonic 
decays caused by $b\to c\,\ell^-\overline{\nu}_\ell$ and 
$b\to u\,\ell^-\overline{\nu}_\ell$ quark-level transitions allow 
extractions of the CKM matrix elements $|V_{cb}|$ and $|V_{ub}|$, 
respectively, as we have seen in Subsection~\ref{subsec:CKM-fits}.

With respect to testing the Standard-Model description of CP violation, 
the major r\^ole is played by non-leptonic $B$ decays. At the quark level, 
they are mediated by the following processes:
\begin{equation}
b\to q_1\,\overline{q_2}\,d\,(s),
\end{equation}
where $q_1,q_2\in\{u,d,c,s\}$ are quark-flavour labels. In this section,
we discuss the basic features of the theoretical framework to deal with 
such non-leptonic transitions.

\subsection{Classification}\label{sec:class}
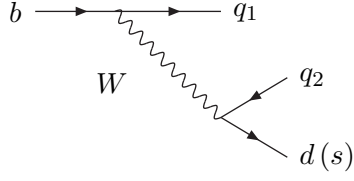
\begin{figure}[ht]
\begin{center}
{\small
\vspace*{-1truecm}\hspace*{4truecm}\begin{picture}(80,50)(80,20)
\Line(10,45)(80,45)
\ArrowLine(27,45)(28,45)
\ArrowLine(62,45)(63,45)
\Photon(40,45)(80,5){2}{10}
\ArrowLine(105,20)(80,5)\ArrowLine(80,5)(105,-10)
\Text(5,45)[r]{$b$}\Text(85,45)[l]{$q_1$}
\Text(110,20)[l]{$q_2$}
\Text(110,-10)[l]{$d\,(s)$}
\Text(45,22)[tr]{$W$}
\end{picture}}
\end{center}
\vspace*{0.8truecm}
\caption[]{Tree diagrams ($q_1,q_2\in\{u,c\}$).}\label{fig:tree-top}
\end{figure}

\begin{figure}[ht]
\begin{center}
{\small
\begin{picture}(140,60)(0,20)
\Line(10,50)(130,50)\Text(5,50)[r]{$b$}\Text(140,50)[l]{$d\,(s)$}
\ArrowLine(24,50)(26,50)\ArrowLine(114,50)(116,50)
\PhotonArc(70,50)(30,0,180){3}{15}
\Text(69,56)[b]{$u,c,t$}\Text(109,75)[b]{$W$}
\Gluon(70,50)(120,10){2}{10}
\ArrowLine(135,23)(120,10)\ArrowLine(120,10)(135,-3)
\Text(85,22)[tr]{$G$}\Text(140,-3)[l]{$q_1$}
\Text(140,23)[l]{$q_2=q_1$}
\end{picture}}
\end{center}
\vspace*{0.5truecm}
\caption[]{QCD penguin diagrams ($q_1=q_2\in\{u,d,c,s\}$).}\label{fig:QCD-top}
\end{figure}
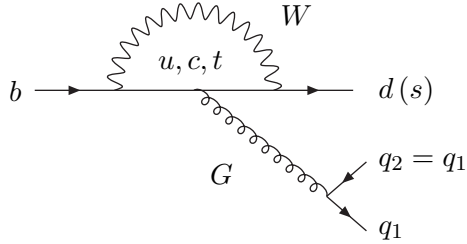

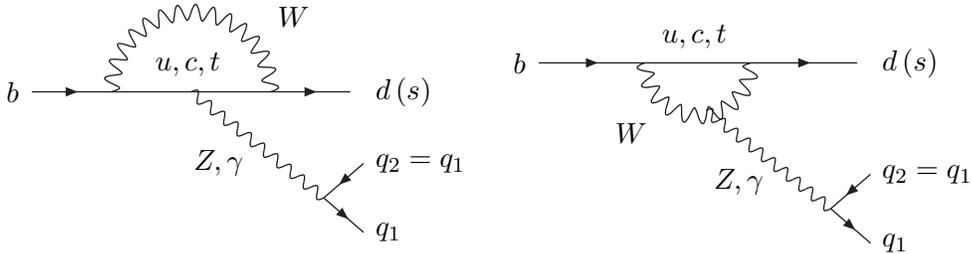
\begin{figure}
\hspace*{1.8truecm}{\small
\begin{picture}(140,60)(0,20)
\Line(10,50)(130,50)\Text(5,50)[r]{$b$}\Text(140,50)[l]{$d\,(s)$}
\ArrowLine(24,50)(26,50)\ArrowLine(114,50)(116,50)
\PhotonArc(70,50)(30,0,180){3}{15}
\Text(69,56)[b]{$u,c,t$}\Text(109,75)[b]{$W$}
\Photon(70,50)(120,10){2}{10}
\ArrowLine(135,23)(120,10)\ArrowLine(120,10)(135,-3)
\Text(90,28)[tr]{$Z,\gamma$}\Text(140,-3)[l]{$q_1$}
\Text(140,23)[l]{$q_2=q_1$}
\end{picture}}

\vspace*{-2truecm}\hspace*{8.4truecm}
{\small
\begin{picture}(140,60)(0,20)
\Line(10,65)(130,65)\Text(5,65)[r]{$b$}\Text(140,65)[l]{$d\,(s)$}
\ArrowLine(29,65)(31,65)\ArrowLine(109,65)(111,65)
\PhotonArc(70,65)(20,180,360){3}{10}
\Text(69,71)[b]{$u,c,t$}\Text(45,35)[b]{$W$}
\Photon(73,47)(120,10){2}{10}
\ArrowLine(135,23)(120,10)\ArrowLine(120,10)(135,-3)
\Text(95,25)[tr]{$Z,\gamma$}\Text(140,-3)[l]{$q_1$}
\Text(140,23)[l]{$q_2=q_1$}
\end{picture}}
\vspace*{0.9truecm}
\caption[]{EW penguin diagrams 
($q_1=q_2\in\{u,d,c,s\}$).}\label{fig:EWP-top}
\end{figure}

There are two kinds of topologies contributing to non-leptonic $B$ decays: 
``tree-diagram-like'' and ``penguin'' topologies. The latter consist of 
gluonic (QCD) and electroweak (EW) penguins, which originate from strong 
and electroweak interactions, respectively. In 
Figs.~\ref{fig:tree-top}--\ref{fig:EWP-top}, the corresponding 
leading-order Feynman diagrams are shown. Depending on their flavour 
content, we may classify $b\to q_1\,\overline{q_2}\,d\,(s)$ transitions 
as follows:
\begin{itemize}
\item $q_1\not=q_2\in\{u,c\}$: only tree diagrams contribute.
\item $q_1=q_2\in\{u,c\}$: tree and penguin diagrams contribute.
\item $q_1=q_2\in\{d,s\}$: only penguin diagrams contribute.
\end{itemize}

\subsection{Low-energy Effective Hamiltonians}\label{subsec:ham}
\subsubsection{General Remarks}
In order to analyse non-leptonic $B$ decays theoretically, we employ
low-energy effective Hamiltonians, which are calculated by making use 
of the operator product expansion~\cite{OPE}. This formalism yields 
transition matrix elements of the following structure:
\begin{equation}\label{ee2}
\langle f|{\cal H}_{\rm eff}|i\rangle=\frac{G_{\rm F}}{\sqrt{2}}\,
\lambda_{\rm CKM}\sum_k C_k(\mu)\langle f|Q_k(\mu)|i\rangle\equiv
\frac{G_{\rm F}}{\sqrt{2}}\,\lambda_{\rm CKM}\left[\vec C^T(\mu)\cdot
\langle \vec Q(\mu)\rangle\right].
\end{equation}
The operator product expansion allows us to disentangle the short-distance
contributions to this transition amplitude from the long-distance 
ones, which are described by perturbative Wilson coefficient 
functions $C_k(\mu)$ and non-perturbative hadronic matrix elements 
$\langle f|Q_k(\mu)|i\rangle$, respectively. As usual, $G_{\rm F}$ is
the Fermi constant, $\lambda_{\rm CKM}$ a CKM factor, and $\mu$ an 
appropriate renormalization scale. This scale separates, roughly speaking,
the short-distance contributions from the long-distance pieces. The $Q_k$ 
are local operators, which are generated by electroweak and strong 
interactions and govern \mbox{``effectively''} the decay in question. 
The Wilson coefficients $C_k(\mu)$ can be considered as scale-dependent 
couplings related to the vertices described by the $Q_k$. 

In general, the Wilson coefficients, which describe essentially the 
whole short-distance dynamics of the transition amplitude (\ref{ee2}), 
include contributions from virtual $W$- and $Z$-bosons, as well as
from virtual top quarks. Mathematically, these contributions are related to
appropriate ``Inami--Lim functions'' \cite{IL}. If extensions of the 
Standard Model are considered, also new particles, such as charged Higgs 
or supersymmetric particles, may contribute to the $C_k(\mu)$. The resulting 
dependence on the masses of these particles can be calculated by evaluating 
penguin and box diagrams with full $W$, $Z$, top-quark and possible 
new-particle exchanges \cite{PB-expansion}. The $\mu$ dependence of the 
Wilson coefficients is governed by the properly included short-distance
QCD corrections. This $\mu$ dependence has to cancel that of the 
non-perturbative hadronic matrix elements $\langle f|Q_k(\mu)|i\rangle$, 
since the physical transition amplitude (\ref{ee2}) cannot depend on 
the choice of the renormalization scale. In this cancellation, usually
several terms on the right-hand side of (\ref{ee2}) are involved. The 
$C_k(\mu)$ depend, in general, also on the employed renormalization 
scheme. This dependence has again to be cancelled by the one of the 
hadronic matrix elements. 

\boldmath
\subsubsection{Tree Decays}\label{subsec:Tree-Ham}
\begin{figure}
\begin{center}
\leavevmode
\epsfysize=4.0truecm 
\epsffile{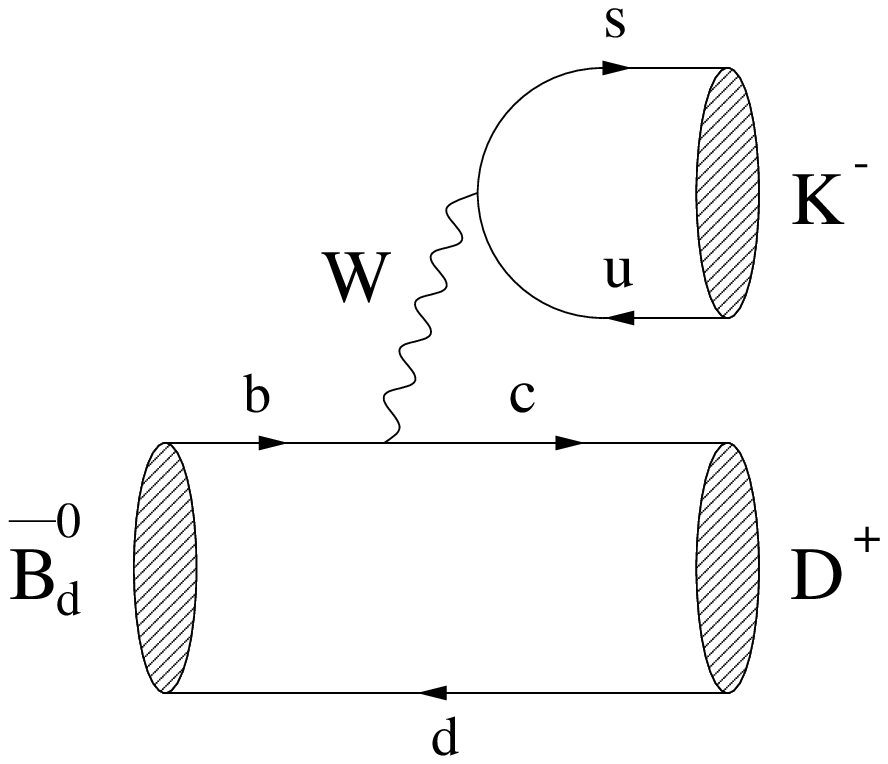} \hspace*{2truecm}
\epsfysize=3.0truecm 
\epsffile{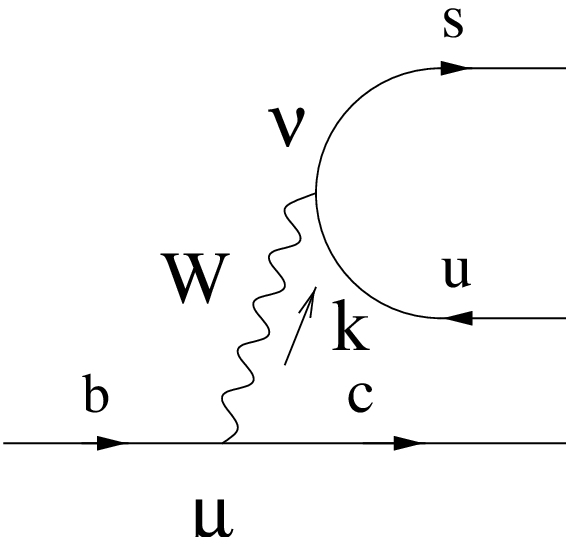}
\caption{Feynman diagram contributing at leading order to 
$\overline{B^0_d}\to D^+K^-$.}\label{fig:feyn-example}
\end{center}
\end{figure}

\begin{figure}[t]
\begin{center}
\leavevmode
\epsfysize=2.8truecm 
\epsffile{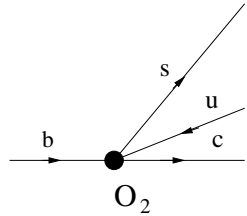}
\caption{The vertex described by the current--current operator 
$O_2$.}\label{fig:feyn-example2}
\end{center}
\end{figure}

\unboldmath
Let us consider the decay $\overline{B^0_d}\to D^+K^-$ as an example.
At the quark level, it originates from $b\to c\overline{u}s$ transitions,
as can be seen in Fig.~\ref{fig:feyn-example}. Consequently, according to 
the classification introduced in Subsection~\ref{sec:class}, it is a pure 
``tree'' decay, receiving no penguin contributions. In order to derive 
the relevant low-energy effective Hamiltonian, we have to consider the 
quark-level process in Fig.~\ref{fig:feyn-example}, yielding the following
transition amplitude:
\begin{equation}\label{trans-ampl}
-\,\frac{g_2^2}{8}V_{us}^\ast V_{cb}
\left[\overline{s}\gamma^\nu(1-\gamma_5)u\right]
\left[\frac{g_{\nu\mu}}{k^2-M_W^2}\right]
\left[\overline{c}\gamma^\mu(1-\gamma_5)b\right].
\end{equation}
Because of $k^2\approx m_b^2\ll M_W^2$, we may write
\begin{equation}
\frac{g_{\nu\mu}}{k^2-M_W^2}\quad\longrightarrow\quad
-\,\frac{g_{\nu\mu}}{M_W^2}\equiv-\left(\frac{8G_{\rm F}}{\sqrt{2}g_2^2}
\right)g_{\nu\mu},
\end{equation}
i.e.\ we may ``integrate out'' the $W$-boson in (\ref{trans-ampl}), 
and arrive at
\begin{eqnarray}
{\cal H}_{\rm eff}&=&\frac{G_{\rm F}}{\sqrt{2}}V_{us}^\ast V_{cb}
\left[\overline{s}_\alpha\gamma_\mu(1-\gamma_5)u_\alpha\right]
\left[\overline{c}_\beta\gamma^\mu(1-\gamma_5)b_\beta\right]\nonumber\\
&&\equiv\frac{G_{\rm F}}{\sqrt{2}}V_{us}^\ast V_{cb}
(\overline{s}_\alpha u_\alpha)_{\mbox{{\scriptsize 
V--A}}}(\overline{c}_\beta b_\beta)_{\mbox{{\scriptsize V--A}}}
\equiv\frac{G_{\rm F}}{\sqrt{2}}V_{us}^\ast V_{cb}O_2\,,\label{O2-def}
\end{eqnarray}
where $\alpha$ and $\beta$ denote $SU(3)_{\rm C}$ colour indices. 
Effectively, the vertex shown in Fig.~\ref{fig:feyn-example} is now
described by the ``current--current'' operator $O_2$, as illustrated in
Fig.~\ref{fig:feyn-example2}.

If we take into account QCD effects, we have to distinguish between
factorizable and non-factorizable QCD vertex corrections. In the former
case, the gluons are exchanged separately {\it within} the $\overline{s}u$
and $\overline{c}b$ quark currents, whereas the gluons are {\it interchanged} 
between these currents in the non-factorizable case. The factorizable
QCD corrections just renormalize the $O_2$ operator. However, the
non-factorizable QCD corrections induce another ``current--current'' 
operator through ``operator mixing''. It is given by
\begin{equation}\label{O1-def}
O_1=\left[\overline{s}_\alpha\gamma_\mu(1-\gamma_5)u_\beta\right]
\left[\overline{c}_\beta\gamma^\mu(1-\gamma_5)b_\alpha\right],
\end{equation}
i.e.\ it differs from $O_2$ through its colour structure. Consequently, 
if we take into account QCD corrections, the low-energy effective 
Hamiltonian takes the following form:
\begin{equation}\label{Heff-example}
{\cal H}_{\rm eff}=\frac{G_{\rm F}}{\sqrt{2}}V_{us}^\ast V_{cb}
\left[C_1(\mu)O_1+C_2(\mu)O_2\right],
\end{equation}
where $C_1(\mu)\not=0$ and $C_2(\mu)\not=1$ are due to QCD renormalization
effects. This Hamiltonian applies to all decay processes, which are mediated 
by $b\to c\overline{u}s$ transitions. The amplitudes of different channels 
belonging to this decay class arise from different hadronic matrix 
elements of $O_1$ and $O_2$. An approach to deal with these quantities, 
referred to as ``factorization'', will be discussed in more detail in 
Subsection~\ref{subsec:fact}.

Let us now have a closer look at the evaluation of the Wilson coefficients 
\cite{CC-REFS}. To this end, we first have to calculate the QCD corrections 
to the vertices shown in Figs.~\ref{fig:feyn-example} and 
\ref{fig:feyn-example2}, yielding the QCD-corrected transition amplitude
\begin{equation}\label{A-QCD}
A_{\rm QCD}=\vec A^T(\mu)\cdot\langle \vec O\rangle_0,
\end{equation}
and the QCD-corrected matrix elements
\begin{equation}\label{ME-QCD}
\langle \vec O(\mu)\rangle=\hat M(\mu)\cdot \langle \vec O\rangle_0,
\end{equation}
respectively. Here the column vector $\langle\vec O\rangle_0$ contains the 
tree-level matrix elements of the operators $O_1$ and $O_2$. The Wilson 
coefficients are now obtained by expressing the QCD-corrected transition
amplitude (\ref{A-QCD}) in terms of the QCD-corrected matrix elements
(\ref{ME-QCD}) as in (\ref{ee2}):
\begin{equation}
A_{\rm QCD}=\vec C^T(\mu)\cdot\langle \vec O(\mu)\rangle,
\end{equation}
yielding
\begin{equation}
\vec C(\mu)=\vec A^T(\mu)\cdot\hat M^{-1}(\mu).
\end{equation}
This procedure is called ``matching'' of the full theory onto an
effective theory, which is described by the low-energy effective Hamiltonian 
${\cal H}_{\rm eff}$. At the one-loop level, the results for the $C_k(\mu)$ 
obtained this way are given by
\begin{equation}\label{CC-expr}
C_1(\mu)=-3\left(\frac{\alpha_s}{4\pi}\right)
\ln\left(\frac{M_W^2}{\mu^2}\right),\quad
C_2(\mu)=1+\frac{3}{N_{\rm C}}
\left(\frac{\alpha_s}{4\pi}\right)\ln\left(\frac{M_W^2}{\mu^2}\right),
\end{equation}
where $N_{\rm C}$ denotes the number of quark colours, i.e.\ $N_{\rm C}=3$ 
for the QCD gauge group $SU(3)_{\rm C}$. 

The characteristic feature of the expressions given in (\ref{CC-expr}) 
are the terms proportional to $\alpha_s\ln(M_W^2/\mu^2)$. Although the
renormalization scale $\mu$ can, in principle, be chosen arbitrarily,
it is natural to use $\mu={\cal O}(m_b)$ in analyses of $B$ decays. 
For such scales, we have to deal with large logarithms in (\ref{CC-expr}).
Sloppily speaking, these large logarithms compensate the smallness of 
the running QCD coupling $\alpha_s(\mu)$. In order to obtain a reliable 
result, these terms have to be resummed to all orders in $\alpha_s$, which 
can be accomplished with the help of renormalization-group techniques. 
Here one exploits that the transition amplitude (\ref{ee2}) cannot depend 
on the chosen renormalization scale $\mu$. The general expression for the 
$C_k(\mu)$ can be written as 
\begin{equation}\label{RG-evol}
\vec C(\mu)=\hat U(\mu, M_W)\cdot\vec C(M_W),
\end{equation}
where $\vec C(M_W)$ denotes the initial conditions, depending on the
short-distance physics at high-energy scales. The evolution matrix 
is given by
\begin{equation}
\hat U(\mu, M_W)=T_g\left\{\exp\left[\int_{g_s(M_W)}^{g_s(\mu)}dg'
\frac{\hat\gamma^T(g')}{\beta(g')}\right]\right\},
\end{equation}
where $g_s$ is the QCD coupling, the operator $T_g$ denotes 
coupling-constant ordering such that the couplings increase from right to 
left, the QCD ``beta function'' $\beta(g)$ governs the renormalization-group
evolution of $g$, 
\begin{equation}
\mu\frac{d}{d\mu}g(\mu)=\beta(g(\mu)),
\end{equation}
and $\hat \gamma$ is the ``anomalous dimension matrix'' of the operators
$\vec O$, entering in 
\begin{equation}
\frac{d}{dg}\vec C(g)=\frac{\hat\gamma^T(g)}{\beta(g)}\cdot\vec C(g).
\end{equation}

Using these renormalization-group techniques, it is possible to sum up the 
following terms of the Wilson coefficients in a systematic way:
\begin{equation}
\alpha_s^n\left[\ln\left(\frac{\mu}{M_W}\right)\right]^n 
\,\,\mbox{(LO)},\quad\,\,\alpha_s^n\left[\ln\left(\frac{\mu}{M_W}\right)
\right]^{n-1}\,\,\mbox{(NLO)},\quad \ldots,
\end{equation}
where the abbreviations LO and NLO refer to the ``leading 
logarithmic'' and ``next-to-leading logarithmic'' approximations, 
respectively. 
For many applications, the LO corrections are not sufficient. 
Important examples are the meaningful use of the QCD scale parameter 
$\Lambda_{\overline{\rm MS}}$ \cite{MS-bar} extracted from various
high-energy processes, and the penguin processes discussed below, 
where the matching procedure required to calculate the top-quark mass 
dependence of the Wilson coefficients can only be performed properly 
by going beyond the LO approximation. In this case, also dependences 
of the Wilson coefficients on the employed renormalization schemes show 
up, which have to be cancelled through the hadronic matrix elements of 
the corresponding operators. Moreover, scale dependences can usually be 
reduced by going beyond LO.

Let us, in order to see the structure of the NLO Wilson coefficients, 
give the result for the current--current operators. Here it is convenient
to introduce operators $O_\pm$ with coefficients $C_\pm(\mu)$, which
do not mix under renormalization and are defined by
\begin{equation}\label{Opm-def}
O_\pm\equiv\frac{1}{2}\left(O_2\pm O_1\right) \quad\mbox{with}\quad
C_\pm(\mu)=C_2(\mu)\pm C_1(\mu).
\end{equation}
The NLO renormalization-group evolution from $M_W$ to $\mu={\cal O}(m_b)$ 
is described by
\begin{equation}
C_\pm(\mu)=U_\pm(\mu,M_W)C_\pm(M_W),
\end{equation}
with
\begin{equation}\label{NLO-evol}
U_\pm(\mu,M_W)=\left[1+\frac{\alpha_s(\mu)}{4\pi}J_{\pm}\right]
\left[\frac{\alpha_s(M_W)}{\alpha_s(\mu)}\right]^{d_\pm}\left[1-
\frac{\alpha_s(M_W)}{4\pi}J_{\pm}\right]
\end{equation}
and
\begin{equation}\label{NLO-init}
C_\pm(M_W)=1+\frac{\alpha_s(M_W)}{4\pi}B_\pm.
\end{equation}
Here 
\begin{equation}\label{coeff1}
J_\pm=\frac{d_\pm}{\beta_0}\beta_1-\frac{\gamma_\pm^{(1)}}{2\beta_0},\quad
d_\pm=\frac{\gamma_\pm^{(0)}}{2\beta_0},\quad
B_\pm=\frac{N_{\rm C}\mp1}{2N_{\rm C}}\left[\pm11+\kappa_\pm\right],
\end{equation}
with
\begin{equation}\label{coeff2}
\beta_0=\frac{11N_{\rm C}-2f}{3},\quad \beta_1=\frac{34}{3}N^2_{\rm C}-
\frac{10}{3}N_{\rm C}f-2\left[\frac{N_{\rm C}^2-1}{2N_{\rm C}}\right]f
\end{equation}
and
\begin{equation}\label{coeff3}
\gamma_\pm^{(0)}=\pm12\left[\frac{N_{\rm C}\mp1}{2N_{\rm C}}\right],\quad
\gamma_\pm^{(1)}=\frac{N_{\rm C}\mp1}{2N_{\rm C}}\left[-21\pm
\frac{57}{N_{\rm C}}\mp\frac{19}{3}N_{\rm C}\pm\frac{4}{3}f-
2\beta_0\kappa_\pm\right].
\end{equation}
The parameter $f$ denotes the number of active quark flavours (5 in
our example, as the top quark is integrated out), and 
$\kappa_\pm$ distinguishes between different renormalization schemes 
\cite{Buras-a1-a2}. The $\beta$ and $\gamma$ coefficients in 
(\ref{coeff1})--(\ref{coeff3}) arise from the expansions
\begin{equation}
\beta(g_s)=-\beta_0\frac{g_s^3}{16\pi^2}-
\beta_1\frac{g^5_s}{\left(16\pi^2\right)^2}+\ldots,\quad
\gamma_\pm(\alpha_s)=\gamma^{(0)}_\pm\left(\frac{\alpha_s}{4\pi}\right)
+\gamma^{(1)}_\pm\left(\frac{\alpha_s}{4\pi}\right)^2+\ldots,
\end{equation}
where the former governs the NLO effective QCD coupling
\begin{equation}
\frac{\alpha_s(\mu)}{4\pi}=\frac{1}{\beta_0\ln(\mu^2/
\Lambda^2_{\overline{\rm MS}})}-\frac{\beta_1}{\beta_0^3}
\Biggl[\frac{\ln\ln(\mu^2/\Lambda^2_{\overline{\rm MS}})}{\ln^2(\mu^2/
\Lambda^2_{\overline{\rm MS}})}\Biggr].
\end{equation}
If we set $J_\pm$ and $B_\pm$ in (\ref{NLO-evol}) and (\ref{NLO-init}) 
equal to zero, we obtain the LO result. 

This example shows that things are getting rather complicated and technical 
at the NLO level. However, for the phenomenological considerations of this 
review, we have just to be familiar with the general structure of the 
appropriate low-energy effective Hamiltonians. For detailed discussions 
of the renormalization-group techniques sketched above, we refer the 
interested reader to the excellent presentations given in 
\cite{BBL-rev,Buras-lect}.

\subsubsection{Decays with Tree and Penguin 
Contributions}\label{subsubsec:Tree-Pen}
In the exploration of CP violation in the $B$ system, non-leptonic decays 
receiving both tree and penguin contributions, i.e.\ $|\Delta B|=1$, 
$\Delta C=\Delta U=0$ modes (see Subsection~\ref{sec:class}), play a
key r\^ole. This feature is due to the fact that CP asymmetries
originate from certain interference effects, which are provided by such 
channels. Because of the penguin topologies, the operator basis is now
much larger than in (\ref{Heff-example}), where we considered a pure 
``tree'' decay. We have to deal with
\begin{equation}\label{e4}
{\cal H}_{\mbox{{\scriptsize eff}}}=\frac{G_{\mbox{{\scriptsize 
F}}}}{\sqrt{2}}\left[\sum\limits_{j=u,c}V_{jr}^\ast V_{jb}\left\{\sum
\limits_{k=1}^2C_k(\mu)\,Q_k^{jr}+\sum\limits_{k=3}^{10}C_k(\mu)\,Q_k^{r}
\right\}\right],
\end{equation}
where the four-quark operators $Q_k^{jr}$ ($j\in\{u,c\}$, $r\in\{d,s\}$) 
can be divided as follows:
\begin{itemize}
\item Current--current operators:
\begin{equation}\label{CC-op-def}
\begin{array}{rcl}
Q_{1}^{jr}&=&(\overline{r}_{\alpha}j_{\beta})_{\mbox{{\scriptsize V--A}}}
(\overline{j}_{\beta}b_{\alpha})_{\mbox{{\scriptsize V--A}}}\\
Q_{2}^{jr}&=&(\overline{r}_\alpha j_\alpha)_{\mbox{{\scriptsize 
V--A}}}(\overline{j}_\beta b_\beta)_{\mbox{{\scriptsize V--A}}}.
\end{array}
\end{equation}
\item QCD penguin operators:
\begin{equation}\label{qcd-penguins}
\begin{array}{rcl}
Q_{3}^r&=&(\overline{r}_\alpha b_\alpha)_{\mbox{{\scriptsize V--A}}}\sum_{q'}
(\overline{q}'_\beta q'_\beta)_{\mbox{{\scriptsize V--A}}}\\
Q_{4}^r&=&(\overline{r}_{\alpha}b_{\beta})_{\mbox{{\scriptsize V--A}}}
\sum_{q'}(\overline{q}'_{\beta}q'_{\alpha})_{\mbox{{\scriptsize V--A}}}\\
Q_{5}^r&=&(\overline{r}_\alpha b_\alpha)_{\mbox{{\scriptsize V--A}}}\sum_{q'}
(\overline{q}'_\beta q'_\beta)_{\mbox{{\scriptsize V+A}}}\\
Q_{6}^r&=&(\overline{r}_{\alpha}b_{\beta})_{\mbox{{\scriptsize V--A}}}
\sum_{q'}(\overline{q}'_{\beta}q'_{\alpha})_{\mbox{{\scriptsize V+A}}}.
\end{array}
\end{equation}
\item EW penguin operators ($e_{q'}$: electrical quark charges):
\begin{equation}\label{ew-penguins}
\begin{array}{rcl}
Q_{7}^r&=&\frac{3}{2}(\overline{r}_\alpha b_\alpha)_{\mbox{{\scriptsize V--A}}}
\sum_{q'}e_{q'}(\overline{q}'_\beta q'_\beta)_{\mbox{{\scriptsize V+A}}}\\
Q_{8}^r&=&
\frac{3}{2}(\overline{r}_{\alpha}b_{\beta})_{\mbox{{\scriptsize V--A}}}
\sum_{q'}e_{q'}(\overline{q}_{\beta}'q'_{\alpha})_{\mbox{{\scriptsize V+A}}}\\
Q_{9}^r&=&\frac{3}{2}(\overline{r}_\alpha b_\alpha)_{\mbox{{\scriptsize V--A}}}
\sum_{q'}e_{q'}(\overline{q}'_\beta q'_\beta)_{\mbox{{\scriptsize V--A}}}\\
Q_{10}^r&=&
\frac{3}{2}(\overline{r}_{\alpha}b_{\beta})_{\mbox{{\scriptsize V--A}}}
\sum_{q'}e_{q'}(\overline{q}'_{\beta}q'_{\alpha})_{\mbox{{\scriptsize V--A}}}.
\end{array}
\end{equation}
\end{itemize}
The current--current operators, which are related to the tree processes
shown in Fig.~\ref{fig:tree-top}, are analogous to the operators $O_1$ and 
$O_2$ we encountered in (\ref{O2-def}) and (\ref{O1-def}); they have 
just a different flavour content. The QCD and EW penguin operators are
associated with the QCD and EW penguin processes shown in 
Figs.~\ref{fig:QCD-top} and \ref{fig:EWP-top}, respectively. In the 
latter case, also certain box diagrams contribute, which are not shown 
explicitly in Fig.~\ref{fig:EWP-top}. Note that the unitarity of the CKM 
matrix, implying
\begin{equation}\label{CKM-UT-REL}
V_{tr}^\ast V_{tb}=-V_{ur}^\ast V_{ub}-V_{cr}^\ast V_{cb},
\end{equation}
has been used in (\ref{e4}) to eliminate the CKM factor $V_{tr}^\ast V_{tb}$,
which is related to penguin topologies with internal top-quark exchanges.

The Wilson coefficient functions, including both LO and NLO QCD corrections 
and LO corrections in the QED coupling $\alpha$, were calculated in 
\cite{Munich,Rome}. In Table~\ref{tab:WC}, we have collected the values 
for these coefficients for up-dated input parameters, as obtained in 
\cite{Buras-lect}, where also a very detailed discussion of the 
technicalities of the corresponding calculations can be found. The 
next-to-leading order results are given for two renormalization schemes, 
the na\"\i ve dimensional regularization scheme (NDR) and the 
't Hooft--Veltman scheme (HV), corresponding to anticommuting 
and non-anticommuting $\gamma_5$ in ${\cal D}\not=4$ \mbox{dimensions}, 
respectively \cite{NDR-HV}. As usual, $\overline{m_b}(m_b)$ denotes the 
running $b$-quark mass in the $\overline{\mbox{MS}}$ scheme \cite{MS-bar}, 
and $\alpha=1/129$ is the QED coupling. 

The results listed in Table~\ref{tab:WC} correspond to the Standard 
Model. In scenarios for new physics, we usually have to deal with
new operators, i.e.\ the operator basis (\ref{CC-op-def})--(\ref{ew-penguins})
is enlarged. The $\Delta F=1$ case, including also the $B$-meson decays with 
tree and penguin contributions discussed here, was considered in \cite{BMU}, 
where the two-loop anomalous dimension matrices for the dim-6 four-quark 
operators were calculated. The corresponding results for the $\Delta F=2$ 
case can be found in \cite{BMU,DF2-NP}.

{\small
\begin{table}[t]
\begin{center}
\begin{tabular}{|c|c|c|c||c|c|c||c|c|c|}
\hline
&\multicolumn{3}{c||}{$\Lambda_{\overline{{\rm MS}}}^{(5)}=160\,\mbox{MeV}$} &
\multicolumn{3}{c||}{$\Lambda_{\overline{{\rm MS}}}^{(5)}=225\,\mbox{MeV}$} &
\multicolumn{3}{c| }{$\Lambda_{\overline{{\rm MS}}}^{(5)}=290\,\mbox{MeV}$} 
\\
\hline
Scheme & LO & NDR & HV & LO & 
NDR & HV & LO & NDR & HV \\
\hline
$C_1$ & -$0.283$ & -$0.171$ & -$0.209$ & -$0.308$ & 
-$0.185$ & -$0.228$ & -$0.331$ & -$0.198$ & -$0.245$ \\
$C_2$ & 1.131 & 1.075 & 1.095 & 1.144 & 
1.082 & 1.105 & 1.156 & 1.089 & 1.114 \\
\hline
$C_3$ & 0.013 & 0.013 & 0.012 & 0.014 & 
0.014 & 0.013 & 0.016 & 0.016 & 0.014 \\
$C_4$ & -$0.028$ & -$0.033$ & -$0.027$ & -$0.030$ & 
-$0.035$ & -$0.029$ & -$0.032$ & -$0.038$ & -$0.032$ \\
$C_5$ & 0.008 & 0.008 & 0.008 & 0.009 & 
0.009 & 0.009 & 0.009 & 0.009 & 0.010 \\
$C_6$ & -$0.035$ & -$0.037$ & -$0.030$ & -$0.038$ & 
-$0.041$ & -$0.033$ & -$0.041$ & -$0.045$ & -$0.036$ \\
\hline
$C_7/\alpha$ & 0.043 & -$0.003$ & 0.006 & 0.045 & 
-$0.002$ & 0.005 & 0.047 & -$0.002$ & 0.005 \\
$C_8/\alpha$ & 0.043 & 0.049 & 0.055 & 0.048 & 
0.054 & 0.060 & 0.053 & 0.059 & 0.065 \\
$C_9/\alpha$ & -$1.268$ & -$1.283$ & -$1.273$ & -$1.280$ & 
-$1.292$ & -$1.283$ & -$1.290$ & -$1.300$ & -$1.293$ \\
$C_{10}/\alpha$ & 0.302 & 0.243 & 0.245 & 0.328 & 
0.263 & 0.266 & 0.352 & 0.281 & 0.284 \\
\hline
\end{tabular}
\end{center}
\caption{The $|\Delta B|=1$, $\Delta C=\Delta U=0$ Wilson coefficients 
at $\mu=\overline{m_b}(m_b)=4.40\,$GeV for $m_t=170\,$GeV as obtained 
in \cite{Buras-lect}.}\label{tab:WC}
\end{table}
}

Let us now return to the Standard Model. Roughly speaking, 
the origin of the penguin operators in (\ref{e4}) is due to 
the penguin topologies in Figs.~\ref{fig:QCD-top} and \ref{fig:EWP-top} 
with internal top-quark exchanges. In the
matching procedure, the top quark is integrated out -- together with the 
$W$- and $Z$-bosons -- at a scale $\mu={\cal O}(M_W)$. Consequently, these 
particles do not show up explicitly in the low-energy effective Hamiltonian 
(\ref{e4}). However, their presence is included in the initial values of 
the renormalization group evolution (\ref{RG-evol}), leading in particular 
to a top-quark mass dependence of the penguin coefficients. As can 
be seen in Figs.~\ref{fig:QCD-top} and \ref{fig:EWP-top}, there are also 
penguin topologies with internal up- and charm-quark exchanges. Within the
language of effective field theory, these contributions show up as
penguin matrix elements of the current--current operators $Q_{1,2}^{jr}$ 
with internal $j$-quark exchanges ($j\in\{u,c\}$), as illustrated
in Fig.~\ref{fig:Pen-ME}. Their proper treatment requires us to go
beyond the LO approximation, representing another example for the 
importance of NLO calculations \cite{RF-1,RF-EWP1}. Following these
lines, we arrive at the renormalization-scheme independent expression
\begin{eqnarray}
\langle\vec Q^T(\mu)\cdot\vec C(\mu)\rangle^{\rm pen}&=&
\sum\limits_{k=3}^{10}\langle Q_k\rangle_0\overline{C}_k(\mu)+
\frac{\alpha_s(\mu)}{8\pi}\left[\frac{10}{9}-G(m_j,k,\mu)\right]
\nonumber\\
&&\times\Biggl\{\left(-\frac{1}{3}\langle Q_3\rangle_0+
\langle Q_4\rangle_0-\frac{1}{3}\langle Q_5\rangle_0+\langle Q_6\rangle_0
\right)\overline{C}_2(\mu)\nonumber\\
&&+\frac{8}{9}\frac{\alpha}{\alpha_s(\mu)}\left(\langle Q_7\rangle_0+
\langle Q_9\rangle_0\right)\left(3\overline{C}_1(\mu)+
\overline{C}_2(\mu)\right)\Biggr\},\label{Pen-ME}
\end{eqnarray}
where
\begin{equation}
G(m,k,\mu)=-4\int\limits_0^1d x\, 
x(1-x)\,\ln\left[\frac{m^2-x(1-x)k^2}{\mu^2}\right]
\end{equation}
describes the loop process shown in Fig.\ \ref{fig:Pen-ME}, with $k$
denoting the four-momentum of the gluons and photons. The 
$\overline{C}_k(\mu)$ are the renormalization-scheme independent Wilson
coefficients introduced in \cite{Munich}. Expression (\ref{Pen-ME}) is 
at the basis of the implementation of the ``Bander--Silverman--Soni (BSS) 
mechanism'' \cite{BSS} in the operator language. It allows us to estimate 
CP-conserving strong phases, which are an essential ingredient for ``direct'' 
CP violation, through the absorptive part of the function $G(m_j,k,\mu)$. 
Several applications of the BSS mechanism can be found in the literature 
(see, for instance, \mbox{\cite{RF-1}--\cite{ward}}). Let us discuss the
impact of penguin topologies with internal charm- and up-quark exchanges in
more detail in the following subsection.

\subsection{Penguins with Internal Charm- and Up-Quark 
Exchanges}\label{subsec:u-c-pens}
\begin{figure}[t]
\vspace*{0.8truecm}
\begin{center}
\begin{picture}(140,60)(0,20)
\Line(10,65)(130,65)\Text(5,65)[r]{$b$}\Text(140,65)[l]{$r$}
\ArrowLine(34,65)(36,65)
\ArrowLine(104,65)(106,65)
\GCirc(70,65){4}{0}
\PhotonArc(70,50)(15,0,360){1}{1}
\ArrowLine(55,49.5)(55,50.5)
\ArrowLine(85,50.5)(85,49.5)
\Text(70,75)[b]{$Q_{1,2}^{jr}$}\Text(48,43)[b]{$j$}\Text(92,43)[b]{$j$}
\Gluon(70,36)(120,10){2}{10}
\ArrowLine(135,23)(120,10)\ArrowLine(120,10)(135,-3)
\Text(95,15)[tr]{$G,\gamma$}\Text(140,-3)[l]{$q$}
\Text(140,23)[l]{$q$}
\end{picture}
\end{center}
\vspace*{0.8truecm}
\caption[]{The one-loop QCD and QED time-like penguin matrix elements of
the current--current operators $Q_{1,2}^{jr}$ 
($j\in\{u,c\}$, $r\in\{d,s\}$, $q\in\{u,d,c,s\}$).}\label{fig:Pen-ME}
\end{figure}
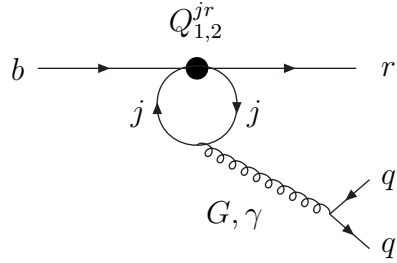

Looking at Figs.~\ref{fig:QCD-top} and \ref{fig:EWP-top}, we observe 
that the amplitude for a generic $b\to r$ penguin process can be
written as follows ($r\in\{d,s\}$):
\begin{equation}
A_{\rm pen}^{(r)}=V_{cr}^\ast V_{cb} P_c^{(r)} + V_{ur}^\ast V_{ub} P_u^{(r)}  
+ V_{tr}^\ast V_{tb} P_t^{(r)},
\end{equation}
were the $P_j^{(r)}$ denote the strong amplitudes of penguin topologies 
with internal $j$-quark exchanges. To be more specific, $P_t^{(r)}$ describes 
the hadronic matrix elements of the penguin operators (\ref{qcd-penguins}) 
and (\ref{ew-penguins}), whereas $P_c^{(r)}$ and $P_u^{(r)}$ are related to 
the matrix elements shown in Fig.~\ref{fig:Pen-ME}, which are associated with 
(\ref{Pen-ME}). Using the unitarity relation (\ref{CKM-UT-REL}) to 
eliminate $V_{ur}^\ast V_{ub}$, we obtain
\begin{equation}\label{APEN}
A_{\rm pen}^{(r)}=V_{tr}^\ast V_{tb}
\left[1+\left(\frac{V_{cr}^\ast V_{cb}}{V_{tr}^\ast V_{tb}}\right)
\Delta P_r\right]\left[P_t^{(r)} -P_u^{(r)}\right],
\end{equation}
where
\begin{equation}
\Delta P_r\equiv\frac{P_c^{(r)}-P_u^{(r)}}{P_t^{(r)}-P_u^{(r)}}.
\end{equation}
Moreover, the generalized Wolfenstein parametrization 
(\ref{NLO-wolf}) with (\ref{CKM-UT-ANGLES}) yields
\begin{equation}
\frac{V_{cd}^\ast V_{cb}}{V_{td}^\ast V_{tb}}=
-\frac{e^{-i\beta}}{R_t}\left[1+{\cal O}(\lambda^4)\right],\quad
\frac{V_{cs}^\ast V_{cb}}{V_{ts}^\ast V_{tb}}=
-\left[1+{\cal O}(\lambda^2)\right].
\end{equation}

In the limit of degenerate up- and charm-quark masses, $\Delta P_r$
would vanish because of the ``Glashow--Iliopoulos--Maiani (GIM) 
mechanism'' \cite{GIM}. However, since $m_u\approx 5\,$MeV, whereas 
$m_c\approx1.3$\,GeV, this GIM cancellation is incomplete, and $\Delta P_r$
may well be sizeable, thereby affecting (\ref{APEN}) significantly. 
Estimates using the BSS mechanism sketched above yield that $|\Delta P_r|$ 
may be as large as ${\cal O}(0.5)$, with a large CP-conserving strong
phase~\cite{BF-u-c-pens}. Consequently, the assumption made in many
analyses that QCD penguin processes are dominated by internal 
top-quark exchanges is not justified, and penguin topologies with 
internal charm and up quarks have also an important phenomenological 
impact \cite{BF-u-c-pens}. An interesting example is the decay 
$B_d\to K^0\overline{K^0}$, where the Standard Model would predict
{\it vanishing} CP violation, if top-quark penguins played the dominant
r\^ole. However, this feature is spoiled by a sizeable value 
of $\Delta P_d$ \cite{RF-BdKK}. Using additional information provided
by $B_s\to K^0\overline{K^0}$, or considering the 
$B_{d,s}\to K^{\ast0}\overline{K^{\ast0}}$ system, weak phases can still
be extracted with the help of $U$-spin flavour-symmetry arguments 
\cite{RF-ang}, as we will see in Subsection~\ref{subsec:BdsDD}. The 
importance of ``charming'' penguins, i.e.\ penguin topologies with 
internal charm-quark exchanges, was also emphasized in 
\cite{charming-pens1}--\cite{charming-pens2}.

\begin{figure}
\vspace*{0.8truecm}
\begin{center}
\leavevmode
\epsfysize=5.0cm 
\epsffile{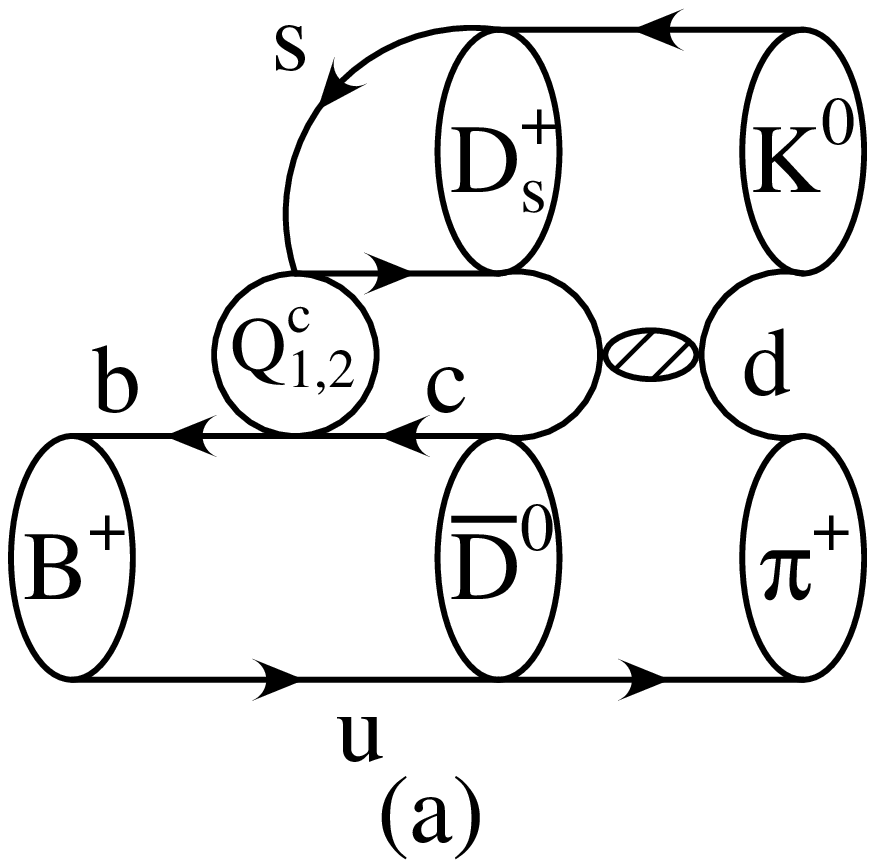} 
  \hspace*{1cm} \boldmath 
  \raisebox{2.3cm}{\large $\in$}  \unboldmath \hspace*{1cm}
\epsfysize=5.0cm 
\epsffile{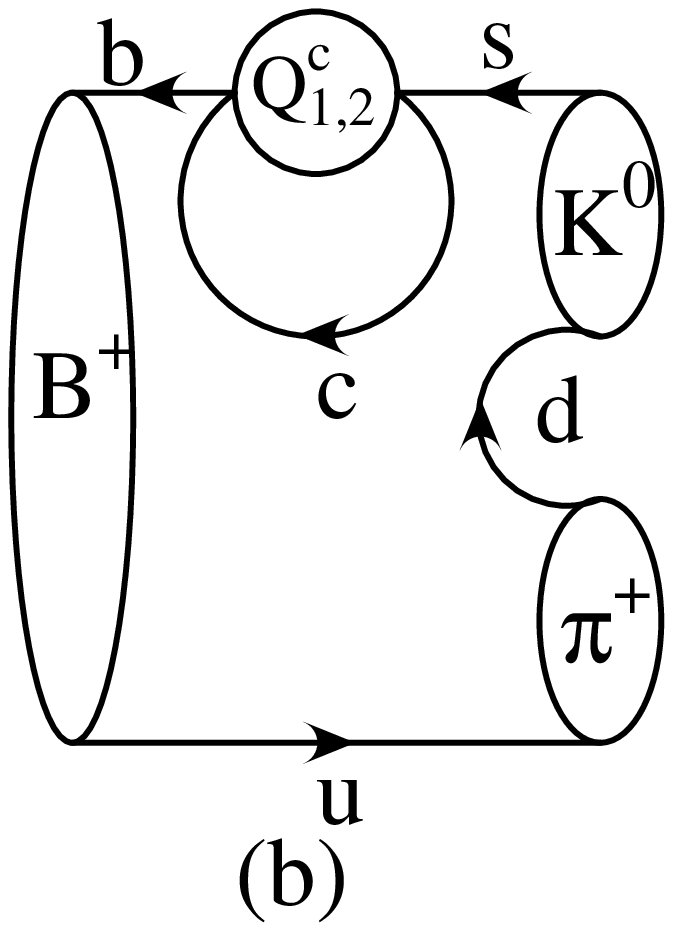}
\end{center}
\vspace*{-0.5truecm}
\caption{Illustration of a rescattering process of the kind 
$B^+\to\{\overline{D^0}D_s^+\}\to\pi^+K^0$ (a), which is contained in 
penguin topologies with 
internal charm-quark exchanges (b).}\label{fig:pen-c}
\end{figure}

The amplitudes $P_u^{(r)}$ and $P_c^{(r)}$ may also receive large 
contributions from rescattering processes \cite{FSI}, which 
can be considered as long-distance penguin topologies with internal 
up and charm quarks \cite{BFM}. An important example is the decay 
$B^+\to \pi^+ K^0$, which receives rescattering contributions of the 
following kind:
\begin{equation}\label{c-rescatter}
B^+\to\{\overline{D^0}D_s^+,\,\overline{D^0}D_s^{\ast+},
\overline{D^{\ast 0}}D_s^{\ast+},\,\ldots\}\to\pi^+K^0
\end{equation}
\begin{equation}\label{u-rescatter}
B^+\to\{\pi^0K^+,\,\pi^0K^{\ast +},\,\rho^0K^{\ast +},\,\ldots\}\to\pi^+K^0.
\end{equation}
Here the dots include also intermediate multibody states. In the
first class of rescattering processes, (\ref{c-rescatter}), we have to 
deal with decays that are caused by the $Q_{1,2}^{c}$ current--current 
operators through insertions into 
tree-diagram-like topologies, and may rescatter into the final state 
$\pi^+K^0$. These final-state interaction effects are related to penguin 
topologies with internal charm-quark exchanges, as can be seen in 
Fig.~\ref{fig:pen-c}. In the case of (\ref{u-rescatter}), we have to
deal both with analogous penguin topologies, where all charm quarks are 
replaced by up quarks, and with annihilation topologies. These issues 
received a lot of attention in the context of extractions of $\gamma$ from 
$B\to\pi K$ decays (see Section~\ref{sec:BpiK}), and require also careful 
attention for several other modes.

\subsection{Electroweak Penguin Effects}
Since the ratio $\alpha/\alpha_s={\cal O}(10^{-2})$ of the QED and QCD 
couplings is very small, we would na\"\i vely expect that EW penguins should 
play a minor r\^ole in comparison with QCD penguins. This would actually 
be the case if the top quark was not heavy. However, the Wilson 
coefficient of the EW penguin operator $Q_9^r$ increases strongly with 
the top-quark mass $m_t$. As can be seen in Table~\ref{tab:WC}, calculated
for $m_t=170\,$GeV, $C_9$ becomes comparable in magnitude with Wilson 
coefficients of QCD penguin operators. 

Because of this feature, we obtain interesting EW penguin effects 
in several non-leptonic $B$ decays \cite{RF-EWP1} (for a review, see
\cite{RF-rev}). A prominent example is $B^-\to \phi K^-$, 
which is governed by QCD penguins \cite{BphiK-old}. However, EW penguins 
may contribute to this channel in ``colour-allowed'' form, thereby reducing 
its branching ratio by ${\cal O}(30\%)$ \cite{RF-EWP1,DH-PhiK}. There are 
even decays, which are {\it dominated} by EW penguins, for example 
$B^-\to\phi \pi^-$ \cite{RF-EWP2} 
and $B_s^0\to\phi \pi^0$ \cite{RF-EWP3}. Unfortunately, the corresponding
branching ratios of ${\cal O}(10^{-8})$ and ${\cal O}(10^{-7})$, 
respectively, are very small. On the other hand, the $B^-\to \phi K^-$ mode
has already been observed by the $B$-factories, with branching ratios at 
the $10^{-5}$ level. We shall come back to this channel in 
Subsection~\ref{subsec:BPhiK}.

EW penguins have also an important impact on the phenomenology of 
$B\to\pi K$ decays \cite{EWP-BpiK,ghlr-ewp}. As we will see in 
Section~\ref{sec:BpiK}, these modes provide several strategies to extract 
the CKM angle $\gamma$. Alternatively, if $\gamma$ is used as an input, 
EW penguin amplitudes can be determined from $B\to\pi K$ branching ratios 
\cite{PAPI,BF-BpiK1}, thereby allowing an interesting comparison with 
the Standard-Model predictions. In \cite{PAPIII}, in addition to strategies 
for extracting the CKM angle $\gamma$ and obtaining experimental insights 
into the world of EW penguins, also a transparent expression for the ratio 
of EW penguin to tree amplitudes was derived.

\subsection{Factorization of Hadronic Matrix Elements}\label{subsec:fact}
In order to discuss ``factorization'', let us consider again the decay 
$\overline{B^0_d}\to D^+K^-$, which we encountered already in the
discussion of the low-energy effective Hamiltonian (\ref{Heff-example}). 
The problem in the evaluation of the $\overline{B^0_d}\to D^+K^-$ 
transition amplitude is the calculation of the hadronic matrix elements 
of the $O_{1,2}$ operators between the $\langle K^-D^+|$ final and 
$|\overline{B^0_d}\rangle$ initial states. Making use of the well-known 
$SU(N_{\rm C})$ colour-algebra relation
\begin{equation}
T^a_{\alpha\beta}T^a_{\gamma\delta}=\frac{1}{2}\left(\delta_{\alpha\delta}
\delta_{\beta\gamma}-\frac{1}{N_{\rm C}}\delta_{\alpha\beta}
\delta_{\gamma\delta}\right)
\end{equation}
to re-write the operator $O_1$, we obtain
\begin{displaymath}
\langle K^-D^+|{\cal H}_{\rm eff}|\overline{B^0_d}\rangle=
\frac{G_{\rm F}}{\sqrt{2}}V_{us}^\ast V_{cb}\Bigl[a_1\langle K^-D^+|
(\overline{s}_\alpha u_\alpha)_{\mbox{{\scriptsize V--A}}}
(\overline{c}_\beta b_\beta)_{\mbox{{\scriptsize V--A}}}
|\overline{B^0_d}\rangle
\end{displaymath}
\vspace*{-0.3truecm}
\begin{equation}
+2\,C_1\langle K^-D^+|
(\overline{s}_\alpha\, T^a_{\alpha\beta}\,u_\beta)_{\mbox{{\scriptsize 
V--A}}}(\overline{c}_\gamma 
\,T^a_{\gamma\delta}\,b_\delta)_{\mbox{{\scriptsize V--A}}}
|\overline{B^0_d}\rangle\Bigr],\nonumber
\end{equation}
with
\begin{equation}\label{a1-def}
a_1=\frac{C_1}{N_{\rm C}}+C_2.
\end{equation}
It is now straightforward to ``factorize'' the hadronic matrix elements:
\begin{eqnarray}
\lefteqn{\left.\langle K^-D^+|
(\overline{s}_\alpha u_\alpha)_{\mbox{{\scriptsize 
V--A}}}(\overline{c}_\beta b_\beta)_{\mbox{{\scriptsize V--A}}}
|\overline{B^0_d}\rangle\right|_{\rm fact}}\nonumber\\
&&=\langle K^-|\left[\overline{s}_\alpha\gamma_\mu(1-\gamma_5)u_\alpha\right]
|0\rangle\langle D^+|\left[\overline{c}_\beta\gamma^\mu
(1-\gamma_5)b_\beta\right]|\overline{B^0_d}\rangle\nonumber\\
&&\propto f_K \,\mbox{[decay constant]}\times 
F_{BD}(M_K^2;0^+) \,\mbox{[form factor]},
\end{eqnarray}
\begin{equation}
\left.\langle K^-D^+|
(\overline{s}_\alpha\, T^a_{\alpha\beta}\,u_\beta)_{\mbox{{\scriptsize 
V--A}}}(\overline{c}_\gamma 
\,T^a_{\gamma\delta}\,b_\delta)_{\mbox{{\scriptsize V--A}}}
|\overline{B^0_d}\rangle\right|_{\rm fact}=0.
\end{equation}
The quantity introduced in (\ref{a1-def}) is a phenomenological
``colour factor'', governing ``colour-allowed'' decays \cite{BSW,NS}. 
The $\overline{B^0_d}\to D^+K^-$ mode belongs to this category, since the
colour indizes of the $K^-$-meson and the $\overline{B^0_d}$--$D^+$ system
in Fig.~\ref{fig:feyn-example} run independently from each other. 
In the case of ``colour-suppressed'' transitions, for instance 
$\overline{B^0_d}\to D^0 \pi^0$, this is not the case, and we have 
to deal with the combination
\begin{equation}\label{a2-def}
a_2=C_1+\frac{C_2}{N_{\rm C}}.
\end{equation}
The coefficients $a_1$ and $a_2$ were analysed beyond the leading 
logarithmic approximation in different renormalization schemes in 
\cite{Buras-a1-a2}. Whereas $a_1=1.01\pm0.02$ is very stable under 
changes of $\Lambda_{\overline{\rm MS}}$ and the renormalization 
scale and scheme, $a_2={\cal O}(0.2)$ is affected sizeably by such
variations.

The phenomenological concept of the ``factorization'' of hadronic matrix 
elements in weak non-leptonic decays has a long history (see, for example,
\cite{BSW}--\cite{Fact-new}). One possibility to justify it is provided 
by the large $N_{\rm C}$ limit \cite{largeN}, where the transition 
amplitudes factorize at leading order in a $1/N_{\rm C}$-expansion, 
which can also be combined with the framework of ``Heavy-Meson Chiral 
Perturbation Theory'' \cite{RF-Nc}. Recently, an interesting approach 
was proposed in \cite{BBNS1}--\cite{BBNS3}, representing an important 
step towards a rigorous theoretical description of a large class of 
non-leptonic two-body $B$ decays. Within this framework, the physical 
idea \cite{fact-idea} that factorization should hold for hadrons 
containing a heavy quark $Q$ with $m_Q\gg \Lambda_{\rm QCD}$ 
is confirmed, and a formalism to calculate the relevant amplitudes at 
the leading order of a $\Lambda_{\rm QCD}/m_Q$-expansion is provided. 
Let us consider a decay $\overline{B}\to M_1M_2$, where $M_1$ picks 
up the spectator quark. If $M_1$ is either a heavy ($D$) or a light 
($\pi$, $K$) meson, and $M_2$ a light \mbox{($\pi$, $K$)} meson, it 
can be shown that the corresponding transition amplitude takes the 
following generic form \cite{BBNS1}:
\begin{equation}\label{QCD-factor}
A(\overline{B}\to M_1M_2)=\left[\mbox{na\"\i ve factorization}\right]
\times\left[1+\mbox{calculable ${\cal O}(\alpha_s)$}+
{\cal O}(\Lambda_{\rm QCD}/m_b)\right].
\end{equation}
While the ${\cal O}(\alpha_s)$ terms, containing also radiative 
non-factorizable corrections to ``na\"\i ve'' factorization, can be 
calculated in a systematic way, the main limitation is due to the 
${\cal O}(\Lambda_{\rm QCD}/m_b)$ terms. A very detailed discussion of 
$B$ decays into heavy--light final states within this ``QCD factorization'' 
approach, including also the $\overline{B^0_d}\to D^+K^-$ mode considered 
above, was given in \cite{BBNS2}, whereas the implications for 
$B\to\pi K, \pi\pi$ modes were studied in \cite{BBNS3}. Another QCD 
approach to deal with non-leptonic $B$ decays into charmless final 
states -- the perturbative hard-scattering (or ``PQCD'') approach -- was 
developed independently in \cite{PQCD}--\cite{PQCD-comp}, and differs from 
the ``QCD factorization'' formalism in some technical aspects. As far as 
phenomenological applications are concerned, the crucial question is 
obviously the importance of the ${\cal O}(\Lambda_{\rm QCD}/m_b)$ terms, 
which are hard to estimate. 
We shall come back to these issues in more detail in Section~\ref{sec:BpiK}.

\boldmath
\section{A Brief Look at the Kaon System}\label{sec:kaon}
\setcounter{equation}{0}
\unboldmath
In 1964, CP violation was discovered in the famous experiment by Christenson 
{\it et al.}~\cite{CP-discovery}, who have observed $K_{\rm L}\to\pi^+\pi^-$ 
decays. If the CP symmetry was conserved by weak interactions, as was
believed until the experimental discovery of CP violation, the mass 
eigenstates $K_{\rm S}$ and $K_{\rm L}$ of the Hamiltonian describing 
$K^0$--$\overline{K^0}$ mixing were eigenstates of the CP operator 
with eigenvalues $+1$ and $-1$, respectively. Consequently, 
since $\pi^+\pi^-$ is a CP-even final state, the detection of 
$K_{\rm L}\to\pi^+\pi^-$ decays signals the non-invariance of weak 
interactions under CP transformations, i.e.\ CP violation. 

Since we are mainly concerned with $B$ decays in this review, the
discussion of CP violation in the kaon system will be rather brief,
and we refer the reader for more detailed presentations to 
\cite{BF-rev,Brev01,buchalla}. The major aspects of kaon physics we have 
to deal with are the CP-violating parameter $\varepsilon$, which plays 
an important r\^ole to constrain the unitarity triangle, as we have already 
seen in Subsection~\ref{subsec:CKM-fits}, and the rare kaon decays 
$K^+\to\pi^+\nu\overline{\nu}$ and $K_{\rm L}\to\pi^0\nu\overline{\nu}$, 
providing interesting relations to the $B$-meson system, as we will discuss 
in Section~\ref{sec:MFV}.

\boldmath
\subsection{CP-violating Observables}\label{subsec:kaon-CP}
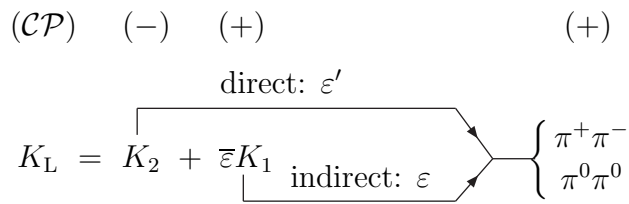
\begin{figure}[b]
\begin{center}
\begin{picture}(360,100)(0,60)
\Text(70,140)[bc]{$({\cal CP})$}
\Text(109,140)[bc]{$(-)$}
\Text(145,140)[bc]{$(+)$}
\Text(276,140)[bc]{$(+)$}
\Text(160,130)[tc]{direct: $\varepsilon'$}
\Text(190,94)[tc]{indirect: $\varepsilon$}
\Text(60,80)[bl]{$K_{\rm L}\,\,=\,\,K_2\,\,+\,\,\overline{\varepsilon}K_1
\qquad\qquad\qquad\qquad \left\{\begin{array}{c}\pi^+\pi^-\\
\pi^0\pi^0\end{array}\right.$}
\Line(105,105)(105,115)
\Line(105,115)(225,115)
\ArrowLine(225,115)(239,96)
\Line(145,90)(145,80)
\Line(145,80)(225,80)
\ArrowLine(225,80)(239,96)
\Line(239,96)(255,96)
\end{picture}
\end{center}
\vspace*{-1truecm}
\caption{Indirect and direct CP violation in $K_{\rm L}\to\pi\pi$ 
decays.}\label{fig:CP-kaon}
\end{figure}

\unboldmath
In the neutral kaon system, CP violation is described by two complex 
quantities, called $\varepsilon$ and $\varepsilon'$, which are defined 
by the following ratios of decay amplitudes:
\begin{equation}\label{defs-eps}
\frac{A(K_{\rm L}\to\pi^+\pi^-)}{A(K_{\rm S}
\to\pi^+\pi^-)}\approx\varepsilon+\varepsilon',\quad
\frac{A(K_{\rm L}\to\pi^0\pi^0)}{A(K_{\rm S}
\to\pi^0\pi^0)}\approx\varepsilon-2\,\varepsilon'.
\end{equation}
These parameters are related to ``indirect'' and ``direct'' CP violation,
as illustrated in Fig.~\ref{fig:CP-kaon}, where $K_1$ and $K_2$ denote the 
CP eigenstates of the neutral kaon system with CP eigenvalues $+1$ and $-1$, 
respectively. Here indirect CP violation is due to the fact that the mass 
eigenstate $K_{\rm L}$ of the neutral kaon system is {\it not} a CP eigenstate 
because of the small admixture of the CP-even $K_1$ state, which may decay 
through a CP-conserving transition into a $\pi\pi$ final state. On the 
other hand, direct CP violation originates from {\it direct} transitions 
of the CP-odd $K_2$ state into the CP-even $\pi\pi$ final state.

After the discovery of indirect CP violation in $K_{\rm L}\to\pi^+\pi^-$, 
it was also observed in $K_{\rm L}\to\pi^0\pi^0$, 
$\pi \ell\overline{\nu}_\ell$, $\pi^+\pi^-\gamma$ modes, and recently in 
$K_{\rm L}\to\pi^+\pi^-e^+e^-$ transitions. All these effects can be 
described by 
\begin{equation}
\varepsilon=(2.280\pm0.013)\times e^{i\frac{\pi}{4}}\times 10^{-3}.
\end{equation}
As we have noted in Subsection~\ref{subsec:CKM-fits}, this observable
fixes a hyperbola in the $\overline{\rho}$--$\overline{\eta}$ plane,
which is given as follows \cite{Brev01}:
\begin{equation}\label{100a}
\overline{\eta} \left[(1-\overline{\rho}) A^2 \eta_2 S_0(x_t)
+ P_c(\varepsilon) \right] A^2 \hat B_K = 0.204,
\end{equation}
and plays an important r\^ole to constrain the unitarity triangle. In this
expression, $\eta_2=0.57$ is a perturbative QCD factor \cite{BJW90}, 
$S_0(x_t)\approx2.38$ with $x_t\equiv m_t^2/M_W^2$ is the Inami--Lim 
function \cite{IL} resulting from box diagrams with $(t, W^\pm)$ 
exchanges (see Fig.~\ref{fig:boxes} for the analogous diagrams contributing
to $B^0_q$--$\overline{B^0_q}$ mixing), and $P_c(\varepsilon)=0.30\pm 0.05$ 
\cite{HN} summarizes the contributions not proportional to $V_{ts}^*V_{td}$.
The renormalization-group invariant quantity 
\begin{equation}\label{BK-def}
\hat B_K\equiv B_K(\mu)\left[\alpha_s^{(3)}(\mu)\right]^{-2/9}
\left[1+\frac{\alpha_s^{(3)}(\mu)}{4\pi}J_3\right]
\end{equation}
parametrizes the relevant hadronic matrix element through
\begin{equation}\label{mix-me}
\langle 
\overline{K^0}|(\overline{s}d)_{\rm V-A}(\overline{s}d)_{\rm V-A}|K^0\rangle
\equiv\frac{8}{3}B_K(\mu)f_K^2M_K^2,
\end{equation}
where $\alpha_s^{(3)}(\mu)$ is the QCD coupling for three active quark
flavours, $J_3=1.895$ in the NDR scheme \cite{BJW90}, $f_K=160$\,MeV is 
the kaon decay constant, and $M_K$ the kaon mass. A reasonable range
for $\hat B_K$ is \cite{Brev01}
\begin{equation}
\hat B_K=0.85\pm0.15,
\end{equation}
which agrees well with recent lattice results (for reviews, see 
\cite{Rome-rev,lellouch}), is slightly above the large--$N_{\rm C}$ estimates 
\cite{large-N-est}, and a bit below the chiral quark-model estimates
\cite{chiral-est}. The numerical value in (\ref{100a}) and the value 
for $P_c(\varepsilon)$ differ slightly from those given in \cite{Brev01} 
due to $\lambda=0.222$ used here instead of $\lambda=0.22$ used in that 
paper \cite{BF-MFV}. This increase of $\lambda$ is made in order to be 
closer to the experimental value of $|V_{ud}|$ \cite{Rome-rev,hoecker}. 

Direct CP violation in $K\to\pi\pi$ transitions is measured by the 
quantity Re$(\varepsilon'/\varepsilon)$, which is governed by the
competition between QCD and EW penguins. For large top-quark masses,
a cancellation between these contributions arises, suppressing
Re$(\varepsilon'/\varepsilon)$ strongly \cite{eps-prime-anat}.
In 1999, new measurements of this observable have demonstrated that 
it is different from zero, thereby establishing direct CP violation and
excluding ``superweak'' models \cite{superweak}. The present results read 
as follows:
\begin{equation}\label{epsprime-res}
\mbox{Re}(\varepsilon'/\varepsilon)=\left\{\begin{array}{ll}
(20.7\pm2.8)\times10^{-4}&\mbox{(KTeV Collaboration \cite{KTeV}),}\\
(15.3\pm2.6)\times10^{-4}&\mbox{(NA48 Collaboration \,\cite{NA48}).}
\end{array}\right.
\end{equation}
Taking into account also the previous measurements yields a world average 
of 
\begin{equation}
\mbox{Re}(\varepsilon'/\varepsilon)=(17.2\pm1.8)\times10^{-4}. 
\end{equation}
Although the short-distance contributions to Re$(\varepsilon'/\varepsilon)$ 
are now fully under control 
\cite{Munich,Rome}, the theoretical predictions are unfortunately 
affected by large uncertainties arising from hadronic matrix elements. 
In units of $10^{-4}$, the various estimates for 
Re$(\varepsilon'/\varepsilon)$ that can be found in the literature range 
from 5 to 40, demonstrating the unsatisfactory present theoretical status of 
this observable (for reviews, see \cite{Brev01,epsprime}). Consequently, 
it does not allow a stringent test of the CP-violating sector of the 
Standard Model, unless better techniques to deal with the hadronic matrix 
elements of the relevant operators are available. A nice representation
of the experimental results on $\mbox{Re}(\varepsilon'/\varepsilon)$
through contours in a plane of hadronic parameters, which may be useful 
in this respect, was recently proposed in \cite{Bu-Ge}.

\boldmath
\subsection{The Rare Decays $K^+\to\pi^+\nu\overline{\nu}$ and 
$K_{\rm L}\to\pi^0\nu\overline{\nu}$}
\unboldmath
In order to test the Standard-Model description of CP violation, the decays 
$K^+\to\pi^+\nu\overline{\nu}$ and $K_{\rm L}\to\pi^0\nu\overline{\nu}$ are 
considerably more promising than Re$(\varepsilon'/\varepsilon)$. Within the 
Standard Model, these loop-induced modes originate from $Z$-penguins and box 
diagrams, and are governed by a single short-distance function 
$X_0(x_t)\approx1.5$. The $K\to\pi\nu\overline{\nu}$ modes are particularly 
interesting since they provide remarkable relations to the $B$ system and 
are very clean from a theoretical point of view. The latter feature
arises from the fact that the required hadronic matrix elements are just 
quark-current matrix elements between pion and kaon states, which can be 
determined from very accessible semi-leptonic $K$ decays. Moreover, it 
was shown that other long-distance contributions are negligible 
\cite{kpinunu-LD}, as well as contributions from higher dimensional 
operators \cite{BI,FLP}. Consequently, the major theoretical uncertainties 
affecting analyses of $K\to\pi\nu\overline{\nu}$ modes originate from the 
scale ambiguities associated with perturbative QCD calculations. The 
inclusion of NLO QCD corrections has reduced these uncertainties 
considerably \cite{Kpi-NLO,BB98}, thereby making $K\to\pi\nu\overline{\nu}$ 
modes an important probe of the Standard Model. 

The derivation of the $K\to\pi\nu\overline{\nu}$ branching ratios
has recently been reviewed in \cite{Brev01}. Within the Standard 
Model, the predictions are given as follows:
\begin{equation}\label{Kpinunu-SM}
{\rm BR}(\kpnn)= (7.5\pm 2.9)\times 10^{-11}, \quad
{\rm BR}(\klnn)= (2.6\pm 1.2)\times 10^{-11}.
\end{equation}
The former channel has already been observed at Brookhaven \cite{Adler00}, 
with 
\begin{equation}\label{Brook}
{\rm BR}(\kpnn)= \left(1.5^{+3.4}_{-1.2}\right)\times 10^{-10}.
\end{equation}
For a review of the experimental prospects concerning these modes, 
see \cite{littenberg}.

Interestingly, the $K\to\pi\nu\overline{\nu}$ branching ratios allow a 
determination of $\sin2\beta$ \cite{BBSIN}. To this end, it is useful
to introduce the following ``reduced'' branching ratios:
\begin{equation}\label{b1b2}
B_1\equiv\frac{1}{\kappa_+}{\rm BR}(\kpnn),\quad
B_2\equiv\frac{1}{\kappa_{\rm L}}{\rm BR}(\klnn),
\end{equation}
where
\begin{equation}\label{kappa-p}
\kappa_+=r_{K^+}\left[\frac{3\,\alpha^2{\rm BR}(K^+\to\pi^0e^+\nu_e)}{2\,\pi^2
\sin^4\Theta_{\rm W}}\right]\lambda^8=4.42\times 10^{-11},
\end{equation}
and
\begin{equation}\label{kappa-L}
\kappa_{\rm L}=\left[\frac{r_{K_{\rm L}}}{r_{K^+}}
\frac{\tau(K_{\rm L})}{\tau(K^+)}\right]\kappa_+=1.93\times 10^{-10}.
\end{equation}
Here $r_{K^+}=0.901$ and $r_{K_{\rm L}}=0.944$ describe isospin-breaking
corrections, which arise in relating $K^+\to\pi^+\nu\overline{\nu}$ and
$K_{\rm L}\to\pi^0\nu\overline{\nu}$ to $K^+\to\pi^0e^+\nu_e$, respectively.
In the Standard Model, we have to an excellent approximation
\begin{equation}\label{sin}
\sin 2\beta=\frac{2\, r_s}{1+r^2_s}
\end{equation}
with
\begin{equation}\label{cbb}
r_s=\sqrt{\sigma}\left[{\sqrt{\sigma(B_1-B_2)}-P_c(\nu\overline{\nu})
\over\sqrt{B_2}}\right],
\end{equation}
where the quantity $P_c(\nu\overline{\nu})=0.40\pm0.06$ describes the 
internal charm-quark contribution to $\kpnn$ \cite{BB98}, and 
\begin{equation}\label{sigma-def}
\sigma= \frac{1}{(1-\lambda^2/2)^2}.
\end{equation} 
In writing (\ref{sin}), we have assumed that $\sin 2\beta>0$, 
as expected in the Standard Model. Note that the numerical values 
in (\ref{kappa-p}) and (\ref{kappa-L}), as well as the value for 
$P_c(\nu\overline{\nu})$, differ slightly from those given in 
\cite{BB98,BBSIN} due to $\lambda=0.222$ used here instead of 
$\lambda=0.22$ used in these papers \cite{BF-MFV}. 

The strength of expression (\ref{sin}) is its theoretical cleanness, 
allowing a precise determination of $\sin 2\beta$ free of hadronic 
uncertainties, and independent of other parameters, such as $|V_{cb}|$, 
$|V_{ub}/V_{cb}|$ and $m_t$. Moreover, it provides a remarkable bridge
to CP violation in the $B$ system, since $\sin2\beta$ can also be
determined in a clean way through CP-violating effects in $B_d\to
J/\psi K_{\rm S}$ decays, as we will see in Subsection~\ref{subsec:BpsiK}. 
The comparison of these two determinations of $\sin 2\beta$ with each 
other is particularly well suited for tests of CP violation in the 
Standard Model, and offers a powerful tool to probe the physics beyond 
it \cite{BBSIN,NIR}. We shall return to this exciting issue in more 
detail in Section~\ref{sec:MFV}, where we discuss various aspects of the 
unitarity triangle, $\sin2\beta$ and $K\to\pi\nu\overline{\nu}$ decays 
in models with minimal flavour violation \cite{BF-MFV}. There we will
also derive a generalization of (\ref{sin}).

In order to obtain the whole picture of CP violation, it is essential
to study this phenomenon outside the kaon system. In this respect,
the $B$ system -- the central topic of this review -- is most promising. 
Let us turn to decays of neutral $B$-mesons first.

\newpage

\boldmath
\section{Time Evolution of Neutral $B$ Decays}\label{sec:neutral}
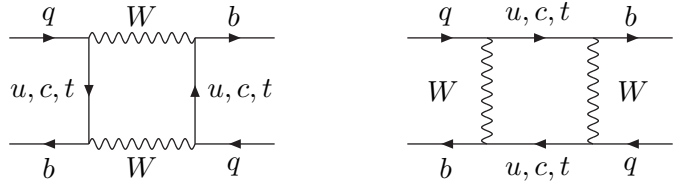
\begin{figure}[h]
\begin{center}
{\small
\begin{picture}(250,70)(0,45)
\ArrowLine(10,100)(40,100)\Photon(40,100)(80,100){2}{8}
\ArrowLine(80,100)(110,100)
\ArrowLine(40,60)(10,60)\Photon(40,60)(80,60){2}{8}
\ArrowLine(110,60)(80,60)
\ArrowLine(40,100)(40,60)\ArrowLine(80,60)(80,100)
\Text(25,105)[b]{$q$}\Text(60,105)[b]{$W$}\Text(95,105)[b]{$b$}
\Text(25,55)[t]{$b$}\Text(60,55)[t]{$W$}\Text(95,55)[t]{$q$}
\Text(35,80)[r]{$u,c,t$}\Text(85,80)[l]{$u,c,t$}
\ArrowLine(160,100)(190,100)\ArrowLine(190,100)(230,100)
\ArrowLine(230,100)(260,100)
\ArrowLine(190,60)(160,60)\ArrowLine(230,60)(190,60)
\ArrowLine(260,60)(230,60)
\Photon(190,100)(190,60){2}{8}\Photon(230,60)(230,100){2}{8}
\Text(175,105)[b]{$q$}\Text(245,105)[b]{$b$}
\Text(175,55)[t]{$b$}\Text(245,55)[t]{$q$}
\Text(210,105)[b]{$u,c,t$}\Text(210,55)[t]{$u,c,t$}
\Text(180,80)[r]{$W$}\Text(240,80)[l]{$W$}
\end{picture}}
\end{center}
\vspace*{-0.4truecm}
\caption{Box diagrams contributing to $B^0_q$--$\overline{B^0_q}$ mixing 
($q\in\{d,s\}$).}\label{fig:boxes}
\end{figure}

\unboldmath
\setcounter{equation}{0}
A characteristic feature of the neutral $B_q$-meson system is 
$B^0_q$--$\overline{B^0_q}$ mixing ($q\in\{d,s\}$). Within the
Standard Model, this phenomenon is induced at lowest order through 
the box diagrams shown in Fig.~\ref{fig:boxes}. The Wigner--Weisskopf 
formalism \cite{WW} yields the following effective Schr{\"o}dinger equation:
\begin{equation}\label{SG-OSZ}
i\,\frac{d}{d\, t}\left(\begin{array}{c} a(t)\\ b(t)
\end{array}
\right)=\left[\left(\begin{array}{cc}
M_{0}^{(q)} & M_{12}^{(q)}\\ M_{12}^{(q)\ast} & M_{0}^{(q)}
\end{array}\right)-
\frac{i}{2}\left(\begin{array}{cc}
\Gamma_{0}^{(q)} & \Gamma_{12}^{(q)}\\
\Gamma_{12}^{(q)\ast} & \Gamma_{0}^{(q)}
\end{array}\right)\right]
\cdot\left(\begin{array}{c}
a(t)\\ b(t)\nonumber
\end{array}
\right),
\end{equation}
which describes the time evolution of the state vector
\begin{equation}
\left\vert \psi_q(t)\right\rangle=a(t)\left\vert B^{0}_q\right\rangle+
b(t)\left\vert\overline{B^{0}_q}\right\rangle.
\end{equation}

\subsection{Solution of the Schr\"odinger Equation}\label{subsec:SG}
It is a straightforward exercise to calculate the eigenstates 
$\left\vert B_{\pm}^{(q)}\right\rangle$ and eigenvalues 
$\lambda_{\pm}^{(q)}$ of the Hamiltonian given in (\ref{SG-OSZ}):
\begin{equation}
\left\vert B_{\pm}^{(q)} \right\rangle  =
\frac{1}{\sqrt{1+\vert \alpha_q\vert^{2}}}
\left(\left\vert B^{0}_q\right\rangle\pm\alpha_q\left\vert
\overline{B^{0}_q}\right\rangle\right)
\end{equation}
\begin{equation}
\lambda_{\pm}^{(q)}  =
\left(M_{0}^{(q)}-\frac{i}{2}\Gamma_{0}^{(q)}
\right)\pm
\left(M_{12}^{(q)}-\frac{i}{2}\Gamma_{12}^{(q)}\right)\alpha_q,
\end{equation}
where
\begin{equation}\label{alpha-q}
\alpha_q e^{+i\left(\Theta_{\Gamma_{12}}^{(q)}+n'
\pi\right)}=
\sqrt{\frac{4\left\vert M_{12}^{(q)}\right\vert^{2}
e^{-i2\delta\Theta_{M/\Gamma}^{(q)}}+\left\vert\Gamma_{12}^{(q)}\right
\vert^{2}}{4\left\vert M_{12}^{(q)}\right\vert^{2}+\left\vert
\Gamma_{12}^{(q)}\right\vert^{2}- 4\left\vert M_{12}^{(q)}\right\vert
\left\vert\Gamma_{12}^{(q)}\right\vert\sin\delta\Theta_{M
/\Gamma}^{(q)}}}.
\end{equation}
Here $M_{12}^{(q)}\equiv e^{i\Theta_{M_{12}}^{(q)}}\vert
M_{12}^{(q)}\vert$, $\Gamma_{12}^{(q)}\equiv
e^{i\Theta_{\Gamma_{12}}^{(q)}}\vert\Gamma_{12}^{(q)}
\vert$ and $\delta\Theta_{M/\Gamma}^{(q)}\equiv\Theta_{M_{12}}^{(q)}-
\Theta_{\Gamma_{12}}^{(q)}$. The $n'\in\{0,1\}$ parametrizes the sign 
of the square root appearing in (\ref{alpha-q}). 

Calculating the dispersive parts of the box diagrams, which are dominated 
to an excellent approximation by the internal top-quark exchanges, 
gives \cite{BuSlSt}
\begin{equation}\label{M12-expr}
M_{12}^{(q)}=\frac{G_{\rm F}^2M_W^2}{12\pi^2}\eta_B M_{B_q}
\hat B_{B_q}f_{B_q}^2(V_{tq}^\ast V_{tb})^2 
S_0(x_t)\,e^{i(\pi-\phi_{\rm CP}(B_q))}.
\end{equation}
Here $\eta_B=0.55$ is a perturbative QCD factor \cite{BJW90,UKJS},
which is common to $M_{12}^{(d)}$ and $M_{12}^{(s)}$, and $M_{B_q}$ 
denotes the $B_q$-meson mass. The non-perturbative parameter $\hat B_{B_q}$ 
is the counterpart of $\hat B_K$ introduced in (\ref{BK-def}), parametrizing
$\langle\overline{B^0_q}|(\overline{b}q)_{\rm V-A}(\overline{b}q)_{\rm V-A}
|B^0_q\rangle$, and $f_{B_q}$ is the $B_q$ decay constant. There is a vast 
literature on the calculations of these parameters, using lattice 
\cite{Rome-rev,FBlat} and QCD sum-rule techniques \cite{QCDSR}; both 
approaches give compatible results. A reasonable range for the combination 
relevant for (\ref{M12-expr}) is \cite{Brev01}
\begin{equation}\label{sqrtBfb}
\sqrt{\hat B_{B_d}}f_{B_d}=(230\pm40)\,{\rm MeV}.
\end{equation}
Finally, the Inami--Lim function $S_0(x_{t})\approx2.38$ appeared already 
in Subsection~\ref{subsec:kaon-CP}, and the convention-dependent phase
$\phi_{\rm CP}(B_q)$ is defined through
\begin{equation}\label{CP-def}
({\cal CP})\left\vert B^{0}_q\right\rangle=
e^{i\phi_{\mbox{{\scriptsize CP}}}(B_q)}
\left\vert\overline{B^{0}_q}\right\rangle.
\end{equation}

Evaluating the absorptive parts of the box diagrams yields
\begin{equation}
\frac{\Gamma_{12}^{(q)}}{M_{12}^{(q)}}\approx
-\frac{3\pi}{2S_0(x_{t})}\frac{m_b^2}{M_W^2}={\cal O}(m_b^2/m_t^2)\ll 1,
\end{equation}
where we have taken into account that $S_0(x_t)\propto x_t=m_t^2/M_W^2$. 
If we expand (\ref{alpha-q}) in powers of this small quantity and neglect 
second-order terms, we arrive at
\begin{equation}
\alpha_q=\left[1+\frac{1}{2}\left|\frac{\Gamma_{12}^{(q)}}{M_{12}^{(q)}}
\right|\sin\delta
\Theta_{M/\Gamma}^{(q)}\right]e^{-i\left(\Theta_{M_{12}}^{(q)}+n'\pi\right)}.
\end{equation}
The deviation of $|\alpha_q|$ from 1 describes CP violation 
in $B^0_q$--$\overline{B^0_q}$ oscillations, and can be probed through
``wrong-charge'' lepton asymmetries:
\begin{equation}\label{wrongcharge}
{\cal A}^{(q)}_{\mbox{{\scriptsize SL}}}\equiv
\frac{\Gamma(B^0_q(t)\to \ell^-\overline{\nu} X)-
\Gamma(\overline{B^0_q}(t)\to\ell^+\nu X)}{\Gamma(B^0_q(t)\to 
\ell^-\overline{\nu} X)+\Gamma(\overline{B^0_q}(t)\to \ell^+\nu X)}
=\frac{|\alpha_q|^4-1}{|\alpha_q|^4+1}\approx
\left|\frac{\Gamma_{12}^{(q)}}{M_{12}^{(q)}}\right|
\sin\delta\Theta^{(q)}_{M/\Gamma}.
\end{equation}
Note that the time dependences cancel in (\ref{wrongcharge}). Because of 
$|\Gamma_{12}^{(q)}|/|M_{12}^{(q)}|\propto m_b^2/m_t^2$ and 
$\sin\delta\Theta^{(q)}_{M/\Gamma}\propto m_c^2/m_b^2$,
the asymmetry ${\cal A}^{(q)}_{\mbox{{\scriptsize SL}}}$ is suppressed by 
a factor $m_c^2/m_t^2={\cal O}(10^{-4})$, and is hence very small in the 
Standard Model. Consequently, it represents an interesting probe to
search for new physics. A recent analysis of the CLEO collaboration 
\cite{cleo-eps-B} yields
\begin{equation}
{\cal A}^{(d)}_{\mbox{{\scriptsize SL}}}/4=-0.0035\pm0.0103\pm0.0015.
\end{equation}

\subsection{Mixing Parameters and Decay Rates}
The time evolution of initially, i.e.\ at time $t=0$, pure $B^0_q$- 
and $\overline{B^0_q}$-meson states is given as follows:
\begin{eqnarray}
\left|B^0_q(t)\right\rangle&=&f_+^{(q)}(t)\left|B^{0}_q\right\rangle
+\alpha_qf_-^{(q)}(t)\left|\overline{B^{0}_q}\right\rangle\label{B0-evol}\\
\left|\overline{B^0_q}(t)\right\rangle&=&\frac{1}{\alpha_q}f_-^{(q)}(t)
\left|B^{0}_q\right\rangle+f_+^{(q)}(t)\left|\overline{B^{0}_q}\right
\rangle,
\end{eqnarray}
where
\begin{equation}
f_{\pm}^{(q)}(t)=\frac{1}{2}\left[e^{-i\lambda_+^{(q)}t}\pm
e^{-i\lambda_-^{(q)}t}\right].
\end{equation}
These time-dependent state vectors allow the calculation of the 
corresponding transition rates. To this end, it is useful to introduce
\begin{equation}
\left|g^{(q)}_{\pm}(t)\right|^2=\frac{1}{4}\left[e^{-\Gamma_{\rm L}^{(q)}t}+
e^{-\Gamma_{\rm H}^{(q)}t}\pm2\,e^{-\Gamma_q t}\cos(\Delta M_qt)\right]
\end{equation}
\begin{equation}
g_-^{(q)}(t)\,g_+^{(q)}(t)^\ast=\frac{1}{4}\left[e^{-\Gamma_{\rm L}^{(q)}t}-
e^{-\Gamma_{\rm H}^{(q)}t}+2\,i\,e^{-\Gamma_q t}\sin(\Delta M_qt)\right],
\end{equation}
and
\begin{equation}\label{xi-def}
\xi_f^{(q)}=e^{-i\Theta_{M_{12}}^{(q)}}
\frac{A(\overline{B_q^0}\to f)}{A(B_q^0\to f)},\quad
\xi_{\overline{f}}^{(q)}=e^{-i\Theta_{M_{12}}^{(q)}}
\frac{A(\overline{B_q^0}\to \overline{f})}{A(B_q^0\to \overline{f})},
\end{equation}
where
\begin{equation}\label{theta-def}
\Theta_{M_{12}}^{(q)}=\pi+2\,\mbox{arg}\left(V_{tq}^\ast V_{tb}\right)-
\phi_{\mbox{{\scriptsize CP}}}(B_q)
\end{equation}
can be read off from (\ref{M12-expr}). Whereas $\Theta_{M_{12}}^{(q)}$ 
depends on the chosen CKM and CP phase conventions, $\xi_f^{(q)}$ and 
$\xi_{\overline{f}}^{(q)}$ are {\it convention-independent} observables.

In (\ref{theta-def}), it has been taken into account that $S_0(x_t)>0$, 
and it has been assumed implicitly that the bag parameter $\hat B_{B_d}$ 
is positive. As emphasized in \cite{GKN}, for $\hat B_{B_d}<0$, 
$\Theta_{M_{12}}^{(q)}$ given in (\ref{theta-def}) would be shifted by 
$\pi$. However, this case appears to be very unlikely. Indeed, all existing 
non-perturbative methods give $\hat B_{B_d}>0$, which we shall also assume 
in this review. A similar comment applies to $\hat B_K$. On the other hand, 
in the case of models with minimal flavour violation, the Inami--Lim function 
$S_0(x_t)$ is replaced by a new parameter $F_{tt}$, which needs not be 
positive. In this case, the shift of (\ref{theta-def}) by $\pi$ has to be 
taken into account \cite{BF-MFV}, and leads to subtleties, as discussed in 
Section~\ref{sec:MFV}.

The $g_{\pm}^{(q)}(t)$ are related to the $f_{\pm}^{(q)}(t)$. However, 
whereas the latter functions depend on $n'$, the $g_{\pm}^{(q)}(t)$ do 
not depend on this parameter. The $n'$-dependence is cancelled by 
introducing the {\it positive} mass difference 
\begin{equation}
\Delta M_q\equiv M_{\rm H}^{(q)}-M_{\rm L}^{(q)}=2\left|M_{12}^{(q)}\right|>0
\end{equation}
of the mass eigenstates $|B_q^{\rm H}\rangle$ (``heavy'') and
$|B_q^{\rm L}\rangle$ (``light''). The quantities $\Gamma_{\rm H}^{(q)}$ and 
$\Gamma_{\rm L}^{(q)}$ denote the corresponding decay widths. 
Their difference can be expressed as
\begin{equation}\label{Expr-DGamma}
\Delta\Gamma_q\equiv\Gamma_{\rm H}^{(q)}-\Gamma_{\rm L}^{(q)}=
\frac{4\mbox{\,Re}\left[M_{12}^{(q)}\Gamma_{12}^{(q)\ast}\right]}{\Delta M_q},
\end{equation}
whereas their average is given by
\begin{equation}
\Gamma_q\equiv\frac{\Gamma^{(q)}_{\rm H}+\Gamma^{(q)}_{\rm L}}{2}=
\Gamma^{(q)}_0.
\end{equation}
There is the following interesting relation:
\begin{equation}\label{DG-MP-rel}
\frac{\Delta\Gamma_q}{\Gamma_q}\approx-\frac{3\pi}{2S_0(x_t)}\frac{m_b^2}
{M_W^2}\,x_q={\cal O}(10^{-2})\times x_q,
\end{equation}
where 
\begin{equation}\label{mix-par}
x_q\equiv\frac{\Delta M_q}{\Gamma_q}=\left\{\begin{array}{cc}
0.75\pm0.02&(q=d)\\
{\cal O}(20)& (q=s)
\end{array}\right.
\end{equation}
denotes the $B^0_q$--$\overline{B^0_q}$ ``mixing parameter''. Consequently,
there may be a sizeable width difference in the $B_s$ system, whereas 
$\Delta\Gamma_d$ is expected to be negligibly small. We shall come back to 
$\Delta\Gamma_s$ in Section~\ref{sec:Bs}.

Combining the formulae listed above, we arrive at the following
transition rates for decays of initially, i.e.\ at $t=0$, present $B^0_q$-
or $\overline{B^0_q}$-mesons:
\begin{equation}\label{rates}
\Gamma(\stackrel{{\mbox{\tiny (---)}}}{B^0_q}(t)\to f)
=\left[\left|g_\mp^{(q)}(t)\right|^2+\left|\xi_f^{(q)}
\right|^2\left|g_\pm^{(q)}(t)\right|^2-
2\mbox{\,Re}\left\{\xi_f^{(q)}
g_\pm^{(q)}(t)g_\mp^{(q)}(t)^\ast\right\}
\right]\tilde\Gamma_f,
\end{equation}
where the time-independent rate $\tilde\Gamma_f$ corresponds to the 
``unevolved'' decay amplitude $A(B^0_q\to f)$, which can be calculated by 
performing the usual phase-space integrations. The rates into the
CP-conjugate final state $\overline{f}$ can be
obtained straightforwardly from (\ref{rates}) through the substitutions
\begin{equation}
\tilde\Gamma_f  \,\,\,\to\,\,\, 
\tilde\Gamma_{\overline{f}},
\quad\,\,\xi_f^{(q)} \,\,\,\to\,\,\, 
\xi_{\overline{f}}^{(q)}.
\end{equation}

\boldmath
\subsection{Determination of $R_t$ from $\Delta M_d$}\label{subsec:Rt-Md}
\unboldmath
As we have noted in our discussion of the constraints on the apex of
the unitarity triangle in the $\overline{\rho}$--$\overline{\eta}$
plane (see Subsection~\ref{subsec:CKM-fits}), the mass difference of 
the eigenstates of the $B_d$ system allows us to determine $R_t$. 
This feature is nicely described by \cite{Brev01,UUT,BB-Bound,BF-MFV}
\begin{equation}\label{RT}
R_t= 1.10\times \frac{ R_0}{A}\frac{1}{\sqrt{\left|S_0(x_{t})\right|}} 
\quad\mbox{with}\quad R_0\equiv\sqrt{\frac{\Delta M_d}{0.50/{\rm ps}}}
\left[\frac{230\,\mbox{MeV}}{\sqrt{\hat B_d} f_{B_d}}\right]
\sqrt{\frac{0.55}{\eta_B}},
\end{equation}
where we have kept the Inami--Lim function $S_0(x_{t})\approx2.38$
explicitly, which will be useful for the discussion of models with
minimal flavour violation in Section~\ref{sec:MFV}. The experimental
range for $\Delta M_d$ is given by \cite{LEPBOSC}
\begin{equation}\label{new-DMd-range}
\Delta M_d=(0.487\pm0.009)\,{\rm ps}^{-1}.
\end{equation}
Alternatively, $R_t$ may be fixed through the ratio $\Delta M_d/\Delta M_s$ 
with the help of $SU(3)$ flavour-symmetry arguments. Concerning
theoretical uncertainties, this approach, which will be discussed together
with the $B_s$ system in Section~\ref{sec:Bs}, is much cleaner.

\subsection{CP-violating Asymmetries}\label{subsec:CPasym}
A particularly simple and interesting situation arises if we restrict 
ourselves to decays of neutral $B_q$-mesons into CP self-conjugate 
final states $|f\rangle$, \mbox{satisfying} the relation 
\begin{equation}
({\cal CP})|f\rangle=\pm\,|f\rangle. 
\end{equation}
Consequently, we have $\xi_f^{(q)}=\xi_{\overline{f}}^{(q)}$ in this
case (see (\ref{xi-def})). Using (\ref{rates}), the corresponding 
time-dependent CP asymmetry can be expressed as follows:
\begin{eqnarray}
\lefteqn{a_{\rm CP}(t)\equiv\frac{\Gamma(B^0_q(t)\to f)-
\Gamma(\overline{B^0_q}(t)\to f)}{\Gamma(B^0_q(t)\to f)+
\Gamma(\overline{B^0_q}(t)\to f)}}\label{ee6}\\
&&=\left[\frac{{\cal A}_{\rm CP}^{\rm dir}(B_q\to f)\,\cos(\Delta M_q t)+
{\cal A}_{\rm CP}^{\rm mix}(B_q\to f)\,\sin(\Delta 
M_q t)}{\cosh(\Delta\Gamma_qt/2)-{\cal A}_{\rm 
\Delta\Gamma}(B_q\to f)\,\sinh(\Delta\Gamma_qt/2)}\right].\nonumber
\end{eqnarray}
In this expression, we have separated the ``direct'' from the 
``mixing-induced'' CP-violating contributions, which are described by
\begin{equation}\label{ee7}
{\cal A}^{\mbox{{\scriptsize dir}}}_{\mbox{{\scriptsize CP}}}(B_q\to f)\equiv
\frac{1-\bigl|\xi_f^{(q)}\bigr|^2}{1+\bigl|\xi_f^{(q)}\bigr|^2}\quad
\mbox{and}\quad
{\cal A}^{\mbox{{\scriptsize mix}}}_{\mbox{{\scriptsize
CP}}}(B_q\to f)\equiv\frac{2\,\mbox{Im}\,\xi^{(q)}_f}{1+
\bigl|\xi^{(q)}_f\bigr|^2}\,,
\end{equation}
respectively. The terminology ``direct CP violation'' refers 
to CP-violating effects, which arise directly at the decay-amplitude level
and are due to interference between different CKM amplitudes; it is the 
same kind of CP violation, which is probed in the neutral kaon system by 
Re$(\varepsilon'/\varepsilon)$.
On the other hand, ``mixing-induced CP violation'' originates
from interference effects between $B_q^0$--$\overline{B_q^0}$ mixing 
and decay processes. The width difference $\Delta\Gamma_q$, which may 
be sizeable in the $B_s$ system, provides another observable 
\begin{equation}\label{ADGam}
{\cal A}_{\rm \Delta\Gamma}(B_q\to f)\equiv
\frac{2\,\mbox{Re}\,\xi^{(q)}_f}{1+\bigl|\xi^{(q)}_f
\bigr|^2},
\end{equation}
which is, however, not independent from ${\cal A}^{\mbox{{\scriptsize 
dir}}}_{\mbox{{\scriptsize CP}}}(B_q\to f)$ and 
${\cal A}^{\mbox{{\scriptsize mix}}}_{\mbox{{\scriptsize CP}}}(B_q\to f)$,
statisfying the following relation:
\begin{equation}\label{Obs-rel}
\Bigl[{\cal A}_{\rm CP}^{\rm dir}(B_q\to f)\Bigr]^2+
\Bigl[{\cal A}_{\rm CP}^{\rm mix}(B_q\to f)\Bigr]^2+
\Bigl[{\cal A}_{\Delta\Gamma}(B_q\to f)\Bigr]^2=1.
\end{equation}

In order to calculate the observable $\xi_f^{(q)}$, which contains 
essentially all the information needed to evaluate the CP asymmetry
(\ref{ee6}), we employ the low-energy effective Hamiltonian (\ref{e4}),
yielding the transition amplitude
\begin{eqnarray}
\lefteqn{A\left(\overline{B^0_q}\to f\right)=\Bigl\langle f\Bigl\vert
{\cal H}_{\mbox{{\scriptsize eff}}}(\Delta B=-1)\Bigr\vert\overline{B^0_q}
\Bigr\rangle=}\\
&&\Biggl\langle f\left|
\frac{G_{\mbox{{\scriptsize F}}}}{\sqrt{2}}\left[
\sum\limits_{j=u,c}V_{jr}^\ast V_{jb}\left\{\sum\limits_{k=1}^2
C_{k}(\mu)\,Q_{k}^{jr}(\mu)
+\sum\limits_{k=3}^{10}C_{k}(\mu)\,Q_{k}^r(\mu)\right\}\right]\right|
\overline{B^0_q}\Biggr\rangle,~~~\mbox{}\nonumber
\end{eqnarray}
where $r\in\{d,s\}$ distinguishes between $b\to d$ and $b\to s$ transitions.
On the other hand, we have also 
\begin{eqnarray}
\lefteqn{A\left(B^0_q\to f\right)=\left\langle f\left|
{\cal H}_{\mbox{{\scriptsize 
eff}}}(\Delta B=-1)^\dagger\right|B^0_q\right\rangle=}\\
&&\hspace*{-0.5truecm}\Biggl\langle f
\left|\frac{G_{\mbox{{\scriptsize F}}}}{\sqrt{2}}
\left[\sum\limits_{j=u,c}V_{jr}V_{jb}^\ast \left\{\sum\limits_{k=1}^2
C_{k}(\mu)\,Q_{k}^{jr\dagger}(\mu)+\sum\limits_{k=3}^{10}
C_{k}(\mu)\,Q_k^{r\dagger}(\mu)\right\}\right]\right|B^0_q
\Biggr\rangle.~~~\mbox{}\nonumber
\end{eqnarray}
If we perform appropriate CP transformations in this expression, i.e.\ 
insert the operator $({\cal CP})^\dagger({\cal CP})=\hat 1$ both 
after $\langle f|$ and in front of $|B^0_q\rangle$, we obtain
\begin{eqnarray}
\lefteqn{A\left(B^0_q\to f\right)=\pm e^{i\phi_{\mbox{{\scriptsize CP}}}
(B_q)}\times}\\
&&\Biggl\langle f\left|
\frac{G_{\mbox{{\scriptsize F}}}}{\sqrt{2}}\left[\sum\limits_{j=u,c}
V_{jr}V_{jb}^\ast\left\{\sum\limits_{k=1}^2
C_{k}(\mu)\,Q_{k}^{jr}(\mu)+\sum\limits_{k=3}^{10}
C_{k}(\mu)\,Q_{k}^r(\mu)\right\}\right]\right|\overline{B^0_q}
\Biggr\rangle,\nonumber
\end{eqnarray}
where we have applied the relation 
\begin{equation}\label{CP-op-trafo}
({\cal CP})\left(Q_k^{jr}\right)^\dagger({\cal CP})^\dagger=Q_k^{jr}, 
\end{equation}
and have furthermore taken into account (\ref{CP-def}). 
Using now (\ref{xi-def}) and (\ref{theta-def}), we finally arrive at
\begin{equation}\label{xi-expr}
\xi_f^{(q)}=\mp\,e^{-i\phi_q}\,
\left[\frac{\sum\limits_{j=u,c}V_{jr}^\ast V_{jb}\bigl\langle 
f\bigl|{\cal Q}^{jr}\bigr|\overline{B^0_q}\bigr\rangle}{\sum\limits_{j=u,c}
V_{jr}V_{jb}^\ast\bigl\langle f\bigl|{\cal Q}^{jr}
\bigr|\overline{B^0_q}\bigr\rangle}\right],
\end{equation}
where 
\begin{equation}
{\cal Q}^{jr}\equiv\sum\limits_{k=1}^2C_k(\mu)\,Q_k^{jr}+
\sum\limits_{k=3}^{10}C_k(\mu)\,Q_k^{r},
\end{equation}
and where
\begin{equation}\label{phiq}
\phi_q\equiv 2\,\mbox{arg} (V_{tq}^\ast V_{tb})=\left\{\begin{array}{cr}
+2\beta&\mbox{($q=d$)}\\
-2\delta\gamma&\mbox{($q=s$)}\end{array}\right.
\end{equation}
is related to the weak $B_q^0$--$\overline{B_q^0}$ mixing phase. 
For the interpretaion of $\beta$ and $\delta\gamma$ in terms of the
unitarity triangles of the CKM matrix, see Fig.~\ref{fig:UT}.
Note that the phase-convention-dependent quantity 
$\phi_{\mbox{{\scriptsize CP}}}(B_q)$ cancels in (\ref{xi-expr}). 

In general, the observable $\xi_f^{(q)}$ suffers from large hadronic 
uncertainties, which are introduced by the hadronic matrix elements in 
(\ref{xi-expr}). However, if the decay $B_q\to f$ is dominated by a 
single CKM amplitude, i.e.
\begin{equation}
A(B^0_q\to f)=e^{-i\phi_f/2}\left(e^{i\delta_f}|M_f|\right),
\end{equation}
the strong matrix element $e^{i\delta_f}|M_f|$ with the CP-conserving
strong phase $\delta_f$ cancels, and $\xi_f^{(q)}$ 
takes the simple form
\begin{equation}\label{ee10}
\xi_f^{(q)}=\mp\exp\left[-i\left(\phi_q-\phi_f\right)\right].
\end{equation}
If the $V_{jr}^\ast V_{jb}$ amplitude plays the dominant r\^ole in 
$\overline{B^0_q}\to f$, we have
\begin{equation}\label{phiD}
\phi_f=2\,\mbox{arg}(V_{jr}^\ast V_{jb})=\left\{\begin{array}{cc}
-2\gamma&\mbox{($j=u$)}\\
0&\,\mbox{($j=c$).}
\end{array}\right.
\end{equation}
In (\ref{phiq}) and (\ref{phiD}), we have employed (\ref{CKM-UT-ANGLES}), 
corresponding to the phase convention chosen in the standard and the
generalized Wolfenstein parametrizations of the CKM matrix. However, the 
results for the observable $\xi_f^{(q)}$ arising from (\ref{ee10}) do of 
course not depend on the choice of the phase convention of the CKM matrix.

Since $B^0_q$--$\overline{B^0_q}$ mixing plays an important r\^ole for
new physics to manifest itself in mixing-induced CP asymmetries of neutral 
$B_q$ decays, let us leave the framework of the Standard Model for a 
moment in the next Subsection, following closely \cite{FM-BpsiK}.

\boldmath
\subsection{Impact of Physics Beyond the Standard Model}\label{subsec:NP-mix}
\unboldmath
The generic way of introducing physics beyond the Standard Model into
analyses of non-leptonic $B$ decays is to use the language of effective 
field theory and to write down all possible dim-6 operators. Of course, 
this has been known already for a long time and lists of the dim-6 operators 
involving all the Standard-Model particles have been published in the 
literature \cite{dim-6}. After having introduced these additional operators, 
we have to construct the generalization of the relevant Standard-Model 
effective Hamiltonian at the scale of the $b$ quark, where we encounter
again dim-6 operators, with Wilson coefficients containing now a 
Standard-Model contribution, and a possible piece of new physics.

The problem with this generic point of view is that the operator basis
is enlarged to such an extent that this general approach becomes almost
useless for phenomenological applications. However, we are dealing with 
non-leptonic decays, in which we are, because of our ignorance of the 
hadronic matrix elements, sensitive neither to the helicity structure 
of the operators nor to their colour structure. The only information 
that is relevant in this case is the flavour structure, and hence we 
introduce the notation
\begin{equation}
\left[(\overline{q}_3 q_2)(\overline{q}_1 Q)\right] = \sum\,
[\mbox{Wilson coeff.}]\times
[\mbox{dim-6 operator mediating $Q \to q_1 \overline{q}_2 q_3$}].
\end{equation}
This sum is renormalization-group invariant, and involves, at  
the scale of the $b$ quark, Standard-Model as well as possible 
non-Standard-Model contributions. In particular, it allows us to estimate 
the relative size of a possible new-physics contribution.

This language can also be applied to the $\Delta B = \pm 2$ operators 
mediating $B^0_d$--$\overline{B^0_d}$ mixing, yielding
\begin{equation}\label{Heff-mix}
{\cal H}_{\rm eff}^{(d)}(\Delta B = + 2) = G_d 
\left[(\overline{b}d)(\overline{b}d) \right]
\end{equation}
as the relevant flavour structure. Within the Standard Model, 
(\ref{Heff-mix}) originates from the box diagrams shown 
in Fig.~\ref{fig:boxes} for $q=d$, which are strongly suppressed by 
the CKM factor $(V_{td} V_{tb}^\ast)^2$, as well as by a loop factor
\begin{equation} \label{loop}
\frac{g^2_2}{64 \pi^2} =
\frac{G_{\rm F} M_W^2}{\sqrt{128} \pi^2} \approx 1 \times 10^{-3},
\end{equation}
making the Standard-Model contribution very small, of the order of
\begin{equation}
G_{\rm SM}^{(d)} = \frac{G_{\rm F}}{\sqrt{2}} \left(\frac{G_{\rm F} 
M_W^2}{\sqrt{128} \pi^2}\right)(V_{td} V_{tb}^\ast)^2.
\end{equation}
Here we did not include the Inami--Lim function $S_0(x_t)$ and
the perturbative QCD corrections for simplicity (see 
Subsection~\ref{subsec:SG}), as we are now only interested in order of 
magnitude estimates. Because of the strongly suppressed Standard-Model 
piece, a new-physics contribution may well be of similar size. If $\Lambda$ 
denotes the characteristic scale of physics beyond the Standard Model, 
we have
\begin{equation}
G_{\rm NP}^{(d)} = \frac{G_{\rm F}}{\sqrt{2}}\left(\frac{G_{\rm F} 
M_W^2}{\sqrt{128} \pi^2}\right)\frac{M_W^2}{\Lambda^2} e^{-i2\psi_d},
\end{equation}
where $\psi_d$ is a possible weak phase, which is induced by the 
new-physics contribution. Finally, we arrive at
\begin{equation}\label{PhiM1}
G_d = \frac{G_{\rm F}}{\sqrt{2}}\left(\frac{G_{\rm F} 
M_W^2}{\sqrt{128} \pi^2}\right)
\left[(V_{td} V_{tb}^\ast)^2+\frac{M_W^2}{\Lambda^2}  
e^{-i2\psi_d}\right]\equiv
|R_d|e^{-i\phi^{\rm NP}_d},
\end{equation}
where the weak phase $\phi^{\rm NP}_d$ is the generalization of
(\ref{phiq}), entering the mixing-induced CP asymmetries. Using 
(\ref{CKM-UT-ANGLES}), we obtain
\begin{equation}\label{PhiM2}
\tan \phi^{\rm NP}_d =
\frac{\sin2 \beta + \varrho^2_d \sin 2\psi_d}
{\cos 2 \beta + \varrho^2_d\cos2\psi_d} 
\end{equation}
with
\begin{equation}\label{rho-NP-def}
\varrho_d=\left(\frac{1}{\lambda^3 A R_t}\right)\left(\frac{M_W}{\Lambda}
\right).
\end{equation}
Since $\varrho_d$ can be of order one even for large $\Lambda$, there can 
be a large phase shift in the mixing phase. If we assume, for example, 
$A R_t=1$, this term equals 1 for a new-physics scale of 
$\Lambda\sim 8$\,TeV. Such contributions affect of course not only the 
CP-violating phase $\phi^{\rm NP}_d$, but also the ``strength'' $|R_d|$ of 
$B^0_d$--$\overline{B^0_d}$ mixing, which would manifest itself as an 
inconsistency in the usual ``standard analysis'' of the unitarity 
triangle discussed in Subsection~\ref{subsec:CKM-fits}. Similar 
considerations can of course also be made for $B^0_s$--$\overline{B^0_s}$ 
mixing, which will be addressed in Subsection~\ref{subsec:Bs-NP}.
 
The formalism developed above has important applications and allows us 
to discuss key modes for the $B$-factories, which will be our next 
topic.

\begin{figure}[b]
\begin{center}
\leavevmode
\epsfysize=4.4truecm 
\epsffile{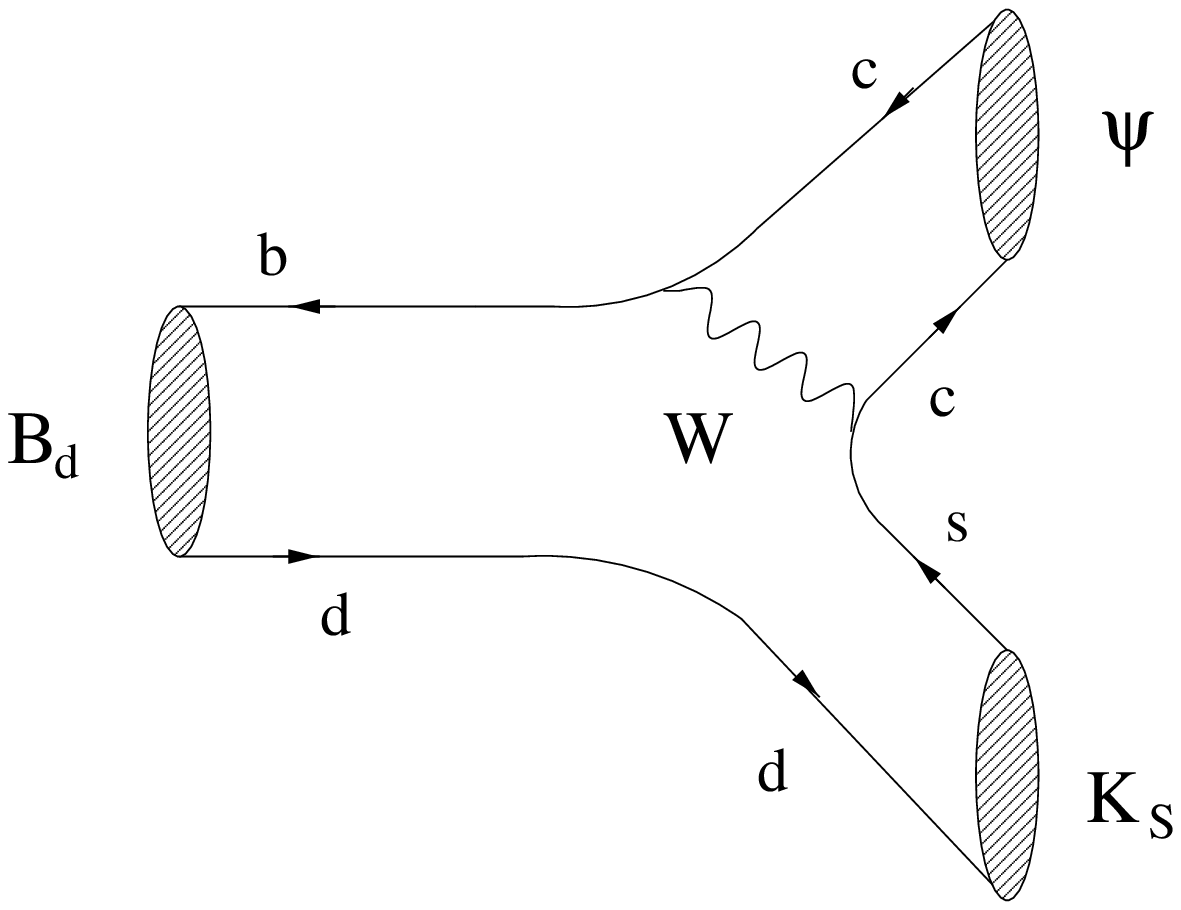} \hspace*{1truecm}
\epsfysize=4.4truecm 
\epsffile{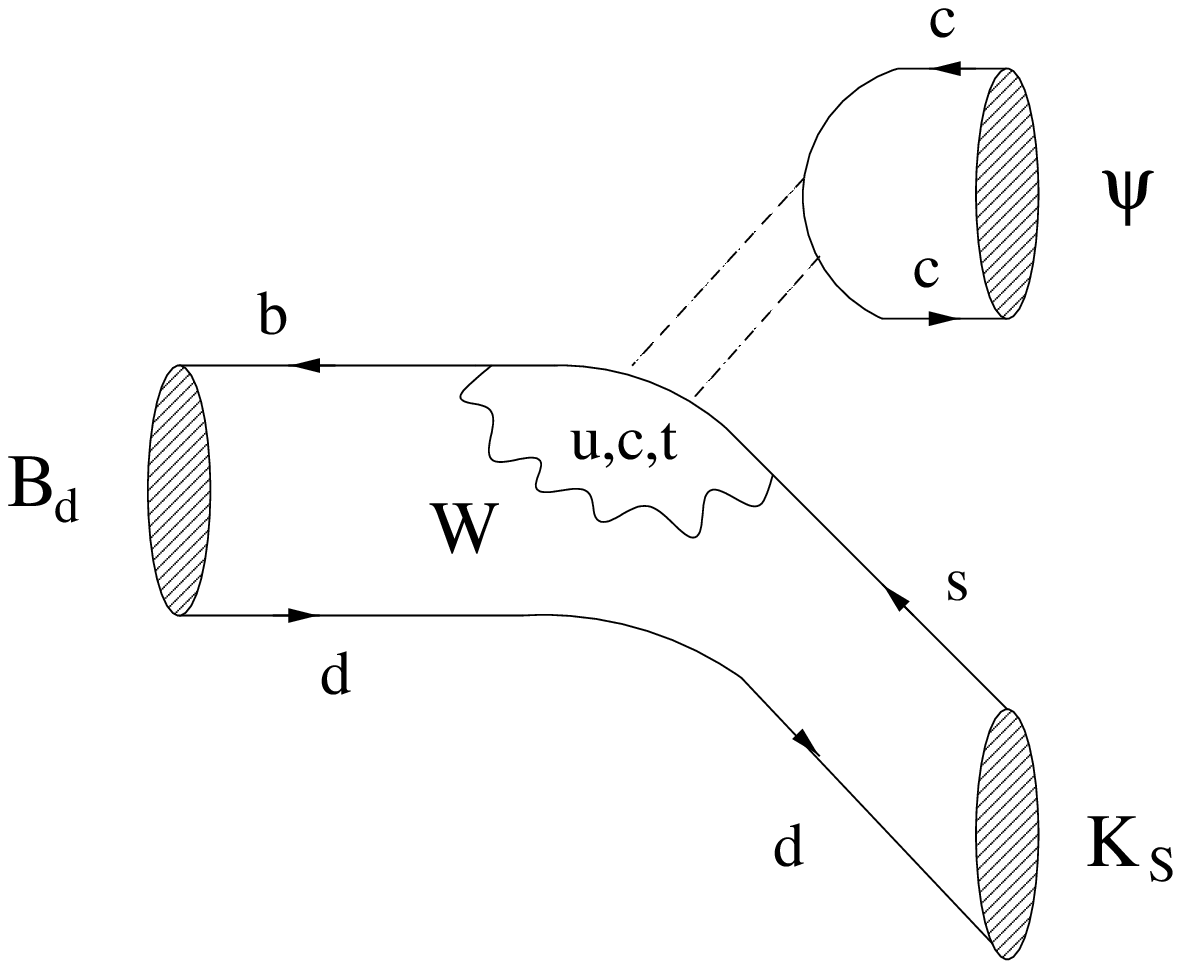}
\end{center}
\vspace*{-0.3truecm}
\caption{Feynman diagrams contributing to $B_d^0\to J/\psi K_{\rm S}$.
The dashed lines in the penguin topology represent a colour-singlet 
exchange.}\label{fig:BdPsiKS}
\end{figure}

\boldmath
\section{Important Benchmark Modes for the $B$-Factories}\label{sec:benchmark}
\unboldmath
\setcounter{equation}{0}
\boldmath
\subsection{The $B\to J/\psi K$ System}\label{subsec:BpsiK}
\unboldmath
\setcounter{equation}{0}
The most important application of the formalism discussed in 
Subsection~\ref{subsec:CPasym} is the extraction of $\beta$ from 
CP-violating effects in the ``gold-plated'' mode 
$B_d\to J/\psi K_{\rm S}$ \cite{bisa}.
\boldmath
\subsubsection{Extracting $\beta$ from 
$B_d\to J/\psi K_{\rm S}$}\label{ssec:beta-extr}
\unboldmath
The decay $B^0_d\to J/\psi K_{\rm S}$ is a transition into a CP eigenstate 
with eigenvalue $-1$ (note that the $J/\psi K_{\rm S}$ system is created 
in a $P$-wave state), and originates from 
$\overline{b}\to \overline{c} c \overline{s}$ quark-level processes. 
As can be seen in Fig.~\ref{fig:BdPsiKS}, we have 
to deal both with tree-diagram-like and with penguin topologies. The 
corresponding amplitude can be written as \cite{RF-BdsPsiK}
\begin{equation}\label{Bd-ampl1}
A(B_d^0\to J/\psi K_{\rm S})=\lambda_c^{(s)}\left(A_{\rm CC}^{c'}+
A_{\rm pen}^{c'}\right)+\lambda_u^{(s)}A_{\rm pen}^{u'}
+\lambda_t^{(s)}A_{\rm pen}^{t'}\,,
\end{equation}
where $A_{\rm CC}^{c'}$ denotes the current--current (CC) contributions,
i.e.\ the ``tree'' processes shown in Fig.\ \ref{fig:BdPsiKS}, and 
the strong amplitudes $A_{\rm pen}^{q'}$ describe the contributions from 
penguin topologies with internal $q$ quarks ($q\in\{u,c,t\})$. These 
penguin amplitudes take into account both QCD and EW penguin contributions. 
The primes in (\ref{Bd-ampl1}) remind us that we are dealing with a 
$\overline{b}\to\overline{s}$ transition, and the
\begin{equation}\label{lamqs-def}
\lambda_q^{(s)}\equiv V_{qs}V_{qb}^\ast
\end{equation}
are CKM factors. If we employ the unitarity of the CKM matrix to eliminate 
$\lambda_t^{(s)}$ (see (\ref{CKM-UT-REL})), and make use of the generalized 
Wolfenstein parametrization (\ref{NLO-wolf}), we obtain 
\begin{equation}\label{Bd-ampl2}
A(B_d^0\to J/\psi K_{\rm S})=\left(1-\frac{\lambda^2}{2}\right){\cal A'}
\left[1+\left(\frac{\lambda^2}{1-\lambda^2}\right)a'e^{i\theta'}e^{i\gamma}
\right],
\end{equation}
where
\begin{equation}\label{Aap-def}
{\cal A'}\equiv\lambda^2A\left(A_{\rm CC}^{c'}+A_{\rm pen}^{ct'}\right)
\end{equation}
with $A_{\rm pen}^{ct'}\equiv A_{\rm pen}^{c'}-A_{\rm pen}^{t'}$, and
\begin{equation}\label{ap-def}
a'e^{i\theta'}\equiv R_b\left(\frac{A_{\rm pen}^{ut'}}{A_{\rm CC}^{c'}+
A_{\rm pen}^{ct'}}\right).
\end{equation}
The quantity $A_{\rm pen}^{ut'}$ is defined in analogy to $A_{\rm pen}^{ct'}$,
and $A$ and $R_b$ are the usual CKM factors, with present experimental
ranges given in (\ref{A-range}) and (\ref{Rb-range}), respectively.

Because of various reasons, it is very difficult to calculate the 
``penguin'' parameter $a'e^{i\theta'}$, introducing $e^{i\gamma}$ into 
the $B_d^0\to J/\psi K_{\rm S}$ amplitude, reliably: the QCD penguins 
contributing to 
$B_d^0\to J/\psi K_{\rm S}$ require a colour-singlet exchange, as indicated 
in Fig.~\ref{fig:BdPsiKS} through the dashed lines, and are 
``Zweig-suppressed''. Such a comment does not apply to the EW penguins, 
which contribute in colour-allowed form. On the other hand, the 
current--current amplitude $A_{\rm CC}^{c'}$ is due to colour-suppressed 
topologies, so that the ratio 
$A_{\rm pen}^{ut'}/(A_{\rm CC}^{c'}+A_{\rm pen}^{ct'})$, which
governs $a'e^{i\theta'}$, may well be sizeable; a plausible estimate
is $a'={\cal O}(0.2)$. Fortunately, this parameter -- and therefore also 
$e^{i\gamma}$ -- enters in (\ref{Bd-ampl2}) in a doubly Cabibbo-suppressed 
way. Consequently, we have to a very good approximation 
$\phi_{\psi K_{\rm S}}=0$, and obtain with the help of (\ref{ee7}) and 
(\ref{ee10})
\begin{equation}\label{e12}
a_{\psi K_{\rm S}}\equiv
-{\cal A}^{\mbox{{\scriptsize mix}}}_{\mbox{{\scriptsize
CP}}}(B_d\to J/\psi K_{\mbox{{\scriptsize S}}})
=-\sin[-(\phi_d-0)]=\sin 2\beta.
\end{equation}
Hence $B_d\to J/\psi K_{\rm S}$ is referred to as the ``gold-plated'' 
mode to determine $\beta$ \cite{bisa}. 

At this point, we have to discuss a subtlety, which arises in $B_q\to f$ 
decays into a final CP eigenstate $|f\rangle$ involving a $K_{\rm S}$-meson,
such as $B_{d(s)}\to J/\psi K_{\rm S}$ or $B_d\to \phi K_{\rm S}$. In this 
case, we have also to take into account the weak $K^0$--$\overline{K^0}$ 
mixing phase $\phi_K=2\,\mbox{arg}(V_{us}^\ast V_{ud})$, modifying the 
convention-independent combination of phases in (\ref{ee10}) as follows:
\begin{equation}
\phi_q+\phi_K-\phi_f.
\end{equation}
However, for the generalized Wolfenstein parametrization (\ref{NLO-wolf}) 
applied in (\ref{phiq}) and (\ref{phiD}), $\phi_K$ is negligibly small
and does not show up explicitly in the mixing-induced CP asymmetries.
Owing to the small value of the CP-violating parameter $\varepsilon$ 
of the neutral kaon system, $\phi_K$ can only affect the mixing-induced
CP asymmetries in very contrived models of new physics \cite{GKN,NiSi}.

Concerning the measurement of $\sin 2\beta$ through (\ref{e12}),
there were already important first steps by the OPAL, CDF and ALEPH
collaborations:
\begin{equation}\label{s2b-old}
\sin2\beta=\left\{\begin{array}{ll}
3.2^{+1.8}_{-2.0}\pm0.5&\mbox{(OPAL \cite{opal})}\\
0.79^{+0.41}_{-0.44}&\mbox{(CDF \cite{cdf})}\\
0.84^{+0.82}_{-1.04}\pm0.16&\mbox{(ALEPH \cite{aleph}).}
\end{array}\right.
\end{equation}
In the summer of 2001, the asymmetric $e^+e^-$ $B$-factories could establish
CP violation in the $B$ system, with the following results for $\sin2\beta$:
\begin{equation}\label{B-factory}
\sin2\beta=\left\{\begin{array}{ll}
0.59\pm 0.14 \pm0.05&\,\mbox{(BaBar \cite{babar-CP-obs})}\\
0.99\pm 0.14 \pm 0.06&\,\mbox{(Belle \cite{belle-CP-obs}).}
\end{array}\right.
\end{equation}
Taking also into account the previous CDF and ALEPH results gives an
average of 
\begin{equation}\label{s2b-world}
\sin2\beta=0.79 \pm 0.10.
\end{equation}

In comparison with $\sin2\beta=0.34\pm0.20\pm0.05$ \cite{babar}
and $0.58^{+0.32+0.09}_{-0.34-0.10}$ \cite{belle} reported by the 
BaBar and Belle collaborations in the spring of 2001, respectively, 
the central values have now moved up considerably. The first results for 
$\sin2\beta$ announced by these experiments in the summer of 2000, 
$0.12\pm0.37\pm0.09$ \cite{BABAR1} and $0.45^{+0.43+0.07}_{-0.44-0.09}$ 
\cite{BELLE1}, gave even smaller central values. Despite the large 
experimental uncertainties, these numbers led already to some excitement 
in the $B$-physics community 
\cite{BB-Bound,BF-MFV,low-sin,BePe,Ali-Lunghi,BCRS}, as they would
have been in disagreement with the ``standard analysis'' of the unitarity 
triangle discussed in Subsection~\ref{subsec:CKM-fits}, yielding
\begin{equation}\label{SM-sin}
0.5\lsim\sin2\beta\lsim0.9.
\end{equation}
In view of the recent Belle result \cite{belle-CP-obs}, the upper bounds
given in (\ref{Rb-bounds}) and (\ref{ubound}) may become important for the
search for new physics.

Since the BaBar and Belle numbers in (\ref{B-factory}) are not fully
consistent with each other, the measurement of $\sin2\beta$ will continue 
to be a very exciting topic. Moreover, it should be noted that the 
$B$-factory results given in (\ref{B-factory}) actually correspond to an 
average over various channels, including $B_d\to J/\psi K_{\rm S}$, 
$\psi(2S) K_{\rm S}$, $\chi_{c1} K_{\rm S}$, $J/\psi K_{\rm L}$ and 
$J/\psi K^\ast$ modes. If just $B_d\to J/\psi K_{\rm S}\,[\to \pi^+\pi^-]$ 
decays are employed, BaBar and Belle obtain $\sin2\beta=0.45\pm0.18$ 
\cite{babar-CP-obs} and $0.81\pm0.20$ \cite{belle-CP-obs}, respectively. On
the other hand, the BaBar and Belle data for $B_d\to J/\psi K_{\rm L}$ 
modes give $\sin2\beta=0.70\pm0.34$ \cite{babar-CP-obs} and $1.31\pm0.23$
\cite{belle-CP-obs}, respectively. We shall come back to these issues in 
\ref{subsubsec:ambig-resol} and Section \ref{sec:MFV}.

After a couple of years of collecting data, an uncertainty of 
$\left.\Delta\sin2\beta\right|_{\rm exp}=0.05$ seems to be achievable at 
the $B$-factories. In the LHC era, this experimental uncertainty should
be reduced further by one order of magnitude \cite{LHC-Report}. In view 
of such a tremendous accuracy, it is crucial to obtain deeper insights 
into the theoretical uncertainties affecting (\ref{e12}). A possibility 
to control them is provided by $B_s\to J/\psi K_{\rm S}$ modes 
\cite{RF-BdsPsiK} (see Subsection~\ref{subsec:BsdPsiKS}). Moreover, 
also direct CP violation allows us to probe the corresponding penguin 
effects \cite{RF-rev}. In order to fully exploit the physics potential of 
$B_d\to J/\psi K_{\rm S}$ decays, their charged counterparts 
$B^\pm\to J/\psi K^\pm$ have to be considered as well \cite{FM-BpsiK}.

\boldmath
\subsubsection{Isospin Relations between $B_d\to J/\psi K_{\rm S}$ 
and $B^\pm\to J/\psi K^\pm$}\label{BpsiK-iso}
\unboldmath
The most general discussion of the $B\to J/\psi K$ system can be 
performed in terms of an isospin decomposition \cite{FM-BpsiK}. Here 
the corresponding initial and final states are grouped in the following 
isodoublets:
\begin{equation}
\left[\begin{array}{c} 
|1/2;+1/2\rangle\\
|1/2;-1/2\rangle\end{array}\right]:
\quad
\underbrace{\left[\begin{array}{c} |B^+\rangle  \\ 
|B^0_d \rangle \end{array}\right],
\quad
\left[\begin{array}{c} |\overline{B^0_d}\rangle  \\ 
-|B^- \rangle  
\end{array}\right]}_{{\cal CP}},
\quad
\underbrace{\left[\begin{array}{c} |J/\psi K^+\rangle  \\ 
|J/\psi K^0 \rangle  
\end{array}\right],
\quad
\left[\begin{array}{c} |J/\psi\overline{K^0}\rangle  \\ 
-|J/\psi K^-  \rangle \end{array}\right]}_{{\cal CP}},
\end{equation}
which are related by CP conjugation. The decays $B^+\to J/\psi K^+$ and 
$B^0_d\to J/\psi K^0$ are described by an effective low-energy Hamiltonian 
of the following structure:
\begin{equation}\label{Heff}
{\cal H}_{\rm eff}=\frac{G_{\rm F}}{\sqrt{2}}\left[V_{cs}V_{cb}^\ast
\left({\cal Q}_{\rm CC}^c-{\cal Q}_{\rm QCD}^{\rm pen}-
{\cal Q}_{\rm EW}^{\rm pen}\right)+V_{us}V_{ub}^\ast
\left({\cal Q}_{\rm CC}^u-{\cal Q}_{\rm QCD}^{\rm pen}-
{\cal Q}_{\rm EW}^{\rm pen}\right)\right],
\end{equation}
where the ${\cal Q}$ denote linear combinations of Wilson coefficients 
and four-quark operators, consisting of CC, QCD penguin and EW penguin 
operators, as listed in (\ref{CC-op-def})--(\ref{ew-penguins}). 
For the following considerations, the flavour structure of these operators 
is crucial:
\begin{equation}\label{CC-def}
{\cal Q}_{\rm CC}^c\sim (\overline{c}c)(\overline{b}s), \quad
{\cal Q}_{\rm CC}^u\sim (\overline{u}u)(\overline{b}s),
\end{equation}
\begin{equation}\label{QCD-def}
{\cal Q}_{\rm QCD}^{\rm pen}\sim \left[(\overline{c}c)+\{(\overline{u}u)
+(\overline{d}d)\}+(\overline{s}s)\right](\overline{b}s),
\end{equation}
\begin{equation}\label{EW-def}
{\cal Q}_{\rm EW}^{\rm pen}\sim \frac{1}{3}\left[2(\overline{c}c)+
\{2(\overline{u}u)-(\overline{d}d)\}-(\overline{s}s)\right](\overline{b}s).
\end{equation}
Since
\begin{equation}
(\overline{u}u)=\frac{1}{2}
\underbrace{\left(\overline{u}u+\overline{d}d\right)}_{I=0}+
\frac{1}{2}
\underbrace{\left(\overline{u}u-\overline{d}d\right)}_{I=1},\quad
2(\overline{u}u)-(\overline{d}d)=\frac{1}{2}
\underbrace{\left(\overline{u}u+\overline{d}d\right)}_{I=0}+
\frac{3}{2}
\underbrace{\left(\overline{u}u-\overline{d}d\right)}_{I=1},
\end{equation}
we conclude that (\ref{Heff}) is a combination of isospin $I=0$ and $I=1$ 
pieces:
\begin{equation}\label{ham-decom}
{\cal H}_{\rm eff} = {\cal H}_{\rm eff}^{I=0} + {\cal H}_{\rm eff}^{I=1},
\end{equation}
where ${\cal H}_{\rm eff}^{I=0}$ receives contributions from all of
the operators in (\ref{CC-def})--(\ref{EW-def}), whereas 
${\cal H}_{\rm eff}^{I=1}$ is only due to ${\cal Q}_{\rm CC}^u$ and
${\cal Q}_{\rm EW}^{\rm pen}$. Using the isospin symmetry, we obtain
\begin{eqnarray}
\langle J/\psi K^+|{\cal H}_{\rm eff}^{I=0}|B^+\rangle&=&
+\langle J/\psi K^0|{\cal H}_{\rm eff}^{I=0}|B^0_d\rangle\\
\langle J/\psi K^+|{\cal H}_{\rm eff}^{I=1}|B^+\rangle&=&
-\langle J/\psi K^0|{\cal H}_{\rm eff}^{I=1}|B^0_d\rangle,
\end{eqnarray}
and arrive at\footnote{For simplicity, the primes of the amplitudes 
introduced in \ref{ssec:beta-extr} are suppressed in the following.}
\begin{equation}\label{AMPLp1}
A(B^+\to J/\psi K^+)=\frac{G_{\rm  
F}}{\sqrt{2}}\left[\lambda_c^{(s)}\left\{
A_c^{(0)}-A_c^{(1)}\right\}+\lambda_u^{(s)}\left\{A_u^{(0)}-
A_u^{(1)}\right\}\right]
\end{equation}
\begin{equation}\label{AMPLd1}
A(B^0_d\to J/\psi K^0)=\frac{G_{\rm F}}{\sqrt{2}}\left[\lambda_c^{(s)}
\left\{A_c^{(0)}+A_c^{(1)}\right\}+\lambda_u^{(s)}
\left\{A_u^{(0)}+A_u^{(1)}\right\}\right],
\end{equation}
where 
\begin{equation}\label{ampl-c}
A_c^{(0)}=A_{\rm CC}^c-A_{\rm QCD}^{\rm pen}-A_{\rm EW}^{(0)},\quad
A_c^{(1)}=-A_{\rm EW}^{(1)}
\end{equation}
\begin{equation}\label{ampl-u}
A_u^{(0)}=A_{\rm CC}^{u (0)}-A_{\rm QCD}^{\rm pen}-A_{\rm EW}^{(0)},\quad
A_u^{(1)}=A_{\rm CC}^{u (1)}-A_{\rm EW}^{(1)}
\end{equation}
can be expressed in terms of the corresponding hadronic matrix elements 
$\langle J/\psi K|{\cal Q}|B\rangle$, i.e.\
are CP-conserving strong amplitudes. Consequently, we may write
\begin{equation}\label{AMPLp2}
A(B^+\to J/\psi K^+)=\frac{G_{\rm F}}{\sqrt{2}}\tilde\lambda^2 A
\left[A_c^{(0)}-A_c^{(1)}\right]
\left[1+\left(\frac{\lambda^2}{1-\lambda^2}\right)R_b
\left\{\frac{A_u^{(0)}-A_u^{(1)}}{A_c^{(0)}-A_c^{(1)}}\right\}
e^{i\gamma}\right]
\end{equation}
\begin{equation}\label{AMPLd2}
A(B^0_d\to J/\psi K^0)=\frac{G_{\rm F}}{\sqrt{2}}\tilde\lambda^2 A
\left[A_c^{(0)}+A_c^{(1)}
\right]\left[1+\left(\frac{\lambda^2}{1-\lambda^2}\right)R_b
\left\{\frac{A_u^{(0)}+A_u^{(1)}}{A_c^{(0)}+A_c^{(1)}}\right\}
e^{i\gamma}\right],
\end{equation}
with 
\begin{equation}\label{lam-tilde}
\tilde\lambda^2\equiv\left(1-\frac{\lambda^2}{2}\right)\lambda^2. 
\end{equation}
The amplitude of the neutral mode takes the same form as (\ref{Bd-ampl2}), 
making, however, its isospin decomposition explicit. As we have already 
noted above, an important observation is that $e^{i\gamma}$ enters in 
(\ref{AMPLp2}) and (\ref{AMPLd2}) in a doubly Cabibbo-suppressed way. 
Moreover, the $A_u^{(0,1)}$ amplitudes are governed by penguin-like 
topologies and annihilation diagrams (see Subsection~\ref{subsec:u-c-pens}), 
and are hence expected to be suppressed with respect to $A_c^{(0)}$, 
which originates also from tree-diagram-like topologies. 
In order to keep track of this feature, we introduce a ``generic'' 
expansion parameter $\overline{\lambda}={\cal O}(0.2)$ 
\cite{PAPI,hierarchy}, which is of the same order as the Wolfenstein 
parameter $\lambda$:
\begin{equation}\label{H1}
\left|A_u^{(0,1)}/A_c^{(0)}\right|={\cal O}(\overline{\lambda}).
\end{equation}
Consequently, the $e^{i\gamma}$ terms in (\ref{AMPLp2}) and (\ref{AMPLd2})
are actually suppressed by ${\cal O}(\overline{\lambda}^3)$. Since the
$A_c^{(1)}$ amplitude is due to dynamically suppressed matrix elements of 
EW penguin operators (see (\ref{ampl-c})), we expect
\begin{equation}\label{H2}
\left|A_c^{(1)}/A_c^{(0)}\right|=\underbrace{{\cal 
O}(\overline{\lambda}^2)}_{{\rm EW\, penguins}}\times
\underbrace{{\cal O}(\overline{\lambda})}_{{\rm Dynamics}}
={\cal O}(\overline{\lambda}^3),
\end{equation}
and obtain the following expression:
\begin{equation}\label{SM-ampl}
A(B^+\to J/\psi K^+)=A_{\rm SM}^{(0)}\left[1+{\cal O}(\overline{\lambda}^3)
\right]=A(B^0_d\to J/\psi K^0),
\end{equation}
with 
\begin{equation}\label{ASM0}
A_{\rm SM}^{(0)}\equiv 
\frac{G_{\rm F}}{\sqrt{2}} \tilde\lambda^2 A\,
A_c^{(0)}.
\end{equation}
Let us note that the plausible hierarchy of strong amplitudes given in 
(\ref{H1}) and (\ref{H2}) may be spoiled by very large rescattering
processes \cite{FSI,BFM}. In the worst case, (\ref{SM-ampl}) may receive 
corrections at the $\overline{\lambda}^2$ level. However, we do not 
consider this a very likely scenario and note that also the 
``QCD factorization'' approach developed in \cite{BBNS1}--\cite{BBNS3} 
is not in favour of such large rescattering effects.


%
%
%
\boldmath
\subsubsection{New Physics in the $B\to J/\psi K$ Decay 
Amplitudes}\label{BpsiK-NP-analysis}
\unboldmath
An important way for new physics to manifest itself in $B_d\to J/\psi 
K_{\rm S}$ decays is through $B^0_d$--$\overline{B^0_d}$ mixing. As 
discussed in Subsection~\ref{subsec:NP-mix}, the mixing-induced CP asymmetry
(\ref{e12}) may be affected significantly this way. An alternative 
mechanism for physics beyond the Standard Model to affect the 
$B\to\psi K$ system is provided by new-physics contributions to the 
decay amplitudes \cite{FM-BpsiK,growo}. The corresponding $\Delta B=\pm1$ 
operators can be treated on the same footing as 
$B^0_d$--$\overline{B^0_d}$ mixing in Subsection~\ref{subsec:NP-mix}. 
In the presence of new physics, the corresponding low-energy effective 
Hamiltonian can still be composed into $I=0$ and $I=1$ pieces, as in 
(\ref{ham-decom}). New physics may affect the Wilson coefficients, 
and may introduce new dim-6 operators, modifying (\ref{SM-ampl}) as 
follows \cite{FM-BpsiK}:
\begin{eqnarray}
A(B^+\to J/\psi K^+)&=&A_{\rm SM}^{(0)}
\left[1+\sum_k r_0^{(k)}e^{i\delta_0^{(k)}}e^{i\varphi_0^{(k)}}-
\sum_j r_1^{(j)}e^{i\delta_1^{(j)}}e^{i\varphi_1^{(j)}}
\right]\label{ampl-NPp}\\
A(B^0_d\to J/\psi K^0)&=&A_{\rm SM}^{(0)}
\left[1+\sum_k r_0^{(k)}e^{i\delta_0^{(k)}}e^{i\varphi_0^{(k)}}+
\sum_j  
r_1^{(j)}e^{i\delta_1^{(j)}}e^{i\varphi_1^{(j)}}\right].\label{ampl-NPd} 
\end{eqnarray}
Here $r_0^{(k)}$ and $r_1^{(j)}$ correspond to the $I=0$ and $I=1$ pieces, 
respectively, $\delta_{0}^{(k)}$ and $\delta_{1}^{(j)}$ are CP-conserving 
strong phases, and $\varphi_{0}^{(k)}$ and $\varphi_{1}^{(j)}$ the 
corresponding CP-violating weak phases. 
The labels $k$ and $j$ distinguish between different new-physics 
contributions to the $I=0$ and $I=1$ sectors.

For the following discussion, we have to make assumptions about the
size of a possible new-physics piece. We shall assume that the new-physics 
contributions to the $I=0$ sector are smaller compared to the  
leading Standard-Model amplitude (\ref{ASM0}) by a factor of order 
$\overline{\lambda}$, i.e.
\begin{equation}\label{gen-strength0}
r_0^{(k)}={\cal O}(\overline{\lambda}).
\end{equation}
Here we have implicitly assumed that there is no flavour suppression 
present. In the case where the new-physics effects are even smaller, 
it is difficult to disentangle them from the Standard-Model contribution. 
This will be addressed -- together with several other scenarios -- below. 
Parametrizing the new-physics amplitudes as in 
Subsection~\ref{subsec:NP-mix} by a scale $\Lambda$, we have
\begin{equation}
\frac{G_{\rm F}}{\sqrt{2}}\frac{M_W^2}{\Lambda^2}\sim\overline{\lambda}
\left[\frac{G_{\rm F}}{\sqrt{2}}\,\lambda^2A\right],
\end{equation}
corresponding to $\Lambda\sim 1$\,TeV. Consequently, as in the example 
given after (\ref{rho-NP-def}), also here we have a generic new-physics 
scale in the TeV regime.

As far as possible new-physics contributions to the $I=1$ sector are 
concerned, we assume a similar ``generic strength'' of the corresponding
operators. However, in comparison with the $I=0$ pieces, the matrix elements 
of the $I=1$ operators, having 
the general flavour structure
\begin{equation}\label{I1-structure}
{\cal Q}_{I=1}\sim (\overline{u}u-\overline{d}d)(\overline{b}s), 
\end{equation}
are expected to suffer from a dynamical suppression. As in (\ref{H1}) 
and (\ref{H2}), we shall assume that this brings another factor of 
$\overline{\lambda}$ into the game, yielding
\begin{equation}\label{gen-strength1}
r_1^{(j)}={\cal O}(\overline{\lambda}^2).
\end{equation}
Employing this kind of counting, the new-physics contributions to the 
$I=1$ sector would be enhanced by a factor of ${\cal O}(\overline{\lambda})$ 
with respect to the $I=1$ Standard-Model pieces. This may actually be the 
case if new physics shows up, for example, in EW penguin processes. 

Consequently, we obtain
\begin{equation}\label{ampl-hier1}
A(B\to J/\psi K)=A_{\rm SM}^{(0)}\biggl[1+
\underbrace{{\cal O}(\overline{\lambda})}_{{\rm NP}_{I=0}}+
\underbrace{{\cal O}(\overline{\lambda}^2)}_{{\rm NP}_{I=1}}+
\underbrace{{\cal O}(\overline{\lambda}^3)}_{{\rm SM}}\biggr].
\end{equation}
In the presence of large rescattering effects, the assumed dynamical 
suppression through a factor of ${\cal O}(\overline{\lambda})$ would no 
longer be effective, thereby modifying (\ref{ampl-hier1}) as follows:
\begin{equation}\label{ampl-hier2}
\left.A(B\to J/\psi K)\right|_{\rm res.}=
\left.A_{\rm SM}^{(0)}\right|_{\rm res.}\times\biggl[1+
\underbrace{{\cal O}(\overline{\lambda})}_{{\rm NP}_{I=0}}+
\underbrace{{\cal O}(\overline{\lambda})}_{{\rm NP}_{I=1}}+
\underbrace{{\cal O}(\overline{\lambda}^2)}_{{\rm SM}}\biggr].
\end{equation}
However, as we have noted above, we do not consider this as a very
likely scenario, and shall use (\ref{ampl-hier1}) in the following 
discussion, neglecting the Standard-Model pieces of 
${\cal O}(\overline{\lambda}^3)$, which are not under theoretical
control.

Concerning the analysis of CP violation, it is obvious that possible 
weak phases appearing in the new-physics contributions play the key r\^ole. 
As was the case for the $\Delta B=\pm2$ operators in 
Subsection~\ref{subsec:NP-mix}, also the $\Delta B = \pm 1$ operators 
could carry such new weak phases, which would then affect the CP-violating 
$B\to J/\psi K$ observables.

\boldmath
\subsubsection{Observables for a General Analysis of New Physics}
\unboldmath
The decays $B^+\to J/\psi K^+$, $B^0_d\to J/\psi K^0$ and their
charge conjugates provide a set of four decay amplitudes $A_i$.
Measuring the corresponding rates, we may determine the $|A_i|^2$.
Since we are not interested in the overall normalization of the
decay amplitudes, we may construct the following three independent 
observables with the help of the $|A_i|^2$:
\begin{eqnarray}
{\cal A}_{\rm CP}^{(+)}&\equiv&
\frac{|A(B^+\to J/\psi K^+)|^2-|A(B^-\to J/\psi K^-)|^2}{|A(B^+\to 
J/\psi K^+)|^2+|A(B^-\to J/\psi K^-)|^2}\label{def-AP}\\
{\cal A}_{\rm CP}^{{\rm dir}}&\equiv&
\frac{|A(B^0_d\to J/\psi K^0)|^2-|A(\overline{B^0_d}\to J/\psi
\overline{K^0})|^2}{|A(B^0_d\to J/\psi K^0)|^2+
|A(\overline{B^0_d}\to J/\psi\overline{K^0})|^2}\label{def-A0}\\
B&\equiv&\frac{\langle|A(B_d\to J/\psi K)|^2\rangle-
\langle|A(B^\pm\to J/\psi K^\pm)|^2\rangle}{\langle|A(B_d\to 
J/\psi K)|^2\rangle+\langle|A(B^\pm\to J/\psi K^\pm)|^2\rangle},\label{def-A}
\end{eqnarray}
where the ``CP-averaged'' amplitudes are generally defined as follows:
\begin{equation}\label{CP-average-ampl}
\left\langle|A(B\to f)|^2\right\rangle\equiv
\frac{1}{2}\left[|A(B\to f)|^2
+|A(\overline{B}\to \overline{f})|^2\right],
\end{equation}
and ${\cal A}_{\rm CP}^{{\rm dir}}$ agrees with the corresponding 
observable introduced in (\ref{ee6}). Mixing-induced CP violation 
provides another observable 
${\cal A}_{\rm CP}^{\rm mix}(B_d\to J/\psi K_{\rm S})$, which is governed by
\begin{equation}\label{xi-NP}
\xi^{(d)}_{\psi K_{\rm S}}=
e^{-i\phi}\left[\frac{1+\sum_k r_0^{(k)}e^{i\delta_0^{(k)}}
e^{-i\varphi_0^{(k)}}+\sum_j r_1^{(j)}e^{i\delta_1^{(j)}}
e^{-i\varphi_1^{(j)}}}{1+\sum_k r_0^{(k)}e^{i\delta_0^{(k)}}
e^{+i\varphi_0^{(k)}}+\sum_j r_1^{(j)}e^{i\delta_1^{(j)}}
e^{+i\varphi_1^{(j)}}}\right],
\end{equation}
as can be seen in (\ref{ee7}). In order to derive (\ref{xi-NP}), we have 
used the parametrization (\ref{ampl-NPd}) to express the corresponding decay 
amplitudes. The phase $\phi$ corresponds to $\phi_d^{\rm NP}+\phi_K$, where
$\phi_d^{\rm NP}$ was introduced in (\ref{PhiM2}), and $\phi_K$ plays usually 
a negligible r\^ole.

In order to search for new-physics effects in the $B\to J/\psi K$ system, 
it is useful to introduce the following combinations of the observables 
(\ref{def-AP}) and (\ref{def-A0}):
\begin{equation}\label{S-D-def}
S\equiv\frac{1}{2}\left[{\cal A}_{\rm CP}^{{\rm dir}}+{\cal A}_{\rm  
CP}^{(+)}
\right],\quad
D\equiv\frac{1}{2}\left[{\cal A}_{\rm CP}^{{\rm dir}}-{\cal A}_{\rm  
CP}^{(+)}
\right].
\end{equation}
Using (\ref{ampl-NPp}) and (\ref{ampl-NPd}), and assuming the hierarchy 
in (\ref{ampl-hier1}), we obtain
\begin{equation}\label{S-expr}
S=-2\left[\sum_k r_0^{(k)}\sin\delta_0^{(k)}\sin\varphi_0^{(k)}\right]
\left[1-2\sum_l r_0^{(l)}\cos\delta_0^{(l)}\cos\varphi_0^{(l)}\right]
={\cal O}(\overline{\lambda}) + {\cal O}(\overline{\lambda}^2)
\end{equation}
\begin{equation}\label{D-expr}
D=-2\sum_j r_1^{(j)}\sin\delta_1^{(j)}\sin\varphi_1^{(j)}=
{\cal O}(\overline{\lambda}^2)
\end{equation}
\begin{equation}\label{B-expr}
B=+2\sum_j r_1^{(j)}\cos\delta_1^{(j)}\cos\varphi_1^{(j)}=
{\cal O}(\overline{\lambda}^2),
\end{equation}
where terms of ${\cal O}(\overline{\lambda}^3)$, including also a
Standard-Model contribution, which is not under theoretical control,
have been neglected. Note that if the dynamical suppression of the
$I=1$ contributions would be larger, $B$ and $D$ would be further
suppressed relative to $S$.

The expression for the mixing-induced CP asymmetry is rather complicated 
and not very instructive. Let us give it for the special case where the 
new-physics contributions to the $I=0$ and $I=1$ sectors involve either 
the same weak or strong phases:
\begin{displaymath}
a_{\psi K_{\rm S}}=
\sin\phi+2\,r_0\cos\delta_0\sin\varphi_0\cos\phi+
2\,r_1\cos\delta_1\sin\varphi_1\cos\phi
\end{displaymath}
\begin{equation}\label{Amix-calc}
~~-\,r_0^2\Bigl[\left(1-\cos2\varphi_0\right)\sin\phi+
\cos2\delta_0\sin2\varphi_0\cos\phi\Bigr]=\sin\phi+
{\cal O}(\overline{\lambda}) + {\cal O}(\overline{\lambda}^2).
\end{equation}
Expressions (\ref{S-expr})--(\ref{B-expr}) also simplify in this case:
\begin{eqnarray}
S&=&-2\,r_0\sin\delta_0\sin\varphi_0+r_0^2\sin2\delta_0\sin2\varphi_0\\
D&=&-2\,r_1\sin\delta_1\sin\varphi_1\\
B&=&2\,r_1\cos\delta_1\cos\varphi_1.
\end{eqnarray}

\subsubsection{Possible Scenarios}\label{BpsiK-scenarios}
As far as CP violation in $B^\pm\to J/\psi K^\pm$ decays is concerned, 
dramatic effects are already excluded by 
\begin{equation}
{\cal A}_{\rm CP}^{(+)}=\left\{\begin{array}{ll}
(-1.8\pm4.3\pm0.4)\% & \mbox{(CLEO \cite{cleo-dir})}\\
(-0.4\pm2.9\pm0.4)\% & \mbox{(BaBar \cite{babar-dir}).}
\end{array}\right.
\end{equation}
Moreover, a recent BaBar analysis yields
$\lambda_{\psi K_{\rm S}}^{(d)}=0.93\pm0.09\pm0.03$ \cite{babar-CP-obs}, where
\begin{equation}
\lambda_{\psi K_{\rm S}}^{(d)}=1/\xi_{\psi K_{\rm S}}^{(d)\ast}.
\end{equation}
This result implies
\begin{equation}
{\cal A}_{\rm CP}^{\rm dir}=(-7\pm10)\%.
\end{equation}
Finally, if we use the most recent BaBar and Belle data for $B\to J/\psi K$, 
we obtain
\begin{equation}\label{B-res}
B=\frac{\mbox{BR}(B_d\to J/\psi K)\,\tau-
\mbox{BR}(B^\pm\to J/\psi K^\pm)}{\mbox{BR}(B_d\to J/\psi K)\,\tau+
\mbox{BR}(B^\pm\to J/\psi K^\pm)}=\left\{\begin{array}{ll}
(-6.2\pm 3.6)\% & \mbox{(BaBar \cite{BABAR-BR-B})}\\
(-10.6\pm 6.8)\% & \mbox{(Belle \cite{BELLE-BR-B})},
\end{array}\right.
\end{equation}
where the numerical values depend rather sensitively on the lifetime ratio
$\tau\equiv\tau_{B^+}/\tau_{B^0_d}$, assumed to be $1.060\pm0.029$ \cite{PDG},
which is consistent with \cite{BABAR-LIFE,BELLE-LIFE}. Because of the large 
uncertainties, we cannot yet draw conclusions. However, the experimental 
situation should improve significantly in the future. 

As can be seen in (\ref{S-expr})--(\ref{B-expr}), the observable $S$ 
provides a ``smoking-gun'' signal for new-physics contributions to the 
$I=0$ sector, while $D$ and $B$ allow us to probe new physics affecting 
the $I=1$ pieces. Since the hierarchy in (\ref{ampl-hier1}) implies 
\begin{equation} 
S=\underbrace{{\cal O}(\overline{\lambda})}_{{\rm NP}_{I=0}}+ 
\underbrace{{\cal O}(\overline{\lambda}^3)}_{\rm SM},\quad 
D=\underbrace{{\cal O}(\overline{\lambda}^2)}_{{\rm NP}_{I=1}} 
+\underbrace{{\cal O}(\overline{\lambda}^3)}_{\rm SM},\quad 
B=\underbrace{{\cal O}(\overline{\lambda}^2)}_{{\rm NP}_{I=1}} 
+\underbrace{{\cal O}(\overline{\lambda}^3)}_{\rm SM}, 
\end{equation} 
we conclude that $S$ may already be accessible at the first-generation 
$B$-factories (BaBar, Belle, Tevatron-II), whereas the latter observables 
will probably be left for second-generation $B$ experiments (BTeV, LHCb). 
However, should $B$ and $D$, in addition to $S$, also be found to be at the
$10\%$ level, i.e.\ should be measured at the first-generation
$B$-factories, we would not only have signals for physics
beyond the Standard Model, but also for large rescattering processes. 

A more pessimistic scenario one can imagine is that $S$ is measured at the 
$\overline{\lambda}^2$ level in the LHC era, whereas no indications for 
non-vanishing values of $D$ and $B$ are found. Then we would still have 
evidence for new physics, which would then correspond to 
$r_0^{(k)}={\cal O}(\overline{\lambda}^2)$ and 
$r_1^{(j)}={\cal O}(\overline{\lambda}^3)$, i.e.
\begin{equation}
A(B\to J/\psi K)=A_{\rm SM}^{(0)}\biggl[1+
\underbrace{{\cal O}(\overline{\lambda}^2)}_{{\rm NP}_{I=0}}+
\underbrace{{\cal O}(\overline{\lambda}^3)}_{{\rm NP}_{I=1}}+
\underbrace{{\cal O}(\overline{\lambda}^3)}_{{\rm SM}}\biggr].
\end{equation}
However, if all three observables are measured to be of 
${\cal O}(\overline{\lambda}^2)$, new-physics effects cannot be 
distinguished from Standard-Model contributions, which could also be 
enhanced to the $\overline{\lambda}^2$ level by large rescattering effects. 
This would be the most unfortunate case for the strategy to search for 
new-physics contributions to the $B\to J/\psi K$ decay amplitudes 
discussed above \cite{FM-BpsiK}. However, further information can be 
obtained with the help of the decay $B_s\to J/\psi K_{\rm S}$, which can 
be combined with $B_d\to J/\psi K_{\rm S}$ through the $U$-spin symmetry of 
strong interactions and may shed light on new physics even in this case.
Within the Standard Model, it allows us to control the -- presumably very 
small -- penguin uncertainties in the determination of $\beta$ from 
$a_{\psi K_{\rm S}}$, and to extract, moreover, the CKM angle $\gamma$
\cite{RF-BdsPsiK}. This mode will be addressed in 
Subsection~\ref{subsec:BsdPsiKS}.

As can be seen in (\ref{Amix-calc}), the mixing-induced CP asymmetry 
$a_{\psi K_{\rm S}}$ is affected both by $I=0$ and by $I=1$ new-physics 
contributions, where the dominant ${\cal O}(\overline{\lambda})$ effects 
are expected to be due to the $I=0$ sector. Neglecting terms of 
${\cal O}(\overline{\lambda}^2)$, we may write
\begin{equation}\label{Amix-simple}
a_{\psi K_{\rm S}}=\sin(\phi+\delta\phi_{\rm NP}^{\rm dir})
\quad\mbox{with}\quad
\delta\phi_{\rm NP}^{\rm dir}=
2\sum_k r_0^{(k)}\cos\delta_0^{(k)}\sin\varphi_0^{(k)}.
\end{equation}
The phase shift $\delta\phi_{\rm NP}^{\rm dir}={\cal O}(\overline{\lambda})$ 
may be as large as ${\cal O}(20^\circ)$. Since the Standard-Model range for 
$\phi$ is given by $30^\circ\lsim\phi=2\beta\lsim 70^\circ$, the 
mixing-induced CP asymmetry  may also be affected significantly by 
new-physics contributions to the $B_d\to J/\psi K_{\rm S}$ 
decay amplitude, and not only in the ``standard'' fashion, through 
$B^0_d$--$\overline{B^0_d}$ mixing, as discussed in  
Subsection~\ref{subsec:NP-mix}. This would be another possibility to 
accommodate ``anomalously'' small or large values of $a_{\psi K_{\rm S}}$. 
In order to gain confidence into such a scenario, it is crucial to 
improve also the measurements of the observables $S$, $D$ and $B$ 
\cite{FM-BpsiK}. 


So far, our considerations were completely general. Let us therefore  
comment briefly on a special case, where the strong phases $\delta_0^{(k)}$ 
and $\delta_1^{(j)}$ take the trivial values $0$ or $\pi$, as in 
factorization. In this case, (\ref{S-expr})--(\ref{B-expr}) 
would simplify as follows:
\begin{equation}
S\approx0,\quad D\approx0,\quad 
B\approx 2\sum_j r_1^{(j)}\sin\varphi_1^{(j)}={\cal  
O}(\overline{\lambda}^2),
\end{equation}
whereas (\ref{Amix-calc}) would yield
\begin{eqnarray}
a_{\psi K_{\rm S}}&=&
\sin\phi+2\,r_0\sin\varphi_0\cos\phi+2\,r_1\sin\varphi_1\cos\phi\nonumber\\
&&-r_0^2\Bigl[\left(1-\cos2\varphi_0\right)\sin\phi+
\sin2\varphi_0\cos\phi\Bigr]=\sin\phi+
{\cal O}(\overline{\lambda}) + {\cal O}(\overline{\lambda}^2).
\end{eqnarray}
The important point is that $S$ and $D$ are governed by sines of the strong 
phases, whereas the new-physics contributions to $B$ and 
$a_{\psi K_{\rm S}}$ involve cosines of the corresponding 
strong phases. Consequently, these terms do {\it not} vanish for 
$\delta\to 0,\pi$. The impact of new physics on 
$a_{\psi K_{\rm S}}$ may still be sizeable in this scenario, 
whereas $B$ could only be measured in the LHC era \cite{LHC-Report}. 
On the other hand, if $S$ 
and $D$ should be observed at the $\overline{\lambda}$ and 
$\overline{\lambda}^2$ levels, respectively, we would not only get a 
``smoking-gun'' signal for new-physics contributions to the $B\to J/\psi K$ 
decay amplitudes, but also for non-factorizable hadronic effects. A 
measurement of all three observables $S$, $D$ and $B$ at the 
$\overline{\lambda}$ level would imply, in addition, large 
rescattering processes, as we have already emphasized above.

\boldmath
\subsubsection{$B_d\to J/\psi[\to \ell^+\ell^-] K^\ast[\to \pi^0 
K_{\rm S}]$ Decays}\label{subsubsec:ambig-resol}
\unboldmath
Let us finally have a look at $B^0_d\to J/\psi K^{\ast0}$ decays, assuming
that direct CP violation vanishes in these channels, as in the Standard 
Model. If the $K^{\ast0}$-meson is observed to decay to the CP eigenstate 
$\pi^0 K_{\rm S}$, the time evolution of the angular distribution of 
the $B_d^0(t)\to J/\psi[\to \ell^+\ell^-] K^\ast[\to \pi^0 K_{\rm S}]$ 
decay products also allows us to probe $\phi$ \cite{DQSTL}, which is given 
by $2\beta$ in the Standard Model. The important feature of the corresponding 
observables is that they do not only allow us to determine $\sin\phi$ -- 
in complete analogy to $a_{\psi K_{\rm S}}$ -- but contain also terms of 
the following form \cite{DDF1}--\cite{DFN}:
\begin{equation}
\cos\delta_f\cos\phi.
\end{equation}
Here $\delta_f$ is a CP-conserving strong phase corresponding to a 
given final-state configuration of the $J/\psi K^{\ast0}$ system. 
Theoretical tools, such as factorization, may be sufficiently accurate 
to determine the sign of $\cos\delta_f$, thereby allowing the direct 
extraction of $\cos\phi$. The knowledge of this quantity, in combination with
$\sin\phi$, allows us then to determine $\phi$ {\it unambiguously}, 
resolving a twofold ambiguity, which arises in the extraction of $\phi$
from the mixing-induced CP asymmetry $a_{\psi K_{\rm S}}=\sin\phi$. 

The resolution of this ambiguity is an important issue and has several 
applications (see, for example, Subsection~\ref{subsec:BDpi} and 
Section~\ref{sec:Uspin}); alternative strategies to accomplish this task 
were proposed in \cite{ambig}. It may also turn out to be crucial for the 
search for new physics. In order to illustrate this point, let us assume
that $\sin\phi$ has been measured to be equal to 0.8 (see (\ref{s2b-world})). 
We would then conclude that $\phi=53^\circ$ or
$127^\circ$, where the former solution would lie perfectly within the 
range $30^\circ\lsim\phi\lsim 70^\circ$ implied by the ``standard analysis''
of the unitarity triangle. The two solutions can be distinguished 
through a measurement of $\cos\phi$, which is equal to $0.6$ and $-0.6$
for $\phi=53^\circ$ and $127^\circ$, respectively. Consequently, a
measurement of $\cos\phi=-0.6$ would imply new physics.

\begin{figure}[t]
\begin{center}
\leavevmode
\epsfysize=4.5truecm 
\epsffile{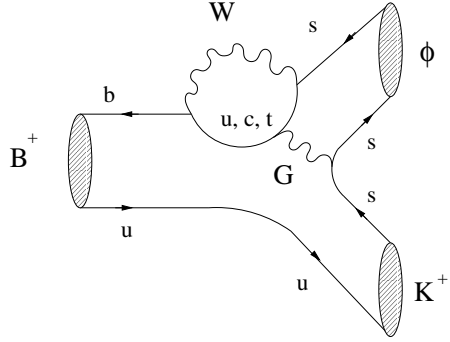} 
\end{center}
\vspace*{-0.6truecm}
\caption{QCD penguin contributions to 
$B^+\to \phi K^+$.}\label{fig:BphiK}
\end{figure}

\boldmath
\subsection{The $B\to\phi K$ System}\label{subsec:BPhiK}
\unboldmath
An important testing ground for the Standard Model is also provided by 
$B\to \phi K$ decays. As can be seen in Fig.~\ref{fig:BphiK}, 
these modes are governed by QCD penguin processes \cite{BphiK-old}, but 
also EW penguins are sizeable \cite{RF-EWP1,DH-PhiK}, and physics beyond 
the Standard Model may have an important impact 
\cite{RF-rev,growo,LoSo,FM-PhiK}. 
In the summer of 2000, the observation of the $B^\pm\to \phi K^\pm$ 
channel was announced by the Belle and CLEO collaborations. The 
experimental situation in the summer of 2001 can be summarized as follows:
\begin{equation}\label{EXP1} 
\mbox{BR}(B^+\to \phi K^+)\times10^6=\left\{\begin{array}{ll} 
11.2^{+2.2}_{-2.0}\pm1.4 & \mbox{(Belle \cite{belle-BphiK})}\\ 
5.5^{+2.1}_{-1.8}\pm0.6 & \mbox{(CLEO \cite{cleo-BphiK})}\\
7.7^{+1.6}_{-1.4}\pm0.8 & \mbox{(BaBar \cite{babar-BphiK})},
\end{array} 
\right. 
\end{equation} 
\begin{equation}\label{EXP2} 
\mbox{}~~~\mbox{BR}(B_d^0\to \phi K^0)\times10^6=\left\{\begin{array}{ll} 
8.9^{+3.4}_{-2.7}\pm1.0 & \mbox{(Belle \cite{belle-BphiK})}\\ 
5.4^{+3.7}_{-2.7}\pm0.7 & \mbox{(CLEO \cite{cleo-BphiK})}\\
8.1^{+3.1}_{-2.5}\pm0.8 & \mbox{(BaBar \cite{babar-BphiK})}.
\end{array} 
\right. 
\end{equation} 
Moreover, a first result for the direct CP asymmetry of the $B^+\to \phi K^+$
transition has already been reported by the BaBar collaboration 
\cite{babar-dir}:
\begin{equation}
{\cal A}_{\rm CP}(B^+\to \phi K^+)=(5\pm20\pm3)\%,
\end{equation}
where ${\cal A}_{\rm CP}(B^+\to \phi K^+)$ is defined in analogy to its
$B^+\to J/\psi K^+$ counterpart in (\ref{def-AP}). Recent calculations
of $B\to\phi K$ modes can be found in \cite{BphiK-recent-calc}. 

In our discussion of the $B\to \phi K$ system, we follow closely 
\cite{FM-PhiK}, and perform an analysis similar to the one for the 
$B\to J/\psi K$ decays given in the previous subsection.

\boldmath
\subsubsection{Decay Amplitudes in the Standard Model}
\unboldmath
If we apply the isospin symmetry and follow \ref{BpsiK-iso}, we obtain 
the following model-independent parametrizations of the $B^+\to \phi K^+$, 
$B^0_d\to \phi K^0$ decay amplitudes:
\begin{equation}\label{AMPLp2-phi} 
A(B^+\to \phi K^+)=\frac{G_{\rm F}}{\sqrt{2}} \tilde\lambda^2
A\left[{\cal A}_c^{(0)}- 
{\cal A}_c^{(1)}\right]\left[1+\left(\frac{\lambda^2}{1-\lambda^2}\right) 
R_b\left\{\frac{{\cal A}_u^{(0)}-{\cal A}_u^{(1)}}{{\cal A}_c^{(0)}- 
{\cal A}_c^{(1)}}\right\}e^{i\gamma}\right] 
\end{equation} 
\begin{equation}\label{AMPLd2-phi} 
A(B^0_d\to \phi K^0)=\frac{G_{\rm F}}{\sqrt{2}} \tilde\lambda^2
A\left[{\cal A}_c^{(0)}+ 
{\cal A}_c^{(1)}\right]\left[1+\left(\frac{\lambda^2}{1-\lambda^2}\right) 
R_b\left\{\frac{{\cal A}_u^{(0)}+{\cal A}_u^{(1)}}{{\cal A}_c^{(0)}+ 
{\cal A}_c^{(1)}}\right\}e^{i\gamma}\right],
\end{equation} 
where ${\cal A}_c^{(0,1)}$ and ${\cal A}_u^{(0,1)}$, which can be expressed
in terms of hadronic matrix elements $\langle \phi K|{\cal Q}|B\rangle$,
correspond to (\ref{ampl-c}) and (\ref{ampl-u}), respectively.

\begin{figure} 
\begin{center} 
\leavevmode 
\epsfysize=4.3truecm  
\epsffile{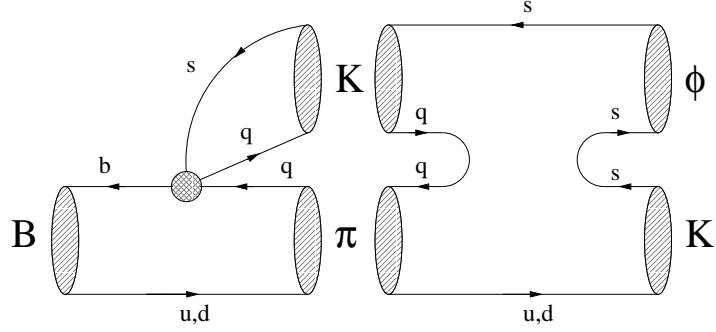} 
\end{center} 
\vspace*{-0.2truecm} 
\caption{Illustration of rescattering processes contributing to $B\to\phi K$ 
through penguin-like topologies with internal $q$-quark exchanges 
($q\in\{u,d\}$). The shaded circle represents insertions of the corresponding 
current--current operators.}\label{fig:rescatter1} 
\end{figure}

\begin{figure} 
\vspace*{0.8truecm} 
\begin{center} 
\leavevmode 
\epsfysize=4.0truecm  
\epsffile{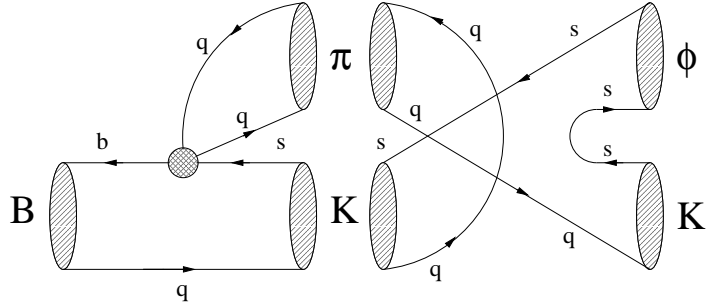} 
\end{center} 
\vspace*{-0.2truecm} 
\caption{Illustration of rescattering processes contributing to $B\to\phi K$  
through annihilation topologies. The shaded circle represents  
insertions of the corresponding current--current operators  
($q\in\{u,d\}$).}\label{fig:rescatter2} 
\end{figure}

At first sight, expressions (\ref{AMPLp2-phi}) and (\ref{AMPLd2-phi}) are 
completely analogous to the ones for the $B^+\to J/\psi K^+$ and  
$B^0_d\to J/\psi K^0$ amplitudes given in (\ref{AMPLp2}) and 
(\ref{AMPLd2}), respectively. However, the dynamics, which is encoded 
in the strong amplitudes ${\cal A}$, is very different. In particular, 
the current--current operators ${\cal Q}_{\rm CC}^c$ cannot contribute 
to $B\to \phi K$ decays, i.e.\ to ${\cal A}_{\rm CC}^c$, through 
tree-diagram-like topologies; they may only do so through penguin topologies 
with internal charm-quark exchanges, which include also  
\begin{equation}\label{rescatter-c} 
B^+\to\{D_s^+\overline{D^0},...\}\to\phi K^+,\quad  
B^0_d\to\{D_s^+D^-,...\}\to\phi K^0 
\end{equation} 
rescattering processes \cite{BFM}, and may actually play an important  
r\^ole (see Subsection~\ref{subsec:u-c-pens}). On the other hand, the 
${\cal A}_{\rm CC}^{u (0,1)}$ amplitudes receive contributions from 
penguin processes with internal up- and down-quark exchanges, as well 
as from annihilation topologies.\footnote{Note that the isospin projection
operators ${\cal Q}\sim (\overline{u}u\pm\overline{d}d)(\overline{b}s)$
involve also $\overline{d}d$ quark currents.}
Such penguins may also be important, in particular in the presence of large 
rescattering processes  \cite{FSI,BFM}; a similar comment applies to 
annihilation topologies. In the $B\to\phi K$ system, the relevant 
rescattering processes are  
\begin{equation}\label{rescatter-u} 
B^+\to\{\pi^0K^+,...\}\to\phi K^+,\quad  
B^0_d\to\{\pi^-K^+,...\}\to\phi K^0, 
\end{equation} 
which are illustrated in Figs.~\ref{fig:rescatter1} and \ref{fig:rescatter2}.
In contrast to (\ref{rescatter-c}), large rescattering effects of this kind 
may affect the search for new physics with $B\to \phi K$ decays, since these 
processes are associated with the weak phase factor $e^{i\gamma}$. Moreover, 
they involve ``light'' intermediate states, and are hence expected to be 
enhanced more easily, dynamically, through long-distance effects than 
(\ref{rescatter-c}), which involve ``heavy'' intermediate states.

Let us now have a closer look at the structure of the $B\to \phi K$ 
decay amplitudes, focusing first on the case corresponding to small  
rescattering effects. From the $B\to\phi K$ counterparts of (\ref{ampl-c}) 
and (\ref{ampl-u}), we expect 
\begin{equation}\label{hier1} 
\left|{\cal A}_u^{(0,1)}/{\cal A}_c^{(0)}\right|={\cal O}(1), 
\end{equation} 
where we have assumed that the CC operators, which contribute only through
penguin or annihilation topologies, yield amplitudes of the same order as 
the QCD penguin operators. In the case of ${\cal A}_c^{(1)}$, 
the situation is different. Here we have to deal with 
an amplitude that is essentially due to EW penguins. Moreover, the 
$B\to \phi K$ matrix elements of $I=1$ operators, having the general 
flavour structure given in (\ref{I1-structure}), are expected to suffer from 
a dynamical suppression. In order to keep track of these features, we 
employ, as in Subsection~\ref{subsec:BpsiK}, again a generic expansion  
parameter $\overline{\lambda}={\cal O}(0.2)$:
\begin{equation}\label{hier2} 
\left|{\cal A}_c^{(1)}/{\cal A}_c^{(0)}\right|= 
\underbrace{{\cal O}(\overline{\lambda})}_{{\rm EW\, penguins}} 
\times\underbrace{{\cal O}(\overline{\lambda})}_{{\rm Dynamics}} 
={\cal O}(\overline{\lambda}^2). 
\end{equation} 
Consequently, we obtain 
\begin{equation}\label{SM-ampl-phi} 
A(B^+\to \phi K^+)={\cal A}_{\rm SM}^{(0)} 
\left[1+{\cal O}(\overline{\lambda}^2)\right]=A(B^0_d\to \phi K^0), 
\end{equation} 
with  
\begin{equation}\label{ASM0-phi} 
{\cal A}_{\rm SM}^{(0)}\equiv  
\frac{G_{\rm F}}{\sqrt{2}}\lambda^2A\,{\cal A}_c^{(0)}. 
\end{equation} 
The terms entering (\ref{SM-ampl-phi}) at the $\overline{\lambda}^2$ level 
contain also pieces that are proportional to the weak phase factor 
$e^{i\gamma}$,  thereby leading to direct CP violation in the 
$B\to \phi K$ system.
 
Let us now consider large rescattering effects of the kind given in 
(\ref{rescatter-u}). Although this case does not appear to be a very
likely scenario,\footnote{Arguments against this possibility, i.e.\
large rescattering effects, were also given in \cite{GIW}.} which is 
also not favoured by the QCD factorization approach 
\cite{BBNS1}--\cite{BBNS3}, it deserves careful attention
to separate possible new-physics effects from those of the Standard Model.
In the worst case, (\ref{hier1}) would be dynamically enhanced as  
\begin{equation}\label{hier1-res} 
\left|{\cal A}_u^{(0,1)}/{\cal A}_c^{(0)}\right|= 
{\cal O}(1/\overline{\lambda}), 
\end{equation} 
and the dynamical suppression in (\ref{hier2}) would no longer be  
effective, i.e. 
\begin{equation}\label{hier2-res} 
\left|{\cal A}_c^{(1)}/{\cal A}_c^{(0)}\right|={\cal O}(\overline{\lambda}). 
\end{equation} 
In such a scenario, (\ref{SM-ampl-phi}) would receive corrections of  
${\cal O}(\overline{\lambda})$, involving also $e^{i\gamma}$. This  
feature may complicate the search for new physics with the help of  
CP-violating effects in $B\to\phi K$ decays. On the other hand, the  
rescattering processes described by (\ref{rescatter-c}) may only affect  
the amplitude ${\cal A}_{\rm SM}^{(0)}$ sizeably through its  
${\cal A}_{\rm CC}^c$ piece, and are not related to a CP-violating 
weak phase factor within the Standard Model.  
 
Before turning to new physics, we would like to emphasize an interesting  
difference between the $B\to \phi K$ and $B\to J/\psi K$ systems.
In the $B\to J/\psi K$ case, the Standard-Model amplitudes corresponding 
to (\ref{SM-ampl-phi}) receive corrections at the  
$\overline{\lambda}^3$ level (see (\ref{SM-ampl})), which may be enhanced 
to ${\cal O}(\overline{\lambda}^2)$ in the presence of large 
rescattering effects. Consequently, within the Standard Model, there may  
be direct CP-violating effects in $B\to J/\psi K$ transitions of at most  
${\cal O}(\overline{\lambda}^2)$, whereas such asymmetries may already 
arise at the $\overline{\lambda}$ level in the $B\to \phi K$ system.  
On the other hand, as $B\to \phi K$ modes are governed by penguin processes, 
their decay amplitudes are more sensitive to new physics.

\boldmath
\subsubsection{Effects of Physics Beyond the Standard 
Model}\label{subsubsec:BphiK-NP}
\unboldmath
In order to analyse the impact of possible new-physics contributions
to the $B\to\phi K$ system, we follow Subsections~\ref{subsec:NP-mix} 
and \ref{BpsiK-NP-analysis}. In analogy to the $B\to J/\psi K$ case, 
the Standard-Model expression (\ref{SM-ampl-phi}) is modified as follows:
\begin{eqnarray} 
A(B^+\to \phi K^+)&=&{\cal A}_{\rm SM}^{(0)} 
\left[1+\sum_k v_0^{(k)}e^{i\Delta_0^{(k)}}e^{i\Phi_0^{(k)}}- 
\sum_j v_1^{(j)}e^{i\Delta_1^{(j)}}e^{i\Phi_1^{(j)}} 
\right]\label{ampl-NPp-phi}\\ 
A(B^0_d\to \phi K^0)&=&{\cal A}_{\rm SM}^{(0)} 
\left[1+\sum_k v_0^{(k)}e^{i\Delta_0^{(k)}}e^{i\Phi_0^{(k)}}+ 
\sum_j v_1^{(j)}e^{i\Delta_1^{(j)}}
e^{i\Phi_1^{(j)}}\right],\label{ampl-NPd-phi}  
\end{eqnarray} 
where $v_0^{(k)}$ and $v_1^{(j)}$ correspond to the $I=0$ and $I=1$ pieces,  
respectively, $\Delta_{0}^{(k)}$ and $\Delta_{1}^{(j)}$ are CP-conserving  
strong phases, and $\Phi_{0}^{(k)}$ and $\Phi_{1}^{(j)}$ the corresponding  
CP-violating weak phases. 

As we have already noted, within the Standard Model, the $B\to \phi K$  
system is governed by QCD penguins. Neglecting, for simplicity,  
EW penguins and the proper renormalization-group evolution, we may  
write 
\begin{equation}\label{SM-pen-expr} 
{\cal A}^{(0)}_{\rm SM}\sim\frac{G_{\rm F}}{\sqrt{2}}\lambda^2 A  
\left[\frac{\alpha_s}{4\pi}\, {\cal C}\right] \langle P_{\rm QCD}\rangle, 
\end{equation} 
where ${\cal C}={\cal O}(1)$ is a perturbative short-distance coefficient, 
which is multiplied by the characteristic loop factor $\alpha_s/(4\pi)$,  
and $P_{\rm QCD}$ denotes an appropriate linear combination of QCD penguin  
operators. Since (\ref{SM-pen-expr}) is a doubly Cabibbo-suppressed loop 
amplitude, new physics could well be of the same order of magnitude. If we 
assume once more that the physics beyond the Standard Model is associated 
with a scale $\Lambda$ and impose that it yields contributions to the 
$B\to \phi K$ amplitudes of the same size as the Standard Model, we obtain 
\begin{equation}\label{generic} 
\frac{G_{\rm F}}{\sqrt{2}}\frac{M_{W}^2}{\Lambda^2}\sim  
\frac{G_{\rm F}}{\sqrt{2}}\lambda^2 A \left[\frac{\alpha_s}{4\pi}\, {\cal C} 
\right], 
\end{equation} 
corresponding to $\Lambda\sim 3$\,TeV.\footnote{In this numerical estimate,
we have assumed $A\times {\cal C}\sim1$ and $\alpha_s=\alpha_s(m_b)
\sim0.2$.} Consequently, for a generic new-physics scale in the TeV regime, 
which we also considered in our $B\to J/\psi K$ analysis, we may well have 
\begin{equation} 
v_0^{(k)}={\cal O}(1).  
\end{equation} 
In deriving (\ref{generic}), we have implicitly assumed that the 
new-physics operators arise at tree level and that there is no 
flavour suppression. In the case where the new-physics effects are less 
pronounced, it may be difficult to disentangle them from the Standard-Model 
contributions. We shall come back to this issue below, discussing 
various scenarios.  
 
Concerning possible new-physics contributions to the $I=1$ sector, we  
assume a ``generic strength'' of the corresponding operators similar to 
(\ref{generic}). However, since these operators have the general flavour 
structure given in (\ref{I1-structure}), their hadronic $B\to\phi K$
matrix elements are expected to suffer from a dynamical suppression. 
As in (\ref{hier2}), we assume that this brings a factor of 
$\overline{\lambda}$ into the game, yielding 
\begin{equation}\label{gen-strength1-phi} 
v_1^{(j)}={\cal O}(\overline{\lambda}). 
\end{equation} 
If we impose such a hierarchy of amplitudes, the new-physics contributions  
to the $I=1$ sector would be enhanced by a factor of  
${\cal O}(\overline{\lambda})$ with respect to the $I=1$ Standard-Model  
pieces. This may actually be the case if new physics shows up, for example,  
in EW penguin processes.  
 
Consequently, we finally arrive at  
\begin{equation}\label{ampl-hier1-phi} 
A(B\to \phi K)={\cal A}_{\rm SM}^{(0)}\biggl[1+ 
\underbrace{{\cal O}(1)}_{{\rm NP}_{I=0}}+ 
\underbrace{{\cal O}(\overline{\lambda})}_{{\rm NP}_{I=1}}+ 
\underbrace{{\cal O}(\overline{\lambda}^2)}_{{\rm SM}}\biggr]. 
\end{equation} 
In deriving this expression, we have assumed that the $B\to \phi K$ decays  
are not affected by rescattering effects. On the other hand, in the presence  
of large rescattering processes of the kind described by (\ref{rescatter-u}),  
the dynamical suppression assumed in (\ref{gen-strength1-phi}) would no 
longer be effective, thereby yielding $v_1^{(j)}={\cal O}(1)$. Analogously, 
the $B\to \phi K$ matrix elements of $I=0$ operators with flavour structure 
\begin{equation}\label{I0-ops1} 
{\cal Q}_{I=0}^{\overline{u}u,\overline{d}d} 
\sim (\overline{u}u+\overline{d}d)(\overline{b}s) 
\end{equation} 
would no longer be suppressed with respect to those of the dynamically  
favoured $I=0$ operators  
\begin{equation}\label{I0-ops2} 
{\cal Q}_{I=0}^{\overline{s}s}\sim (\overline{s}s)(\overline{b}s), 
\end{equation} 
and would also contribute to $v_0^{(k)}$ at ${\cal O}(1)$. A similar  
comment applies to the matrix elements of the $I=0$ operators with 
the following flavour content: 
\begin{equation}\label{I0-ops3} 
{\cal Q}_{I=0}^{\overline{c}c}\sim (\overline{c}c)(\overline{b}s), 
\end{equation} 
whose dynamical suppression in $B\to\phi K$ decays may be reduced through 
rescattering effects of the kind given in (\ref{rescatter-c}), which may 
also affect the ${\cal A}_{\rm SM}^{(0)}$ amplitude, as we have noted 
above. Consequently, in the presence of large rescattering effects,  
the decay amplitude (\ref{ampl-hier1-phi}) is modified as follows: 
\begin{equation}\label{ampl-hier2-phi} 
\left.A(B\to \phi K)\right|_{\rm res.}= 
\left.{\cal A}_{\rm SM}^{(0)}\right|_{\rm res.}\times 
\biggl[1+\underbrace{{\cal O}(1)}_{{\rm NP}_{I=0}}+ 
\underbrace{{\cal O}(1)}_{{\rm NP}_{I=1}}+ 
\underbrace{{\cal O}(\overline{\lambda})}_{{\rm SM}}\biggr]. 
\end{equation} 
Let us emphasize that the rescattering contributions to the  
prefactor on the right-hand side of this equation are due to  
(\ref{rescatter-c}), whereas the hierarchy in square brackets is  
governed by large rescattering processes of the type described  
by (\ref{rescatter-u}).

\boldmath 
\subsubsection{Observables for a General Analysis of 
New Physics}\label{sec:obs} 
\unboldmath 
The observables provided by the $B\to\phi K$ system are completely
analogous to the set of $B\to J/\psi K$ observables introduced in
(\ref{def-A}) and (\ref{S-D-def}). The corresponding observables 
${\cal S}$, ${\cal D}$ and ${\cal B}$ allow us to separate the $I=0$ 
contributions from the $I=1$ sector, and play a key r\^ole in the search 
for new physics. Moreover, they may provide valuable insights into the 
$B\to \phi K$ hadron dynamics. In the case of the decay
$B_d\to \phi K_{\rm S}$, mixing-induced CP violation yields an additional 
observable ${\cal A}_{\rm CP}^{\rm mix}(B_d\to \phi K_{\rm S})$. Explicit
expressions for these quantities, which are rather complicated, 
can be found in \cite{FM-PhiK}; 
let us here just note that they are governed by
\begin{equation}\label{S-D-simple}
{\cal S}\sim v_0,\quad {\cal D}\sim v_1,\quad
{\cal B}\sim v_1.
\end{equation}

\subsubsection{Possible Scenarios}
An interesting probe to search for new physics is provided by a 
comparison of the mixing-induced CP asymmetries in $B_d\to \phi K_{\rm S}$
and $B_d\to J/\psi K_{\rm S}$ \cite{RF-rev,growo,LoSo}. Using the 
hierarchies in (\ref{ampl-hier1}) and (\ref{ampl-hier1-phi})
yields the following relation \cite{FM-PhiK}:
\begin{equation}\label{mix-diff1} 
{\cal A}^{\mbox{{\scriptsize mix}}}_{\mbox{{\scriptsize CP}}} 
(B_d\to \phi K_{\rm S})- 
{\cal A}^{\mbox{{\scriptsize mix}}}_{\mbox{{\scriptsize CP}}} 
(B_d\to J/\psi K_{\rm S})=\underbrace{{\cal O}(1)}_{{\rm NP}_{I=0}} + 
\underbrace{{\cal O}(\overline{\lambda})}_{{\rm NP}_{I=1}} 
+\underbrace{{\cal O}(\overline{\lambda}^2)}_{\rm SM}\, , 
\end{equation} 
where the $\sin\phi$ terms, which may also deviate from the Standard-Model  
expectation, cancel. The contributions entering at the  
$\overline{\lambda}$ and $\overline{\lambda}^2$ levels may also contain  
new-physics effects from $B_d\to J/\psi K_{\rm S}$, whereas the  
${\cal O}(1)$ term would essentially be due to new physics in the
$B_d\to \phi K_{\rm S}$ channel. 

As can be seen in (\ref{S-D-simple}), ${\cal S}$ provides a ``smoking-gun'' 
signal for new-physics contributions to the $I=0$ amplitude, whereas 
${\cal D}$ and ${\cal B}$ probe new-physics effects in the $I=1$ sector. 
If we employ the hierarchy arising in (\ref{ampl-hier1-phi}), we obtain
\begin{equation}\label{obs-hier1} 
{\cal S}=\underbrace{{\cal O}(1)}_{{\rm NP}_{I=0}}+ 
\underbrace{{\cal O}(\overline{\lambda}^2)}_{\rm SM},\quad 
{\cal D}=\underbrace{{\cal O}(\overline{\lambda})}_{{\rm NP}_{I=1}} 
+\underbrace{{\cal O}(\overline{\lambda}^2)}_{\rm SM},\quad 
{\cal B}=\underbrace{{\cal O}(\overline{\lambda})}_{{\rm NP}_{I=1}} 
+\underbrace{{\cal O}(\overline{\lambda}^2)}_{\rm SM}, 
\end{equation} 
where the Standard-Model contributions are not under theoretical control.  
If the dynamical suppression of the $I=1$ contributions were larger 
than ${\cal O}(\overline{\lambda})$, ${\cal D}$ and ${\cal B}$ would be  
further suppressed with respect to ${\cal S}$. On the other hand, if  
the rescattering effects described by (\ref{rescatter-u}) were very  
large -- and not small, as assumed in (\ref{obs-hier1}) -- {\it all} three
observables would be of ${\cal O}(1)$. In such a situation, we would not  
only have signals for physics beyond the Standard Model, but also for  
large rescattering processes.  

The discussion given above corresponds to the most optimistic scenario 
concerning the generic strength of possible new-physics effects in the  
$B\to \phi K$ system. Let us now consider a more pessimistic case,  
where the new-physics contributions are smaller by a factor of  
${\cal O}(\overline{\lambda})$: 
\begin{equation}\label{ampl-hier3-phi} 
A(B\to \phi K)={\cal A}_{\rm SM}^{(0)}\biggl[1+ 
\underbrace{{\cal O}(\overline{\lambda})}_{{\rm NP}_{I=0}}+ 
\underbrace{{\cal O}(\overline{\lambda}^2)}_{{\rm NP}_{I=1}}+ 
\underbrace{{\cal O}(\overline{\lambda}^2)}_{{\rm SM}}\biggr]. 
\end{equation} 
Now the new-physics contributions to the $I=1$ sector can no longer be  
separated from the Standard-Model contributions. However, we  
would still get an interesting pattern for the $B\to \phi K$ observables, 
providing evidence for new physics: whereas (\ref{mix-diff1}) and  
${\cal S}$ would both be sizeable, i.e.\ of ${\cal O}(10\%)$ and within  
reach of the $B$-factories, ${\cal D}$ and ${\cal B}$ would be strongly  
suppressed. However, if these two observables, in addition to  
(\ref{mix-diff1}) and ${\cal S}$, are found to be also at the $10\%$  
level, new physics cannot be distinguished from Standard-Model  
contributions, which could also be enhanced to the $\overline{\lambda}$  
level by large rescattering effects. This would be the most unfortunate 
case for the search for new-physics contributions to the $B\to \phi K$ 
decay amplitudes \cite{FM-PhiK}. As we have seen in 
Subsection~\ref{subsec:BpsiK}, there is an analogous case in the
$B\to J/\psi K$ system, which does also not appear to be a very likely 
scenario.

In general, $B\to J/\psi K$ and $B\to \phi K$ modes offer powerful
tools to test the Standard-Model description of CP violation. Hopefully,
the experimental data for these decays, which are very accessible at
the $B$-factories, will shed light on the physics beyond the Standard Model.
A decay lying in some sense between these channels is the transition 
$B_d\to\pi^+\pi^-$, which is our next topic: whereas $B\to J/\psi K$ and 
$B\to \phi K$ are governed by tree-diagram-like and penguin contributions, 
respectively, both topologies play an important r\^ole in $B_d\to\pi^+\pi^-$. 
This feature leads to serious problems in the extraction of $\alpha$ from 
the mixing-induced CP asymmetry of $B_d\to\pi^+\pi^-$, introducing large
hadronic uncertainties into the corresponding Standard-Model expression.

\boldmath
\subsection{The $B\to\pi\pi$ System}\label{subsec:Bpipi}
\unboldmath
\boldmath
\subsubsection{Probing $\alpha$ through $B_d\to\pi^+\pi^-$}
\begin{figure}
\begin{center}
\leavevmode
\epsfysize=4.0truecm 
\epsffile{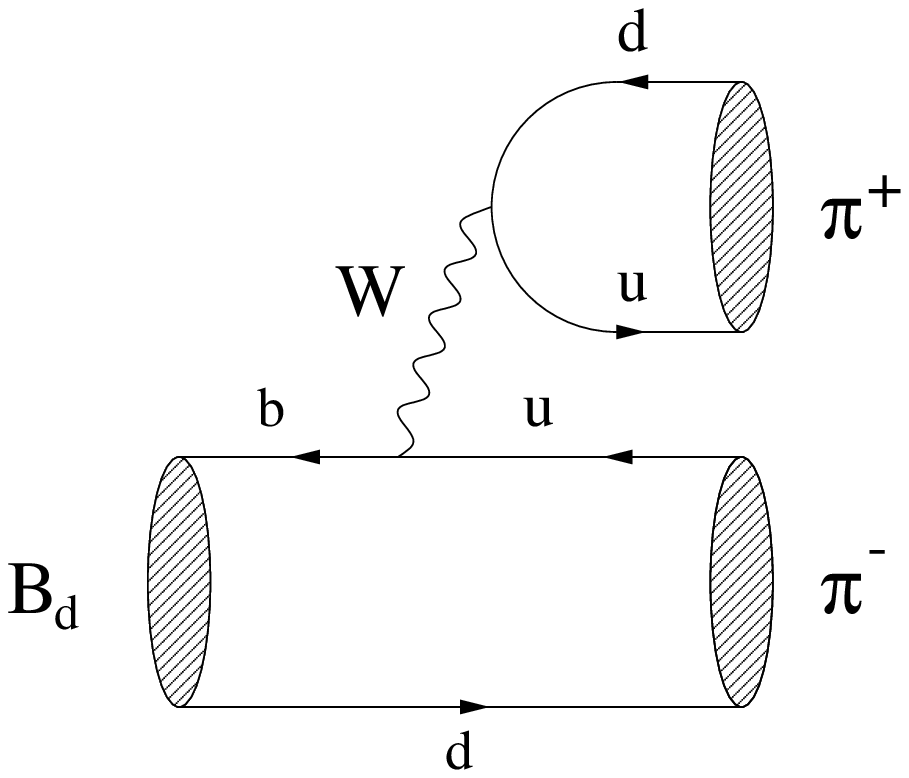} \hspace*{1truecm}
\epsfysize=4.5truecm 
\epsffile{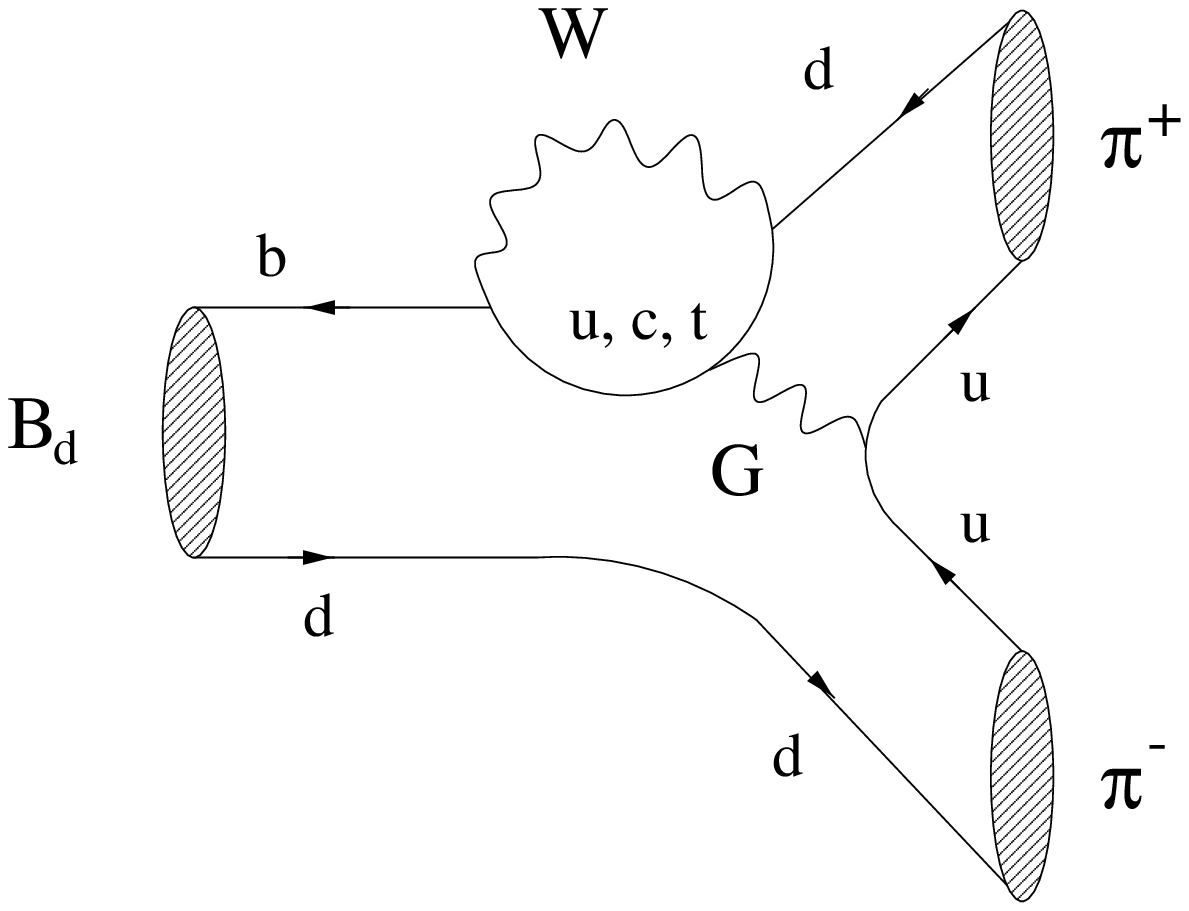}
\end{center}
\vspace*{-0.3truecm}
\caption{Feynman diagrams contributing to 
$B_d^0\to\pi^+\pi^-$.}\label{fig:bpipi}
\end{figure}

\unboldmath
The transition $B_d^0\to\pi^+\pi^-$ is a decay into a CP eigenstate 
with eigenvalue $+1$, and originates from 
$\overline{b}\to\overline{u}u\overline{d}$ quark-level processes, as 
can be seen in Fig.\ \ref{fig:bpipi}. Within the Standard Model, the 
corresponding decay amplitude can be expressed -- in analogy to 
(\ref{Bd-ampl1}) -- in the following way \cite{RF-BsKK}:
\begin{equation}\label{Bpipi-ampl}
A(B_d^0\to\pi^+\pi^-)=\lambda_u^{(d)}\left(A_{\rm CC}^{u}+
A_{\rm pen}^{u}\right)+\lambda_c^{(d)}A_{\rm pen}^{c}+
\lambda_t^{(d)}A_{\rm pen}^{t}
={\cal C}\left(e^{i\gamma}-d e^{i\theta}\right),
\end{equation}
where
\begin{equation}\label{C-DEF}
{\cal C}\equiv\lambda^3A\,R_b\left(A_{\rm CC}^{u}+A_{\rm pen}^{ut}\right)
\quad\mbox{with}\quad A_{\rm pen}^{ut}\equiv A_{\rm pen}^{u}-A_{\rm pen}^{t},
\end{equation}
and
\begin{equation}\label{D-DEF}
d e^{i\theta}\equiv\frac{1}{R_b}
\left(\frac{A_{\rm pen}^{ct}}{A_{\rm CC}^{u}+A_{\rm pen}^{ut}}\right).
\end{equation}
In contrast to the $B_d^0\to J/\psi K_{\rm S}$ amplitude (\ref{Bd-ampl2}), 
the ``penguin'' parameter $d e^{i\theta}$ does {\it not} enter in
(\ref{Bpipi-ampl}) in a doubly Cabibbo-suppressed way. If we assume, for 
a moment, that $d=0$, the formalism discussed in 
Subsection~\ref{subsec:CPasym} yields
\begin{equation}\label{bpipi-ideal}
{\cal A}^{\mbox{{\scriptsize mix}}}_{\mbox{{\scriptsize
CP}}}(B_d\to\pi^+\pi^-)=-\sin[-(2\beta+2\gamma)]=-\sin2\alpha,
\end{equation}
which would allow a determination of $\alpha$. Unfortunately, it is 
expected that this relation is strongly affected by penguin effects, 
which were analysed by many authors over the last couple of years
\cite{charming-pens1,charming-pens2,BBNS1,BBNS3,alpha-uncert,SiWo}.

Large penguin effects in $B_d\to\pi^+\pi^-$ are also indicated by the 
$B$-factory data. In the summer of 1999, the observation of this 
channel was announced by the CLEO collaboration. The present results 
for its CP-averaged branching ratio read as follows:
\begin{equation}
\mbox{BR}(B_d\to\pi^+\pi^-)\times 10^{6}=\left\{\begin{array}{ll}
4.3^{+1.6}_{-1.4}\pm0.5 & \mbox{(CLEO \cite{CLEO-BpiK})}\\
4.1\pm1.0\pm0.7 & \mbox{(BaBar \cite{babar-BpiK})}\\
5.6^{+2.3}_{-2.0}\pm 0.4 & \mbox{(Belle \cite{belle-BpiK}).}
\end{array}\right.
\end{equation}
Using $SU(3)$ flavour-symmetry arguments and plausible dynamical assumptions,
the CP-averaged $B_d\to\pi^+\pi^-$ branching ratio can be combined with 
that of $B_d\to\pi^\mp K^\pm$  \cite{SiWo} to derive constraints on the 
penguin parameter $d$ \cite{RF-bpipi}. The present $B$-factory results 
imply $d\gsim 0.2$, as we will discuss in more detail in 
\ref{subsubsec:U-spin-Replace} (see Fig.~\ref{fig:d}). Moreover, 
theoretical considerations are also in favour of sizeable values of
$d$ (see (\ref{d-QCD-fact})). Consequently, we have 
already strong evidence that the approximation $d=0$, i.e.\ the neglect 
of penguins in $B_d\to\pi^+\pi^-$, is not justified. 

Concerning the direct and mixing-induced CP-violating observables of
$B_d\to\pi^+\pi^-$, first results are already available from 
the BaBar collaboration \cite{BABAR-Bpipi-CP}:
\begin{equation}\label{Bpipi-CP-asym-res}
{\cal A}_{\rm CP}^{\rm dir}(B_d\to\pi^+\pi^-)=(-25^{+45}_{-47}\pm14)\%,
\quad
{\cal A}_{\rm CP}^{\rm mix}(B_d\to\pi^+\pi^-)=(-3^{+56}_{-53}\pm11)\%.
\end{equation}
Needless to note, because of the large experimental uncetainties, no
conclusions can be drawn at present, although a direct CP asymmetry
at the level of $25\%$ would also imply large penguin effects.

\begin{figure}
\centerline{
\epsfxsize=8.0truecm
\epsffile{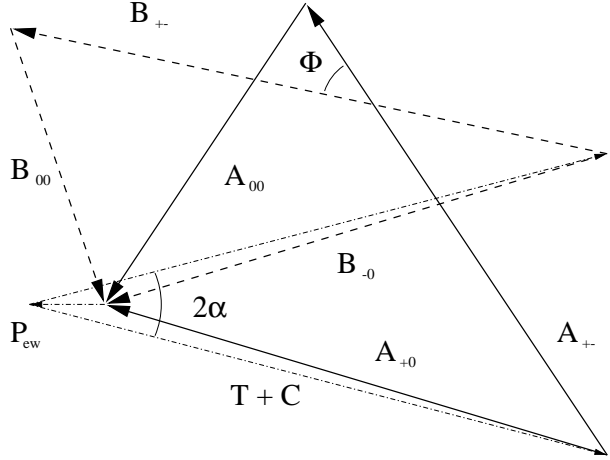}}
\caption{Illustration of the $B\to\pi\pi$ isospin triangles. 
Here the amplitudes $A$ correspond to (\ref{su21}); the amplitudes 
$B$ correspond to the ones in (\ref{su22}), rotated by 
$e^{-2i\beta}$.}\label{fig:bpipi-isospin}
\end{figure}

\boldmath
\subsubsection{Isospin Relations between $B\to\pi\pi$ Amplitudes}
\unboldmath
There are various strategies to control the penguin uncertainties affecting
the extraction of $\alpha$ with the 
help of additional experimental data. The best known approach was proposed 
by Gronau and London~\cite{GL}, employing isospin relations between the
$B\to\pi\pi$ decay amplitudes. Since $B^-\to\pi^-\pi^0$ is a 
$\Delta I=3/2$ transition, the QCD penguin operators (\ref{qcd-penguins}) 
with $r=d$, which mediate $\Delta I=1/2$ transitions, do not contribute. 
Consequently, if we neglect EW penguins for a moment, this channels 
receives only contributions from colour-allowed and colour-suppressed 
tree-diagram-like topologies, which are described by strong amplitudes 
$T$ and $C$, respectively, and we obtain
\begin{equation}
A(B^\pm\to\pi^\pm\pi^0)=e^{\pm i\gamma}e^{i\delta{T+C}}|T+C|,
\end{equation}
yielding
\begin{equation}\label{rel1}
\frac{A(B^+\to\pi^+\pi^0)}{A(B^-\to\pi^-\pi^0)}=e^{2i\gamma}.
\end{equation}
Moreover, the isospin symmetry implies the following amplitude 
relations \cite{GL}:
\begin{eqnarray}
\sqrt{2}\,A(B^+\to\pi^+\pi^0)&=&A(B^0_d\to\pi^+\pi^-)+
\sqrt{2}\,A(B^0_d\to\pi^0\pi^0)\label{su21}\\
\sqrt{2}\,A(B^-\to\pi^-\pi^0)&=&A(\overline{B^0_d}\to\pi^+\pi^-)+
\sqrt{2}\,A(\overline{B^0_d}\to\pi^0\pi^0),\label{su22}
\end{eqnarray}
which can be represented as two triangles in the complex plane. These
triangles, i.e.\ their sides, can be fixed through the six $B\to\pi\pi$ 
branching ratios. In order to determine also their relative orientation, 
we rotate the CP-conjugate triangle by $e^{-2i\beta}$. The resulting 
situation is illustrated in Fig.~\ref{fig:bpipi-isospin}, where the angle 
$\Phi$ can be extracted from mixing-induced CP violation \cite{PAPI}: 
applying the formalism discussed in Subsection~\ref{subsec:CPasym},
we obtain
\begin{equation}
\xi_{\pi^+\pi^-}^{(d)}=-e^{-i2\beta}\left[\frac{A(\overline{B^0_d}\to
\pi^+\pi^-)}{A(B_d\to\pi^+\pi^-)}\right]=-\left|\frac{B_{+-}}{A_{+-}}\right|
e^{i\Phi},
\end{equation}
which implies
\begin{equation}
{\cal A}^{\mbox{{\scriptsize mix}}}_{\mbox{{\scriptsize
CP}}}(B_d\to\pi^+\pi^-)=-\,\frac{2|A_{+-}||B_{+-}|}{|A_{+-}|^2+
|B_{+-}|^2}\sin\Phi\,.
\end{equation}
If we use (\ref{rel1}), and take into account that the CP-conjugate 
triangle was rotated by $e^{-2i\beta}$, we conclude that the angle 
between the $B^+\to\pi^+\pi^0$ and $B^-\to\pi^-\pi^0$ amplitudes is 
given by $2\alpha$. For simplicity, we have chosen $\phi_{\rm CP}=0$ 
in the discussion given above. The corresponding determination of 
$\alpha$ does, of course, not depend on the choice of this CP phase. 
It is an easy exercise to convince ourselves from this feature. 

On the other hand, the EW penguin amplitude $P_{\rm ew}$, which we have
neglected so far, does affect this extraction of $\alpha$, as can be seen 
in Fig.~\ref{fig:bpipi-isospin}. Although the determination of the sides of
the isospin triangles and of their relative orientation remains unchanged, 
the angle between $A_{+0}$ and $B_{-0}$ is shifted from $\alpha$ by
$\Delta\alpha={\cal O}(|P_{\rm ew}|/|T+C|)$, which corresponds to a
small correction of at most a few degrees \cite{ghlr-ewp,PAPIII}. As 
was noticed recently \cite{BF-BpiK1,GPY}, also the EW penguin contribution 
can be taken into account with the help of the $SU(2)$ isospin symmetry, 
yielding
\begin{equation}\label{q-pipi-expr}
\left[\frac{P_{\rm ew}}{T+C}\right]=
-\,1.3\times 10^{-2}\times\left|\frac{V_{td}}{V_{ub}}\right|e^{i\alpha},
\end{equation}
where the numerical factor depends only on Wilson coefficients of EW 
penguin and CC operators, i.e.\ does not involve hadronic matrix elements.

Isospin is broken not only by the quark charges, as in the EW 
penguin operators, but also by the up- and down-quark mass difference, which 
generates $\pi^0$--$\eta,\eta'$ mixing and converts the isospin triangle 
relations (\ref{su21}) and (\ref{su22}) between the $B\to\pi\pi$ amplitudes 
into quadrilaterals. The impact of these isospin-violating effects
on the extraction of $\alpha$ was analysed in \cite{gardner}, and was 
found to be significant if $\sin 2\alpha$ is small.

Unfortunately, the $B\to\pi\pi$ triangle approach is very challenging from
an experimental point of view, since it requires a measurement of the
$B_d^0\to\pi^0\pi^0$ branching ratio and its CP conjugate. Because of the
two neutral pions in the final state, this mode is very difficult to
reconstruct. Moreover, the corresponding CP-averaged branching ratio is
expected to be very small, i.e.\ BR$\left.(B_d\to\pi^0\pi^0)\right|_{\rm TH}
\mathrel{\hbox{\rlap{\hbox{\lower4pt\hbox{$\sim$}}}\hbox{$<$}}}
{\cal O}(10^{-6})$. In a recent analysis, the CLEO collaboration has 
obtained the upper limit BR$(B_d\to\pi^0\pi^0)<5.7\times10^{-6}$ 
(90\% C.L.) \cite{cleo-pi0pi0}. Such upper bounds may be useful to constrain
the QCD penguin uncertainties affecting the determination of $\alpha$
from $B_d\to\pi^+\pi^-$ decays \cite{GrQu}--\cite{GLSS}.

\boldmath
\subsubsection{Extracting $\alpha$ from $B\to\rho\pi$ Modes}
\unboldmath
Because of the $B_d\to\pi^0\pi^0$ problem of the Gronau--London approach
to extract $\alpha$ from $B\to\pi\pi$ isospin relations, alternative 
strategies are very desirable. An interesting one is provided by 
$B\to\rho\pi$ modes \cite{Brhopi}. These decays are more complicated than 
the $B\to\pi\pi$ system, since their final states consist of the three 
different isospin configurations $I=0,1,2$ instead of $I=0,2$. Performing
an isospin analysis of the decays $B^+\to\rho^+\pi^0$, 
$B^+\to\rho^0\pi^+$, $B^0_d\to\rho^+\pi^-$, $B^0_d\to\rho^-\pi^+$,
$B^0_d\to\rho^0\pi^0$ and of their charge conjugates 
yields the following two pentagonal relations:
\begin{equation}
\sqrt{2}\left(A^{+0}+A^{0+}\right)=A^{+-}+A^{-+}+2\,A^{00}
\end{equation}
\begin{equation}
\sqrt{2}\left(\overline{A}^{+0}+\overline{A}^{0+}\right)=
\overline{A}^{+-}+\overline{A}^{-+}+2\,\overline{A}^{00},
\end{equation}
which correspond to the $B\to\pi\pi$ triangle relations (\ref{su21}) and 
(\ref{su22}).
The ten $B\to\rho\pi$ rates allow us to fix the sides of both isospin 
pentagons. Measuring in addition mixing-induced CP violation 
in $B_d\to \rho^+\pi^-, \rho^-\pi^+, \rho^0\pi^0$, it is also possible 
to determine $\alpha$.

\begin{figure}
\begin{center}
\leavevmode
\epsfysize=3.0truecm 
\epsffile{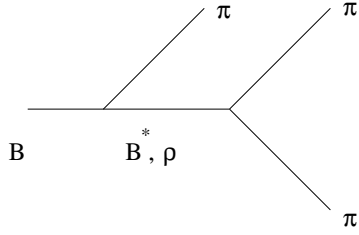}
\end{center}
\vspace*{-0.6truecm}
\caption{Polar diagrams contributing to $B\to\pi\pi\pi$.}\label{fig:polar}
\end{figure}

In practice, this approach is obviously quite complicated and suffers from
multiple discrete ambiguities. In order to avoid them, Quinn 
and Snyder have considered a maximum-likelihood fit to the parameters of
the full Dalitz plot distribution, and found that it is possible to extract
$\alpha$, as well as other parameters, with ${\cal O}(10^{3})$ 
Monte--Carlo-generated events \cite{qs}. Here the basic assumption is 
that the $B\to3\pi$ events are fully dominated by $B\to\rho\pi$. Taking
into account detector efficiencies and background effects, one finds that
${\cal O}(10^{4})$ $B\to\rho\pi$ events are required, corresponding to 
${\cal O}(10^{9})$ $B \overline{B}$ events, which may be beyond of what 
can be achieved in first-generation $B$-factory experiments. The approach 
appears even to be challenging for the LHC era. 

To attack this problem, strategies to use ``early'' data on 
$B\to\rho\pi$ modes were recently proposed in \cite{qusi}, employing Dalitz 
plot analyses of $B_d \to \rho \pi \to \pi^+ \pi^- \pi^0$ decays. 
In that paper, it was pointed out that important parameters -- unfortunately 
not $\alpha$ -- can be determined from untagged and tagged time-integrated 
measurements, suggesting that the extraction of $\alpha$ from the 
time-dependent data sample can be accomplished with a smaller data sample 
than would be required if all parameters were to be obtained from that 
time-dependent data sample alone. 

Another recent development concerning $B\to\rho\pi$ modes are studies of  
the impact of the ``polar diagrams'' shown in Fig.~\ref{fig:polar}.
As was pointed out in \cite{dea}, these processes may lead to significant 
effects in $B^\mp\to\pi^\mp\pi^\mp\pi^\pm$ decays, whereas they are 
expected to be negligible in other charged decays. In the neutral case, only 
$B_d\to\rho^0\pi^0$ transitions may be affected. Since the processes induced 
by such polar diagrams represent an irreducible background in the Dalitz 
plot, charged $B\to\rho\pi$ and neutral $B_d\to\rho^0\pi^0$ channels should 
be discarded from the extraction of $\alpha$ \cite{dea}. Further studies of 
these issues are desirable.

\boldmath
\subsubsection{Other Approaches to Extract $\alpha$}
\unboldmath
Another strategy to extract $\alpha$ is provided by 
$B^\pm\to\rho^\pm\rho^0(\omega)\to\rho^\pm\pi^+\pi^-$ decays, where 
$\rho^0(\omega)$ denotes the $\rho^0\,$--$\,\omega$ interference 
region \cite{et-got}. In this case, experimental data on 
$e^+e^-\to\pi^+\pi^-$ processes can be used to constrain the hadronic 
uncertainties affecting the corresponding direct CP asymmetry, which 
is related to $\sin\alpha$ and may well be as large as ${\cal O}(20\%)$ 
at the $\omega$ invariant mass. In this context, it is worth mentioning 
that also direct CP violation in three-body decays such as 
$B^\pm\to K^\pm\pi^+\pi^-$, involving various intermediate 
resonances, was considered \cite{eos}. Here the Dalitz plot distributions 
may provide information on the CKM angle $\gamma$ (see also \cite{bbgm}). 

A possibility to eliminate the penguin uncertainties in the extraction 
of $\alpha$ from mixing-induced CP violation in $B_d\to\pi^+\pi^-$ is also
provided by the $B_d\to K^0\overline{K^0}$ channel, which can be related 
to $B_d\to\pi^+\pi^-$ through the $SU(3)$ flavour symmetry of strong 
interactions \cite{BF-alpha}. 

Another -- rather simple -- strategy to determine $\alpha$ was proposed 
in \cite{FM-alpha}. Using the unitarity of the CKM matrix to eliminate
$\lambda_c^{(d)}$, we may rewrite (\ref{Bpipi-ampl}) as follows:
\begin{equation}\label{j_tp}
A(B_d^0\to\pi^+\pi^-)=e^{i\gamma}\,T+e^{-i\beta}\,P\,,
\end{equation}
where 
\begin{equation}
T\equiv\left|\lambda_u^{(d)}\right|\left(A_\mathrm{CC}^u+A_\mathrm{pen}^u-
A_\mathrm{pen}^c\right) \quad\mbox{and}\quad 
P\equiv\left|\lambda_t^{(d)}\right|\left(A_\mathrm{pen}^t-
A_\mathrm{pen}^c\right).
\end{equation}
For the following considerations, also the CP-conserving 
strong phase $\delta\equiv{\rm arg}(PT^\ast)$
plays an important r\^ole. Since the $B^0_d$--$\overline{B^0_d}$ mixing
phase is given by $2\beta$ in the Standard Model, the unitarity relation
$\alpha+\beta+\gamma=180^\circ$ allows us to express the CP-violating
observables ${\cal A}_{\rm CP}^{\rm dir}(B_d\to\pi^+\pi^-)$ and
${\cal A}_{\rm CP}^{\rm mix}(B_d\to\pi^+\pi^-)$ as functions
of the CKM angle $\alpha$ and the two hadronic parameters $|P/T|$ and 
$\delta$. Consequently, we have two observables at our disposal, depending 
on three ``unknowns''. The corresponding expressions simplify considerably, 
if we keep only the leading-order terms in $x\equiv|P/T|$, yielding
\begin{eqnarray}
{\cal A}_{\rm CP}^{\rm dir}(B_d\to\pi^+\pi^-)&=&2x\sin\delta\sin\alpha+
{\cal O}(x^2)\label{Adir-simp}\\
{\cal A}_{\rm CP}^{\rm mix}(B_d\to\pi^+\pi^-)&=&
-\sin2\alpha-2x\cos\delta\cos2\alpha\sin\alpha+
{\cal O}(x^2).\label{Amix-simp}
\end{eqnarray}
If the parameter $x$ is fixed through an additional input, both $\alpha$
and the strong phase $\delta$ can be determined from (\ref{Adir-simp})
and (\ref{Amix-simp}). To this end, we may use, for example, the $SU(3)$ 
flavour symmetry and certain dynamical assumptions \cite{FM-alpha}:
\begin{equation}
x \approx \lambda R_t\left(\frac{f_{\pi}}{f_K}\right)
\sqrt{\frac{\mbox{BR}(B^+\to\pi^+K^0)}{2\,\mbox{BR}(B^+\to\pi^+\pi^0)}}.
\end{equation}

Refinements of this approach were discussed in \cite{charles}. It 
should be emphasized that 
${\cal A}_{\rm CP}^{\rm dir}(B_d\to\pi^+\pi^-)$ and
${\cal A}_{\rm CP}^{\rm mix}(B_d\to\pi^+\pi^-)$ allow us to fix 
contours in the $\alpha$--$x$ plane in a {\it theoretically clean} way. 
Unfortunately, it appears very challenging to determine also $x$,
which would allow the extraction of $\alpha$, in a theoretically reliable 
manner. In order not to comprise the LHC year-1 statistics, $x$ would be 
required with a theoretical uncertainty smaller than $10\%$ \cite{LHC-Report}. 
Despite the recent theoretical progress made in \cite{BBNS1}--\cite{PQCD-comp},
it appears questionable whether such an accuracy can eventually be 
achieved (see \cite{charming-pens2} and Subsection \ref{subsec:BpiK-calc}). 
Moreover, any QCD-based approach to 
calculate $x$ requires also the CKM factor $|V_{td}/V_{ub}|$. This input 
can be avoided, if all weak phases are expressed in terms of the generalized 
Wolfenstein parameters $\overline{\rho}$ and $\overline{\eta}$, allowing us 
to fix contours in the $\overline{\rho}$--$\overline{\eta}$ plane through the 
$B_d\to\pi^+\pi^-$ observables~\cite{charles}.

Let us finally note that a particularly promising strategy is 
provided by the decay $B_s\to K^+K^-$, which is related to 
$B_d\to\pi^+\pi^-$ by interchanging all down and strange quarks, 
i.e.\ through the $U$-spin flavour symmetry of strong interactions. 
A combined analysis of these two channels allows a simultaneous 
determination of $\beta$ and $\gamma$ \cite{RF-BsKK}, which has 
certain theoretical advantages, appears to be promising for Tevatron-II 
\cite{wuerth}, and is ideally suited for the LHC era \cite{LHC-Report}. 
This approach will be discussed in Subsection~\ref{subsec:BsKK-Uspin}.

\boldmath
\subsection{$B_d\to D^{(\ast)\pm}\pi^{\mp}$ Decays}\label{subsec:BDpi}
\begin{figure}
\vspace*{-0.3truecm}
\begin{center}
\leavevmode
\epsfysize=4.0truecm 
\epsffile{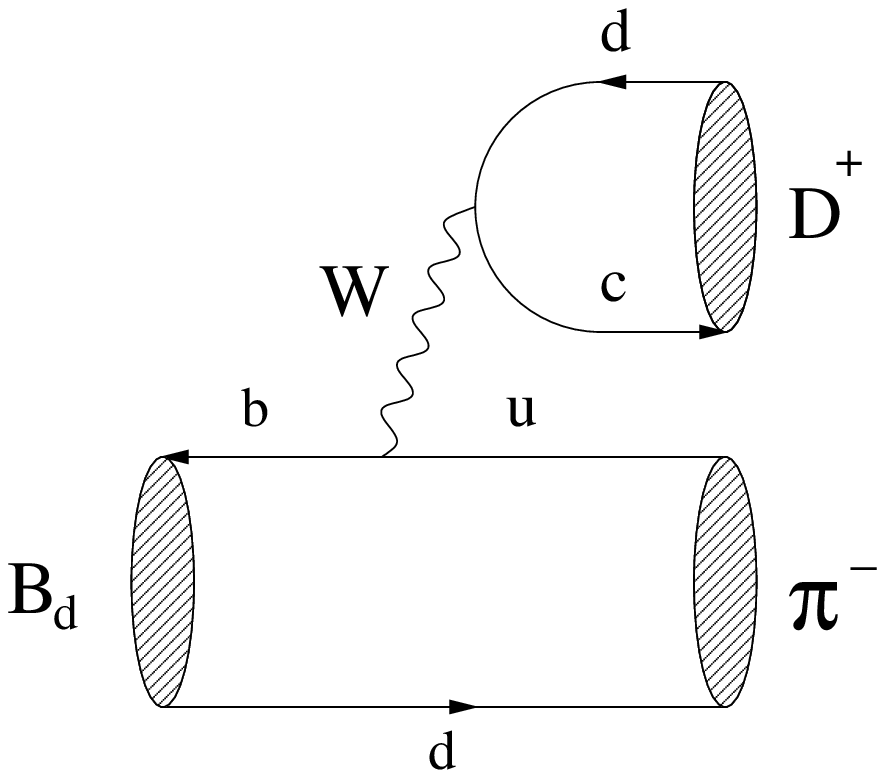} \hspace*{1truecm}
\epsfysize=4.0truecm 
\epsffile{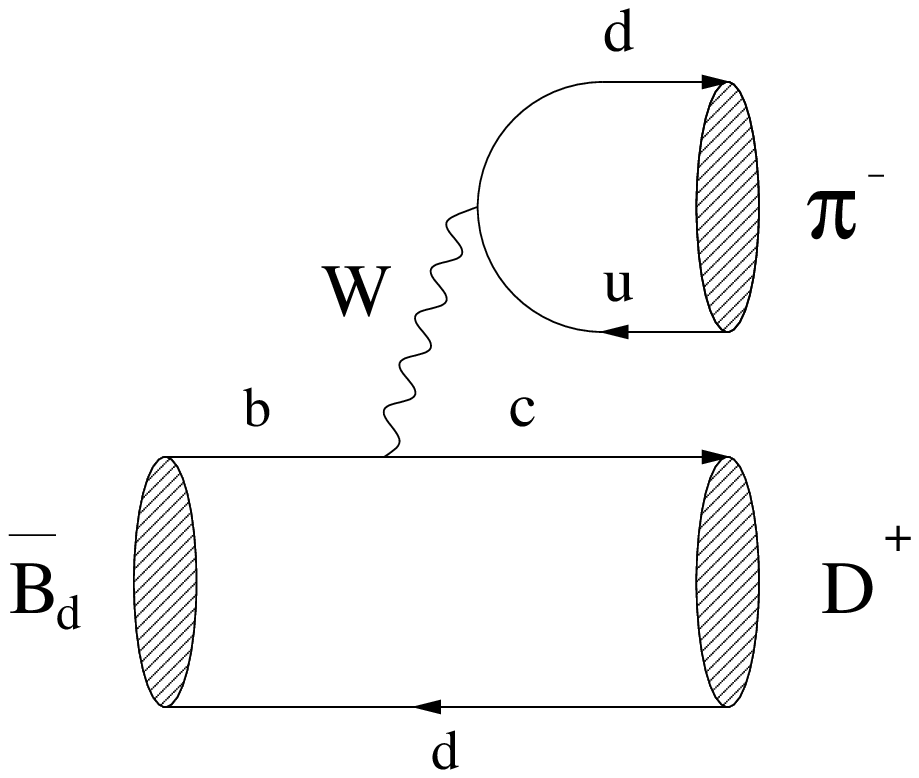}
\end{center}
\vspace*{-0.6truecm}
\caption[]{Feynman diagrams contributing to $B^0_d, \overline{B^0_d}\to 
D^{(\ast)+}\pi^-$ decays.}\label{fig:BDpi}
\end{figure}

\unboldmath
So far, we have put a strong emphasis on neutral $B$ decays into  
CP eigenstates. However, in order to extract angles of the unitarity 
triangle, there are also interesting decays into final states that are 
{\it not} eigenstates of the CP operator. An important example is given 
by $B_d\to D^{(\ast)\pm}\pi^{\mp}$ decays, which are mediated by 
$\overline{b}\to\overline{u}c\overline{d}$ ($b\to c\overline{u}d$) 
quark-level processes and receive hence only contributions from 
tree-diagram-like topologies. As can be seen in Fig.~\ref{fig:BDpi}, 
$B^0_d$- and $\overline{B^0_d}$-mesons may both decay into 
$D^{(\ast)+}\pi^-$, thereby leading to interference effects between 
$B^0_d$--$\overline{B^0_d}$ mixing and decay processes.
These interference effects allow an interesting -- theoretically clean -- 
determination of $2\beta+\gamma$.

The relevant transition amplitudes can be expressed as hadronic matrix 
elements of 
\begin{eqnarray}
{\cal H}_{\mbox{{\scriptsize eff}}}(\overline{B^0_d}\to f)&=&
\frac{G_{\mbox{{\scriptsize F}}}}{\sqrt{2}}\,\overline{v}
\left[\overline{{\cal O}}_1\,{ C}_1(\mu)+
\overline{{\cal O}}_2\,{ C}_2(\mu)\right]\\
{\cal H}_{\mbox{{\scriptsize eff}}}(B^0_d\to f)&=&
\frac{G_{\mbox{{\scriptsize F}}}}{\sqrt{2}}\,v^\ast
\left[{\cal O}_1^\dagger\,{ C}_1(\mu)+{\cal O}_2^\dagger\,
{ C}_2(\mu)\right],
\end{eqnarray}
where $f$ is a final state with valence-quark content 
$c\overline{d}\,d\overline{u}$, for example $D^{\ast+}\pi^-$,
and $\overline{{\cal O}}_k$ and ${\cal O}_k$ denote current--current 
operators, which are given by
\begin{equation}
\begin{array}{rclrcl}
\overline{{\cal O}}_1&=&(\overline{d}_\alpha u_\beta)_{\mbox{{\scriptsize 
V--A}}}\left(\overline{c}_\beta b_\alpha\right)_{\mbox{{\scriptsize V--A}}},&
~~\overline{{\cal O}}_2&=&(\overline{d}_\alpha u_\alpha)_{\mbox{{\scriptsize 
V--A}}}\left(\overline{c}_\beta b_\beta\right)_{\mbox{{\scriptsize V--A}}},\\
{\cal O}_1&=&(\overline{d}_\alpha c_\beta)_{\mbox{{\scriptsize V--A}}}
\left(\overline{u}_\beta b_\alpha\right)_{\mbox{{\scriptsize V--A}}},&
~~{\cal O}_2&=&(\overline{d}_\alpha c_\alpha)_{\mbox{{\scriptsize V--A}}}
\left(\overline{u}_\beta b_\beta\right)_{\mbox{{\scriptsize 
V--A}}},
\end{array}
\end{equation}
and are analogous to the ones we encountered in \ref{subsec:Tree-Ham};
their NLO Wilson coefficients can be calculated with the help of 
(\ref{Opm-def})--(\ref{coeff3}). Using (\ref{NLO-wolf}) and
(\ref{CKM-UT-ANGLES}) yields
\begin{equation}
\overline{v}\equiv V_{ud}^\ast V_{cb}=\left(1-\frac{\lambda^2}{2}\right)
A\lambda^2,\quad v\equiv V_{cd}^\ast V_{ub}=-A\lambda^4\left(
\frac{R_b}{1-\lambda^2/2}\right)e^{-i\gamma}.
\end{equation} 
On the other hand, taking into account CP relations (\ref{CP-def}) and 
(\ref{CP-op-trafo}), we obtain
\begin{eqnarray}
\lefteqn{\left\langle f\left|{\cal O}_1^\dagger(\mu){ C}_1(\mu)+
{\cal O}_2^\dagger(\mu){ C}_2(\mu)\right|B^0_d\right\rangle}
\nonumber\\
&&=\left\langle f\left|({\cal CP})^\dagger({\cal CP})\left[
{\cal O}_1^\dagger(\mu){ C}_1(\mu)+{\cal O}_2^\dagger(\mu){ C}_2(\mu)
\right]({\cal CP})^\dagger({\cal CP})\right|B^0_d\right\rangle\\
&&=e^{i\phi_{\mbox{{\scriptsize CP}}}(B_d)}\,\Bigl\langle\overline{f}\Bigl|
{\cal O}_1(\mu){ C}_1(\mu)+{\cal O}_2(\mu){ C}_2(\mu)\Bigr|
\overline{B^0_d}\Bigr\rangle.
\nonumber
\end{eqnarray}
Consequently, the relevant decay amplitudes take the following form:
\begin{eqnarray}
A(\overline{B^0_d}\to f)&=&\left\langle f\left|{\cal 
H}_{\mbox{{\scriptsize eff}}}(\overline{B^0_d}\to f)\right|
\overline{B^0_d}\right\rangle=\frac{G_{\mbox{{\scriptsize F}}}}{\sqrt{2}}
\,\overline{v}\,\,\overline{M}_f\\
A(B^0_d\to f)&=&\left\langle f\left|{\cal H}_{\mbox{{\scriptsize eff}}}
(B^0_d\to f)\right|B^0_d\right\rangle\,=\,e^{i\phi_{\mbox{{\scriptsize 
CP}}}(B_d)}\frac{G_{\mbox{{\scriptsize F}}}}{\sqrt{2}}\,v^\ast
M_{\overline{f}},
\end{eqnarray}
with hadronic matrix elements
\begin{eqnarray}
\overline{M}_f&\equiv&\Bigl\langle f\Bigl|\overline{{\cal O}}_1(\mu)
{ C}_1(\mu)+\overline{{\cal O}}_2(\mu){ C}_2(\mu)
\Bigr|\overline{B^0_d}\Bigr\rangle\\
M_{\overline{f}}&\equiv&\Bigl\langle\overline{f}\Bigl|{\cal O}_1(\mu)
{ C}_1(\mu)+{\cal O}_2(\mu){ C}_2(\mu)
\Bigr|\overline{B^0_d}\Bigr\rangle.
\end{eqnarray}

We are now in a position to calculate the observable $\xi_f^{(d)}$,
governing the time evolution of the $B^0_d(t),\overline{B^0_d}(t)
\to f$ decay processes. Using (\ref{xi-def}) and (\ref{theta-def}), 
we arrive at
\begin{equation}
\xi_f^{(d)}=e^{-i\Theta_{M_{12}}^{(d)}}
\frac{A(\overline{B^0_d}\to f)}{A(B^0_d\to f)}
=-\,e^{-i(\phi_d+\gamma)}\left(\frac{1-\lambda^2}{\lambda^2R_b}\right)
\frac{\overline{M}_f}{M_{\overline f}}.
\end{equation}
Performing an 
analogous calculation for $\overline{f}\equiv D^{(\ast)-}\pi^+$ yields
\begin{equation}
\xi_{\overline{f}}^{(d)}=e^{-i\Theta_{M_{12}}^{(d)}}
\frac{A(\overline{B^0_d}\to \overline{f})}{A(B^0_d\to \overline{f})}=
-\,e^{-i(\phi_d+\gamma)}\left(\frac{\lambda^2R_b}{1-\lambda^2}\right)
\frac{M_{\overline f}}{\overline{M}_f}.
\end{equation}

If the time-dependent rates corresponding to $B^0_d(t),\overline{B^0_d}(t)
\to f$ and $B^0_d(t), \overline{B^0_d}(t)\to \overline{f}$ decay processes 
are measured, the observables $\xi_f^{(d)}$ and $\xi_{\overline{f}}^{(d)}$ 
can be determined (see (\ref{rates})). Since the hadronic matrix elements 
$\overline{M}_f$ and $M_{\overline{f}}$, as well as the rather poorly
known CKM factor $R_b$, cancel in the following combination \cite{RF-rev}:
\begin{equation}\label{Prod}
\xi_f^{(d)}\times\xi_{\overline{f}}^{(d)}=e^{-2i(\phi_d+\gamma)},
\end{equation}
we may extract the weak phase $\phi_d+\gamma$ in a {\it theoretically clean} 
way \cite{BDpi}. Moreover, as $\phi_d$, i.e.\ $2\beta$, can be determined 
straightforwardly with the help of the ``gold-plated'' mode 
$B_d\to J/\psi K_{\rm S}$ (see also \ref{subsubsec:ambig-resol}), we may 
extract $\gamma$ from (\ref{Prod}). 

Unfortunately, the $\overline{b}\to\overline{u}c\overline{d}$ transition 
in Fig.\ \ref{fig:BDpi} is doubly Cabibbo-suppressed by 
$\lambda^2R_b\approx0.02$ with respect to $b\to c\overline{u}d$, so that 
the interference effects are tiny. However, the approach is nevertheless 
experimentally interesting, if decays into $D^{\ast\pm}\pi^\mp$ states 
are considered, where the $D^{\ast\pm}$-mesons continue to decay through 
strong interactions, $D^{\ast\pm}\to D\pi^\pm$. Here the branching ratios 
are large, i.e.\ ${\cal O}(10^{-3})$, and the $D^{\ast\pm}\pi^\mp$ states 
can be reconstructed with a good efficiency and modest backgrounds. In order 
to boost statistics, a partial reconstruction technique can be applied. 
Experimental feasibility studies for BaBar and the LHC were performed in 
\cite{Babar-book} and \cite{LHC-Report}, respectively. The analyses can 
also be extended to $B_d\to D^{\ast\pm} a_1^\mp$ modes, exhibiting a 
branching ratio that is about three times larger than that of 
$B_d\to D^{\ast\pm}\pi^{\mp}$ \cite{LHC-Report}; advantages of other 
resonances were emphasized in \cite{diehl}. However, the two spin-1 particles 
arising in $B_d\to D^{\ast\pm} a_1^\mp$ complicate the extraction of 
$\phi_d+\gamma$, requiring an angular analysis (for an analogous
problem, see \cite{FD2}).

\subsection{Summary}
In this section, we have discussed $B$-factory benchmark modes. 
The key features of these transitions are interference effects 
between $B^0_d$--$\overline{B^0_d}$ mixing and decay processes, 
providing observables for the extraction of weak phases, where 
hadronic matrix elements cancel.

The most important channel is the ``gold-plated'' decay 
$B_d\to J/\psi K_{\rm S}$, allowing a determination of $\sin2\beta$.
Using this and similar modes, CP violation in the $B$ system
could recently be observed by the BaBar and Belle collaborations.
Since their results for $\sin2\beta$ are not fully consistent with 
each other, the measurement of this quantity 
will continue to be a very exciting topic. Taking into account also 
previous results from the CDF and ALEPH collaborations, the resulting 
average is now in good agreement with the range implied by the ``standard 
analysis'' of the unitarity triangle. However, as we have illustrated in the 
context of $B_d\to J/\psi[\to \ell^+\ell^-] K^\ast[\to \pi^0 K_{\rm S}]$ 
decays, new physics may even hide in such a situation, making it important 
to determine also $\cos2\beta$. 

The preferred mechanism for physics beyond the Standard Model to manifest 
itself in these measurements is through contributions to 
$B^0_d$--$\overline{B^0_d}$ mixing. In order to obtain the whole picture, 
$B^\pm\to J/\psi K^\pm$ modes, which are related to $B_d\to J/\psi K_{\rm S}$ 
through the isospin symmetry of strong 
interactions, should be considered as well. Then we may -- in addition to 
the usual mixing-induced CP asymmetry $a_{\psi K_{\rm S}}$ -- introduce a 
set of three observables, allowing a general analysis of possible new-physics 
contributions to the different isospin sectors of the $B\to J/\psi K$ system. 
Imposing a plausible dynamical hierarchy of amplitudes, we have seen that one 
of these observables may already be accessible at the first-generation 
$B$-factories, whereas the remaining ones will probably be left for the 
LHC era. However, in the presence of large rescattering effects, all three 
new-physics observables may be sizeable.

A similar analysis can also be performed for the $B\to\phi K$ system,
which is very sensitive to new-physics effects at the amplitude level, 
since these modes originate from $\overline{b}\to\overline{s}$ penguin 
processes. Within the Standard Model, mixing-induced CP violation in 
$B_d\to \phi K_{\rm S}$ is related to $\sin2\beta$, as in the
$B_d\to J/\psi K_{\rm S}$ case. A difference between these measurements 
would probably be due to new-physics contributions to the $B\to\phi K$ 
decay amplitudes. In order to get the full picture, three additional 
observables have to be measured, providing not only ``smoking-gun'' signals 
for new-physics contributions to the two different $B\to\phi K$ isospin 
channels, but also valuable insights into hadron dynamics. Whereas the 
$B\to\phi K$  system is, in general, a powerful tool to search for 
indications of new physics, there is also an unfortunate case, where 
such effects cannot be distinguished from those of the Standard Model. 

If penguin effects played a negligible r\^ole in the decay 
$B_d\to\pi^+\pi^-$, the angle $\alpha$ of the unitarity triangle
could be determined from the corresponding mixing-induced CP asymmetry.
However, both experimental data and theoretical considerations 
indicate that the penguin effects cannot be neglected. Using isospin 
relations between $B\to\pi\pi$ amplitudes, the corresponding hadronic
uncertainties could in principle be eliminated, thereby allowing a 
determination of $\alpha$. Since this approach requires a 
measurement of $B_d\to\pi^0\pi^0$, it is unfortunately very difficult 
in practice. An alternative to solve the penguin problem in the extraction 
of $\alpha$ is given by $B\to\rho\pi$ modes, and a particularly promising 
way to make use of the CP-violating $B_d\to\pi^+\pi^-$ observables is 
offered by $B_s\to K^+K^-$ modes, as we will see in 
Subsection~\ref{subsec:BsKK-Uspin}.

Finally, we have also considered the pure ``tree'' decays 
$B_d\to D^{(\ast)\pm}\pi^\mp$, providing a theoretically clean 
determination of the weak phase $2\beta+\gamma$. Although the relevant 
interference effects between $B^0_d$--$\overline{B^0_d}$ mixing and 
decay processes are doubly Cabibbo-suppressed, this approach is 
nevertheless experimentally interesting. It has also a counterpart in 
the $B_s$ system, which is our next topic.

\boldmath
\section{A Closer Look at the $B_s$ System}\label{sec:Bs}
\unboldmath
\setcounter{equation}{0}
\boldmath
\subsection{General Remarks and Differences to the $B_d$ System}
\unboldmath
Unfortunately, at the $e^+e^-$ $B$-factories operating at the 
$\Upsilon(4S)$ resonance (BaBar, Belle, CLEO), no $B_s$-mesons are 
accessible, since $\Upsilon(4S)$ states decay only to $B_{u,d}$-mesons, 
but not to $B_s$.\footnote{Operating these machines at the $\Upsilon(5S)$
resonance would also make $B_s$-mesons accessible \cite{petrak}.} On the 
other hand, the physics potential of the $B_s$ system is very promising 
for hadron machines (Tevatron, LHC), where plenty of $B_s$-mesons are 
produced \cite{LHC-Report}. In some sense, $B_s$ physics is therefore 
the ``El Dorado'' for $B$ experiments at hadron colliders.
There are important differences between the $B_d$ and $B_s$ systems:
\begin{itemize}
\item Within the Standard Model, the $B^0_s$--$\overline{B^0_s}$ mixing 
phase probes the tiny angle $\delta\gamma$ in the unitarity triangle
shown in Fig.\ \ref{fig:UT} (b), and is hence negligibly small:
\begin{equation}
\phi_s=-2\delta\gamma=-2\lambda^2\eta={\cal O}(-0.03)={\cal O}(-2^\circ),
\end{equation}
whereas $\phi_d=2\beta={\cal O}(45^\circ)$.
\item A large mixing parameter $x_s={\cal O}(20)$ is expected in the 
Standard Model, whereas $x_d=0.75\pm0.02$ (see (\ref{mix-par})). The 
present lower bound is given as follows \cite{LEPBOSC}:
\begin{equation}\label{Ms-bound}
\Delta M_s>15.0\,\mbox{ps}^{-1},\quad
x_s>21.3\,\, (95\% \,\,{\rm C.L.}).
\end{equation}
\item There may be a sizeable width difference 
$\Delta\Gamma_s/\Gamma_s={\cal O}(-10\%)$ between the mass eigenstates 
of the $B_s$ system that is due to CKM-favoured $b\to c\overline{c}s$ 
quark-level transitions into final states common to $\overline{B^0_s}$ 
and $B^0_s$ \cite{BuSlSt,DG-hist}, whereas $\Delta\Gamma_d$ is negligibly 
small.\footnote{Note that $\Delta\Gamma_s$ is negative in the Standard 
Model, as can be seen in (\ref{DG-MP-rel}).} The present CDF and LEP 
average is given as follows \cite{stocchi}:
\begin{equation}
\Delta\Gamma_s/\Gamma_s=-0.16^{+0.16}_{-0.13},\quad
|\Delta\Gamma_s|/\Gamma_s<0.31\,\, (95\% \,\,{\rm C.L.}).
\end{equation}
\end{itemize}
Let us next discuss interesting phenomenological implications of the 
mixing parameters $\Delta M_s$ and $\Delta\Gamma_s$ in more detail.

\begin{figure}
\centerline{\rotate[r]{
\epsfysize=11.1truecm
{\epsffile{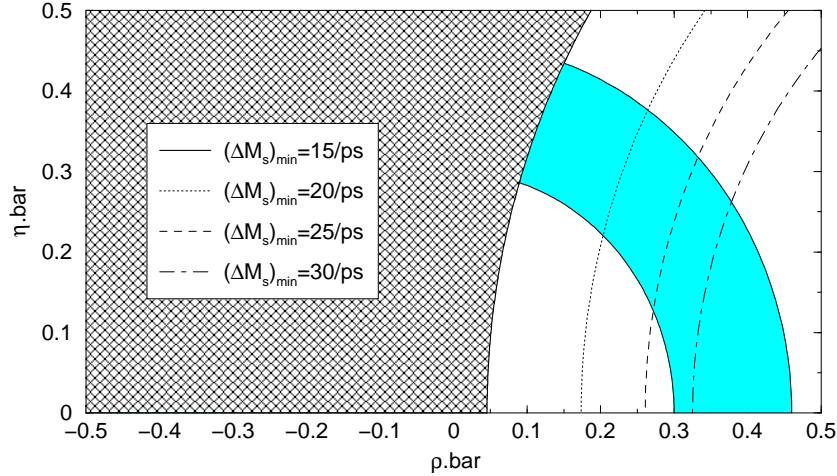}}}}
\caption{The impact of the upper limit $(R_t)_{\rm max}$
on the allowed range in the 
$\overline{\rho}$--$\overline{\eta}$ plane for $\xi=1.15$. The
shaded region corresponds to $R_b=0.38\pm0.08$.}\label{fig:UT-constr}
\end{figure}

\boldmath
\subsection{$\Delta M_s$ and Constraints in the 
$\overline{\rho}$--$\overline{\eta}$ Plane}\label{Bs-gen}
\unboldmath
As we have already noted in Subsections~\ref{subsec:CKM-fits} and 
\ref{subsec:Rt-Md}, the mass difference $\Delta M_d$ plays an important 
r\^ole to constrain the apex of the unitarity triangle. In particular, 
it allows us to fix a circle in the $\overline{\rho}$--$\overline{\eta}$ 
plane around $(1,0)$ with radius $R_t$ through (\ref{RT}). The theoretical 
cleanliness of this determination of $R_t$ is limited by the 
non-perturbative parameter $\sqrt{\hat B_{B_d}}f_{B_d}$, with its current
range given in (\ref{sqrtBfb}). In view of these uncertainties, it is
actually more advantageous to employ
\begin{equation}
\left|\frac{V_{td}}{V_{ts}}\right|=\xi\sqrt{\frac{M_{B_s}}{M_{B_d}}}
\sqrt{\frac{\Delta M_d}{\Delta M_s}}.
\end{equation}
Using $|V_{ts}|=|V_{cb}|$, which holds up to corrections of 
${\cal O}(\lambda^2)$, we arrive at \cite{Brev01,UUT,BB-Bound,BF-MFV}
\begin{equation}\label{Rt-Ms}
R_t=0.83\times\xi\times\sqrt{\frac{\Delta M_d}{0.50\,\mbox{ps}^{-1}}}
\sqrt{\frac{15.0\,\mbox{ps}^{-1}}{\Delta M_s}},
\end{equation}
which is much cleaner than (\ref{RT}), since the $SU(3)$-breaking parameter
\begin{equation}
\xi\equiv\frac{\sqrt{\hat B_{B_s}}f_{B_s}}{\sqrt{\hat B_{B_d}}f_{B_d}}=
1.15\pm0.06
\end{equation}
can be obtained with an accuracy that is considerably higher than those 
of the $\sqrt{\hat B_{B_q}}f_{B_q}$ 
\cite{Rome-rev,FBlat}. Interestingly, the presently available experimental 
lower bound on $\Delta M_s$ can be transformed into an upper bound on 
$R_t$ \cite{Buras-Rt}. Using $(\Delta M_d)_{\rm max}=0.50\,\mbox{ps}^{-1}$,
we obtain
\begin{equation}\label{Rtmax}
(R_t)_{\rm max}=0.83\times\xi\times
\sqrt{\frac{15.0\,\mbox{ps}^{-1}}{(\Delta M_s)_{\rm min}}}.
\end{equation}
In Fig.~\ref{fig:UT-constr}, we show the impact of this relation on the 
allowed range in the $\overline{\rho}$--$\overline{\eta}$ plane. The 
strong experimental lower bound $\Delta M_s>15.0\,\mbox{ps}^{-1}$ 
($95\%$ C.L.) excludes already a large part in the 
$\overline{\rho}$--$\overline{\eta}$ plane (crossed region for $\xi=1.15$), 
implying in particular $\gamma<90^\circ$.

The present experiments searching for $B^0_s$--$\overline{B^0_s}$ mixing, 
which give the lower bound (\ref{Ms-bound}), yield also a local minimum of 
the log-likelihood function around $\Delta M_s=17.7\,\mbox{ps}^{-1}$, 
which is $2.5\,\sigma$ away from being zero. The consequences of a possible
future measurement of $\Delta M_s=(17.7\pm1.4)\,\mbox{ps}^{-1}$ motivated
by this observation were recently studied both within the Standard Model 
and within supersymmetric models with minimal flavour violation \cite{AL}. 
In this analysis, it was argued that if $\Delta M_s$ should actually be 
found in this range, a large class of supersymmetric models would be 
disfavoured, and of all models considered in \cite{AL}, the best fit to 
the data would be obtained for the Standard Model. It is expected 
that run II of the Tevatron will provide a measurement of 
$\Delta M_s$ in the near future.

\boldmath
\subsection{$\Delta\Gamma_s$ and Untagged Decay Rates}
\unboldmath
The most appropriate tool to analyse $\Delta\Gamma_s$ is provided by the
heavy-quark expansion, where $1/m_b$ corrections were determined
\cite{BBD}, and next-to-leading order QCD corrections were calculated 
\cite{DGamma-cal}. In order to predict $\Delta\Gamma_s$, certain
hadronic matrix elements, parametrized through non-perturbative
bag parameters and $f_{B_s}$, are required. Recent lattice analyses give
\begin{equation}\label{DG-lat-pred}
-\left(\frac{\Delta\Gamma_s}{\Gamma_s}\right)=\left\{
\begin{array}{ll}
(9.7^{+1.4}_{-3.5}\pm2.5\pm2.0\pm1.6)\% & 
\mbox{(Hashimoto {\it et al.}~\cite{hashimoto})}\\
(4.7\pm1.5\pm1.6)\% & \mbox{(Becirevic {\it et al.}~\cite{rome}).}
\end{array}\right.
\end{equation}
Although the relevant bag parameters are in rather good agreement in 
these papers, the results in (\ref{DG-lat-pred}) differ since 
$\Delta\Gamma_s$ was expressed in \cite{rome} using $\Delta M_d$ as a 
normalization. The advantage of this approach is that hadronic uncertainties 
enter only in ratios. On the other hand, $|V_{ts}/V_{td}|^2$ is required as 
an input, making the prediction of $\Delta\Gamma_s$ sensitive to the fits of 
the unitarity triangle discussed in Subsection~\ref{subsec:CKM-fits}, which 
depend also on theoretical assumptions and may be affected by new physics. 
A detailed discussion of the present theoretical status of $\Delta\Gamma_s$ 
was recently given in \cite{BeLe}, where the result
\begin{equation}
\Delta\Gamma_s/\Gamma_s=-\left(9.3^{+3.4}_{-4.6}
\right)\% 
\end{equation}
was obtained. The main uncertainties are due to residual scale dependences
and $1/m_b$ corrections. As argued in \cite{BeLe}, further improvements
appear to be very challenging.

A non-vanishing width difference $\Delta\Gamma_s$ would allow the 
extraction of CP-violating weak phases from the following ``untagged'' 
$B_s$ rates \cite{FD2,dunietz,FD1,RF-Bs,Hawaii}:
\begin{equation}\label{untag-def}
\Gamma_s[f(t)]\equiv\Gamma(B^0_s(t)\to f)+\Gamma(\overline{B^0_s}(t)
\to f).
\end{equation}
If we use (\ref{rates}), we obtain
\begin{equation}\label{untagged}
\Gamma_s[f(t)]\propto\left[\left(1+\left|\xi_f^{(s)}\right|^2\right)
\left(e^{-\Gamma_{\rm L}^{(s)}t}+
e^{-\Gamma_{\rm H}^{(s)}t}\right)
-2\,\mbox{Re}\,\xi_f^{(s)}\left(e^{-\Gamma_{\rm L}^{(s)}t}-
e^{-\Gamma_{\rm H}^{(s)}t}\right)\right],
\end{equation}
providing the observable ${\cal A}_{\Delta\Gamma}(B_s\to f)$ 
introduced in (\ref{ADGam}). Interestingly, the $\Delta M_st$ terms
appearing in the ``tagged'' rates $\Gamma(B^0_s(t)\to f)$ and 
$\Gamma(\overline{B^0_s}(t)\to f)$ cancel in their ``untagged'' 
combination (\ref{untag-def}). Because of 
the large mixing parameter $x_s$, the $\Delta M_st$ terms oscillate 
very rapidly and are hard to resolve. Although it should nevertheless be 
feasible to keep track of these oscillations, studies of untagged 
rates are also interesting in terms of efficiency, acceptance and purity. 
The following channels are particularly promising in the context of
extracting weak phases from untagged $B_s$ data samples:
\begin{itemize}
\item $B_s\to J/\psi\phi$, allowing us to probe $\phi_s=-2\delta\gamma$ 
\cite{FD1}. 
\item The $B_s\to K^{\ast+}K^{\ast-}, K^{\ast0}\overline{K^{\ast0}}$
system, providing a strategy to extract $\gamma-2\delta\gamma$ \cite{FD1};
bounds implied by $B_s\to K^{+}K^{-}, K^{0}\overline{K^{0}}$ were
studied in \cite{RF-Bs}.
\item  $B_s$ decays caused by $\overline{b}\to\overline{u}c\overline{s}$ 
($b\to c\overline{u}s$) quark-level processes \cite{FD2}, for example
$B_s\to D_s^{\ast\pm}K^{\ast\mp}$ or $B_s\to D^{\ast0}\phi$, allowing 
also a determination of $\gamma-2\delta\gamma$.
\end{itemize}
Here $-2\delta\gamma$ is negligibly small in the Standard Model, as we
have noted above. In the case of $B\to VV$ decays into two vector mesons, 
the untagged angular distributions 
of their decay products have to be considered, which provide many more 
observables than $B\to PP$ modes into two pseudoscalar mesons (for a general 
discussion of the observables of $B\to VV$ decays, see \cite{ChiWo}). This 
feature is exploited in the untagged strategies listed above. Since the 
$B_s\to J/\psi\phi$ mode is of central interest for hadron colliders, we 
shall come back to this channel in Subsection~\ref{subsec:Bspsiphi}.

\subsection{Impact of New Physics}\label{subsec:Bs-NP}
As we have seen in Subsection~\ref{subsec:NP-mix}, 
$B^0_d$--$\overline{B^0_d}$ and $B^0_s$--$\overline{B^0_s}$ mixing may be 
strongly affected by new physics, since these are highly CKM-suppressed
loop-induced fourth order weak interaction processes. In particular --
in addition to $\Delta M_s=2|M_{12}^{(s)}|$ -- also the 
$B^0_s$--$\overline{B^0_s}$ mixing phase 
$\phi_s\sim\mbox{arg}M_{12}^{(s)}$
may be modified in the presence of new physics as follows:
\begin{equation}\label{PhiM2s}
\tan \phi^{\rm NP}_s \approx
\frac{-2\delta\gamma + \varrho^2_s \sin 2\psi_s}
{1 + \varrho^2_s\cos2\psi_s}\approx
\frac{\varrho^2_s \sin 2\psi_s}
{1 + \varrho^2_s\cos2\psi_s},
\end{equation}
where
\begin{equation}\label{rho-NPs-def}
\varrho_s=\left(\frac{1}{\lambda^2 A}\right)\left(\frac{M_W}{\Lambda}
\right).
\end{equation}
For a new-physics scale $\Lambda$ in the TeV regime, $\rho_s$ would be
of order one. The impact on $\Delta M_s$ would affect the determination 
of $R_t$ with the help of (\ref{Rt-Ms}), provided new physics enters 
differently in $\Delta M_d$. In contrast to $B^0_d$--$\overline{B^0_d}$ 
mixing (see (\ref{PhiM2}) and (\ref{rho-NP-def})), the Standard-Model
``background'' $\phi^{\rm SM}_s=-2\delta\gamma=-2\lambda^2\eta$ is negligibly 
small in (\ref{PhiM2s}). Consequently, CP-violating effects in $B_s$ 
decays represent a very sensitive probe for new physics \cite{NiSi,DFN}. 
To simplify the following discussion, we shall write from now on generically 
$\phi_s$ for the $B^0_s$--$\overline{B^0_s}$ mixing phase.

Interestingly, new-physics contributions to $B^0_s$--$\overline{B^0_s}$ 
mixing may also affect the width difference $\Delta\Gamma_s$ \cite{gro-DG}. 
Using expression (\ref{Expr-DGamma}) for $\Delta\Gamma_s$, we obtain
\begin{equation}\label{DG-NP}
\Delta\Gamma_s=\Delta\Gamma_s^{\rm SM}\cos\phi_s,
\end{equation}
which yields a reduction of $|\Delta\Gamma_s|$ for $\phi_s\not=0$ or $\pi$. 
This relation is of course convention-independent, as shown 
explicitly in \cite{DFN}. In (\ref{DG-NP}), we have employed -- as we 
do throughout this review -- the phase convention of the standard 
and generalized Wolfenstein parametrizations of the CKM matrix, where 
the phase of $\Gamma_{12}^{(s)}$ vanishes to an excellent 
approximation, as $b\to c\overline{c}s$ processes play the key r\^ole for
this off-diagonal element of the $B^0_s$--$\overline{B^0_s}$
decay matrix.\footnote{Since $\Gamma_{12}^{(s)}$ is governed by
CKM-favoured tree-level decays, it is practically insensitive to
new physics.} Consequently, we may identify $\phi_s$ with 
\begin{equation}
\mbox{arg}\left(M_{12}^{(s)}\right)-
\mbox{arg}\left(-\Gamma_{12}^{(s)}\right)=\phi_s-0,
\end{equation}
which actually enters in (\ref{DG-NP}).

In a recent paper \cite{DFN}, various strategies to detect a sizeable phase 
$\phi_s$ were proposed, providing ``smoking-gun'' signals for new-physics 
contributions to $B^0_s$--$\overline{B^0_s}$ mixing. For the untagged 
case, also a new approach was presented, allowing the determination of 
$\phi_s$ from simple measurements of lifetimes and branching ratios. 
Here no two-exponential fits as in (\ref{untagged}) are involved, 
thereby requiring considerably less statistics. Moreover, also methods 
to determine $\phi_s$ {\it unambiguously} were suggested, which play -- 
among other things -- an important r\^ole for the extraction of $\gamma$
from the strategies discussed in the next subsection. In the corresponding
decays, as well as in $B_s\to J/\psi \phi$, $B^0_s$--$\overline{B^0_s}$ 
mixing represents the preferred mechanism for new physics to manifest
itself.

\boldmath
\subsection{Strategies using Pure Tree Decays of $B_s$-Mesons}
\unboldmath
An interesting class of $B_s$ decays is due to 
$\overline{b}\to\overline{u}c\overline{s}$ ($b\to c\overline{u}s$)
quark-level transitions, providing the $B_s$ variant of the 
$B_d\to D^{(\ast)\pm}\pi^\mp$ approach to extract $\phi_d+\gamma$
discussed in Subsection~\ref{subsec:BDpi}. Here we have also to deal 
with pure ``tree'' decays, where both $B_s^0$- and $\overline{B_s^0}$-mesons
may decay into the same final state $f$. The resulting interference effects 
between decay and mixing processes allow a {\it theoretically clean} 
extraction of $\phi_s+\gamma$ from 
\begin{equation}
\xi_f^{(s)}\times\xi_{\overline{f}}^{(s)}=e^{-2i(\phi_s+\gamma)}.
\end{equation}
An interesting difference to the $B_d\to D^{(\ast)\pm}\pi^\mp$ approach
is that both decay paths of $B_s^0, \overline{B_s^0}\to f$ are of the same 
order in $\lambda$, thereby leading to larger interference effects.
There are several strategies making use of these features: 
\begin{itemize}
\item We may consider the colour-allowed decays $B_s\to D_s^\pm K^\mp$ 
\cite{ADK}, or the colour-suppressed modes $B_s\to D^0\phi$ \cite{GL0}. 
In these strategies, ``tagged'' analyses of the corresponding time-dependent
rates have to be performed, i.e.\ the $\Delta M_st$ terms have to be 
resolved. The selection of $B_s\to D_s^\pm K^\mp$ transitions is 
unfortunately experimentally challenging, since $B_s\to D_s^\pm\pi^\mp$ 
events, which come with a 20 times larger branching ratio, have to be 
rejected \cite{LHC-Report}.
\item As we have seen above, in the case of 
$B_s\to D_s^{\ast\pm} K^{\ast\mp}$ or $B_s\to D^{\ast0}\phi$, the 
observables of the corresponding angular distributions provide
sufficient information to extract $\phi_s+\gamma$ from ``untagged''
analyses \cite{FD2}, requiring a sizeable $\Delta\Gamma_s$. A
``tagged'' strategy involving $B_s\to D_s^{\ast\pm} K^{\ast\mp}$ modes
was proposed in \cite{LSS}.
\item Recently, strategies making use of ``CP-tagged'' $B_s$ decays were 
proposed \cite{FP}, which require a symmetric $e^+e^-$ collider operated 
at the $\Upsilon(5S)$ resonance. In this approach, initially present CP 
eigenstates $B_s^{\rm CP}$ are employed, which can be tagged through
the fact that the $B_s^0/\overline{B_s^0}$ mixtures have anti-correlated 
CP eigenvalues at the $\Upsilon(5S)$ resonance. Here $B_s\to D_s^\pm K^\mp, 
D_s^\pm K^{\ast\mp}, D_s^{\ast\pm} K^{\ast\mp}$ modes may be used. 
\end{itemize}
The extraction of $\gamma$ from the weak phase $\phi_s+\gamma$ provided
by these approaches requires $\phi_s$ as an additional input, which is
negligibly small in the Standard Model. Whereas it appears to be
quite unlikely that the amplitudes of the pure tree decays listed above 
are affected significantly by new physics, as they involve no 
flavour-changing neutral-current processes, this is not the case for
the $B^0_s$--$\overline{B^0_s}$ mixing phase $\phi_s$. In order to probe 
this quantitiy, $B_s\to J/\psi\,\phi$ transitions offer interesting 
strategies.

\boldmath
\subsection{The Golden Mode $B_s\to J/\psi\phi$}\label{subsec:Bspsiphi}
\unboldmath
This decay is the $B_s$ counterpart of $B_d\to J/\psi K_{\rm S}$ and
provides interesting strategies to extract $\Delta M_s$ and $\Delta\Gamma_s$, 
and to probe $\phi_s$ \cite{DDF1,DFN}; experimental feasibility studies
for the LHC can be found in \cite{LHC-Report}. The corresponding Feynman 
diagrams are analogous to those shown in Fig.~\ref{fig:BdPsiKS}. Since the 
final state of $B_s\to J/\psi\phi$ is an admixture of different CP 
eigenstates, we have to use the angular distribution of the 
$J/\psi\to \ell^+\ell^-$ and $\phi\to K^+K^-$ decay products to disentangle 
them \cite{DDLR}. 
\subsubsection{Structure of the Angular Distribution}
\begin{figure}
\centerline{
\epsfysize=6.0truecm
{\epsffile{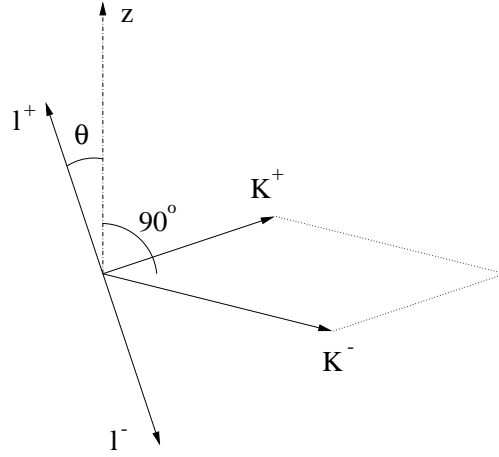}}}
\caption{The kinematics of 
$B_s^0\to J/\psi[\to\ell^+\ell^-]\phi\,[\to K^+K^-]$ in the
$J/\psi$ rest frame.}\label{fig:kin}
\end{figure}

The time-dependent angular distribution of the 
$B_s^0\to J/\psi[\to \ell^+\ell^-] \phi[\to K^+K^-]$ decay products 
can be written generically as follows~\cite{DDF1}:
\begin{equation}\label{ang}
f(\Theta,\Phi,\Psi;t)=\sum_k b^{(k)}(t)\,
g^{(k)}(\Theta,\Phi,\Psi),
\end{equation}
where we have denoted the angles describing the kinematics of the decay
products of $J/\psi\to \ell^+\ell^-$ and $\phi\to K^+K^-$ by $\Theta$, 
$\Phi$ and $\Psi$. For instance, $\Theta$ describes the angle between 
the direction of the $\ell^+$ and the $z$ axis in the $J/\psi$ rest frame,
where the $z$ axis is definded to be perpendicular to the decay plane of 
$\phi\to K^+K^-$, as shown in Fig.~\ref{fig:kin}

The functions $b^{(k)}(t)$ describe the time evolution 
of the angular distribution (\ref{ang}). They can be expressed in terms of 
real or imaginary parts of the following bilinear combinations of decay 
amplitudes:
\begin{equation}
A^\ast_{\tilde f}(t)A_f(t)
=\left\langle(J/\psi\phi)_{\tilde f}|{\cal H}_{\rm eff}|
B^0_s(t)\right\rangle^\ast\left\langle(J/\psi\phi)_{f}|{\cal H}_{\rm eff}|
B^0_s(t)\right\rangle,
\end{equation}
where ${\cal H}_{\rm eff}$ is the hermitian conjugate of (\ref{e4})
with $r=s$, and $|B^0_s(t)\rangle$ is given by (\ref{B0-evol}).
The labels $f$ and $\tilde f$ specify the relative polarization of the 
$J/\psi$- and $\phi$-mesons in given final-state configurations 
$(J/\psi\phi)_f$ and $(J/\psi\phi)_{\tilde f}$, respectively. It is 
convenient to introduce linear polarization amplitudes $A_0(t)$, 
$A_\parallel(t)$ and $A_\perp(t)$ \cite{pol}, corresponding to linear
polarization states of the vector mesons, which are either longitudinal
($0$) or transverse to their directions of motion. In the latter case,
the polarization states may either be parallel ($\parallel$) or 
perpendicular ($\perp$) to one another. Whereas $A_\perp(t)$ describes 
a CP-odd final-state configuration, both $A_0(t)$ and $A_\parallel(t)$ 
correspond to CP-even final-state configurations, i.e.\ to the CP 
eigenvalues $-1$ and $+1$, respectively. The observables $b^{(k)}(t)$ 
are then given by
\begin{equation}\label{obs1}
\left|A_f(t)\right|^2\quad\mbox{with}\quad f\in\{0,\parallel,\perp\},
\end{equation}
as well as by the interference terms
\begin{equation}\label{obs2}
\mbox{Re}\{A_0^\ast(t)A_\parallel(t)\}\quad\mbox{and}\quad
\mbox{Im}\{A_f^\ast(t)A_\perp(t)\} \quad\mbox{with}\quad f\in\{0,\parallel\}.
\end{equation}

\boldmath
\subsubsection{Structure of the Observables}
\unboldmath
Applying once more the formalism discussed in Section~\ref{sec:neutral}, 
we find that the time evolution of the $b^{(k)}(t)$ is governed by  
\begin{equation}
\xi^{(s)}_{\psi\phi}\,\propto\, e^{-i2\phi_s}\left[\frac{\lambda_u^{(s)*}
\tilde A_{\rm pen}^{ut'}+\lambda_c^{(s)*}\left(\tilde A_{\rm CC}^{c'}+
\tilde A_{\rm pen}^{ct'}\right)}{\lambda_u^{(s)}\tilde A_{\rm pen}^{ut'}+
\lambda_c^{(s)}\left(\tilde A_{\rm CC}^{c'}+\tilde A_{\rm pen}^{ct'}\right)}
\right],
\end{equation}
where the amplitudes are analogous to the ones introduced in 
(\ref{Bd-ampl1}); for simplicity, we have suppressed the label $f$ of
the $(J/\psi\phi)_f$ final-state configuration in this expression. We 
expect -- as in our discussion of the $B_d\to J/\psi K_{\rm S}$ case -- 
that 
\begin{equation}
\left|\frac{\lambda_u^{(s)}\tilde A_{\rm pen}^{ut}}{\lambda_c^{(s)}
\left(\tilde A_{\rm CC}^c+\tilde A_{\rm pen}^{ct}
\right)}\right|={\cal O}(\overline{\lambda}^3),
\end{equation}
yielding
\begin{equation}\label{xispsiphi}
\xi^{(s)}_{\psi\phi}\,\propto\, e^{-i\phi_s}
\left[1-\,i\,\sin\gamma\times{\cal O}(\overline{\lambda}^3)\right].
\end{equation}
Since $\phi_s={\cal O}(0.03)$ in the Standard Model, there may well 
be -- in contrast to the determination of $\phi_d$ from 
$B_d\to J/\psi\,K_{\rm S}$ -- significant hadronic uncertainties 
in the extraction of $\phi_s$ from the 
$B_s\to J/\psi[\to \ell^+\ell^-] \phi[\to K^+K^-]$ angular distribution.
These hadronic uncertainties, which may become an important issue for the 
LHC, can be controlled with the help of the decay $B_d\to J/\psi\rho^0$, 
which has also some other interesting features \cite{RF-ang}.

In the following discussion, we assume 
\begin{equation}\label{SIMP}
\xi^{(s)}_{\psi\phi}\,\propto\, e^{-i\phi_s},
\end{equation} 
i.e.\ that the $B_s\to J/\psi \phi$ decay amplitudes do not involve a 
CP-violating weak phase, which implies vanishing direct CP violation. 
The general formalism, where this assumption is not made and also penguin 
effects are taken into account, was presented in \cite{RF-ang}. Moreover, 
we also assume, for simplicity, that new physics does not affect the
structure of (\ref{SIMP}), thereby manifesting itself only through a 
{\it sizeable} value of $\phi_s$. Since the decays $B_s\to J/\psi \phi$ 
and $B_d\to J/\psi K_{\rm S}$ originate from the same quark-level 
transitions, the general new-physics analysis discussed in 
Subsection~\ref{subsec:BpsiK} allows us to test this 
assumption already in the $B\to J/\psi K$ system.

\subsubsection{CP Asymmetries and Manifestation of New Physics}
An important implication of (\ref{SIMP}) is that $B_s\to J/\psi\phi$
exhibits very small CP violation in the Standard Model, making this 
channel an interesting probe to search for indications of new physics 
\cite{NiSi,DFN}. The quantity $\sin\phi_s$ governs CP violation in 
$B_s\to J/\psi\phi$, as can be seen in the following expression:
\begin{equation}\label{CP1}
\frac{\Gamma(t)-\overline{\Gamma}(t)}{\Gamma(t)+\overline{\Gamma}(t)}=
\left[\frac{1-D}{F_+(t)+D F_-(t)}\right]
\sin(\Delta M_s t)\,\sin\phi_s.
\end{equation}
Here $\Gamma(t)$ and $\overline{\Gamma}(t)$  denote the time-dependent
rates for decays of initially, i.e.\ at time $t=0$, present $B^0_s$- and
$\overline{B^0_s}$-mesons into $J/\psi\phi$ final states, respectively,
\begin{equation}\label{dilut}
D\equiv\frac{|A_{\perp}(0)|^2}{|A_0(0)|^2 + |A_{\|}(0)|^2}=
0.1\,\ldots\,0.5
\end{equation}
is a hadronic factor, and 
\begin{equation}\label{Fpm-def}
F_{\pm}(t)\equiv\frac{1}{2}\left[\left(1\pm\cos\phi_s\right)
e^{+\Delta\Gamma_s t/2}+\left(1\mp\cos\phi_s\right)
e^{-\Delta\Gamma_s t/2}\right].
\end{equation}
The range in (\ref{dilut}) corresponds to factorization \cite{DDF1}, 
and is in good agreement with a recent analysis of the $B_s\to J/\psi\phi$ 
polarization amplitudes $A_0(0)$, $A_{\|}(0)$, $A_{\perp}(0)$ performed 
by the CDF collaboration \cite{CDF-angular}. 

In fact, $\sin\phi_s$ may be sizeable in extensions of the 
Standard Model. An example is the symmetrical 
$SU_{\rm L}(2)\times SU_{\rm R}(2)\times U(1)$ model with spontaneous CP 
violation (SB--LR) \cite{LR-model1}--\cite{LR-model2}, 
which was applied to the $B_s\to J/\psi\phi$ observables in \cite{BF-LR}. 
In this model, $\sin\phi_s$ may be as large as ${\cal O}(-40\%)$. On the 
other hand, small values of $a_{\psi K_{\rm S}}$ are favoured, which 
are no longer compatible with the recently updated $B$-factory results,
yielding the average given in (\ref{s2b-world}).
Another scenario for new physics, where large values of $\sin\phi_s$ 
may arise, is provided by models allowing mixing to a new isosinglet 
down quark, as in $E_6$ \cite{sil}. Let us note that the new-physics 
contributions to the $B_s\to J/\psi\phi$ decay amplitudes are negligible 
in these and the SB--LR  model, as assumed above.

Although the CP asymmetry (\ref{CP1}) may provide a ``smoking-gun'' 
signal for new-physics contributions to $B^0_s$--$\overline{B^0_s}$, 
mixing, it does not allow a clean determination of $\sin\phi_s$ because 
of the hadronic parameter $D$. This feature is due to the fact that
the $J/\psi\phi$ final states are admixtures of different CP eigenstates.
As we have noted above, in order to solve this problem, an angular analysis 
of the $B_s\to J/\psi[\to\ell^+\ell^-]\phi\,[\to\ K^+K^-]$ decay products 
has to be performed. Since the full three-angle distribution is quite
complicated,\footnote{It is given in \cite{DDF1}, together with
appropriate weighting functions to extract the observables.} let us 
consider here the one-angle distribution
\begin{equation}\label{one-angle}
\frac{d \Gamma (t)}{d \cos \Theta} \propto
\left(|A_0(t)|^2 + |A_{\|}(t)|^2\right)\,\frac{3}{8}\,(1 + \cos ^2 \Theta)
+ |A_{\perp}(t)|^2\,\frac{3}{4} \sin^2 \Theta,
\end{equation}
where the kinematics and the definition of the polar angle $\Theta$ is 
illustrated in Fig.~\ref{fig:kin}. The one-angle distribution 
(\ref{one-angle}) allows us now to extract the observables 
\begin{equation}
P_+(t)\equiv|A_0(t)|^2 + |A_{\|}(t)|^2,\quad P_-(t)\equiv|A_{\perp}(t)|^2, 
\end{equation}
as well as their CP conjugates, thereby providing the CP asymmetries
\begin{equation}\label{ASYM}
\frac{P_\pm(t)-\overline{P}_\pm(t)}{P_\pm(t)+\overline{P}_\pm(t)}=
\pm\,\frac{1}{F_\pm(t)}\,\sin(\Delta M_s t)\,\sin\phi_s.
\end{equation}
On the other hand, untagged data samples are sufficient to determine 
\begin{equation}\label{UNTAG}
P_\pm(t)+\overline{P}_\pm(t)\propto
\left[(1\pm\cos\phi_s)e^{-\Gamma_{\rm L}^{(s)} t}
+(1\mp\cos\phi_s)e^{-\Gamma_{\rm H}^{(s)} t}\right].
\end{equation}
New-physics effects would be indicated by the following features:
\begin{itemize}
\item Sizeable values of the CP-violating asymmetries (\ref{ASYM}).
\item The untagged observables (\ref{UNTAG}) depend on two exponentials.
\end{itemize}
In contrast to (\ref{CP1}), these observables do not involve the hadronic
parameter $D$ and allow a clean determination of $\phi_s$. A detailed 
discussion of other strategies to search for new physics with $B_s$ decays 
is given in \cite{DFN}, where also the general time-dependent expressions
for the observables of the $B_s\to J/\psi\phi$ three-angle distribution
can be found.

\subsection{Summary}
The $B_s$-meson system has several interesting features and offers a nice 
play ground for $B$ experiments at hadron machines (Tevatron-II, LHC and 
BTeV). Already the present experimental lower bound on $\Delta M_s$ has 
an important impact on the allowed range for the apex of the unitarity 
triangle in the $\overline{\rho}$--$\overline{\eta}$ plane, implying in 
particular $\gamma<90^\circ$. It is expected that $B^0_s$--$\overline{B^0_s}$ 
mixing will be observed at the Tevatron in the near future, which would
provide much more stringent constraints on the unitarity triangle. 

In contrast to the $B_d$-meson system, the width difference between
the mass eigenstates may be sizeable in the $B_s$ case, i.e.\ at
the level of $10\%$. This width difference may allow extractions of
CP-violating weak phases from ``untagged'' $B_s$ data samples.

Since the $B^0_s$--$\overline{B^0_s}$ mixing phase $\phi_s$ is negligibly
small in the Standard Model, CP violation is tiny in $B_s\to J/\psi\phi$, 
which is the $B_s$ counterpart of $B_d\to J/\psi K_{\rm S}$. 
In addition to strategies to determine the $B^0_s$--$\overline{B^0_s}$ 
mixing parameters, this mode is therefore a very sensitive tool to 
search for new-physics contributions, allowing in particular an extraction
of $\phi_s$. Further $B_s$ benchmark modes are the
$B_s\to D_s^{(\ast)\pm} K^{(\ast)\mp}$ channels, which are pure ``tree'' 
decays and offer various strategies to determine the weak phase 
$\phi_s+\gamma$. 

In the approaches to explore CP violation we have considered so far, 
interference effects between $B^0_{d,s}$--$\overline{B^0_{d,s}}$ 
mixing and decay processes played a key r\^ole to get rid of poorly known 
hadronic matrix elements, thereby allowing the extraction of CP-violating 
weak phases. Let us now turn to decays of charged $B$-mesons, where amplitude 
relations have to be used to accomplish this task.

\boldmath
\section{CP Violation in Charged $B$ Decays}\label{sec:charged}
\unboldmath
\setcounter{equation}{0}
\subsection{General Remarks}
Since there are no mixing effects present in the charged $B$-meson 
system, non-vanishing CP asymmetries of the kind 
\begin{equation}\label{CP-charged}
{\cal A}_{\mbox{{\scriptsize CP}}}(B^+\to f)
\equiv\frac{\Gamma(B^+\to f)-\Gamma(B^-\to\overline{f})}{\Gamma(B^+
\to f)+\Gamma(B^-\to\overline{f})},
\end{equation}
which correspond to ${\cal A}_{\rm CP}^{\rm dir}(B_q\to f)$ defined in
(\ref{ee7}), would give us immediate and unambiguous evidence for direct 
CP violation in the $B$ system. As we have seen in 
Subsection~\ref{subsec:kaon-CP}, this kind of CP violation has already
been established in the neutral kaon system through the measurement of
$\mbox{Re}(\varepsilon'/\varepsilon)\not=0$. Recently, the NA48 
collaboration has reported the following CP asymmetry
\cite{NA48}:
\begin{equation}
\frac{\Gamma(K^0\to\pi^+\pi^-)-
\Gamma(\overline{K^0}\to\pi^+\pi^-)}{\Gamma(K^0\to\pi^+\pi^-)+
\Gamma(\overline{K^0}\to\pi^+\pi^-)}=\left(5.0\pm0.9\right)\times 10^{-6},
\end{equation}
which is the neutral $K$-meson analogue of 
${\cal A}_{\rm CP}^{\rm dir}(B_q\to f)$, and makes direct CP violation 
in the kaon system more apparent than the observable 
$\mbox{Re}(\varepsilon'/\varepsilon)$. 

Whereas direct CP violation is extremely small in kaon decays, 
the CP-violating asymmetries (\ref{CP-charged}) may be as large as
${\cal O}(30\%)$ in the most fortunate cases, for example in 
$B^+\to K^+\overline{K^0}$ \cite{RF-BdKK}, because of the different 
CKM structure of the relevant $B$-decay amplitudes. These CP asymmetries 
arise from the interference between amplitudes with different CP-violating 
weak and CP-conserving strong phases. Due to the unitarity of 
the CKM matrix, any non-leptonic $B$-decay amplitude can be expressed, 
within the Standard Model, in the following way:
\begin{eqnarray}
A(B^+\to f)&=&|A_1|e^{i\delta_1}e^{+i\varphi_1}+
|A_2|e^{i\delta_2}e^{+ i\varphi_2}\label{ampl-dec}\\
A(B^-\to\overline{f})&=&|A_1|e^{i\delta_1}e^{-i\varphi_1}+
|A_2|e^{i\delta_2}e^{-i\varphi_2}.\label{ampl-dec-CP}
\end{eqnarray}
Here the $\delta_{1,2}$ are CP-conserving strong phases, which are
induced by final-state interaction processes, whereas the $\varphi_{1,2}$ 
are CP-violating weak phases, which originate from the CKM matrix. Using 
(\ref{ampl-dec}) and (\ref{ampl-dec-CP}), we obtain
\begin{equation}\label{ACP-dir-expr}
{\cal A}_{\mbox{{\scriptsize CP}}}(B^+\to f)=
\frac{-2|A_1||A_2|\sin(\varphi_1-\varphi_2)\sin(\delta_1-\delta_2)}{|A_1|^2+
2|A_1||A_2|\cos(\varphi_1-\varphi_2)\cos(\delta_1-\delta_2)+|A_2|^2}.
\end{equation}
Consequently, a non-vanishing direct CP asymmetry 
${\cal A}_{\mbox{{\scriptsize CP}}}(B^+\to f)$ requires both a non-trivial 
strong and a non-trivial weak phase difference. In addition to the hadronic 
amplitudes $|A_{1,2}|$, the strong phases $\delta_{1,2}$ lead to particularly 
large hadronic uncertainties in (\ref{ACP-dir-expr}), thereby destroying 
the clean relation to the CP-violating phase difference 
$\varphi_1-\varphi_2$, which is usually related to the angle $\gamma$ of
the unitarity triangle.

\begin{figure}
\begin{center}
\leavevmode
\epsfysize=3.8truecm 
\epsffile{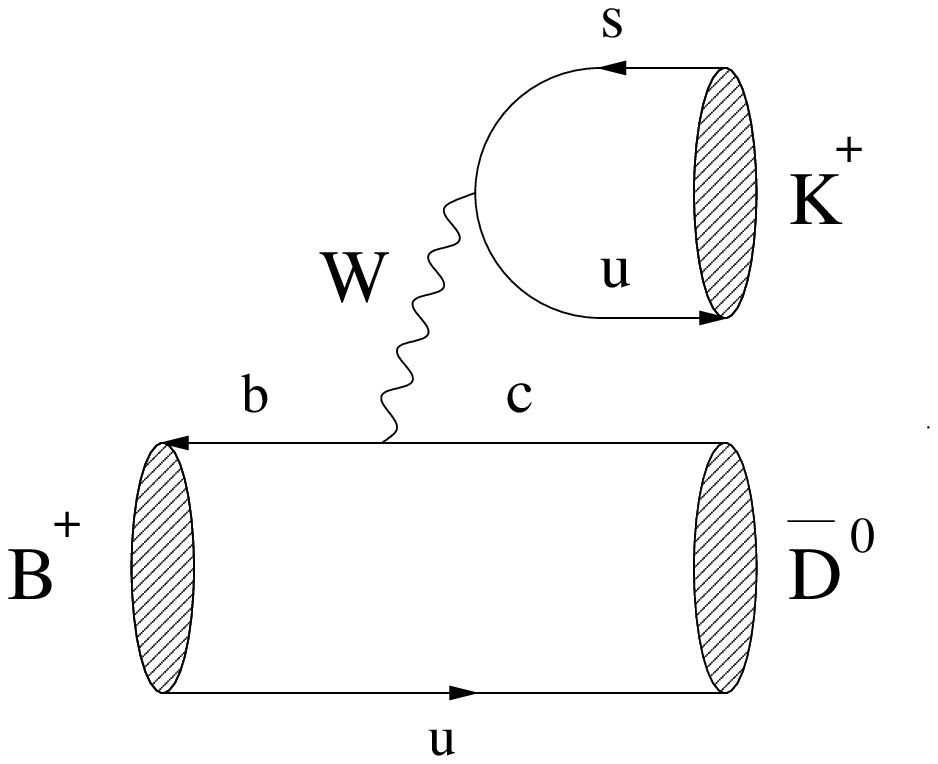} \hspace*{1truecm}
\epsfysize=4.5truecm 
\epsffile{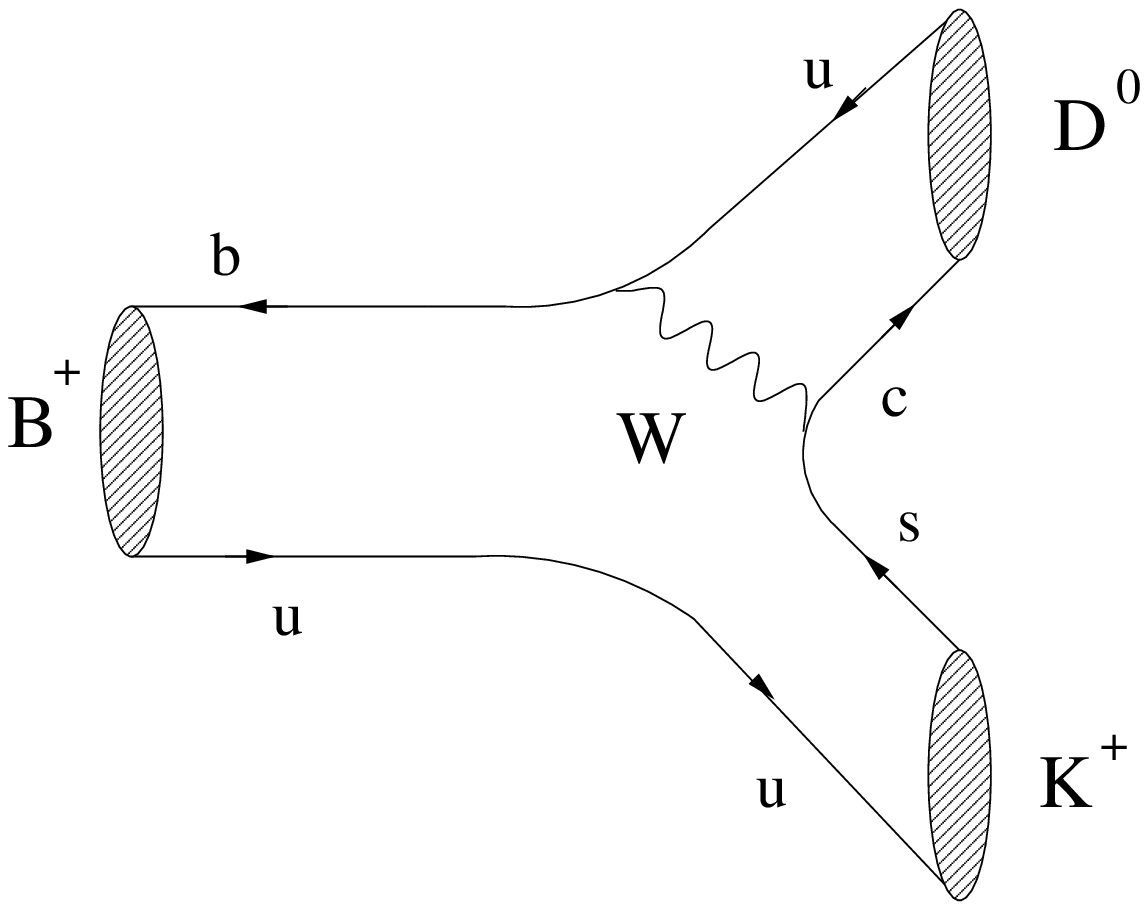}
\end{center}
\vspace*{-0.4truecm}
\caption{Feynman diagrams contributing to $B^+\to K^+\overline{D^0}$ and 
$B^+\to K^+D^0$. }\label{fig:BDK}
\end{figure}

\boldmath
\subsection{Extracting $\gamma$ from $B^\pm\to K^\pm D$ Decays}
\unboldmath
\subsubsection{Triangle Relations and Experimental Problems}
\begin{figure}
\begin{center}
\begin{picture}(320,75)(0,0)
\Line(50,10)(208,10) \ArrowLine(207,10)(209,10)
\DashLine(50,10)(240,40){6}\ArrowLine(237,39.5)(239,40)
\DashLine(210,10)(240,40){6}\ArrowLine(238,38)(239,39)
\Line(50,10)(180,40)\ArrowLine(177,39.5)(179,40.2)
\Line(210,10)(180,40)\ArrowLine(181,39)(180,40)
\Text(130,2)[t]{$A(B^+_u\to K^+\overline{D^0})=A(B^-_u\to K^-D^0)$}
\Text(82,40)[r]{$\sqrt{2}\,A(B^+_u\to K^+ D_+^0)$}
\Text(245,20)[l]{$A(B^-_u\to K^-\overline{D^0})$}
\Text(315,60)[br]{$\sqrt{2}\,A(B^-_u\to K^- D_+^0)$}
\Text(105,60)[lb]{$A(B^+_u\to K^+D^0)$}
\Text(210,26.5)[c]{$2\gamma$}\CArc(210,10)(9,50,130)
\end{picture}
\end{center}
\caption{The extraction of $\gamma$ from 
$B_u^\pm\to K^\pm\{D^0,\overline{D^0},D^0_+\}$ 
decays.}\label{fig:BDK-triangle}
\end{figure}
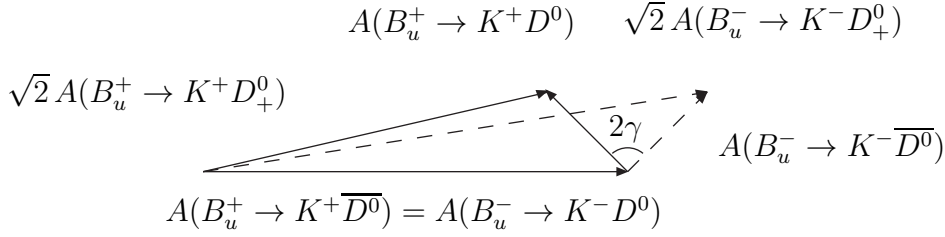

An important tool to eliminate the hadronic uncertainties in charged
$B$ decays is given by amplitude relations. The prototype of this 
approach, which was proposed by Gronau and Wyler \cite{gw}, uses 
$B^\pm \to K^\pm D$ decays and allows a {\it theoretically clean} extraction 
of the CKM angle $\gamma$. The decays $B^+\to K^+\overline{D^0}$ and 
$B^+\to K^+D^0$ are pure ``tree'' decays, as can be seen in 
Fig.~\ref{fig:BDK}. If we make, in addition, use of the transition 
$B^+\to D^0_+K^+$, where $D^0_+$ denotes the CP eigenstate of the 
neutral $D$-meson system with eigenvalue $+1$,
\begin{equation}\label{ED85}
|D^0_+\rangle=\frac{1}{\sqrt{2}}\left(|D^0\rangle+
|\overline{D^0}\rangle\right),
\end{equation}
we obtain
\begin{eqnarray}
\sqrt{2}A(B^+\to K^+D^0_+)&=&A(B^+\to K^+D^0)+
A(B^+\to K^+\overline{D^0})\\
\sqrt{2}A(B^-\to K^-D^0_+)&=&A(B^-\to K^-\overline{D^0})+
A(B^-\to K^-D^0).
\end{eqnarray}
These relations can be represented as two triangles in 
the complex plane. Since we have only to deal with tree-diagram-like 
topologies, we obtain moreover
\begin{eqnarray}
A(B^+\to K^+\overline{D^0})&=&A(B^-\to K^-D^0)\\
A(B^+\to K^+D^0)&=&A(B^-\to K^-\overline{D^0})\times e^{2i\gamma},
\end{eqnarray}
allowing a clean extraction of $\gamma$, as shown in 
Fig.~\ref{fig:BDK-triangle}. Unfortunately, these triangles are 
very squashed, since $B^+\to K^+D^0$ is colour-suppressed 
with respect to $B^+\to K^+\overline{D^0}$:
\begin{equation}\label{BDK-suppr}
\tilde r\equiv\left|\frac{A(B^+\to K^+D^0)}{A(B^+\to K^+\overline{D^0}}\right|=
\left|\frac{A(B^-\to K^-\overline{D^0})}{A(B^-\to K^-D^0}\right|\approx
\frac{1}{\lambda}\frac{|V_{ub}|}{|V_{cb}|}\times\frac{a_2}{a_1}
\approx0.1,
\end{equation}
where the phenomenological ``colour'' factors were introduced in
Subsection~\ref{subsec:fact}.

In 1998, the CLEO collaboration reported the observation of the colour-allowed
decay $B^+\to K^+\overline{D^0}$ with the following branching ratio
\cite{cleo-BDK}:
\begin{equation}
\mbox{BR}(B^+\to K^+\overline{D^0})=
(0.257\pm0.065\pm0.032)\times10^{-3}.
\end{equation}
Meanwhile, this decay has also been seen by Belle \cite{belle-BDK}. 
Using (\ref{BDK-suppr}), we expect
\begin{equation}
\mbox{BR}(B^+\to K^+D^0)\approx 10^{-2}\times
\mbox{BR}(B^+\to K^+\overline{D^0}).
\end{equation}
While $\mbox{BR}(B^+\to K^+\overline{D^0})$ can be 
determined using conventional methods, the measurement of 
$\mbox{BR}(B^+\to K^+D^0)$ suffers from considerable experimental 
problems \cite{ads}:
\begin{itemize}
\item If this colour-suppressed branching ratio is measured through 
Cabibbo-favoured hadronic decays of the $D^0$, e.g.\ through 
$B^+\to K^+D^0[\to K^-\pi^+]$, we obtain large interference effects 
with the colour-allowed decay chain $B^+\to K^+\overline{D^0}[\to K^-\pi^+]$, 
where the $\overline{D^0}$ decay is doubly Cabibbo-suppressed. 
\item All possible hadronic tags of the $D^0$-meson in $B^+\to K^+D^0$ will 
be affected by such interference effects.
\item These problems could be circumvented through semi-leptonic 
tags $D^0\to \ell^+\nu_\ell X_s$. However, here we have to deal with 
large backgrounds due to $B^+\to \ell^+\nu_\ell X_c$, which are
difficult to control.
\end{itemize}
Moreover, decays of neutral $D$-mesons into CP eigenstates, such as
$D^0_+\to\pi^+\pi^-$ or $D^0_+\to K^+K^-$, involve small efficiencies 
and are experimentally challenging.

\boldmath
\subsubsection{Alternative Approaches and Constraints on $\gamma$}
\unboldmath
As we have just seen, the original method proposed by Gronau and Wyler 
\cite{gw} will unfortunately be very difficult in practice. A variant of 
this approach was proposed by Atwood, Dunietz and Soni \cite{ads}. In order 
to overcome the problems listed above, these authors consider the 
following decay chains:
\begin{equation}
B^+\to K^+\overline{D^0}\,[\to f_i],\quad B^+\to K^+D^0 \,[\to f_i],
\end{equation}
where $f_i$ denotes doubly Cabibbo-suppressed (Cabibbo-favoured) non-CP 
modes of the $\overline{D^0}$ ($D^0$), for instance, $f_i=K^-\pi^+$, 
$K^-\pi^+\pi^0$. In order to determine $\gamma$, at least two different 
final states $f_i$ have to be employed. In this method, one makes 
use of the large interference effects, which spoil the hadronic tag 
of the $D^0$-meson in the original Gronau--Wyler method. In contrast to the
case of $B^+\to K^+D^0_+$ discussed above, here both contributing decay 
amplitudes should be of comparable size, thereby leading to potentially 
large CP-violating effects. Furthermore, the difficult to measure 
branching ratio $\mbox{BR}(B^+\to K^+D^0)$ is not required, but can 
rather be determined as a by-product. Unfortunately, this approach is 
also challenging, since many channels are involved, with total branching 
ratios of ${\cal O}(10^{-7})$ or even smaller. An accurate determination 
of the relevant $D$-meson branching ratios BR$(D^0\to f_i)$ and 
BR$(\overline{D^0}\to f_i)$ is also essential for this method. For
further refinements of this approach, see \cite{ADS2}.

The experimental problems related to the measurement of
$\mbox{BR}(B^+\to K^+D^0)$ were also avoided in 
\cite{gro-BDK,xing-BDK}, where the colour-allowed mode 
$B^+\to K^+\overline{D^0}$ and the two CP eigenstates $|D^0_\pm\rangle$ 
were considered, providing the observables
\begin{equation}\label{R-pm}
R_\pm\equiv2\left[\frac{\mbox{BR}(B^+\to K^+D_\pm^0)+
\mbox{BR}(B^-\to K^-D_\pm^0)}{\mbox{BR}(B^+\to K^+\overline{D^0})+
\mbox{BR}(B^-\to K^-D^0)}\right]=1\pm2\,\tilde r
\cos\tilde \delta\cos\gamma+\tilde r^2,
\end{equation}
where $\tilde r$ was introduced in (\ref{BDK-suppr}) and $\tilde \delta$ 
denotes a strong phase. As in the case of certain CP-averaged
$B\to\pi K$ branching ratios \cite{FM} (see \ref{subsec:FM-bound}),
the ratios in (\ref{R-pm}) imply
\begin{equation}
\sin^2\gamma\leq \mbox{min}\,(R_+,R_-),
\end{equation}
constraining $\gamma$ in the case of $\mbox{min}\,(R_+,R_-)<1$. 
Moreover, we have
\begin{equation}\label{R-sum}
R_++R_-=2(1+\tilde r^2),
\end{equation}
allowing in principle the determination of $\tilde r$, and the 
CP-violating observables
\begin{equation}
{\cal A}_\pm\equiv\frac{\mbox{BR}(B^+\to K^+D_\pm^0)-
\mbox{BR}(B^-\to K^-D_\pm^0)}{\mbox{BR}(B^+\to K^+\overline{D^0})+
\mbox{BR}(B^-\to K^-D^0)},
\end{equation}
which are equal in magnitude and have opposite signs, yielding the
combined asymmetry
\begin{equation}
{\cal A}\equiv {\cal A}_--{\cal A}_+=2\,\tilde r\sin\tilde \delta\sin\gamma.
\end{equation}
Consequently, we have three observables at our disposal, $R_+$, $R_-$ and 
${\cal A}$, allowing the extraction of $\gamma$, $\tilde \delta$ and 
$\tilde r$. However, because of $\tilde r\approx0.1$, this approach will 
also be very challenging in practice, in particular the resolution of the 
$\tilde r^2$ terms in (\ref{R-sum}) and of the 
CP-violating effects will be very difficult. It is possible to gain the 
knowledge of $\gamma$ by using, in addition, isospin-related neutral 
$B\to K^{(\ast)}D$ decays and neglecting certain annihilation topologies 
\cite{GroRo-BDK,JaKo}. This dynamical assumption can be tested through 
$B^+\to K^{(\ast)0}D^+$ modes, requiring branching ratios at the $10^{-7}$ 
level or even smaller \cite{GroRo-BDK}. 

Another alternative for the extraction of $\gamma$ is provided by 
$B_d\to K^{\ast0}D$ decays \cite{dun}, where the triangles are more 
equilateral and the interference effects associated with the 
hadronic tags of the $D^0$-mesons are less pronounced. 
But all sides are small, i.e.\ colour-suppressed, so that these decays 
are also not perfectly suited for the ``triangle'' approach.

\begin{figure}
\begin{center}
\leavevmode
\epsfysize=4.4truecm 
\epsffile{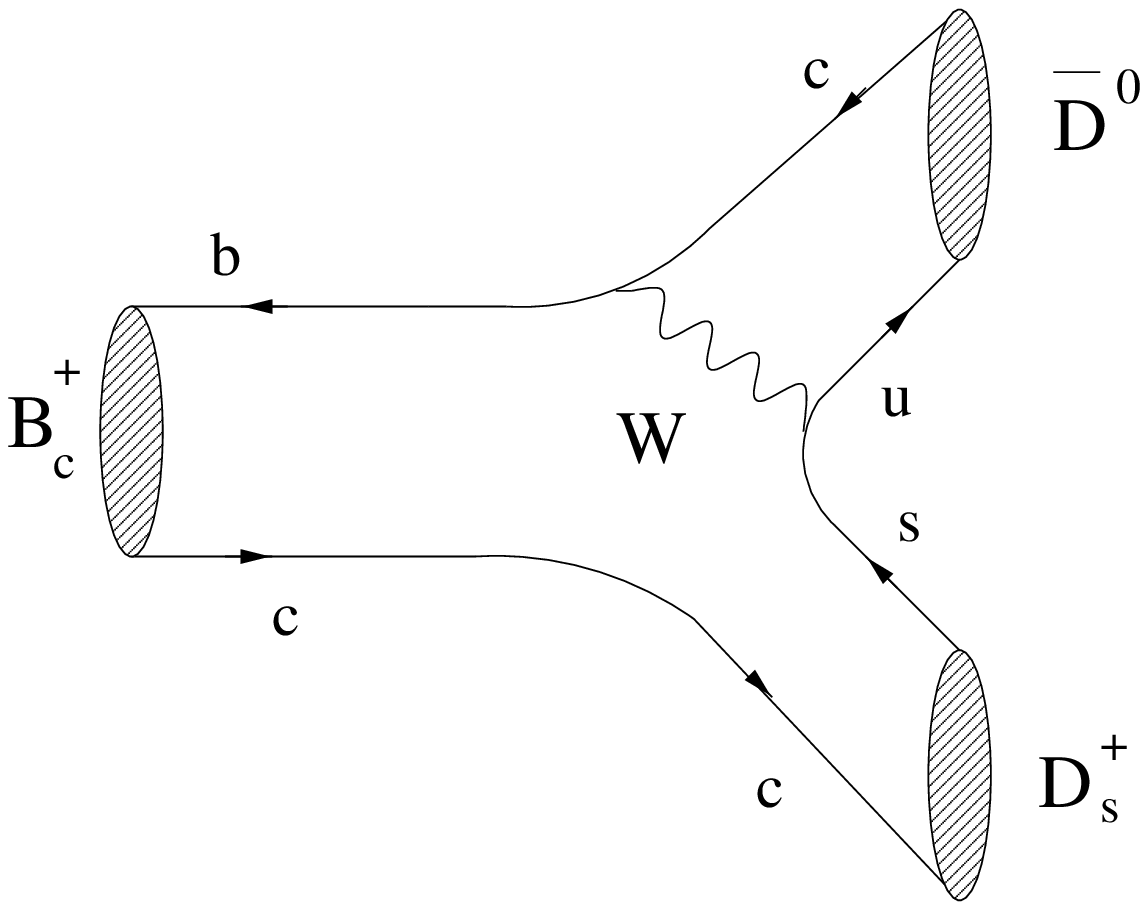} \hspace*{1.3truecm}
\epsfysize=4.0truecm 
\epsffile{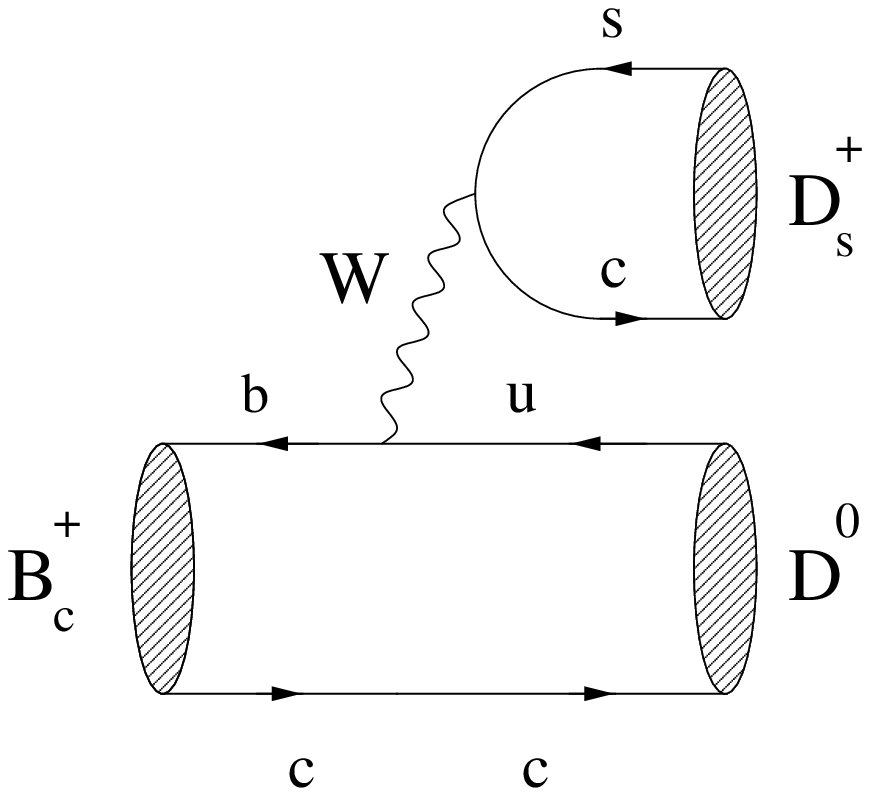}
\end{center}
\vspace*{-0.4truecm}
\caption{Feynman diagrams contributing to $B^+_c\to D_s^+\overline{D^0}$ and 
$B^+\to D_s^+D^0$. }\label{fig:BcDsD}
\end{figure}

\boldmath
\subsection{Extracting $\gamma$ from $B^\pm_c\to D_s^\pm D$ Decays}
\unboldmath
\boldmath
\subsubsection{Ideal Realization of Triangle Relations}
\begin{figure}
\begin{center}
\begin{picture}(320,150)(0,0)
\Line(50,10)(208,10) \ArrowLine(207,10)(209,10)
\DashLine(50,10)(290,130){6}\ArrowLine(288,129)(290,130)
\DashLine(210,10)(290,130){6}\ArrowLine(289,129)(290,130)
\Line(50,10)(130,130)\ArrowLine(128.3,128)(129,129)
\Line(210,10)(130,130)\ArrowLine(131.5,128)(131,129)
\Text(130,2)[t]{$A(B^+_c\to D_s^+\overline{D^0})=A(B^-_c\to D^-_sD^0)$}
\Text(82,80)[r]{$\sqrt{2}\,A(B^+_c\to D_s^+ D_+^0)$}
\Text(255,60)[l]{$A(B^-_c\to D^-_s\overline{D^0})$}
\Text(310,140)[br]{$\sqrt{2}\,A(B^-_c\to D_s^- D_+^0)$}
\Text(95,140)[lb]{$A(B^+_c\to D^+_sD^0)$}
\Text(210,29)[c]{$2\gamma$}\CArc(210,10)(33,57,123)
\end{picture}
\end{center}
\caption{The extraction of $\gamma$ from 
$B_c^\pm\to D^\pm_s\{D^0,\overline{D^0},D^0_+\}$ decays.}\label{fig:triangles}
\end{figure}
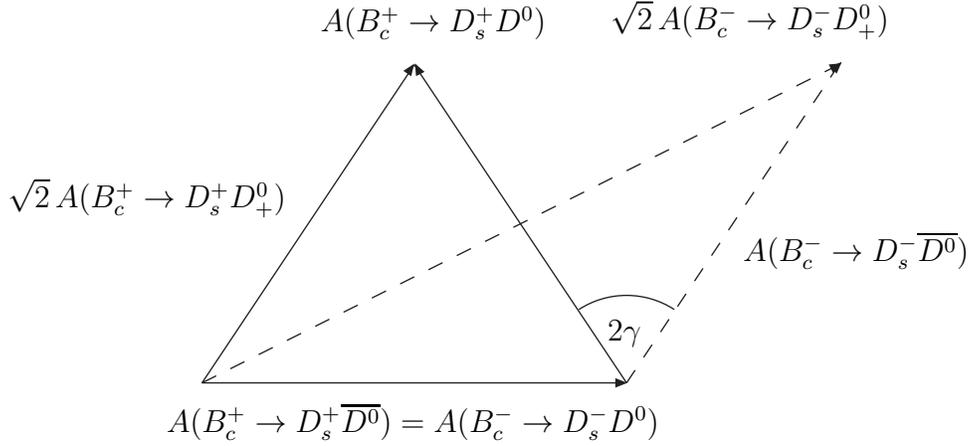

\unboldmath
The decays $B_c^\pm\to D_s^\pm D$ are the $B_c$-meson counterparts of 
the $B_u^\pm\to K^\pm D$ modes and allow also an extraction of $\gamma$ 
\cite{masetti}, which relies on the amplitude relations 
\begin{eqnarray}
\sqrt{2}A(B_c^+\to D_s^+D^0_+)&=&A(B_c^+\to D_s^+D^0)+
A(B_c^+\to D_s^+\overline{D^0})\\
\sqrt{2}A(B_c^-\to D_s^-D^0_+)&=&A(B_c^-\to D_s^-\overline{D^0})+
A(B_c^-\to D_s^-D^0),
\end{eqnarray}
with
\begin{eqnarray}
A(B^+_c\to D_s^+\overline{D^0})&=&A(B^-_c\to D_s^-D^0)\\
A(B_c^+\to D_s^+D^0)&=&A(B_c^-\to D_s^-\overline{D^0})\times e^{2i\gamma}.
\end{eqnarray}
At first sight, everything is completely analogous to the $B_u^\pm\to K^\pm D$
case. However, there is an important difference \cite{fw}, 
which becomes obvious by comparing the Feynman diagrams shown in 
Figs.\ \ref{fig:BDK} and \ref{fig:BcDsD}: in the $B_c^\pm\to D_s^\pm D$ 
system, the amplitude with the rather small CKM matrix element $V_{ub}$ 
is not colour suppressed, while the larger element $V_{cb}$ comes with 
a colour-suppression factor. Therefore, we obtain
\begin{equation}\label{Bc-ratio1}
\left|\frac{A(B^+_c\to D_s^+ D^0)}{A(B^+_c\to D_s^+ 
\overline{D^0})}\right|=
\left|\frac{A(B^-_c\to D_s^-\overline{D^0})}{A(B^-_c\to D_s^- 
D^0)}\right|\approx\frac{1}{\lambda}\frac{|V_{ub}|}{|V_{cb}|}
\times\frac{a_1}{a_2}={\cal O}(1),
\end{equation}
and conclude that the two amplitudes are similar in size. In contrast 
to this favourable situation, in the decays $B_u^{\pm}\to K^{\pm}D$, 
the matrix element $V_{ub}$ comes with the colour-suppression factor, 
resulting in a very stretched triangle. The extraction of $\gamma$ from 
the $B_c^\pm\to D_s^\pm D$ triangles is illustrated in 
Fig.~\ref{fig:triangles}, which should be compared with the
squashed $B^\pm_u\to K^\pm D$ triangles shown in 
Fig.\ \ref{fig:BDK-triangle}.

\boldmath
\subsubsection{$U$-Spin Relations and Experimental Remarks}
\unboldmath
A situation similar to the $B^\pm_u\to K^\pm D$ system arises in  
$B_c^\pm\to D^\pm D$ decays, which are obtained from the 
$B_c^\pm\to D^\pm_sD$ channels by interchanging all down and strange 
quarks, i.e.\ through the $U$-spin symmetry of strong interactions. 
These modes satisfy the amplitude relations
\begin{eqnarray}
\sqrt{2}A(B_c^+\to D^{+}D^0_+)&=&A(B^+_c\to D^{+}D^0)+
A(B^+_c\to D^+\overline{D^0})\\
\sqrt{2}A(B_c^-\to D^-D^0_+)&=&
A(B^-_c\to D^-\overline{D^0})+ A(B^-_c\to D^- D^0),
\end{eqnarray}
as well as
\begin{eqnarray}
A(B^+_c\to D^+ \overline{D^0})&=&A(B^-_c\to D^- D^0)\\
A(B^+_c\to D^+ D^0)&=&e^{i2\gamma}A(B^-_c\to D^- \overline{D^0}).
\end{eqnarray}
Because of CKM factors different from the $B_c^\pm\to D_s^\pm D$ case, 
we obtain
\begin{equation}\label{BcU-ratio}
\left|\frac{A(B^+_c\to D^+ D^0)}{A(B^+_c\to D^+ \overline{D^0})}\right|=
\left|\frac{A(B^-_c\to D^- \overline{D^0}}{A(B^-_c\to D^- D^0)}\right|
\approx\lambda^2\times\left(\frac{1}{\lambda}\frac{|V_{ub}|}{|V_{cb}|}
\times\frac{a_1}{a_2}\right)\lsim 0.1,
\end{equation}
implying triangles of the same shape as in the $B_u^\pm\to K^\pm D$ approach. 
The decays $B_d\to K^{\ast 0}D$ \cite {dun}, whose amplitudes 
are all colour suppressed and proportional to $\lambda^3 R_b$, have no 
analogue in the $B_c$ system.

The $B_c^\pm\to D^\pm D$ strategy is obviously affected by 
interference problems of the same kind as the original Gronau--Wyler approach,
i.e.\ we expect amplitudes of the same order of magnitude for the decay 
chains $B^+_c\to D^+D^0[\to K^-\pi^+]$ and 
$B^+_c\to D^+\overline{D^0}[\to K^-\pi^+]$. In order to extract $\gamma$, 
we could employ the same idea as in \cite{ads}. However, in the case of 
the $B_c$ system, an alternative is provided by the follwing $U$-spin 
relations:
\begin{eqnarray}
A(B^+_c\to D^+D^0)&=&-\lambda/(1-\lambda^2/2)\, 
A(B^+_c\to D^+_sD^0)\label{U-spin1}\\
A(B^+_c\to D^+\overline{D^0})&=&(1-\lambda^2/2)/\lambda\, 
A(B^+_c\to D^+_s\overline{D^0}).\label{U-spin2}
\end{eqnarray}
Since the amplitudes on the right-hand sides of these equations are
of the same order of magnitude (see (\ref{Bc-ratio1})), the
interference effects due to $D^0,\overline{D^0}\to \pi^\pm K^\mp$ are
practically unimportant in their measurement and in the associated 
$B_c^\pm\to D^\pm_s D$ strategy to determine $\gamma$. Consequently, 
this is the preferred $B_c$ approach to extract $\gamma$. 
Nevertheless, the Cabibbo-enhanced decay $B^+_c\to D^+\overline{D^0}$ 
plays an important r\^ole to increase the statistics for the measurement 
of the basis of the triangles shown in Fig.~\ref{fig:triangles} with the 
help of (\ref{U-spin2}). 

At the LHC, one expects ${\cal O}(10^{10})$ untriggered $B_c$-mesons 
per year of running \cite{LHC-Bc}. Estimates of $B_c$ branching ratios 
can already be found in the literature \cite{Bc-BRs}, yielding, however, 
in several cases conflicting results. The following values seem 
reasonable \cite{fw}:
\begin{equation}\label{BR-Bc-est}
\mbox{BR}(B^+_c\to D_s^+D^0)\approx 10^{-5},\quad
\mbox{BR}(B^+_c\to D_s^+ \overline{D^0})\approx
10^{-5}\mbox{--}10^{-6}.
\end{equation}
Moreover, we expect
\begin{equation}
\mbox{BR}(B^+_c\to D^+ \overline{D^0})\approx10^{-4}\mbox{--}10^{-5},
\end{equation}
allowing the measurement of the $B^+_c\to D_s^+ \overline{D^0}$ branching
ratio with the help of (\ref{U-spin2}). A very crude feasibility study
using (\ref{BR-Bc-est}) gives around 20 events per year at the LHC, 
demonstrating that the $B_c$ system may well contribute to our understanding
of CP violation. More refined experimental feasibility studies 
of the $B_c^\pm\to D_{(s)}^\pm D$ system are strongly encouraged.
From a theoretical point of view, it provides an ideal realization of 
the ``triangle'' approach to extract $\gamma$ \cite{fw}. 

%
%
%
\boldmath
\subsection{Impact of New Physics: $D^0$--$\overline{D^0}$ Mixing enters
the Stage} 
\unboldmath
Since the charged $B$ decays considered in the previous two subsections 
receive only contributions from tree-diagram-like topologies, involving
no flavour-changing neutral-current processes, it appears to be quite
unlikely that their amplitudes are affected significantly through 
new-physics contributions. However, we have to keep in mind that 
$D^0$--$\overline{D^0}$ mixing has been neglected in the corresponding
strategies to extract $\gamma$. This phenomenon is usually characterized by 
a CP-violating weak phase $\theta_D$, and by 
\begin{equation}
x_D\equiv \frac{\Delta M_D}{\Gamma_D},\quad
y_D\equiv \frac{\Delta\Gamma_D}{2\Gamma_D},
\end{equation}
where $\Delta M_D$ and $\Delta \Gamma_D$ denote the mass and width differences
of the neutral $D$-meson mass eigenstates, respectively, and $\Gamma_D$ 
is the average width. These mixing parameters are very small in the 
Standard Model, typical estimates give values for $x_D$ and $y_D$ at the 
$10^{-3}$ level. The parameter $x_D$ is governed by virtual transitions 
and may hence well be enhanced by one order of magnitude through new 
physics, whereas this appears to be rather unlikely for $y_D$, as this 
quantity is closely related to $D$ decay rates, where large new-physics 
effects are quite unlikely. However, $y_D$ may be enhanced by certain 
resonance effects in the Standard Model (for a recent review of these 
issues, see \cite{bigi-D}). 

A recent result $y_D=(3.42\pm1.39\pm0.74)\%$ from the FOCUS collaboration 
\cite{FOCUS}, corresponding to a signal at the 2\,$\sigma$ level, led 
already to some excitement \cite{BGLNP}. However, this effect has 
still to be confirmed by other experiments -- the results reported in 
\cite{E791,CLEO-yD} are consistent with zero -- and it is crucial to 
investigate whether such a value could also be accommodated in the Standard 
Model. Moreover, there is a problem of compatibility between the FOCUS and 
a recent CLEO result \cite{CLEO-D}, which may point towards a large 
CP-conserving strong phase difference. The $B$-factories will also provide 
very valuable insights into charm physics \cite{Babar-book}; a recent Belle 
study finds $y_D=(1.16^{+1.67}_{-1.65})\%$ \cite{tanaka}.

Apart from being a powerful tool to search for physics beyond the Standard 
Model, charm physics is also an important ingredient for $B^\pm\to K^\pm D$ 
strategies to extract $\gamma$. The impact of $D^0$--$\overline{D^0}$ mixing 
on these approaches was analysed in \cite{ADS2,MeSi,Silva-Soffer}. Since we 
have to deal with interference effects of ${\cal O}(10^{-1})$ in these modes, 
mixing parameters $x_D$ and $y_D$ at the $10^{-2}$ level may affect the 
extraction of $\gamma$ significantly \cite{MeSi,Silva-Soffer}. Strategies 
to include $D^0$--$\overline{D^0}$ mixing in the determination of $\gamma$ 
from $B^\pm\to K^\pm D$ decays can be found in \cite{ADS2,Silva-Soffer}. 
Since all amplitudes are of the same order of magnitude in the 
$B_c^\pm\to D_s^\pm D$ approach , the sensitivity to $D^0$--$\overline{D^0}$ 
mixing is considerably smaller in this case, which is another advantage of
the $B_c$ decays.

\subsection{Summary}
Since charged $B$-mesons do not exhibit mixing-induced CP violation, 
we have to search for fortunate decays, where the hadronic uncertainties
affecting the extraction of weak phases from direct CP asymmetries
can be eliminated through amplitude relations. The prototype 
of this approach employs $B^\pm_u\to K^\pm \{D^0,\overline{D^0},D^0_+\}$ 
modes, allowing a clean determination of $\gamma$ with the 
help of certain triangle relations, which are due to the CP eigenstate
$D^0_+$ of the neutral $D$-meson system. Unfortunately, the corresponding 
amplitude triangles are very squashed, and further problems related to 
hadronic tags of the $D$-mesons arise; variants of the $B^\pm_u\to K^\pm D$
strategy were proposed to deal with these problems. From a theoretical point 
of view, the triangle approach is realized in an ideal way in the 
$B^\pm_c\to D^\pm_s \{D^0,\overline{D^0},D^0_+\}$ system, where all 
amplitudes are of similar size. On the other hand, $B_c$-mesons are 
not as accessible as ``ordinary'' $B_u$- or $B_d$-mesons. The preferred
mechanism for new physics to manifest itself in these strategies is
$D^0$--$\overline{D^0}$ mixing, which may lead to significant effects 
in the $B^\pm_u\to K^\pm D$ approaches. Amplitude relations will also play 
a key r\^ole in the following $B\to\pi K$ discussion.

\begin{table}[t]
\begin{center}
\begin{tabular}{|c|c|c|c|}
\hline
Decay Mode & CLEO \cite{CLEO-BpiK} & BaBar \cite{babar-BpiK} & Belle 
\cite{belle-BpiK}\\
\hline
$B_d\to\pi^\mp K^\pm$ & $17.2^{+2.5}_{-2.4}\pm1.2$ & 
$16.7\pm1.6\pm1.3$ & $19.3^{+3.4+1.5}_{-3.2-0.6}$\\
$B^\pm\to\pi^0K^\pm$ & $11.6^{+3.0+1.4}_{-2.7-1.3}$ & $10.8^{+2.1}_{-1.9}
\pm1.0$ & $16.3^{+3.5+1.6}_{-3.3-1.8}$ \\
$B^\pm\to\pi^\pm K$ & $18.2^{+4.6}_{-4.0}\pm1.6$ & $18.2^{+3.3}_{-3.0}\pm2.0$& 
$13.7^{+5.7+1.9}_{-4.8-1.8}$\\
$B_d\to\pi^0 K$ & $14.6^{+5.9+2.4}_{-5.1-3.3}$ & $8.2^{+3.1}_{-2.7}\pm1.2$ &
$16.0^{+7.2+2.5}_{-5.9-2.7}$ \\
\hline
$B_d\to\pi^+ \pi^-$ & $4.3^{+1.6}_{-1.4}\pm0.5$ &
$4.1\pm1.0\pm0.7$ & $5.6^{+2.3}_{-2.0}\pm0.4$\\
$B^\pm\to\pi^\pm \pi^0$  & $5.6^{+2.6}_{-2.3}\pm1.7$ & 
$5.1^{+2.0}_{-1.8}\pm0.8$ & $7.8^{+3.8+0.8}_{-3.2-1.2}$\\
\hline
$B_d\to K^+ K^-$ & $<1.9$ (90\% C.L.) & $<2.5$ (90\% C.L.) 
& $<2.7$ (90\% C.L.)\\
$B^\pm\to K^\pm K$ & $<5.1$ (90\% C.L.) & $<2.4$ (90\% C.L.) 
& $<5.0$ (90\% C.L.)\\
\hline
\end{tabular}
\caption{CP-averaged $B\to \pi K,\pi\pi, K K$ branching ratios 
in units of $10^{-6}$.}\label{tab:BPIK}
\end{center}
\end{table}

\boldmath
\section{Phenomenology of $B\to\pi K$ Decays}\label{sec:BpiK}
\unboldmath
\setcounter{equation}{0}
\subsection{General Remarks and Experimental Status}
The physics potential of $B\to\pi K$ modes is very interesting for
the $B$-factories, as these channels provide promising strategies
to probe the angle $\gamma$ of the unitarity triangle. In contrast 
to the $B^\pm\to K^\pm D$ and $B^\pm_c\to D_s^\pm D$ decays discussed 
in the previous section, $B\to\pi K$ transitions are very accessible at 
these machines, thereby providing experimentally feasible constraints or
determinations of $\gamma$. On the other hand, the $B\to\pi K$ strategies 
are not theoretically clean, and require flavour-symmetry arguments and
certain plausible dynamical assumptions. In 1994, Gronau, Hern\'andez, 
London and Rosner pointed out in a series of pioneering papers that such 
a theoretical input allows extractions of weak phases from 
$B\to \pi K,\pi\pi, K K$ modes \cite{GRL,GHLR-SU3}.

The first observation of $B\to\pi K$ decays was announced by the
CLEO collaboration in 1997 \cite{CLEO97}, and has triggered a lot of 
theoretical work. In this context, the bound on $\gamma$ derived in 
\cite{FM}, making use of an approach to determine this angle with the help 
of the $B^0_d\to\pi^-K^+$, $B^+\to\pi^+K^0$ system \cite{PAPIII}, received 
a lot of attention. Since 1997, the experimental situation has improved 
considerably due to the efforts by the CLEO, BaBar and Belle collaborations. 
Now we have not only results available for $B^0_d\to\pi^-K^+$ and 
$B^+\to\pi^+K^0$, but the whole set of $B\to\pi K$ decays has been observed. 
The present status of the measurements of the corresponding CP-averaged 
branching ratios, which are defined in accordance with 
(\ref{CP-average-ampl}), is summarized in Table~\ref{tab:BPIK}. Since the 
$B\to\pi K$ strategies to probe $\gamma$ require also some input from the 
$B\to\pi\pi$ system, we have included these modes as well, whereas the 
decays $B_d\to K^+K^-$ and $B^\pm\to K^\pm K$ represent sensitive probes 
for rescattering effects. First analyses of CP-violating asymmetries in 
$B\to\pi K$ modes have also been performed. The corresponding results are 
listed in Table~\ref{tab:BPIK-Asym}, and do not yet provide signals for
direct CP violation.\footnote{The CP asymmetry for $B_d\to\pi^\mp K^\pm$
obtained by the BaBar Collaboration has recently been updated at the 
Lepton Photon 01 Conference, with the result $0.07\pm0.08\pm0.02$ 
\cite{BABAR-Bpipi-CP}.} In the future, 
the experimental situation should continue to improve significantly through 
the $B$-factories.

During the last couple of years, there was of course also theoretical 
progress. As far as the $B_d\to\pi^\mp K^\pm$, $B^\pm\to\pi^\pm K$ 
system is concerned, more refined strategies to probe $\gamma$ were 
proposed in \cite{defan,BpiK-mixed}. Moreover, it was noticed that 
the charged $B^\pm\to\pi^0 K^\pm$, $B^\pm\to \pi^\pm K$ \cite{NR,neubert-BpiK} 
and neutral $B_d\to \pi^\mp K^\pm$, $B_d\to\pi^0 K$ \cite{BF-BpiK1,BF-neut} 
modes are particularly well suited to extract $\gamma$, where in the 
latter case also mixing-induced CP violation in $B_d\to\pi^0 K_{\rm S}$ 
can be employed as an additional ingredient. The charged and neutral
approaches have certain theoretical advantages and are less sensitive to
possible rescattering effects, which can be probed -- and even taken 
into account -- through additional data on $B^\pm\to K^\pm K$ modes 
\cite{BFM,BF-BpiK1,defan,FKNP,RF-FSI,GR-FSI,FSI-strat}.

\begin{table}[t]
\begin{center}
\begin{tabular}{|c|c|c|c|}
\hline
Decay Mode & CLEO \cite{CLEO-BpiK-asym} & BaBar \cite{babar-BpiK}
& Belle \cite{Belle-BpiK-asym}\\
\hline
$B_d\to\pi^\mp K^\pm$ & $0.04\pm0.16$ & $0.19\pm0.10\pm0.03$
& $-0.044_{-0.186-0.018}^{+0.167+0.021}$ \\
$B^\pm\to\pi^0K^\pm$ & $0.29\pm0.23$ & $0.00\pm0.18\pm0.04$ &
$0.059_{-0.222-0.055}^{+0.196+0.017}$\\
$B^\pm\to\pi^\pm K$ & $-0.18\pm0.24$ & $0.21\pm0.18\pm0.03$ &  
$-0.098_{-0.430-0.020}^{+0.343+0.063}$ \\
$B_d\to\pi^0 K$ & n.a.\ & n.a.\ & n.a.\ \\
\hline
\end{tabular}
\caption{CP asymmetries ${\cal A}_{\rm CP}$ as defined in 
(\ref{CP-charged}) for $B\to\pi K$ modes.}\label{tab:BPIK-Asym}
\end{center}
\end{table}

The philosophy of these strategies is to extract a maximal amount of 
information, including $\gamma$ and hadronic parameters, with a 
``minimal'' input from theory, using only flavour-symmetry and plausible 
dynamical arguments. Concerning the description of the dynamics 
of $B\to\pi K,\pi\pi$ decays, important theoretical progress has recently 
been made through the development of the ``QCD factorization'' 
\cite{BBNS1}--\cite{BBNS3} and the perturbative hard-scattering (or ``PQCD'') 
approaches \cite{PQCD}--\cite{PQCD-comp}, allowing the systematic calculation 
of the relevant hadronic parameters at the leading order of a 
$\Lambda_{\rm QCD}/m_b$-expansion, and estimates of some of the 
contributions entering at the next-to-leading order level.

A very detailed analysis was recently performed within the 
QCD factorization approach in \cite{BBNS3}, providing valuable 
insights and allowing a reduction of the $SU(3)$-breaking corrections, 
which affect the strategies to probe $\gamma$. Moreover, rescattering 
processes are found to play a very minor r\^ole. If one is willing 
to use a stronger input from \mbox{QCD factorization}, concerning mainly 
CP-conserving strong phases, more stringent constraints on $\gamma$ can 
be obained. Exploiting QCD factorization in a maximal way, $\gamma$ 
can be determined by fitting simultaneously all calculated CP-averaged 
$B\to\pi K,\pi\pi$ branching ratios to the corresponding experimental 
results.\footnote{Using ``na\"\i ve'' factorization, such an approach 
was advocated in \cite{HSW}.} Here the central question is of course 
whether power corrections in $\Lambda_{\rm QCD}/m_b$ can really be controlled 
reliably. In a recent paper \cite{charming-pens2}, it was argued that 
non-perturbative penguin topologies with internal charm- and up-quark 
exchanges \cite{BF-u-c-pens,charming-pens1} (see 
Subsection~\ref{subsec:u-c-pens}) may play an important r\^ole in this 
context, thereby precluding the extraction of $\gamma$ from the calculated 
$B\to\pi K,\pi\pi$ branching ratios. Moreover, large CP asymmetries in 
certain modes are found in this analysis, in contrast to the predictions 
of QCD factorization. 

In this review, we shall focus on the former kind of strategies, using
a ``minimal'' theoretical input. These approaches do not only allow the 
extraction of $\gamma$, but also of strong phases and certain ratios of 
tree to penguin amplitudes. The hadronic parameters thus determined can 
then be compared with the theoretical predictions, provided, for example, 
by the QCD factorization or PQCD approaches.

\begin{figure}
\begin{center}
\leavevmode
\epsfysize=4.3truecm 
\epsffile{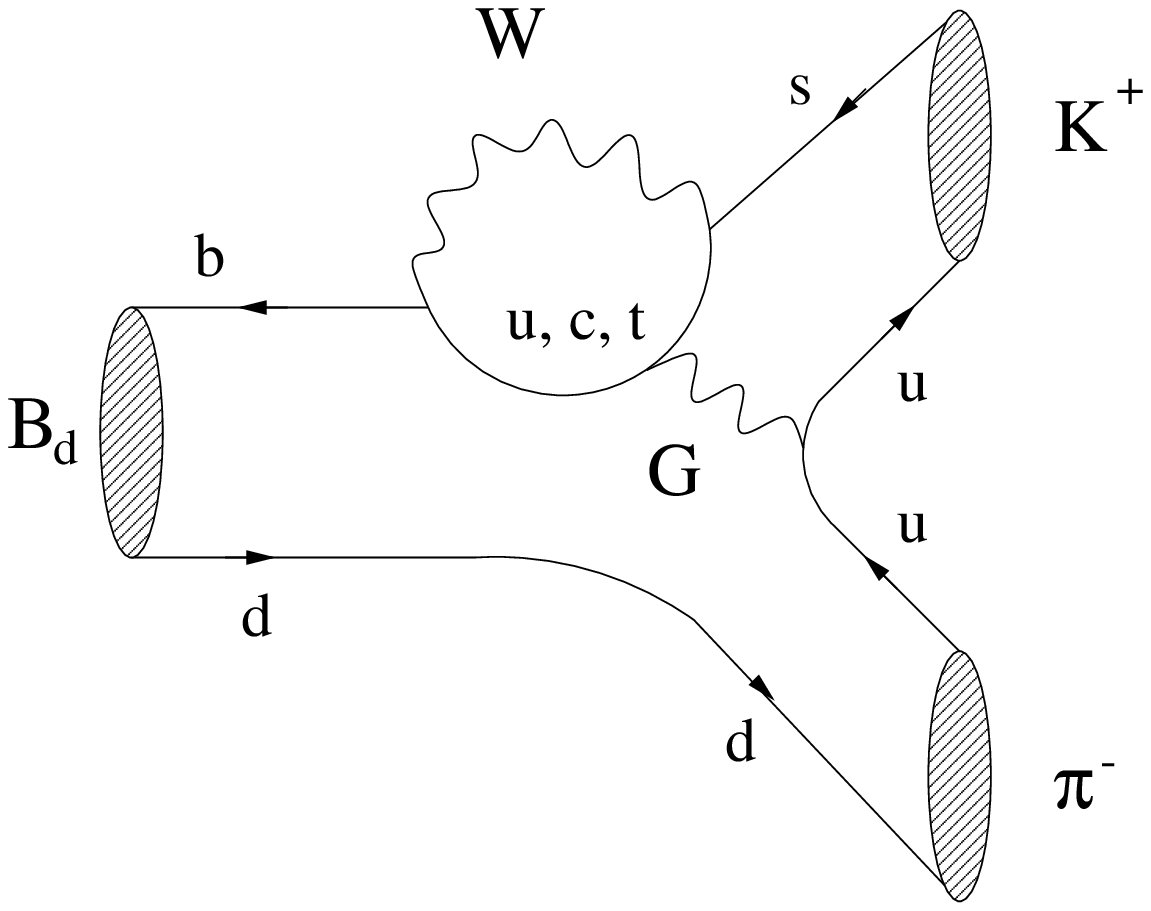} \hspace*{1.4truecm}
\epsfysize=3.8truecm 
\epsffile{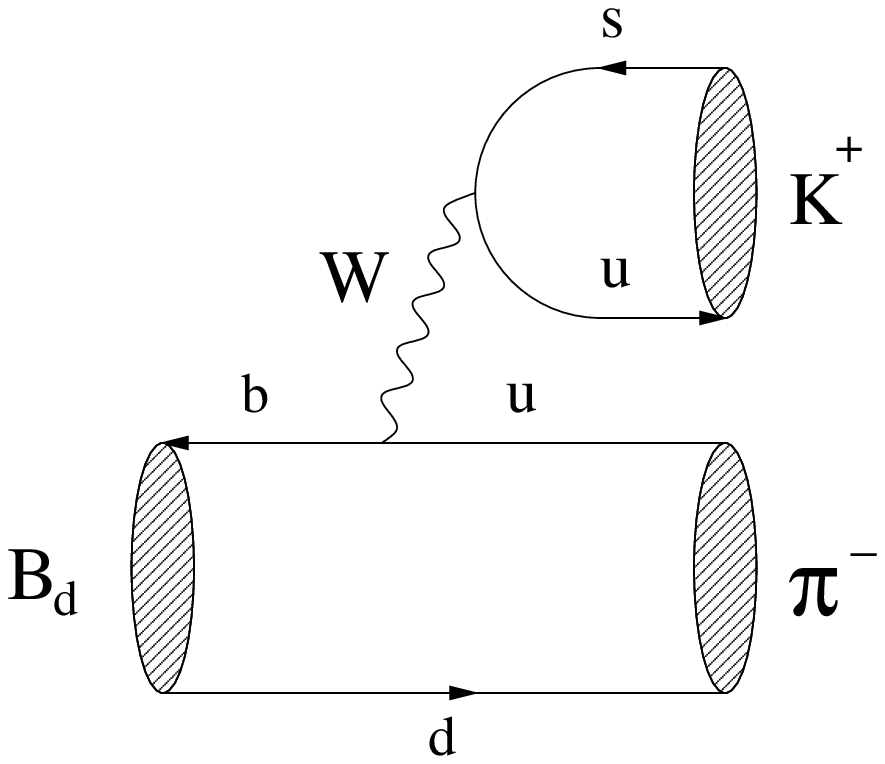}
\end{center}
\vspace*{-0.3truecm}
\caption{Feynman diagrams contributing to 
$B^0_d\to\pi^-K^+$.}\label{fig:BpiK-neutral}
\end{figure}

\boldmath
\subsection{The $B_d\to\pi^\mp K^\pm$, $B^\pm\to\pi^\pm K$ 
System}\label{sec:BPIK-gen}
\unboldmath
\subsubsection{Amplitude Structure and Isospin Relations}
In order to get more familiar with $B\to\pi K$ modes, let us consider 
the $B_d\to\pi^\mp K^\pm$, $B^\pm\to\pi^\pm K$ system first
\cite{PAPIII,FM,defan}. The decay $B^0_d\to\pi^-K^+$
receives contributions both from penguin and from
colour-allowed tree-diagram-like topologies, as can be seen in 
Fig.~\ref{fig:BpiK-neutral}. Within the
framework of the Standard Model, we may write
\begin{equation}\label{mixed-ampl-neut}
A(B^0_d\to\pi^-K^+)=-\lambda^{(s)}_u\left[\tilde P_u+\tilde 
P_{\rm ew}^{(u){\rm C}}+\tilde{\cal T}\right]-\lambda^{(s)}_c\left[\tilde P_c+
\tilde P_{\rm ew}^{(c){\rm C}}\right]-
\lambda^{(s)}_t\left[\tilde P_t+\tilde P_{\rm ew}^{(t){\rm C}}\right],
\end{equation}
where the overall minus sign is related to the definition of $SU(3)$ meson 
states adopted in \cite{GRL,GHLR-SU3}, $\tilde P_q$ and 
$\tilde P_{\rm ew}^{(q){\rm C}}$ denote contributions from QCD and EW penguin 
topologies with internal $q$-quark exchanges $(q\in\{u,c,t\})$, respectively, 
$\tilde{\cal T}$ is due to the colour-allowed 
$\overline{b}\to\overline{u}u\overline{s}$ tree-diagram-like topologies, 
and the CKM factors $\lambda^{(s)}_q$ were introduced in (\ref{lamqs-def}).
The label ``C'' reminds us that only colour-suppressed EW penguin topologies
contribute to $B^0_d\to\pi^-K^+$. Because of the tiny ratio 
$|\lambda^{(s)}_u/\lambda^{(s)}_t|\approx0.02$, QCD penguins play the 
dominant r\^ole in (\ref{mixed-ampl-neut}), despite their loop suppression. 
This interesting feature applies to all $B\to\pi K$ modes. Consequently, 
these decays are governed by flavour-changing neutral-current processes 
and are sensitive probes for new-physics effects 
\cite{FM-BpiK-NP}--\cite{matias}. 

Performing a similar exercise for the decay $B^+\to\pi^+K^0$ yields
\begin{equation}\label{BpiK-charged-ampl}
A(B^+\to\pi^+K^0)=\lambda^{(s)}_u\left[P_u+P_{\rm ew}^{(u){\rm C}}+
{\cal A}\right]+\lambda^{(s)}_c\left[P_c+P_{\rm ew}^{(c){\rm C}}\right]+
\lambda^{(s)}_t\left[P_t+P_{\rm ew}^{(t){\rm C}}\right],
\end{equation}
where the notation is as in (\ref{mixed-ampl-neut}), and ${\cal A}$ 
describes annihilation topologies. If we make use of the unitarity of 
the CKM matrix to eliminate $\lambda^{(s)}_t$ and apply the generalized 
Wolfenstein parametrization (\ref{NLO-wolf}), we obtain 
\begin{equation}\label{Bpampl}
A(B^+\to\pi^+K^0)\equiv P_{\rm c}=-\left(1-\frac{\lambda^2}{2}\right)\lambda^2A
\left(1+\rho_{\rm c} e^{i\theta_{\rm c}}e^{i\gamma}\right)
{\cal P}_{tc}^{\rm c},
\end{equation}
where
\begin{equation}\label{Ptc}
{\cal P}_{tc}^{\rm c}\equiv\left|{\cal P}_{tc}^{\rm c}\right|
e^{i\delta_{tc}^{\rm c}}=
P_t-P_c+P_{\rm ew}^{(t){\rm C}}-P_{\rm ew}^{(c){\rm C}},
\end{equation}
and
\begin{equation}\label{rho-def}
\rho_{\rm c} e^{i\theta_{\rm c}}=\left(\frac{\lambda^2}{1-\lambda^2}\right) 
R_b\left[1-\left(\frac{{\cal P}_{uc}^{\rm c}+
{\cal A}}{{\cal P}_{tc}^{\rm c}}\right)\right].
\end{equation}
In these expressions, $\delta_{tc}^{\rm c}$ and 
$\theta_{\rm c}$ denote CP-conserving strong phases, and 
${\cal P}_{uc}^{\rm c}$ is defined in analogy to (\ref{Ptc}). 
The quantity $\rho_{\rm c} e^{i\theta_{\rm c}}$, where the label ``c''
reminds us that we are dealing with a charged $B\to\pi K$ decay, is a 
measure of the strength of certain rescattering effects, as we will see below.
This notation will be useful in Subsection~\ref{sec:BpiK-cn}.

If we apply the $SU(2)$ isospin symmetry of strong interactions, implying 
\begin{equation}\label{PEN-rel}
\tilde P_c=P_c\quad\mbox{and}\quad\tilde P_t=P_t\,, 
\end{equation}
the QCD penguin topologies with internal top- and charm-quark exchanges 
contributing to $B_d^0\to\pi^-K^+$ and $B^+\to\pi^+K^0$ can be related 
to each other, yielding 
\begin{equation}\label{iso-mixed}
A(B_d^0\to\pi^-K^+)+A(B^+\to\pi^+K^0)=-\left(T+P_{\rm ew}^{\rm C}\right),
\end{equation}
where 
\begin{equation}
P_{\rm ew}^{\rm C}=-\left(1-\frac{\lambda^2}{2}\right)\lambda^2A 
\left[\left\{\tilde P_{\rm ew}^{(t){\rm C}}-\tilde
P_{\rm ew}^{(c){\rm C}}\right\}-\left\{P_{\rm ew}^{(t){\rm C}}-
P_{\rm ew}^{(c){\rm C}}\right\}\right]
\end{equation}
is essentially due to EW penguins, and 
\begin{equation}\label{T-def}
T=\lambda^4A\,R_b\left[
\tilde{\cal T}-{\cal A}+\left\{\tilde P_u-P_u\right\}+\left\{\tilde
P_{\rm ew}^{(u){\rm C}}-\tilde P_{\rm ew}^{(t){\rm C}}\right\}-
\left\{P_{\rm ew}^{(u){\rm C}}-P_{\rm ew}^{(t){\rm C}}\right\}\right]
e^{i\gamma}
\end{equation}
is usually referred to as a ``tree'' amplitude. However, owing to a 
subtlety in the implementation of the isospin symmetry, $T$ does not 
only receive contributions from colour-allowed tree-diagram-like 
topologies, but also from penguin and annihilation topologies 
\cite{BFM,defan}. Since the amplitudes $T$ and $P_{\rm ew}^{\rm C}$ 
have the following phase structure:
\begin{equation}
T\equiv|T|e^{i\delta_T}e^{i\gamma},\quad
P_{\rm ew}^{\rm C}\equiv-|P_{\rm ew}^{\rm C}|e^{i\delta_{\rm ew}^{\rm C}},
\end{equation}
where $\delta_T$ and $\delta_{\rm ew}^{\rm C}$ denote CP-conserving
strong phases, we obtain
\begin{equation}\label{mixed-AR}
A(B_d^0\to\pi^-K^+)+A(B^+\to\pi^+K^0)=
-\left(e^{i\gamma}-q_{\rm C}e^{i\omega_{\rm C}}\right)|T|e^{i\delta_T},
\end{equation}
with
\begin{equation}\label{qC-def}
q_{\rm C}\equiv \left|P_{\rm ew}^{\rm C}/T\right| \quad\mbox{and}\quad 
\omega_{\rm C}\equiv \delta_{\rm ew}^{\rm C}-\delta_T.
\end{equation}
Relation (\ref{mixed-AR}) allows us to probe $\gamma$ through the 
$B_d\to\pi^\mp K^\pm$, $B^\pm\to\pi^\pm K$ system. As we will see below,
the charged $B^\pm\to\pi^0 K^\pm$, $B^\pm\to \pi^\pm K$ and neutral 
$B_d\to \pi^\mp K^\pm$, $B_d\to\pi^0 K$ modes also satisfy isospin 
relations of the same structure.

\boldmath
\subsubsection{A Simple Bound on $\gamma$}\label{subsec:FM-bound}
\unboldmath
Interestingly, already the CP-averaged $B_d\to\pi^\mp K^\pm$, 
$B^\pm\to\pi^\pm K$ branching ratios may imply interesting 
constraints on $\gamma$ \cite{FM}. To simplify the discussion,
we make the following plausible dynamical assumptions:
\begin{itemize}
\item The parameter 
$\rho_{\rm c} e^{i\theta_{\rm c}}\propto \lambda^2 R_b\approx 0.02$ 
plays a negligible r\^ole in (\ref{Bpampl}), i.e.  
\begin{equation}\label{BpKpKo-simple}
A(B^+\to\pi^+K^0)=-|P_{\rm c}|e^{i\delta_{tc}^{\rm c}}=
A(B^-\to\pi^-\overline{K^0}).
\end{equation}
\item The colour-suppressed EW penguins play also a negligible r\^ole.
\end{itemize}
Using isospin relation (\ref{mixed-AR}), we may then re-write the 
$B^0_d\to\pi^-K^+$ decay amplitude as follows:
\begin{equation}\label{BdKmKp-simple}
A(B^0_d\to\pi^-K^+)=|P_{\rm c}|e^{i\delta_{tc}^{\rm c}}
\left(1-re^{i\delta}e^{i\gamma}\right),
\end{equation}
where 
\begin{equation}\label{simple-r-def}
r\equiv|T/P_{\rm c}| \quad\mbox{and}\quad 
\delta\equiv\delta_T-\delta_{tc}^{\rm c}. 
\end{equation}
Taking into account (\ref{BpKpKo-simple}), we observe that
(\ref{BdKmKp-simple}) implies amplitude triangles in the complex plane 
that are analogous to those shown in Fig.~\ref{fig:BDK-triangle}, allowing 
an extraction of $\gamma$ if $r$ is known \cite{PAPIII}. From (\ref{Bpampl}), 
(\ref{PEN-rel}) and (\ref{T-def}) we expect
\begin{equation}
r\approx\lambda^2 R_b \left|\tilde{\cal T}/\tilde{\cal P}_{tc}\right|
={\cal O}(\overline{\lambda}),
\end{equation}
where we have employed once more the generic expansion parameter 
$\overline{\lambda}$ introduced in Subsection~\ref{subsec:BpsiK}. 
This relation demonstrates nicely the dominance of the QCD penguin 
topologies in a quantitative way.

In order to probe $\gamma$, we consider the CP-averaged decay amplitude
\begin{equation}
\left\langle|A(B_d\to\pi^\mp K^\pm)|^2\right\rangle=
|P_{\rm c}|^2\left(1-2r\cos\delta\cos\gamma+r^2\right),
\end{equation}
where the overall normalization can be fixed through
\begin{equation}
\left\langle|A(B^\pm\to\pi^\pm K)|^2\right\rangle=|P_{\rm c}|^2.
\end{equation}
Finally, we arrive at the simple expression
\begin{equation}\label{R-param}
R\equiv
\left[\frac{\mbox{BR}(B_d^0\to\pi^- K^+)+\mbox{BR}(\overline{B_d^0}\to\pi^+ 
K^-)}{\mbox{BR}(B^+\to\pi^+ K^0)+\mbox{BR}(B^-\to\pi^-\overline{K^0})}
\right]\frac{\tau_{B^+}}{\tau_{B^0_d}}=
1-2r\cos\delta\cos\gamma+r^2.
\end{equation}
If we take a very conservative point of view, the hadronic quantities 
$\delta$ and $r$ are unknown parameters. Let us, therefore, treat them
as free parameters, which yields the following minimal value for $R$:
\begin{equation}\label{FM-bound}
\left.R_{\rm min}\right|_{\delta,r}=\sin^2\gamma\leq R,
\end{equation}
implying the allowed range \cite{FM}
\begin{equation}
0^\circ\leq\gamma\leq\gamma_0 \quad\lor\quad 
180^\circ-\gamma_0\leq\gamma\leq180^\circ,
\end{equation}
with
\begin{equation}
\gamma_0=\arccos(\sqrt{1-R}).
\end{equation}
This constraint on $\gamma$ is only effective if $R$ is found to be smaller 
than one. In 1997, when CLEO reported the first result on the CP-averaged
$B_d\to\pi^\mp K^\pm$, $B^\pm\to\pi^\pm K$ branching ratios, the result
was $R=0.65\pm0.40$. The central value $R=0.65$ would imply 
$\gamma_0=54^\circ$ \cite{FM}, thereby excluding a large range in the 
$\overline{\rho}$--$\overline{\eta}$ plane \cite{GNPS}. The present
experimental status of $R$ is summarized in Table~\ref{tab:BPIK-obs},
where $R_{\rm c}$ and $R_{\rm n}$ are the counterparts of $R$ in the
charged and neutral $B\to\pi K$ systems, respectively, which will be
discussed in Subsection~\ref{sec:BpiK-cn}. Unfortunately, 
the present experimental uncertainties are still too large to draw any 
conclusions and to decide whether $R<1$. The experimental situation 
should, however, improve significantly in the next couple of years.

\begin{table}[t]
\begin{center}
\begin{tabular}{|c|c|c|c|}
\hline
Observable & CLEO \cite{CLEO-BpiK} & BaBar \cite{babar-BpiK} & Belle 
\cite{belle-BpiK}\\
\hline
$R$ & $1.00\pm0.30$ & $0.97\pm0.23$ & $1.50\pm0.66$ \\
$R_{\rm c}$ & $1.27\pm0.47$ & $1.19\pm0.35$ & $2.38\pm1.12$ \\
$R_{\rm n}$ & $0.59\pm0.27$ & $1.02\pm0.40$ & $0.60\pm0.29$ \\
\hline
\end{tabular}
\caption{Ratios of CP-averaged $B\to \pi K$ branching 
ratios as defined in (\ref{R-param}), (\ref{Rc-def}) and (\ref{Rn-def}). 
For the evaluation of $R$, we have used $\tau_{B^+}/\tau_{B^0_d}=
1.060\pm0.029$.}\label{tab:BPIK-obs}
\end{center}
\vspace*{-0.3truecm}
\end{table}

\subsubsection{Impact of Rescattering Processes}\label{subsubsec:BpiK-res}
An important limitation of the theoretical accuracy of the
$B_d\to\pi^\mp K^\pm$, $B^\pm\to\pi^\pm K$ strategies to probe $\gamma$ 
is due to rescattering effects, which may affect the two dynamical 
assumptions made in \ref{subsec:FM-bound}. In particular, we expect 
na\"\i vely from (\ref{rho-def}) that
\begin{equation}\label{rho-naive}
\rho_{\rm c} e^{i\theta_{\rm c}}={\cal O}(\overline{\lambda}^2),
\end{equation}
and conclude that this parameter should play a negligible r\^ole in 
$B^+\to\pi^+K^0$. Moreover, this channel should hence exhibt tiny CP 
violation at the few percent level. These expectations, and the smallness
of the colour-suppressed EW penguins, may in principle be affected 
by very large rescattering effects \cite{FSI,BFM}.

As we have seen in Subsection~\ref{subsec:u-c-pens}, there are basically
two different kinds of rescattering processes, originating from 
$\overline{b}\to\overline{c}c\overline{s}$ and 
$\overline{b}\to\overline{u}u\overline{s}$ quark-level transitions. 
In the case of the $B^+\to\pi^+K^0$ mode, we have to deal with 
$B^+\to\{\overline{D^0}D_s^+,\,\ldots\}\to\pi^+K^0$ and
$B^+\to\{\pi^0K^+,\,\ldots\}\to\pi^+K^0$ rescattering processes.
Similar contributions arise also in the $B^0_d\to\pi^-K^+$ channel. 
The former kind of rescattering effects can be considered as penguin 
topologies with internal charm-quark exchanges (see Fig.~\ref{fig:pen-c}), 
and is included in the amplitudes $\tilde P_c$ and $P_c$ in 
(\ref{mixed-ampl-neut}) and (\ref{BpiK-charged-ampl}), respectively. These 
processes may contribute significantly to the $B_d\to\pi^\mp K^\pm$, 
$B^\pm\to\pi^\pm K$ branching ratios. However, as $P_c$ appears both 
in the numerator and in the denominator in (\ref{rho-def}), the parameter 
$\rho_{\rm c} e^{i\theta_{\rm c}}$ is unlikely to be enhanced significantly 
from the $\overline{\lambda}^2$ level by this kind of rescattering. On 
the other hand, the latter kind of rescattering processes is related to 
penguin topologies with internal up-quark exchanges and annihilation 
topologies (for analogous processes, see Figs.~\ref{fig:rescatter1} and 
\ref{fig:rescatter2}), affecting $\tilde P_u$, $P_u$ and 
${\cal A}$. Whereas these amplitudes play a minor r\^ole for the branching 
ratios, they affect the parameter $\rho_{\rm c} e^{i\theta_{\rm c}}$. 
In the presence of dramatic rescattering processes of this kind, i.e.\ of
\begin{equation}\label{BpiK-u-res}
B^+\to\{\pi^0K^+,\pi^0K^{\ast+},\ldots\}\to\pi^+K^0, 
\end{equation}
the parameter $\rho_{\rm c}$ may be enhanced to the $\overline{\lambda}$ level:
\begin{equation}\label{rho-FSI}
\left.\rho_{\rm c} e^{i\theta_{\rm c}}\right|_{\rm res.}=
{\cal O}(\overline{\lambda}),
\end{equation}
thereby implying CP-violating asymmetries in $B^\pm\to\pi^\pm K$
of ${\cal O}(10\%)$. In several of the analyses listed in \cite{FSI},
assuming elastic rescattering as a toy model or making use of Regge 
phenomenology, such an enhancement was actually found. 

Fortunately, we may also obtain experimental insights into these issues.
In this respect, the decay $B^+\to K^+ \overline{K^0}$, which is related 
to $B^+\to\pi^+K^0$ by interchanging all down and strange quarks, i.e.\ 
through the $U$-spin flavour symmetry of strong interactions, plays a key 
r\^ole \cite{BFM,defan,FKNP,RF-FSI}.  Using a notation similar to that in 
(\ref{Bpampl}) yields
\begin{equation}
A(B^+\to K^+ \overline{K^0})=\lambda^3A\left[1-
\left(\frac{1-\lambda^2}{\lambda^2}\right)\rho^{(d)} 
e^{i\theta^{(d)}}e^{i\gamma}\right]{\cal P}_{tc}^{(d)},
\end{equation}
where $\rho^{(d)} e^{i\theta^{(d)}}$ corresponds to 
$\rho_{\rm c} e^{i\theta_{\rm c}}$ given in (\ref{rho-def}). Since 
this parameter enters -- in contrast to (\ref{Bpampl}) -- with 
$1/\lambda^2$ in the $B^+\to K^+ \overline{K^0}$ amplitude, the 
corresponding branching ratio and CP asymmetry represent a very 
sensitive probe for an enhancement of $\rho^{(d)}$ through rescattering
processes. The $U$-spin symmetry of strong interactions implies
\begin{equation}\label{U-spin-KK}
\rho^{(d)}=\rho_{\rm c},\quad \theta^{(d)}=\theta_{\rm c},
\end{equation}
allowing us to determine $\rho_{\rm c}$ and $\theta_{\rm c}$ as functions 
of $\gamma$ through the observables of the $B^\pm\to \pi^\pm K$, 
$B^\pm\to K^\pm K$ system \cite{defan,RF-FSI}. Moreover, we obtain the 
following relation between the corresponding CP asymmetries and CP-averaged 
branching ratios:
\begin{equation}\label{CP-BR-rel}
\frac{{\cal A}_{\rm CP}(B^+\to \pi^+ K^0)}{{\cal A}_{\rm CP}(B^+\to 
K^+\overline{K^0})}=-R_{SU(3)}^2\left[\frac{\mbox{BR}(B^\pm\to 
K^\pm K)}{\mbox{BR}(B^\pm\to \pi^\pm K)}\right],
\end{equation}
where $R_{SU(3)}$ describes $SU(3)$ breaking, with
\begin{equation}
\left.R_{SU(3)}\right|_{\rm fact}=
\left[\frac{M_B^2-M_\pi^2}{M_B^2-M_K^2}\right]
\left[\frac{F_{B\pi}(M_K^2;0^+)}{F_{BK}(M_K^2;0^+)}\right].
\end{equation}
Here the form factors $F_{B\pi}(M_K^2;0^+)$ and $F_{BK}(M_K^2;0^+)$ 
parametrize the quark--current matrix elements 
$\langle\pi|(\overline{b}d)_{\rm V-A}|B\rangle$ and
$\langle K|(\overline{b}s)_{\rm V-A}|B\rangle$, respectively. Using
the model of Bauer, Stech and Wirbel (BSW) \cite{BSW} yields
$R_{SU(3)}={\cal O}(0.7)$.

The presently available upper bounds on the CP-averaged 
$B^\pm\to K^\pm K$ branching ratio imply already interesting upper bounds 
on $\rho_{\rm c}$. To this end, we employ the $U$-spin relations in 
(\ref{U-spin-KK}), and consider the quantity \cite{BF-neut}
\begin{equation}\label{K-def}
K\equiv\left[\frac{1}{\epsilon\,R_{SU(3)}^2}\right]\left[
\frac{\mbox{BR}(B^\pm\to\pi^\pm K)}{\mbox{BR}(B^\pm\to K^\pm K)}\right]=
\frac{1+2\,\rho_{\rm c}\cos\theta_{\rm c}\cos\gamma+
\rho_{\rm c}^2}{\epsilon^2-2\,\epsilon\,\rho_{\rm c}
\cos\theta_{\rm c}\cos\gamma+\rho_{\rm c}^2},
\end{equation}
where $\epsilon\equiv\lambda^2/(1-\lambda^2)$. The 
expression on the right-hand side implies the following allowed range 
for $\rho_{\rm c}$ (for detailed discussions, see \cite{RF-bpipi,pirjol} 
and \ref{subsubsec:U-spin-Replace}):
\begin{equation}\label{rho-range}
\frac{1-\epsilon\,\sqrt{K}}{1+\sqrt{K}}\leq\rho_{\rm c}\leq
\frac{1+\epsilon\,\sqrt{K}}{|1-\sqrt{K}|}.
\end{equation}
The present CLEO data give $\mbox{BR}(B^\pm\to K^\pm K)/
\mbox{BR}(B^\pm\to\pi^\pm K)<0.3$ ($90\%$ C.L.) \cite{CLEO-BpiK}. 
Using (\ref{rho-range}), this upper bound implies $\rho_{\rm c}<0.15$ 
for $R_{SU(3)}=0.7$, and is not in favour of dramatic rescattering effects,
although the bound on $\rho_{\rm c}$ is still one order of magnitude above the 
na\"\i ve ${\cal O}(\overline{\lambda}^2)$ expectation.\footnote{If we
assume $\mbox{BR}(B^\pm\to K^\pm K)/\mbox{BR}(B^\pm\to\pi^\pm K)<0.15$, as
indicated by the BaBar bound on $\mbox{BR}(B^\pm\to K^\pm K)$ in 
Table~\ref{tab:BPIK}, we obtain $\rho_{\rm c}<0.12$.} Rescattering 
effects can also be probed through $B_d\to K^+K^-$ modes \cite{GR-FSI},
which also do not seem to show an anomalous enhancement.

The second dynamical assumption in \ref{subsec:FM-bound} was related to
the colour-suppressed EW penguins. Using the expressions for the 
corresponding four-quark operators (see (\ref{ew-penguins})), neglecting 
$Q_7^s$ and $Q_8^s$ because of their tiny Wilson coefficients, and 
performing appropriate Fierz transformations, we may write 
\cite{BF-BpiK1,defan} (see also \cite{PAPIII})
\begin{equation}\label{qc-expr}
q_{\rm C}e^{i\omega_{\rm C}}=
\frac{3}{2\lambda^2R_b}\left[\frac{C_1'(\mu)C_{10}(\mu)-
C_2'(\mu)C_9(\mu)}{C_2'^2(\mu)-C_1'^2(\mu)}\right]
a_{\rm C} e^{i\omega_{\rm C}},
\end{equation}
where 
\begin{equation}
C_1'(\mu)\equiv C_1(\mu)+\frac{3}{2}\,C_9(\mu),\quad C_2'(\mu)\equiv C_2(\mu)+
\frac{3}{2}\,C_{10}(\mu), 
\end{equation}
and 
\begin{equation}
a_{\rm C} e^{i\omega_{\rm C}}\equiv\frac{a_2^{\rm eff}}{a_1^{\rm eff}}
=\frac{C_1'(\mu)\,\zeta(\mu)+C_2'(\mu)}{C_1'(\mu)+C_2'(\mu)\,\zeta(\mu)}
\end{equation}
with
\begin{equation}\label{zeta-def}
\zeta(\mu)\equiv\frac{\langle K^0\pi^+|Q_2^u(\mu)|B^+\rangle+\langle 
K^+\pi^-|Q_2^u(\mu)|B^0_d\rangle}{\langle K^0\pi^+|Q_1^u(\mu)|B^+\rangle+
\langle K^+\pi^-|Q_1^u(\mu)|B^0_d\rangle}
\end{equation}
denotes a hadronic parameter, representing a measure of ``colour 
suppression'' in the $B_d\to\pi^\mp K^\pm$, $B^\pm\to\pi^\pm K$ system. 
The combination of Wilson coefficients in (\ref{qc-expr}) is essentially 
renormalization-scale independent and changes only by 
${\cal O}(1\%)$ when evolving from $\mu=M_W$ down to $\mu=m_b$.
Using the leading-order coefficients in Table~\ref{tab:WC} for 
$\Lambda_{\overline{{\rm MS}}}^{(5)}=225\,\mbox{MeV}$ yields
\begin{equation}
q_{\rm C}=0.71\times\left[\frac{0.38}{R_b}\right]\times a_{\rm C}. 
\end{equation}
It is plausible to assume that $a_{\rm C}$ is associated with a suppression
of ${\cal O}(\overline{\lambda})$, thereby yielding
\begin{equation}\label{qC-naive}
q_{\rm C}={\cal O}(\overline{\lambda}).
\end{equation}
Note that this implies $|P_{\rm ew}^{\rm C}/P|={\cal O}(\overline{\lambda}^2)$
because of $r={\cal O}(\overline{\lambda})$. Consequently, EW penguins
contribute at the same level to the $B_d\to\pi^\mp K^\pm$, $B^\pm\to\pi^\pm K$
amplitudes as the parameter $\rho_{\rm c} e^{i\theta_{\rm c}}$. In the 
presence of dramatic rescattering processes of the kind given in 
(\ref{BpiK-u-res}), the dynamical suppression in (\ref{qC-naive}) would no 
longer be effective, i.e.
\begin{equation}
\left.q_{\rm C}\right|_{\rm res.}={\cal O}(1),
\end{equation}
yielding contributions to the decay amplitudes at the $\overline{\lambda}$
level. The parameter $r$ would be of ${\cal O}(\overline{\lambda})$ also
in the presence of large rescattering effects, although its value may
be affected significantly. 

Consequently, in the case of large rescattering effects mediated by
$\overline{b}\to\overline{u}u\overline{s}$ quark-level processes,
both assumptions made in \ref{subsec:FM-bound} would no longer hold.
As we have noted above, the present experimental upper bounds on 
$B\to KK$ modes are not pointing towards such a picture. Moreover, 
it is disfavoured theoretically by the QCD factorization approach, 
where the predictions for the relevant parameters read as follows 
\cite{BBNS3}:\footnote{The first errors are the sums of all theoretical
errors added in quadrature, whereas the second errors are due to 
$R_b$; the second lines show the results without 
certain weak annihilation contributions.}
\begin{equation}\label{rho-QCD-fact}
\rho_{\rm c}=\left\{\begin{array}{l}
(2.0\pm0.1\pm0.4)\times 10^{-2}\\
(1.9\pm0.1\pm0.4)\times 10^{-2},
\end{array}\right.
\quad 
\theta_{\rm c}=\left\{\begin{array}{l}
(13.6\pm4.4)^\circ\\
(16.6\pm5.2)^\circ
\end{array}\right.
\end{equation}
\begin{equation}\label{qC-QCD-fact}
q_{\rm C}=\left\{\begin{array}{l}
(8.3\pm4.5\mp1.7)\times 10^{-2}\\
(8.9\pm4.9\mp1.8)\times 10^{-2},
\end{array}\right.
\quad
\omega_{\rm C}=\left\{\begin{array}{l}
(-60.2\pm49.5)^\circ\\
(-54.2\pm44.2)^\circ.
\end{array}\right.
\end{equation}
In particular, it is found that the bound in (\ref{FM-bound}) is violated 
through these parameters by at most $1.5\%$, and only in the region 
$72^\circ<\gamma<86^\circ$. This analysis suggests that these effects can 
be safely neglected until $R$ is measured with an accuracy better than 
$10\%$.

\boldmath
\subsubsection{More Refined Strategies to Probe 
$\gamma$}\label{ref-gamma-strat}
\unboldmath
In addition to the ratio $R$ of CP-averaged branching ratios, 
also the ``pseudo-asymmetry'' 
\begin{equation}\label{A0-def}
A_0\equiv\left[\frac{\mbox{BR}(B^0_d\to\pi^-K^+)-\mbox{BR}(\overline{B^0_d}\to
\pi^+K^-)}{\mbox{BR}(B^+\to\pi^+K^0)+\mbox{BR}(B^-\to\pi^-\overline{K^0})}
\right]\frac{\tau_{B^+}}{\tau_{B^0_d}}=
{\cal A}_{\rm CP}(B^0_d\to\pi^-K^+)R
\end{equation}
plays an important r\^ole in the probing of $\gamma$ \cite{BpiK-mixed}. 
In order to derive parametrizations for $R$ and $A_0$, it is useful
to introduce the following generalization of $r$:
\begin{equation}
r\equiv\frac{|T|}{\sqrt{\langle|A(B^\pm\to\pi^\pm K)|^2\rangle}},
\end{equation}
which reduces to (\ref{simple-r-def}) for $\rho_{\rm c}=0$. Using 
(\ref{Bpampl}) and isospin relation (\ref{iso-mixed}),
the observables can be expressed in terms of $\gamma$ and the
hadronic parameters $r$, $\delta$, $q_{\rm C}$, $\omega_{\rm C}$, 
$\rho_{\rm c}$, $\theta_{\rm c}$. With the help of the corresponding 
parametrizations, which are given in \cite{defan}, we may 
now proceed as in \ref{subsec:FM-bound}. If we treat both $\delta$ and $r$ 
as unknown quantities, we arrive at 
\cite{defan} 
\begin{equation}\label{Rmin-refined}
\left.R_{\rm min}\right|_{\delta,r}=
\left[\frac{1+2\,q_{\rm C}\, \rho_{\rm c} \cos(\theta_{\rm c}+\omega_{\rm C})
+q_{\rm C}^2\,\rho_{\rm c}^2}{\left(1-2\,q_{\rm C} \cos\omega_{\rm C}
\cos\gamma + q_{\rm C}^2\right)\left(1+2\,\rho_{\rm c} \cos\theta_{\rm c} 
\cos\gamma+\rho_{\rm c}^2\right)}\right]\sin^2\gamma,
\end{equation}
representing the generalization of (\ref{FM-bound}). On the other
hand, if we  make also use of the asymmetry $A_0$, contours in the 
$\gamma$--$r$ plane can be determined, which provide stronger
constraints on $\gamma$. The sensitivity on 
$\rho_{\rm c} e^{i\theta_{\rm c}}$ and 
$q_{\rm C}e^{i\omega_{\rm C}}$ was studied in great detail
in \cite{defan}. It should be emphasized that the former parameter
can be taken into account in (\ref{Rmin-refined}) and the contours
in the $\gamma$--$r$ plane through the $B^\pm\to K^\pm K$ observables.
As we have seen above, they can be combined with those of 
$B^\pm\to \pi^\pm K$ through the $U$-spin symmetry, thereby allowing us
to determine $\rho_{\rm c}$ and $\theta_{\rm c}$ as functions of $\gamma$ 
\cite{defan,RF-FSI}. 

If the parameter $r$, i.e.\ the magnitude of the ``tree'' amplitude $T$, 
can be determined, we are in a position to extract $\gamma$ and the
strong phase $\delta$. At this stage, a certain model dependence enters. 
An approximate way to fix this amplitude is to neglect the 
colour-suppressed CC operator contributions to $B^+\to\pi^+\pi^0$, 
and to use the $SU(3)$ flavour symmetry to relate the colour-allowed 
CC amplitude of that decay to $T$, yielding \cite{PAPIII}
\begin{equation}
r\approx\sqrt{2}\,\lambda\frac{f_K}{f_\pi}
\sqrt{\frac{\mbox{BR}(B^\pm\to\pi^\pm\pi^0)}{\mbox{BR}(B^\pm\to\pi^\pm K)}}.
\end{equation}
Using this expression, the experimental results in Table~\ref{tab:BPIK} 
imply $r\approx0.2$. Another approach is provided by ``na\"\i ve'' 
factorization:
\begin{equation}\label{T-fact}
\left.|T|\right|_{{\rm fact}}=
\frac{G_{\rm F}}{\sqrt{2}}\,\lambda\,|V_{ub}|\,a_1 \left(M_{B_d}^2-
M_\pi^2\right) f_{K} F_{B\pi}(M_K^2;0^+).
\end{equation}
The BSW model \cite{BSW} and QCD sum rules \cite{QCDSrules} give 
$F_{B\pi}(M_K^2;0^+)=0.3$, leading to
\begin{equation}\label{T-fact2}
\left.|T|\right|_{{\rm fact}}=a_1\times\left[\frac{|V_{ub}|}{3.2\times
10^{-3}}\right]\times7.8\times10^{-9}\,\mbox{GeV}.
\end{equation}
As noted in \cite{BpiK-mixed}, also analyses of semi-leptonic 
$B^0_d\to \pi^-\ell^+\nu_\ell$ decays may be very useful to fix $|T|$ 
through arguments based on factorization. Using (\ref{T-fact2}), we 
obtain \cite{FM}
\begin{equation}
\left.r\right|_{{\rm fact}}=0.18\times a_1\times
\left[\frac{|V_{ub}|}{3.2\times10^{-3}}\right]
\sqrt{\left[\frac{1.8\times10^{-5}}{\mbox{BR}(B^\pm\to\pi^\pm K)}\right]
\times\left[\frac{\tau_{B_u}}{1.6\,\mbox{ps}}\right]}.
\end{equation}
The most refined theoretical analysis employing QCD factorization
gives \cite{BBNS3}
\begin{equation}
\overline{r}\equiv u_{\rm c}\, r=\left\{\begin{array}{l}
(20.6\pm3.5\pm4.1)\times 10^{-2}\\
(22.0\pm3.6\pm4.4)\times 10^{-2},
\end{array}\right.
\quad 
\delta=\left\{\begin{array}{l}
(-5.7\pm4.4)^\circ\\
(-6.2\pm4.6)^\circ,
\end{array}\right.
\end{equation}
where 
\begin{equation}\label{u-def}
u_{\rm c}=\sqrt{1+2\,\rho_{\rm c}\cos\theta_{\rm c}\cos\gamma+\rho_{\rm c}^2}.
\end{equation}
Since $\rho_{\rm c}$ is very small in the QCD factorization approach 
(see (\ref{rho-QCD-fact})), $\overline{r}$ and $r$ take very similar
values, which are also consistent with the other approaches to fix $r$.
We shall come back to the determination of $r$ in 
Subsection~\ref{subsec:BspiK-Uspin}, using $B_s\to\pi^\pm K^\mp$ modes. 

An advantage of the $B_d\to\pi^\mp K^\pm$, $B^\pm\to\pi^\pm K$ strategies 
to probe $\gamma$ is that the correponding decays do not involve neutral
pions and are hence also very accessible at hadron machines, in particular 
at the LHC. There one expects measurements of $R$ and $A_0$ with uncertainties
at the $3\%$ level \cite{LHC-Report}. On the other hand, this approach 
suffers from the theoretical disadvantage that $r$ requires some additional 
input and that the treatment of EW penguins is based on colour-suppression 
arguments. Moreover, $r$ may in principle be affected by rescattering 
processes of the kind given in (\ref{BpiK-u-res}), which may also enhance 
the EW penguin contributions. The charged and neutral $B\to\pi K$ systems 
have interesting theoretical advantages in this respect.

\boldmath
\subsection{The Charged and Neutral $B\to\pi K$ Systems}\label{sec:BpiK-cn}
\unboldmath
In 1994, Gronau, Rosner and London proposed an $SU(3)$ strategy to determine 
$\gamma$ with the help of the charged decays $B^{\pm}\to\pi^{\pm} K$, 
$\pi^0K^{\pm}$, $\pi^0\pi^{\pm}$ \cite{GRL}. However, as was pointed out by 
Deshpande and He \cite{EWP-BpiK}, this elegant approach is spoiled by EW 
penguins \cite{ghlr-ewp}, which play an important r\^ole because of the 
large top-quark mass \cite{RF-EWP1}. In 1998, the Gronau--Rosner--London 
strategy was resurrected by Neubert and Rosner \cite{NR}, who showed that 
the EW penguin contributions can be controlled in this case theoretically 
with the help of the general expressions for the corresponding four-quark 
operators, appropriate Fierz transformations, and the $SU(3)$ flavour 
symmetry of strong interactions (see also \cite{PAPIII}). The neutral 
$B\to\pi K$ system is completely analogous in this respect, providing, 
however, an additional observable, if $B_d\to\pi^0K_{\rm S}$ modes are 
considered \cite{BF-BpiK1}.

\subsubsection{Parametrization of Decay Amplitudes and Observables}
The starting point of our discussion is the following isospin relation:
\begin{displaymath}
\sqrt{2}\,A(B^+\to\pi^0K^+)\,+\,A(B^+\to\pi^+K^0)=
\sqrt{2}\,A(B^0_d\to\pi^0K^0)\,+\,A(B^0_d\to\pi^-K^+)
\end{displaymath}
\begin{equation}\label{iso1}
=-\left[(T+C)\,+\,P_{\rm ew}\right]\equiv 3\,A_{3/2},
\end{equation}
where the combination $(T+C)$ originates from colour-allowed and 
colour-suppressed $\overline{b}\to\overline{u}u\overline{s}$ 
tree-diagram-like topologies, $P_{\rm ew}$ is due to colour-allowed 
and colour-suppressed EW penguin constributions, and $A_{3/2}$ reminds 
us that there is only an $I=3/2$ isospin component in (\ref{iso1}). 
In the Standard Model, these amplitudes can be parametrized as 
\begin{equation}\label{Ampl-def}
T+C\equiv|T+C|\,e^{i\delta_{T+C}}\,e^{i\gamma},\quad
P_{\rm ew}\equiv-\,|P_{\rm ew}|e^{i\delta_{\rm ew}},
\end{equation}
where $\delta_{T+C}$ and $\delta_{\rm ew}$ denote CP-conserving strong
phases. Consequently, we may write 
\begin{displaymath}
\sqrt{2}\,A(B^+\to\pi^0K^+)\,+\,A(B^+\to\pi^+K^0)=
\sqrt{2}\,A(B^0_d\to\pi^0K^0)\,+\,A(B^0_d\to\pi^-K^+)
\end{displaymath}
\begin{equation}\label{iso-rewritten}
=-\left(e^{i\gamma}-q e^{i\omega}\right)|T+C|e^{i\delta_{T+C}},
\end{equation}
with
\begin{equation}\label{q-def}
q\equiv \left|P_{\rm ew}/(T+C)\right| \quad\mbox{and}\quad 
\omega\equiv \delta_{\rm ew}-\delta_{T+C}.
\end{equation}

Let us focus on the charged $B\to\pi K$ system first. Since 
(\ref{iso-rewritten}) has the same phase structure as the
$B_d^0\to\pi^- K^+$, $B^+\to\pi^+K^0$ isospin relation 
(\ref{mixed-AR}), we have just to replace $A(B_d^0\to\pi^- K^+)$
by $\sqrt{2}A(B^+\to\pi^0K^+)$, where the factor of $\sqrt{2}$
is related to the wavefunction of the neutral pion. The observables 
corresponding to $R$ and $A_0$, which were introduced in (\ref{R-param}) 
and (\ref{A0-def}), respectively, are hence given as follows:
\begin{eqnarray}
R_{\rm c}&\equiv&
2\left[\frac{\mbox{BR}(B^+\to\pi^0 K^+)+
\mbox{BR}(B^-\to\pi^0 K^-)}{\mbox{BR}(B^+\to\pi^+ K^0)+
\mbox{BR}(B^-\to\pi^- \overline{K^0})}\right]\label{Rc-def}\\
A_0^{\rm c}&=&
2\left[\frac{\mbox{BR}(B^+\to\pi^0 K^+)-
\mbox{BR}(B^-\to\pi^0 K^-)}{\mbox{BR}(B^+\to\pi^+ K^0)+
\mbox{BR}(B^-\to\pi^- \overline{K^0})}\right]=
{\cal A}_{\rm CP}(B^+\to\pi^0K^+)R_{\rm c}.
\end{eqnarray}
Completely general parametrizations for $R_{\rm c}$ and $A_0^{\rm c}$ can be 
obtained with the help of the formalism discussed in \ref{ref-gamma-strat} 
by just making the following replacements \cite{BF-BpiK1}:\footnote{For 
an alternative parametrization, see \cite{neubert-BpiK}.}
\begin{equation}\label{replace-c}
r\to r_{\rm c}\equiv\frac{|T+C|}{\sqrt{\langle|P_{\rm c}|^2\rangle}},\quad
\delta\to \delta_{\rm c}\equiv\delta_{T+C}-\delta_{tc}^{\rm c},\quad
q_{\rm C}\to q,\quad \omega_{\rm C}\to \omega.
\end{equation}
Since the parameters $\rho_{\rm c}$ and $\theta_{\rm c}$ are related to 
the charged mode $B^+\to\pi^+K^0$, they do not have to be replaced. These 
substitutions can of course also be performed in (\ref{Rmin-refined}), 
yielding then the expression for 
$\left.R_{\rm min}^{\rm c}\right|_{\delta_{\rm c},r_{\rm c}}$.

In the case of the neutral $B\to\pi K$ system, $\sqrt{2}A(B^0_d\to\pi^0K^0)$
takes the r\^ole of $A(B^+\to\pi^+ K^0)$. In analogy to (\ref{Bpampl}), 
we may write
\begin{equation}\label{Bnampl}
\sqrt{2}\,A(B^0_d\to\pi^0K^0)\equiv P_{\rm n}=
-\left(1-\frac{\lambda^2}{2}\right)\lambda^2A
\left(1+\rho_{\rm n} e^{i\theta_{\rm n}}e^{i\gamma}\right)
{\cal P}_{tc}^{\rm n},
\end{equation}
where
\begin{equation}\label{rho-n-def}
\rho_{\rm n} e^{i\theta_{\rm n}}=\left(\frac{\lambda^2}{1-\lambda^2}
\right)R_b\left[1-\left(\frac{{\cal P}_{uc}^{\rm n}-
{\cal C}}{{\cal P}_{tc}^{\rm n}}\right)\right].
\end{equation}
Here ${\cal C}$ is due to colour-suppressed tree-diagram-like topologies. 
The neutral $B\to\pi K$ observables corresponding to $R$ and $A_0$ are 
therefore given by
\begin{eqnarray}
R_{\rm n}&\equiv&\frac{1}{2}
\left[\frac{\mbox{BR}(B^0_d\to\pi^- K^+)+
\mbox{BR}(\overline{B^0_d}\to\pi^+K^-)}{\mbox{BR}(B^0_d\to\pi^0K^0)+
\mbox{BR}(\overline{B^0_d}\to\pi^0\overline{K^0})}\right]\label{Rn-def}\\
A_0^{\rm n}&=&\frac{1}{2}
\left[\frac{\mbox{BR}(B^0_d\to\pi^- K^+)-
\mbox{BR}(\overline{B^0_d}\to\pi^+K^-)}{\mbox{BR}(B^0_d\to\pi^0K^0)+
\mbox{BR}(\overline{B^0_d}\to\pi^0\overline{K^0})}\right]=
{\cal A}_{\rm CP}(B^0_d\to\pi^-K^+)R_{\rm n}.
\end{eqnarray}
As in the charged $B\to\pi K$ case, general parametrizations of these 
quantities can be obtained straightforwardly from the expressions for 
the $B_d\to\pi^\mp K^\pm$, $B^\pm\to\pi^\pm K$ observables derived in
\cite{defan} by performing appropriate substitutions \cite{BF-BpiK1}:
\begin{equation}\label{replace-n}
r\to r_{\rm n}\equiv\frac{|T+C|}{\sqrt{\langle|P_{\rm n}|^2\rangle}},\quad
\delta\to \delta_{\rm n}\equiv\delta_{T+C}-\delta_{tc}^{\rm n},\quad
q_{\rm C}\to q,\quad \omega_{\rm C}\to \omega.
\end{equation}
Moreover, the parameters of the $B^+\to\pi^+ K^0$ mode have to 
be replaced as follows:
\begin{equation}
\rho_{\rm c}\to\rho_{\rm n},\quad \theta_{\rm c}\to\theta_{\rm n}.
\end{equation}

\subsubsection{Theoretical Advantages}
The theoretical advantages of the charged and neutral $B\to\pi K$ systems
to probe $\gamma$ are related to the following two features:
\begin{itemize}
\item The amplitude $T+C$ can be fixed through the $B^+\to\pi^+\pi^0$
decay with the help of the $SU(3)$ flavour symmetry of strong 
interactions \cite{GRL}:
\begin{equation}\label{T-C-det}
T+C=-\,\sqrt{2}\,\frac{V_{us}}{V_{ud}}\,
\frac{f_K}{f_{\pi}}\,A(B^+\to\pi^+\pi^0),
\end{equation}
where the ratio $f_K/f_{\pi}=1.2$ of the kaon and pion decay constants
takes into account factorizable $SU(3)$-breaking corrections. This
relation allows us to determine the parameters $r_{\rm c}$ and $r_{\rm n}$. 
\item Performing similar tricks as in the derivation of (\ref{qc-expr}), 
the EW penguin parameter $q e^{i\omega}$ can be fixed through the $SU(3)$ 
flavour symmetry \cite{NR} (see also \cite{BF-BpiK1,PAPIII}):
\begin{equation}\label{q-expr}
q e^{i\omega}=
\frac{3}{2\lambda^2R_b}\left[\frac{C_1'(\mu)C_{10}(\mu)-
C_2'(\mu)C_9(\mu)}{C_2'^2(\mu)-C_1'^2(\mu)}\right]=
0.71\times\left[\frac{0.38}{R_b}\right].
\end{equation}
In contrast to (\ref{qc-expr}), this expression does {\it not} involve a
hadronic parameter. Taking into account factorizable $SU(3)$ breaking,
the central value of $0.71$ is shifted to $0.68$. Expression (\ref{q-expr}) 
has also a counterpart in the $B\to\pi\pi$ system (see (\ref{q-pipi-expr})), 
which is based on the $SU(2)$ isospin symmetry and allows us to take 
into account the EW penguin effects in the $B\to\pi\pi$ triangle approach 
to extract $\alpha$ \cite{BF-BpiK1,GPY}.
\end{itemize}
It should be emphasized that (\ref{T-C-det}) and (\ref{q-expr}) are
consequences of the $SU(3)$ flavour symmetry. Therefore, these relations
cannot be affected by rescattering effects. In the formalism discussed
above, such processes may only manifest themselves through anomalously
large parameters $\rho_{\rm c} e^{i\theta_{\rm c}}$ and 
$\rho_{\rm n} e^{i\theta_{\rm n}}$, which are na\"\i vely expected at the 
$\overline{\lambda}^2$ level. Using additional experimental information 
provided by $B^\pm \to K^\pm K$ modes, the parameter 
$\rho_{\rm c} e^{i\theta_{\rm c}}$ can be taken into account through 
the $U$-spin symmetry, as we have seen in \ref{subsubsec:BpiK-res}. In the 
case of the neutral $B\to\pi K$ strategy, we have an additional observable 
${\cal A}_{\rm CP}^{\rm mix}(B_d\to\pi^0K_{\rm S})$ at our disposal, allowing 
us to take into account the parameter $\rho_{\rm n} e^{i\theta_{\rm n}}$ in
an exact manner, i.e.\ without making use of flavour-symmetry 
arguments~\cite{BF-BpiK1}. A sizeable value of 
$\rho_{\rm n} e^{i\theta_{\rm n}}$ would be signalled both by a violation of
the relation \cite{PAPIII}
\begin{equation}
{\cal A}_{\rm CP}^{\rm mix}(B_d\to\pi^0K_{\rm S})=
{\cal A}_{\rm CP}^{\rm mix}(B_d\to J/\psi K_{\rm S}),
\end{equation}
and by large direct CP violation in the $B_d\to\pi^0K_{\rm S}$ channel.

\boldmath
\subsubsection{Strategies to Probe $\gamma$}\label{subsubsec:gamma-det}
\unboldmath
If the observables $R_{\rm c}$, $A_0^{\rm c}$ and $R_{\rm n}$, $A_0^{\rm n}$
are measured, $\gamma$ and the strong phases $\delta_{\rm c}$ and 
$\delta_{\rm n}$ can be extracted from the charged and neutral $B\to\pi K$ 
decays in a similar manner as discussed in \ref{ref-gamma-strat}. Looking 
at Tables~\ref{tab:BPIK-Asym} and \ref{tab:BPIK-obs}, it becomes obvious 
that this approach cannot yet be performed in practice. In particular, CP 
violation in $B^\pm\to\pi^0K^\pm$ and $B_d\to\pi^\mp K^\pm$ has not yet been 
observed. However, non-trivial bounds on $\gamma$ may already be obtained 
from the ratios $R_{\rm c,n}$ of the CP-averaged decay rates. 

The most simple constraints on $\gamma$ arise from expression 
(\ref{Rmin-refined}) for $\left.R_{\rm min}\right|_{\delta,r}$,
if we perform the replacements of variables specified above. 
In the derivation of this minimal value for $R$, both $\delta$ and $r$ 
were treated as free variables. However, in the case of the charged and 
neutral $B\to\pi K$ decays, the parameters $r_{{\rm c,n}}$ can be fixed 
with the help of the $SU(3)$ flavour symmetry through $B^\pm\to\pi^\pm \pi^0$ 
decays. Using the experimental results listed in Table~\ref{tab:BPIK} and 
adding the uncertainties in quadrature gives the numbers summarized in 
Table~\ref{tab:r}. Consequently, it is actually more appropriate to treat 
only $\delta_{{\rm c,n}}$ as an ``unknown'' hadronic parameter, thereby 
varying it within the range $[0,2\pi]$. This procedure yields the following 
extremal, i.e.\ minimal and maximal, values for $R_{\rm c,n}$ \cite{BF-BpiK1}:
\begin{equation}\label{Rext}
\left.R_{\rm c,n}^{\rm ext}
\right|_{\delta_{\rm c,n}}=1 \pm 2\frac{r_{\rm c,n}}{u_{\rm c,n}}
\sqrt{h_{\rm c,n}^2+k_{\rm c,n}^2}+v^2r_{\rm c,n}^2,
\end{equation}
where
\begin{eqnarray}
h_{\rm c,n}&=&\cos\gamma+\rho_{\rm c,n}\cos\theta_{\rm c,n}-q\left[
\,\cos\omega+\rho_{\rm c,n}\cos(\theta_{\rm c,n}-\omega)\cos\gamma\,
\right]\label{h-def}\\
k_{\rm c,n}&=&\rho_{\rm c,n}\sin\theta_{\rm c,n}+q\left[\,\sin\omega
-\rho_{\rm c,n}\sin(\theta_{\rm c,n}-\omega)\cos\gamma\,\right]
\end{eqnarray}
\begin{equation}
v=\sqrt{1-2\,q\cos\omega\cos\gamma+q^2},\label{v-def}
\end{equation}
and $u_{\rm c}$ was already introduced in (\ref{u-def}). If $r$ is fixed
through an additional theoretical input and the corresponding straightforward
replacements of variables are performed, these general formulae apply of 
course also to the $B_d\to\pi^\mp K^\pm$ $B^\pm\to\pi^\pm K$ system. 
The rather complicated expressions simplify considerably, if we assume that 
$\rho_{\rm c}$ and $\rho_{\rm n}$ are negligibly small, and take into account 
that (\ref{q-expr}) implies $\omega=0$.

\begin{table}[t]
\begin{center}
\begin{tabular}{|c|c|c|c|}
\hline
Parameter & CLEO \cite{CLEO-BpiK} & BaBar \cite{babar-BpiK} & Belle 
\cite{belle-BpiK}\\
\hline
$r_{\rm c}$ & $0.21\pm0.06$ & $0.21\pm0.05$ & $0.30\pm0.09$ \\
$r_{\rm n}$ & $0.17\pm0.06$ & $0.21\pm0.06$ & $0.19\pm0.12$ \\
\hline
\end{tabular}
\caption{Present experimental results for $r_{\rm c}$ and 
$r_{\rm n}$.}\label{tab:r}
\end{center}
\end{table}

\begin{figure}
\centerline{\rotate[r]{
\epsfysize=11.9truecm
{\epsffile{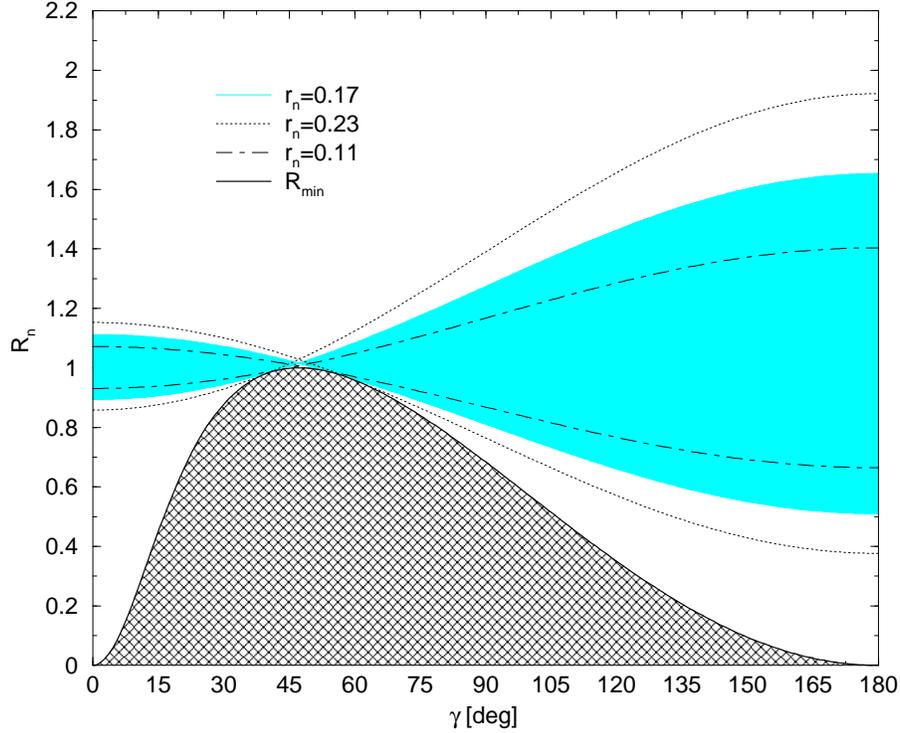}}}}
\caption{The dependence of the extremal values of $R_{\rm n}$ (neutral
$B\to\pi K$ system) described by
(\ref{Rmin-refined}) and (\ref{Rext}) on the CKM angle $\gamma$ for 
$q e^{i\omega}=0.68$ and $\rho_{\rm n}=0$.}\label{fig:Rn}
\end{figure}

Let us now illustrate the corresponding bounds on $\gamma$ in more detail
\cite{BF-neut}. The present experimental status of $R_{\rm (c,n)}$ is 
summarized in Table~\ref{tab:BPIK-obs}. We observe that both the CLEO and the 
Belle data point towards $R_{\rm c}>1$ and $R_{\rm n}<1$, whereas the central 
values of the BaBar collaboration are close to one, with a small preferrence 
of $R_{\rm c}>1$. In Fig.~\ref{fig:Rn}, we show the dependence of the 
extremal values of $R_{\rm n}$ on $\gamma$, using the formulae given above. 
Here the crossed region below the $R_{\rm min}$ curve is {\it excluded}. 
As in the case of the bound in (\ref{FM-bound}), this feature can only 
be transformed into constraints on $\gamma$ if $R_{\rm n}$ is found to 
be smaller than one. Interestingly, the present data from the CLEO and 
Belle collaborations may actually indicate that $R_{\rm n}<1$, although 
the uncertainties are still too large to draw any definite conclusions 
on this exciting possibility. Both measurements agree very well, having 
the same central value $R_{\rm n}=0.6$. If we consider this number as an 
example, the $R_{\rm min}$ curve in Fig.~\ref{fig:Rn} implies the allowed 
range $0^\circ\leq\gamma\leq19^\circ\,\lor\,97^\circ\leq\gamma
\leq180^\circ$. If we use additional information on the parameter 
$r_{\rm n}$, we may put even stronger constraints on $\gamma$. 
For $r_{\rm n}=0.17$, the {\it allowed} range (\ref{Rext}) for
$R_{\rm n}$ is described by the shaded region in Fig.~\ref{fig:Rn},
implying $134^\circ\leq\gamma\leq180^\circ$ for $R_{\rm n}=0.6$.

The curves for the charged $B\to\pi K$ system are very similar to the
ones shown in Fig.~\ref{fig:Rn}. The only difference arises in
the contours related to $\left.R_{\rm c}^{\rm ext}\right|_{\delta_{\rm c}}$,
since $r_{\rm c}$ is expected to be slightly different from $r_{\rm n}$. 
If $R_{\rm c}$ should be found to be larger than one, as 
favoured by the present $B$-factory data, the 
$\left.R_{\rm min}^{\rm c}\right|_{\delta_{\rm c},r_{\rm c}}$ curve would 
not be effective, and the parameter $r_{\rm c}$ has to be fixed in order 
to constrain $\gamma$. If we take the central values of the present
CLEO data to illustrate this bound, $r_{\rm c}=0.21$ and $R_{\rm c}=1.3$, 
we obtain $84^\circ\leq\gamma\leq180^\circ$. 

The allowed ranges for $\gamma$ arising in these examples would be of 
great phenomenological interest, as they are complementary to the range 
implied by the usual indirect fits of the unitarity triangle discussed 
in Subsection~\ref{subsec:CKM-fits}. In particular, as the second quadrant 
for $\gamma$ would be favoured, there would be essentially no overlap 
between these ranges, which could be interpreted as a manifestation 
of new physics \cite{BF-neut}. Other arguments for $\cos\gamma<0$ using 
$B\to PP$, $PV$ and $VV$ decays were given in \cite{RF-bpipi,HY} (see
also \ref{subsubsec:U-spin-Replace}). The central values of the Belle 
collaboration, $R_{\rm c}=2.4$ and $r_{\rm c}=0.3$, would be an even more 
striking signal for new physics, as all values of $\gamma$ would be 
excluded.\footnote{For slightly smaller $R_{\rm c}$ or larger $r_{\rm c}$, 
only values of $\gamma$ close to $180^\circ$ would be allowed.} However, 
the large experimental uncertainties do not yet allow us to draw definite 
conclusions. 

Before we can speculate on physics beyond the Standard Model, it is of 
course crucial to explore the hadronic uncertainties. For the formalism 
discussed above, this was done in \cite{BF-BpiK1}; within a different 
framework, similar considerations were made in \cite{neubert-BpiK}.
The theoretical accuracy of the bounds on $\gamma$ arising from the
charged and neutral $B\to\pi K$ systems is limited both by non-factorizable 
$SU(3)$-breaking corrections and by rescattering processes. The former may 
affect the determination of the parameters $qe^{i\omega}$ and $r_{\rm c,n}$, 
whereas the latter may lead to sizeable values of $\rho_{\rm c,n}$. 

As we will discuss in the next subsection, the QCD factorization 
approach provides valuable insights into the $SU(3)$-breaking corrections, 
thereby allowing a reduction of the corresponding uncertainties \cite{BBNS3}. 
Although the parameters $\rho_{\rm c,n} e^{i\theta_{\rm c,n}}$ are of 
${\cal O}(\overline{\lambda}^2)$ in this framework, it should be emphasized
that they can be taken into account through additional data, as we have
noted above. 
In order to illustrate their impact on the bounds on $\gamma$, let us take 
$\rho_{\rm n}=0.05$ and $\theta_{\rm n}\in\{0^\circ,180^\circ\}$. 
For the example given above, we obtain then the allowed ranges 
$0^\circ\leq\gamma\leq\left(19\pm1\right)^\circ\,\lor\,
\left(97\pm4\right)^\circ\leq\gamma\leq180^\circ$, and
$\left(134\pm2\right)^\circ\leq\gamma\leq180^\circ$.
The feature that the uncertainty due to $\rho_{\rm n}$ is larger in the 
case of $\left.R_{\rm min}^{\rm n}\right|_{\delta_{\rm n},r_{\rm n}}$ 
can be understood easily by performing an expansion of (\ref{Rmin-refined}) 
and (\ref{Rext}) in powers of $\rho_{\rm n}$, and neglecting second-order 
terms of ${\cal O}(\rho_{\rm n}^2)$, ${\cal O}(r_{\rm n}\,\rho_{\rm n})$ 
and ${\cal O}(r_{\rm n}^2)$:
\begin{eqnarray}
\left.R^{\rm n}_{\rm min}\right|_{\delta_{\rm n},r_{\rm n}}^{\rm L.O.}&=&
\left[\frac{1+2\,\rho_{\rm n}\cos\theta_{\rm n}\left(q-\cos\gamma
\right)}{1-2\,q\,\cos\gamma+q^2}\right]\sin^2\gamma\label{Rmin-approx}\\
\left.R_{\rm n}^{\rm ext}
\right|_{\delta_{\rm n}}^{\rm L.O.}&=&1\,\pm\,2\,r_{\rm n}
\left|\cos\gamma-q\right|.\label{Rext-approx}
\end{eqnarray}
Here we have moreover made use of (\ref{q-expr}), which gives
$\omega=0$. Interestingly, as was noted for the charged $B\to\pi K$ system 
in \cite{NR}, there are no terms of ${\cal O}(\rho_{\rm n})$ present in 
(\ref{Rext-approx}), in contrast to (\ref{Rmin-approx}). Consequently, 
the bounds on $\gamma$ related to (\ref{Rmin-refined}) are affected more 
strongly by $\rho_{\rm n}$ then those implied by (\ref{Rext}).

\boldmath
\subsubsection{Constraints in the $\overline{\rho}$--$\overline{\eta}$
Plane}
\unboldmath
In addition to the theoretical uncertainties due to non-factorizable
$SU(3)$-breaking corrections and rescattering effects, another 
uncertainty of the constraints on $\gamma$ is related to the 
CKM factor $R_b$, which arises in the expression (\ref{q-expr}) for 
the EW penguin parameter $qe^{i\omega}$. Because of this feature, 
it is actually more appropriate to consider constraints in 
the $\overline{\rho}$--$\overline{\eta}$ plane instead of the bounds
on $\gamma$ \cite{BF-neut,neubert-proc}. A similar ``trick'' was also 
employed for $B_d\to\pi^+\pi^-$ decays in \cite{charles}, and recently 
for the $B\to\pi K$ system in \cite{BBNS3}.

\begin{figure}
\centerline{\rotate[r]{
\epsfysize=11.1truecm
{\epsffile{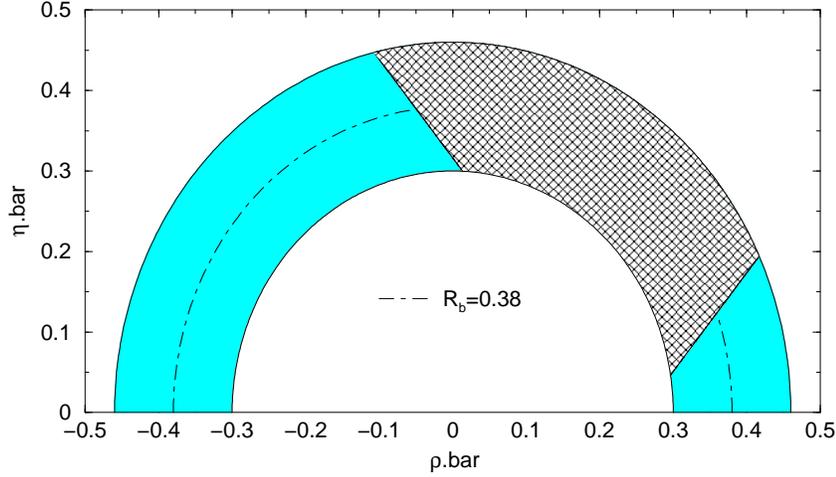}}}}
\caption{The constraints in the $\overline{\rho}$--$\overline{\eta}$ 
plane implied by (\ref{Rmin-refined}) for $R_{\rm n}=0.6$,
$q e^{i\omega}=0.68\times[0.38/R_b]$, and $\rho_{\rm n}=0$. The shaded 
region is the allowed range for the apex of the unitarity triangle, whereas 
the ``crossed'' region is excluded through 
$\left.R_{\rm min}^{\rm n}\right|_{\delta_{\rm n},r_{\rm n}}$.
}\label{fig:rho-eta-const1}
\end{figure}

\begin{figure}
\centerline{\rotate[r]{
\epsfysize=11.1truecm
{\epsffile{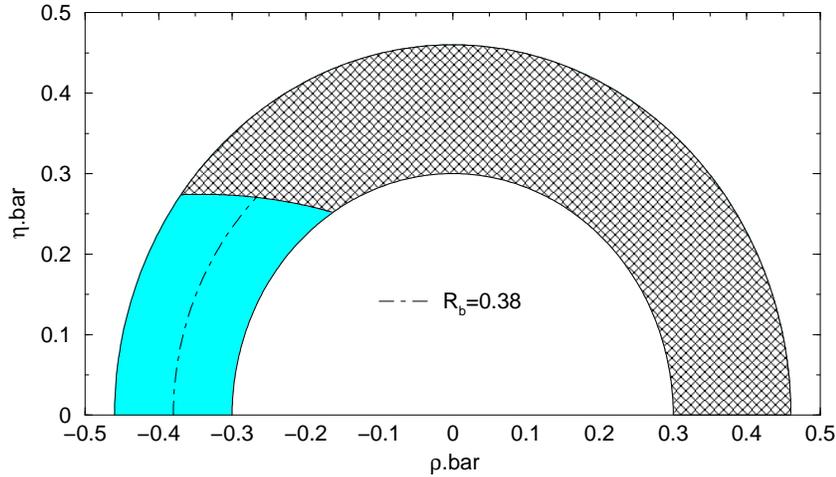}}}}
\caption{The constraints in the $\overline{\rho}$--$\overline{\eta}$ 
plane implied by (\ref{Rext}) for $R_{\rm n}=0.6$, $r_{\rm n}=0.17$,
$q e^{i\omega}=0.68\times[0.38/R_b]$, and $\rho_{\rm n}=0$. The shaded 
region is the allowed range for the apex of the unitarity triangle, whereas 
the ``crossed'' region is excluded through 
$\left.R_{\rm n}^{\rm ext}\right|_{\delta_{\rm n}}$.
}\label{fig:rho-eta-const2}
\end{figure}

The constraints in the $\overline{\rho}$--$\overline{\eta}$
plane can be obtained straightforwardly from (\ref{Rmin-refined}) and 
(\ref{Rext}). In the former case, we obtain \cite{BF-neut}
\begin{equation}\label{constr1}
\cos\gamma=R_{\rm c,n}q\pm\sqrt{\left(1-R_{\rm c,n}\right)\left(
1-R_{\rm c,n}q^2\right)},
\end{equation}
whereas we have in the latter case
\begin{equation}\label{constr2}
\cos\gamma=\frac{1-R_{\rm c,n}\pm2\,q\,r_{\rm c,n}+\left(1+
q^2\right)r_{\rm c,n}^2}{2\,r_{\rm c,n}\left(q\,r_{\rm c,n}\pm1\right)}.
\end{equation}
In these expressions, we have assumed, for simpliciy, $\rho_{\rm c,n}=0$ 
and $\omega=0$. The right-hand sides of these formulae depend implicitly 
on the CKM factor $R_b$ through the EW penguin parameter $qe^{i\omega}$, 
which is given by (\ref{q-expr}). In order to get rid of $R_b$, we consider 
contours in the $\overline{\rho}$--$\overline{\eta}$ plane. They can be 
obtained with the help of (\ref{constr1}) and (\ref{constr2}) by taking 
into account (see Fig.\ \ref{fig:UT} (a))
\begin{equation}
\overline{\rho}=R_b\cos\gamma,\quad\overline{\eta}=R_b\sin\gamma,
\end{equation}
and are illustrated in Figs.~\ref{fig:rho-eta-const1} and 
\ref{fig:rho-eta-const2} for the neutral $B\to\pi K$ examples 
given in \ref{subsubsec:gamma-det}. The corresponding allowed ranges 
in the $\overline{\rho}$--$\overline{\eta}$ plane should be compared 
with the situation in Fig.~\ref{fig:UT-constr}, excluding essentially
the whole second quadrant.

\boldmath
\subsubsection{Bounds on Strong Phases}\label{subsubsec:delta}
\unboldmath
The general expressions for $R_{\rm c,n}$ allow us to determine 
$\cos\delta_{\rm c,n}$ as functions of $\gamma$ \cite{BF-neut}:
\begin{eqnarray}
\cos\delta_{\rm c,n}&=&\frac{1}{h_{\rm c,n}^2+k_{\rm c,n}^2}\left[
\frac{\left(1-R_{\rm c,n}+v^2r_{\rm c,n}^2\right)u_{\rm c,n}
h_{\rm c,n}}{2\,r_{\rm c,n}}\right.\nonumber\\
&&\left.\pm k_{\rm c,n}\sqrt{h_{\rm c,n}^2+
k_{\rm c,n}^2-\left[\frac{\left(1-R_{\rm c,n}+v^2r_{\rm c,n}^2\right)
u_{\rm c,n}}{2\,r_{\rm c,n}}\right]^2}\right].\label{cosdelta}
\end{eqnarray}
Taking into account that $|\cos\delta_{\rm c,n}|\leq1$, we obtain again 
the allowed range for $\gamma$ related to (\ref{Rext}). In addition,
there arise also constraints on the strong phases $\delta_{\rm c,n}$. 
In the case of $R_{\rm n}=0.6$ and $r_{\rm n}=0.17$, we obtain 
$-1\leq\cos\delta_{\rm n}\leq-0.84$, whereas $R_{\rm c}=1.3$ and 
$r_{\rm c}=0.21$ would imply $+0.25\leq\cos\delta_{\rm c}\leq+1$.

As can be seen in (\ref{replace-c}) and (\ref{replace-n}), we have 
$\delta_{\rm n}-\delta_{\rm c}=\delta_{tc}^{\rm c}-\delta_{tc}^{\rm n}$, 
where $\delta_{tc}^{\rm c}$ and $\delta_{tc}^{\rm n}$ are the strong 
phases of the amplitudes ${\cal P}_{tc}^{\rm c}$ and ${\cal P}_{tc}^{\rm n}$,
which describe the differences of the penguin topologies with internal top- 
and charm-quark exchanges contributing to $B^+\to\pi^+K^0$ and 
$B_d^0\to\pi^0K^0$, respectively. These topologies consist of 
QCD and EW penguins, where the latter contribute to $B^+\to\pi^+K^0$ 
only in colour-suppressed form. In contrast, $B_d^0\to\pi^0K^0$ 
receives contributions both from colour-allowed and from colour-suppressed 
EW penguins. Nevertheless, they are expected to be at most 
${\cal O}(20\%)$ of the QCD penguin contributions.
If we neglect the EW penguins and make use of isospin flavour-symmetry 
arguments, we obtain ${\cal P}_{tc}^{\rm n}\approx{\cal P}_{tc}^{\rm c}$, 
yielding $\delta_{\rm n}\approx\delta_{\rm c}$ and $\cos\delta_{\rm n}\approx
\cos\delta_{\rm c}$. Employing moreover factorization, these cosines
are expected to be close to $+1$. 

Consequently, the present CLEO and Belle data point towards a 
``puzzling'' situation, whereas no such discrepancies arise for 
the results of the BaBar collaboration. It is of course too early 
to draw any definite conclusions. However, if future data should 
confirm this ``discrepancy'', it may be an indication for new-physics 
contributions to the EW penguin sector, or a manifestation of large 
flavour-symmetry-breaking effects \cite{BF-neut}. Obviously, further 
studies are required to distinguish between these possibilities. In this 
context, it should be kept in mind that there may also be ``unconventional'' 
sources for flavour-symmetry-breaking effects. An example is 
$\pi^0$--$\eta$,\,$\eta'$ mixing. As was emphasized in \cite{gardner}, 
the isospin violation arising from such effects could mock new physics 
in the extraction of the CKM angle $\alpha$ from $B\to\pi\pi$ isospin 
relations. The impact on the isospin relations involving $B\to \pi^0 K$
decays may also be sizeable.

\boldmath
\subsection{Towards Calculations of $B\to\pi K, \pi\pi$ 
Decays}\label{subsec:BpiK-calc}
\unboldmath
For many years, calculations of $B\to\pi K, \pi\pi$ modes employed the
Bander--Silverman--Soni mechanism \cite{BSS} (see \ref{subsubsec:Tree-Pen}), 
where the strong phases are obtained from absorptive parts of time-like
penguin diagrams of the kind shown in Fig.~\ref{fig:Pen-ME}. Concerning 
analyses of exclusive modes (for examples, see \cite{RF-1}--\cite{AKL-calc2}), 
a problem of this approach is that the transition amplitudes depend, if one 
applies ``na\"\i ve'' factorization to deal with the hadronic matrix elements, 
on a variable $k^2$, where $k$ denotes the four-momentum of the gluons
and photons appearing in Fig.~\ref{fig:Pen-ME}. A conceptual improvement 
was proposed in \cite{BL,ward,RF-1} by applying the Brodsky--Lepage model 
\cite{BrLe}. In this approach, the spectator quark is ``boosted'' through 
the exchange of an ``extra'' gluon and the $k^2$ dependence of the decay 
rates is removed by folding the perturbatively calculated loop amplitudes 
with appropriate wavefunctions.

Recently, important theoretical progress could be made through the 
observation that a rigorous ``factorization'' formula of the structure 
given in (\ref{QCD-factor}) holds also for non-leptonic $B$-decays into 
two light mesons \cite{BBNS1}. In the corresponding formalism, the 
``QCD factorization'' approach, soft non-factorizable contributions and 
final-state interaction effects are suppressed by $\Lambda_{\rm QCD}/m_b$. 
At the leading order in $\Lambda_{\rm QCD}/m_b$, CP-conserving strong phases 
arise essentially from perturbatively calculable QCD corrections. Since 
integrations over the light-cone momentum fractions of the constituent 
quarks inside the mesons are performed, no dependences of the transition 
amplitudes on unphysical parameters such as $k^2$ appear in this approach.

A very comprehensive analysis of $B\to\pi K, \pi\pi$ decays within this
framework was performed in \cite{BBNS3}, where also the $SU(3)$-breaking 
corrections affecting the strategies to probe $\gamma$ discussed in 
\ref{subsubsec:gamma-det} were explored. Following these lines, it is 
possible to reduce the corresponding hadronic uncertainties to the level 
of non-factorizable corrections that violate simultaneously the $SU(3)$ 
flavour symmetry and are power suppressed in the heavy-quark limit. 
Consequently, these corrections are parametrically suppressed by a product 
of the three small quantities $1/N_{\rm C}$, $(m_s-m_d)/\Lambda_{\rm QCD}$
and $\Lambda_{\rm QCD}/m_b$. The relevant equations, where these
corrections enter, are (\ref{T-C-det}) and (\ref{q-expr}), which are
required to fix $r_{\rm c,n}$ and $qe^{i\omega}$, respectively. 
Concerning the determination of $r_{\rm c}$, we may write 
\begin{equation}\label{rc-refined}
r_{\rm c}=R_{\rm th} r^{\rm c}_{\rm exp},
\end{equation}
where
\begin{equation}
r^{\rm c}_{\rm exp}=\sqrt{2} \,\tan\theta_{\rm C}\, \frac{f_K}{f_\pi}
\sqrt{\frac{\mbox{BR}(B^\pm\to\pi^\pm\pi^0)}{\mbox{BR}(B^\pm\to\pi^\pm K)}}
\end{equation}
corresponds to the determination of $T+C$ through (\ref{T-C-det}). The
analysis of \cite{BBNS3} gives
\begin{equation}\label{Rth}
R_{\rm th}=0.98\pm0.05.
\end{equation}
Combining this with the present experimental values for $r^{\rm c}_{\rm exp}$,
averaged over all present $B$-factory results, Beneke {\it et al.}\ obtain 
\begin{equation}\label{rc-final}
r_{\rm c}=0.218\pm0.034_{\rm exp}\pm0.011_{\rm th}. 
\end{equation}
If both $SU(3)$-breaking corrections and small electromagnetic contributions
are taken into account, expression (\ref{q-expr}) for the EW 
penguin parameter $qe^{i\omega}$ is rescaled by the following factor 
\cite{BBNS3}:
\begin{equation}\label{Rq}
R_q=(0.84\pm0.10)e^{-i(2.5\pm2.8)^\circ},
\end{equation}
where about half of the deviation from one is due to mostly factorizable
$SU(3)$ breaking. Taking also into account the uncertainty due to the CKM 
factor $R_b$, the final result reads 
\begin{equation}\label{q-final}
q=\left(58.8\pm6.7\mp11.8\right)\times 10^{-2},
\end{equation}
where the first error is theoretical and the second error arises from
$R_b$. The remaining uncertainties of $R_{\rm th}$ and
$R_q$ are due to terms of 
\begin{equation}
{\cal O}\left(\frac{1}{N_{\rm C}}\times \frac{m_s-m_d}{\Lambda_{\rm QCD}}
\times \frac{\Lambda_{\rm QCD}}{m_b}\right)=
{\cal O}\left(\frac{1}{N_{\rm C}}\times\frac{m_s-m_d}{m_b}\right),
\end{equation}
which are na\"\i vely estimated not to exceed the few percent level.
Consequently, it may eventually be possible to determine $\gamma$ 
from the $B\to\pi K$ strategies reviewed above with a theoretical 
accuracy of about $10^\circ$. 

If the parameter $r_{\rm c}$ is not determined through (\ref{rc-refined})
but calculated directly in the QCD factorization approach, we encounter
much larger theoretical uncertainties. The predictions for $r_{\rm c}$
and $\delta_{\rm c}$ are given as follows \cite{BBNS3}:
\begin{equation}\label{rc-QCD-fact}
\overline{r}_{\rm c}\equiv u_{\rm c}\,r_{\rm c}=\left\{\begin{array}{l}
\left(23.9\pm4.5\pm4.8\right)\times 10^{-2}\\
\left(25.7\pm4.8\pm5.1\right)\times 10^{-2},
\end{array}\right.
\quad
\delta_{\rm c}=\left\{\begin{array}{l}
\left(-9.6\pm3.8\right)^\circ\\
\left(-10.2\pm4.1\right)^\circ,
\end{array}\right.
\end{equation}
where the notation is as in (\ref{rho-QCD-fact}) and (\ref{qC-QCD-fact}), 
$u_{\rm c}$ was introduced in (\ref{u-def}), and the CP-conserving strong 
phase is so small because of 
\begin{equation}\label{QCD-fact-phase}
\delta_{\rm c}={\cal O}\left(\alpha_s(m_b),\Lambda_{\rm QCD}/m_b\right).
\end{equation}
Since $u_{\rm c}$ is very close to one within QCD factorization, 
$\overline{r}_{\rm c}$ and $r_{\rm c}$ take very similar values. It should 
be noted that the comparison between (\ref{rc-final}) and (\ref{rc-QCD-fact}) 
represents a non-trivial test of the QCD factorization approach, which 
could lead to a reduction of the uncertainties related to the modelling of
power corrections to the heavy-quark limit. In this context, potentially 
large corrections may arise from certain annihilation topologies, as 
first noted in \cite{PQCD-anni}. Moreover, we have to care about ``chirally
enhanced'' corrections of the following structure:
\begin{equation}
r_\chi^\pi=\frac{2\,M_\pi^2}{(m_u+m_d)\,m_b},\quad
r_\chi^K=\frac{2\,M_K^2}{(m_{u,d}+m_s)\,m_b},
\end{equation}
which are formally of ${\cal O}(\Lambda_{\rm QCD}/m_b)$, but numerically
close to unity. Phenomenological analyses of $B\to\pi K, \pi\pi$ decays
require definitely estimates of these corrections. In \cite{BBNS3}, a
certain prescription was used to get some handle on them. 

Using an input from QCD factorization that is stronger than 
(\ref{Rth}) and (\ref{Rq}), more stringent constraints on $\gamma$
and the allowed range in the $\overline{\rho}$--$\overline{\eta}$ plane 
can be obtained. As a first step, we may employ that the CP-conserving 
strong phase $\delta_{\rm c}$ is predicted to be very small due to
(\ref{QCD-fact-phase}), so that $\cos\delta_{\rm c}$ governing $R_{\rm c}$ 
is close to one. As a second step, information on $\gamma$ can be obtained 
from the predictions for the branching ratios and the observables 
$R_{\rm (c,n)}$. Finally, the information from all CP-averaged 
$B\to\pi K,\pi\pi$ branching ratios can be combined into a single global 
fit for the generalized Wolfenstein parameters $\overline{\rho}$ and 
$\overline{\eta}$ \cite{BBNS3}. For these approaches, it is of course 
crucial that the contributions entering at the 
${\cal O}(\Lambda_{\rm QCD}/m_b)$ level can be controlled reliably. 

In a recent paper \cite{charming-pens2}, it was argued that 
non-perturbative penguin topologies with internal charm- and 
up-quark exchanges may play an important r\^ole in this context. 
If the analysis of the unitarity triangle performed in \cite{Rome-rev}
is used to determine $\gamma$, thereby fixing it to the range in the
first quadrant given in (\ref{Rome-gamma}), the $B\to\pi K$ branching 
ratios calculated within QCD factorization \cite{BBNS1} are found to be 
systematically smaller than the measured values, whereas the branching 
ratio for $B_d\to \pi^+ \pi^-$ is about a factor of two larger than the 
experimental result. On the other hand, if $\gamma$ is treated as a 
free parameter, a fit to the data yields $\gamma=(163 \pm 12)^\circ$, 
which is in strong disagreement to the range given in (\ref{Rome-gamma})
(see also \ref{subsubsec:U-spin-Replace}). In the case of 
$B^0_d\to\pi^-K^+$ and $B^+\to\pi^0K^+$, the CP-averaged branching 
ratios are now enhanced by almost a factor of two, since the
interference between penguins and trees is constructive for values
of $\gamma$ in the second quadrant. On the other hand, 
$\mbox{BR}(B_d\to\pi^+\pi^-)$ is reduced significantly, as the 
interference between trees and penguins is now destructive. Since the
decays $B^+\to\pi^+K^0$ and $B^0_d\to\pi^0K^0$ are essentially independent 
of the angle $\gamma$, their branching ratios -- in particular the one
of $B^0_d\to\pi^0K^0$ -- are still much smaller than the experimental 
values. 

These features could be interpreted either as a manifestation of new 
physics, or as an indication of large $\Lambda_{\rm QCD}/m_b$ corrections. 
The authors of \cite{charming-pens2} focus on the latter possibility 
and explore the impact of non-perturbative penguin topologies with 
internal charm- and up-quark exchanges. Actually, as far as the $B\to\pi K$
branching ratios are concerned, the most important effects are due to the 
former ``charming'' penguins, as we have seen in \ref{subsubsec:BpiK-res};
they may also affect the strong phases $\delta_{tc}^{\rm (c,n)}$
significantly, which are a key ingredient for CP violation. In the 
$B\to\pi\pi$ case, also the latter topologies may become important, 
since they are not Cabibbo-suppressed in $\overline{b}\to\overline{d}$ 
transitions. 

If $\gamma$ is assumed to lie again within the range in (\ref{Rome-gamma}), 
and the non-perturbative penguin contributions are taken into account 
by adding appropriate amplitudes, which contain also 
other ${\cal O}(\Lambda_{\rm QCD}/m_b)$ terms with the same 
quantum numbers, for instance the annihilation diagrams considered in 
\cite{PQCD-anni}, the predictions for the branching ratios can be brought 
closer to the experimental results. Concerning the $B_d^0\to\pi^0K^0$ mode,
the branching ratio is now in perfect agreement with the central value of 
BaBar, but still much smaller than those of CLEO and Belle. Since the 
non-perturbative penguin amplitudes cannot (yet) be calculated 
from first principles, they are determined in this approach through a 
fit to the data, where $SU(3)$ flavour-symmetry arguments are used
to simplify the analysis. It should be noted that also here 
no anomalous enhancement of the amplitude corresponding to penguin 
topologies with internal up-quark exchanges shows up, as would be the 
case in the presence of dramatic rescattering processes of the kind 
given in (\ref{BpiK-u-res}).

Another interesting feature of this approach is that large direct CP 
violation arises in the $B^0_d\to\pi^-K^+$, $B^+\to\pi^0K^+$ and 
$B_d^0\to\pi^+\pi^-$ channels, which would be in contrast to the picture of 
QCD factorization. On the other hand, the CP-violating asymmetries 
in $B^+\to\pi^+ K^0$ and $B_d^0\to \pi^0K^0$ would be very small, i.e.\ at
the few percent level, as expected ``na\"\i vely'' from (\ref{Bpampl}) 
and (\ref{Bnampl}). It should be noted that the large CP asymmetries of
the $B\to\pi K$ modes are exactly those that are required for the strategies 
to extract $\gamma$ discussed in \ref{ref-gamma-strat} and 
\ref{subsubsec:gamma-det}. In the decay $B_d\to\pi^+\pi^-$, a large 
sensitivity to $\Lambda_{\rm QCD}/m_b$ effects is found, thereby making the 
extraction of $\alpha$ from ${\cal A}_{\rm CP}^{\rm mix}(B_d\to\pi^+\pi^-)$ 
through a calculation of the hadronic parameters in (\ref{Amix-simp}) 
questionable. 

Because of the non-perturbative penguin topologies, the sensitivity of the 
calculated branching ratios on $\gamma$ is now lost. In particular, if one 
tries to fit also $\gamma$, one finds that this angle remains essentially 
undetermined with the present experimental accuracy \cite{charming-pens2}. 
An enhancement of the $B\to\pi K$ penguin amplitudes through intrinsically 
non-perturbative charm rescattering was also discussed recently in 
\cite{Rosner-charm}. The terms entering at ${\cal O}(\Lambda_{\rm QCD}/m_b)$ 
in non-leptonic decays will certainly continue to be one of the hot topics 
in $B$ physics. Probably it will take some time until this issue will
have been settled.

Further progress on the theoretical description of $B\to\pi K,\pi\pi$ 
decays will hopefully be made. For a detailed review of other recent 
analyses using QCD factorization, we refer the reader to \cite{BBNS3}, 
where all technicalities are discussed. Detailed comparisons between 
QCD factorization and PQCD can be found in \cite{BBNS3,PQCD-comp}. 
Another approach to calculate the $B\to\pi\pi$ hadronic matrix elements
using QCD light-cone sum rules was proposed in \cite{khod}. As far as 
the determination of $\gamma$ is concerned, the strategies considered in 
\ref{subsubsec:gamma-det} will eventually exhibit the highest theoretical 
accuracy. Moreover, in addition to $r_{\rm c,n}$, they will also provide 
the strong phases $\delta_{\rm c,n}$ and the parameters 
$\rho_{\rm c,n}e^{i\theta_{\rm c,n}}$. The insights into hadron 
dynamics thus obtained will be a very fertile and decisive testing 
ground for the theoretical approaches to deal with $B\to\pi K,\pi\pi$ 
modes.

\boldmath
\subsection{Impact of New Physics}
\unboldmath
In the theoretically clean strategies to measure $\gamma$ with the 
help of pure ``tree'' decays, such as $B_d\to D^{\ast\pm}\pi^\mp$ or 
$B_s\to D_s^\pm K^\mp$, there are no contributions from flavour-changing 
neutral-current (FCNC) processes. Consequently, new physics is expected to 
play a minor r\^ole in these approaches. On the other hand, penguins 
are crucial for $B\to\pi K$ decays, so that these modes are very 
sensitive to new physics \cite{FM-BpiK-NP}--\cite{matias} (see also 
the model-independent analysis of the $B\to\phi K$ system \cite{FM-PhiK}
discussed in \ref{subsubsec:BphiK-NP}). 

In order to analyse new-physics effects in the $B\to\pi K$ strategies 
to probe $\gamma$, it is useful to distinguish between scenarios of 
physics beyond the Standard Model, where the isospin symmetry is either 
preserved or violated \cite{FM-BpiK-NP}. The point is that isospin 
relations are at the basis of these approaches, as we have seen above. 
An interesting example for isospin-conserving new physics is given by 
models with enhanced chromomagnetic dipole operators \cite{kagan}, 
where $B^\pm\to\pi^\pm K$ decays may well exhibit direct CP violation 
at the level of $10\%$. On the other hand, the preferred place for 
isospin-violating scenarios of new physics to manifest themselves is 
the EW penguin sector, which plays a particularly important r\^ole 
in the charged and neutral $B\to\pi K$ approaches, employing the 
isospin relation (\ref{iso1}), where EW penguins enter also in colour-allowed 
form. If we consider, for example, models with tree-level FCNC couplings 
of the $Z$-boson, extended gauge models with an extra $Z'$-boson, or SUSY 
models with broken R-parity, EW penguin-like contributions may actually 
be much larger than those of the Standard-Model EW penguins (for a 
detailed study, see \cite{GNK}). Finally, it should not be forgotten
that we have used the unitarity of the CKM matrix in 
the parametrizations of the $B\to\pi K$ decay amplitudes. The chances 
to encounter discrepancies in the $B\to\pi K$ strategies are therefore 
very good. In fact, as we have seen in Subsection~\ref{sec:BpiK-cn}, 
the present data may already point towards puzzling constraints on 
$\gamma$ and strong phases in the charged and neutral $B\to\pi K$ decays.
Better data are needed to clarify these exciting issues.

\begin{figure}
\vspace*{-2.7truecm}
\begin{center}
\begin{tabular}{ll}
\epsfysize=9.8cm
\epsffile{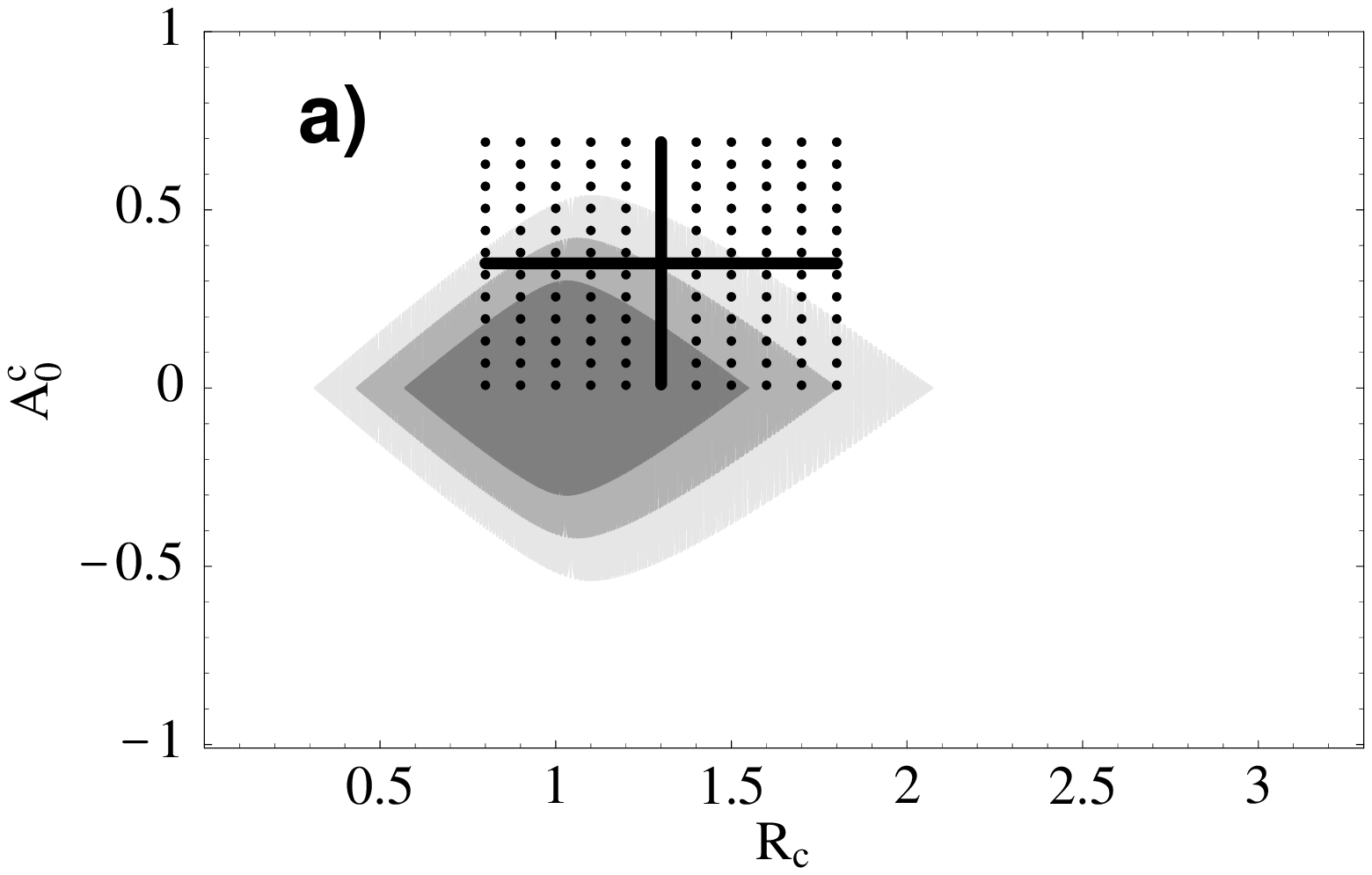}
&
\epsfysize=9.8cm
\epsffile{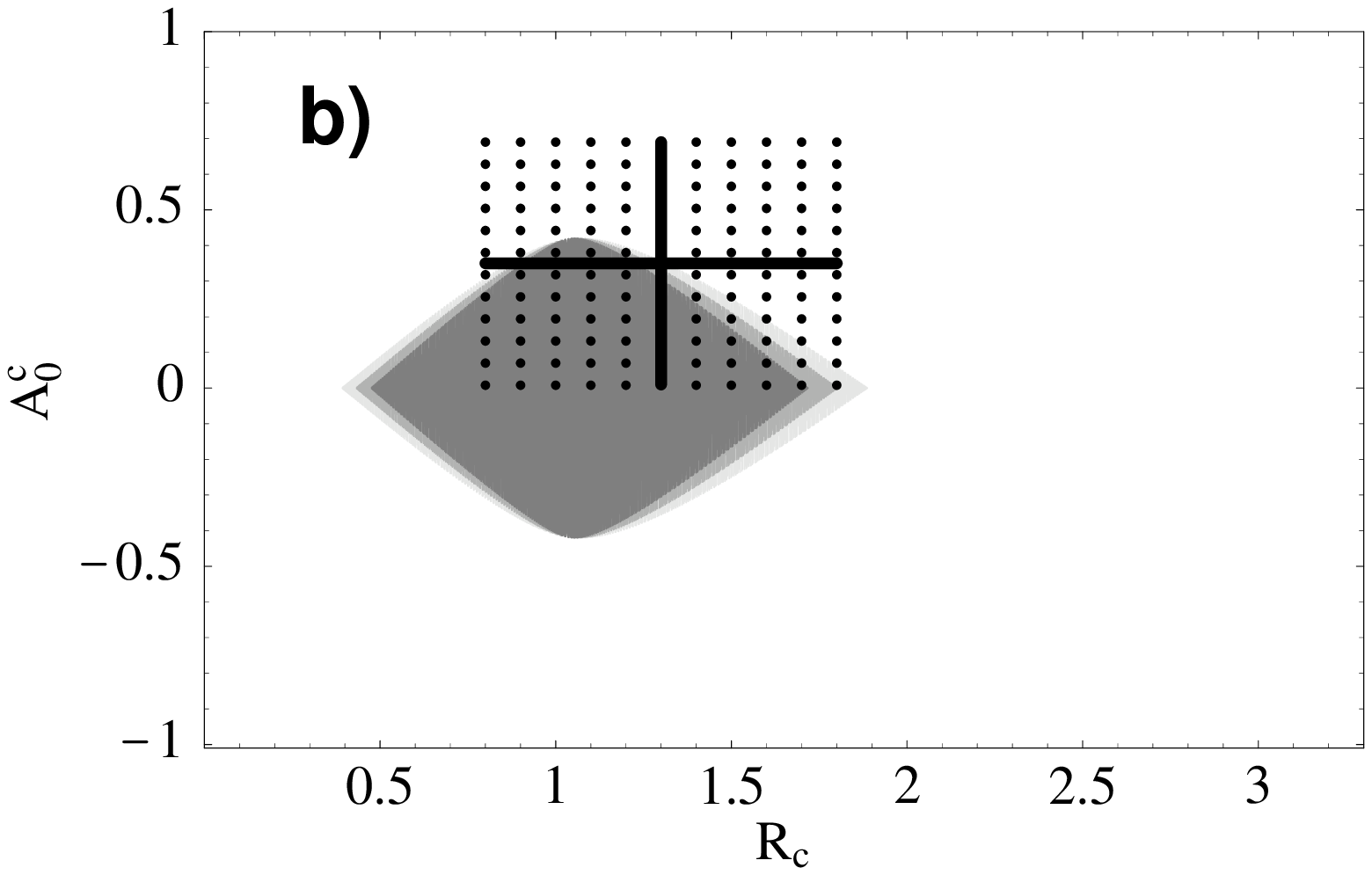}
\end{tabular}
\end{center}
\vspace*{-3truecm}
\caption{Allowed regions characterizing the $B^\pm\to\pi^0K^\pm$, $\pi^\pm K$ 
system in the Standard Model: (a) $0.15\leq r_{\rm c}\leq0.27$, $q=0.63$; 
(b) $r_{\rm c}=0.21$, $0.48\leq q\leq0.78$ ($\rho_{\rm c}=0$).}
\label{fig:BpiK-char-cont}
\end{figure}

We may also arrive at a situation, where {\it no} solutions for $\gamma$ 
and strong phases are found. This would be the most striking signal 
for new-physics effects in the $B\to\pi K$ system. Since the Standard 
Model predicts strong correlations between the corresponding observables,
such a case may be indicated immediately \cite{FMat}. In 
Fig.~\ref{fig:BpiK-char-cont}, we show the allowed range for the charged 
$B\to\pi K$ system in the $R_{\rm c}$--$A_0^{\rm c}$ plane. Here the 
dotted regions correspond to the CLEO results published in 
\cite{CLEO-BpiK,CLEO-BpiK-asym}. If future measurements should lie 
significantly outside the allowed ranges shown in this figure, we would 
have an indication for new physics. On the other hand, if we should find 
values lying within these regions, this would not automatically imply a 
``confirmation'' of the Standard Model. In this case, we would then be 
in a position to extract a value for $\gamma$ by following the strategies 
described above, which may also lead to discrepancies. Similar allowed 
regions can of course also be obtained in the observable spaces of the 
$B_d\to\pi^\mp K^\pm$, $B^\pm\to\pi^\pm K$ and $B_d\to\pi^\mp K^\pm$, 
$B_d\to\pi^0 K$ systems \cite{FMat}. A set of sum rules relating the 
CP-averaged branching ratios and CP asymmetries of $B\to\pi K$ decays 
was recently derived in \cite{matias}, which is useful to analyse the 
data in view of possible isospin-breaking effects.

\subsection{Summary}
The phenomenology of $B\to\pi K$ decays provides interesting strategies 
to determine $\gamma$. To this end, three different combinations of 
$B\to\pi K$ decays may be considered: the $B_d\to\pi^\mp K^\pm$, 
$B^\pm\to\pi^\pm K$ system, the charged $B^\pm\to\pi^0K^\pm$, 
$B^\pm\to\pi^\pm K$ system, and the neutral $B_d\to\pi^\mp K^\pm$, 
$B_d\to\pi^0 K$ system. In the first case, dynamical assumptions about 
rescattering and colour-suppressed EW penguin processes have to be made, 
in addition to flavour-symmetry arguments. Recent theoretical developments 
are supporting these plausible assumptions, as well as experimental bounds 
on $B\to KK$ branching ratios. In the case of the charged and neutral 
$B\to\pi K$ strategies, EW penguin topologies, which contribute there also 
in colour-allowed form, can be taken into account through the $SU(3)$ 
flavour symmetry, and the sensitivity to rescattering effects is smaller. 
Using data on $B^\pm\to K^\pm K$ decays, the rescattering processes can 
in principle be taken into account in the charged $B\to\pi K$ strategy, 
whereas mixing-induced CP violation in $B_d\to\pi^0K_{\rm S}$ allows 
us to include these effects exactly in the neutral $B\to \pi K$ approach. 

Appropriate ratios of CP-averaged $B\to\pi K$ decay rates provide
valuable contraints both on $\gamma$ and on CP-conserving strong phases. 
In order to determine these quantities, also certain CP 
asymmetries have to be measured. Interestingly, the present data may 
point towards a ``puzzling'' situation, although the experimental 
uncertainties are still too large to draw definite conclusions. 
Since $B\to\pi K$ modes are governed by penguin topologies, they
are sensitive probes for new physics. Moreover, the unitarity of the
CKM matrix is employed in the strategies to probe $\gamma$. 

Important theoretical progress concerning the description of the
$B\to\pi K, \pi\pi$ hadron dynamics has recently been made, 
providing a framework for systematic calculations in the 
$m_b\gg\Lambda_{\rm QCD}$ limit. Concerning the $B\to\pi K$ strategies 
to probe $\gamma$, this interesting development allows a reduction of 
$SU(3)$-breaking corrections, and gives confidence into dynamical 
assumptions related to rescattering effects. Further theoretical issues
concern mainly the importance of $\Lambda_{\rm QCD}/m_b$ corrections. 

The experimental situation will soon improve significantly. Eventually, 
it may be possible to extract $\gamma$ through the $B\to\pi K$ strategies 
with an accuracy of ${\cal O}(10^\circ)$.

\boldmath
\section{Phenomenology of $U$-spin-related $B$ Decays}\label{sec:Uspin}
\unboldmath
\setcounter{equation}{0}
Let us now focus on strategies to extract the angles of the unitarity
triangle -- in particular $\gamma$ -- from pairs of $B$-meson decays, 
which are related to each other through the $U$-spin flavour symmetry 
of strong interactions. In analogy to the well-known isospin symmetry, 
$U$ spin is also described by an $SU(2)$ subgroup of the full 
flavour-symmetry group $SU(3)_{\rm F}$. Whereas isospin relates down 
and up quarks, $U$ spin relates down and strange quarks. Consequently, 
$U$-spin-breaking effects are generally expected to be more significant than 
isospin breaking. On the other hand, an important advantage of the $U$-spin 
symmetry is that it is satisfied -- in addition to QCD penguins -- also by 
EW penguin topologies, since the down and strange quarks have the same 
electrical charges. In this review, we have already encountered $U$-spin 
arguments several times \cite{BFM,defan,FKNP,pirjol,Lipkin}, and first
approaches to extract CKM phases were pointed out in 1993 \cite{snowmass}. 
However, the great power of the $U$-spin symmtery to determine weak
phases and hadronic parameters was noticed just recently in the 
strategies proposed in \cite{RF-BdsPsiK,RF-BsKK,BspiK}, which are our 
next topic. Since these approaches involve also decays of $B_s$-mesons, 
$B$ experiments at hadron colliders are required to implement them in 
practice. At Tevatron-II, we will have first access to the corresponding 
modes and interesting results are expected \cite{wuerth}. In the era of 
BTeV and the LHC, the $U$-spin strategies can then be fully exploited 
\cite{LHC-Report}.

\boldmath
\subsection{The $B_{s(d)}\to J/\psi K_{\rm S}$ 
System}\label{subsec:BsdPsiKS}
\unboldmath
As we have seen in Subsection~\ref{subsec:BpsiK}, the ``gold-plated'' mode 
$B_d\to J/\psi K_{\rm S}$ plays an outstanding r\^ole in the determination 
of the $B^0_d$--$\overline{B^0_d}$ mixing phase $\phi_d$, i.e.\ of the 
CKM angle $\beta$. In this subsection, we consider the decay 
$B_s\to J/\psi K_{\rm S}$ \cite{RF-BdsPsiK}, which is related to 
$B_d\to J/\psi K_{\rm S}$ by interchanging all down and strange quarks
(see Fig.~\ref{fig:BdPsiKS}), and may allow an interesting extraction of 
$\gamma$.

\subsubsection{Amplitude Structure}
In analogy to (\ref{Bd-ampl2}), the $B_s^0\to J/\psi K_{\rm S}$ decay 
amplitude can be expressed as follows: 
\begin{equation}\label{Bs-ampl}
A(B_s^0\to J/\psi K_{\rm S})=-\lambda\,{\cal A}\left(1-a e^{i\theta}
e^{i\gamma}\right),
\end{equation}
where
\begin{equation}
{\cal A}\equiv\lambda^2A\left(A_{\rm CC}^{c}+A_{\rm pen}^{ct}\right)
\end{equation}
and
\begin{equation}\label{a-def}
a e^{i\theta}\equiv R_b\left(\frac{A_{\rm pen}^{ut}}{A_{\rm CC}^{c}+
A_{\rm pen}^{ct}}\right)
\end{equation}
correspond to (\ref{Aap-def}) and (\ref{ap-def}), respectively. It should
be emphasized that the Standard-Model expressions (\ref{Bd-ampl2}) 
and (\ref{Bs-ampl}) rely only on the unitarity of the CKM matrix.

Comparing (\ref{Bs-ampl}) with (\ref{Bd-ampl2}), we observe that 
$a e^{i\theta}$ enters in the $B_s^0\to J/\psi K_{\rm S}$ decay 
amplitude in a Cabibbo-allowed way, whereas $a' e^{i\theta'}$ is 
doubly Cabibbo-suppressed in the $B_d^0\to J/\psi K_{\rm S}$ 
amplitude. Consequently, there may be sizeable CP violation in 
$B_s\to J/\psi K_{\rm S}$, which provides {\it two} independent 
observables, ${\cal A}_{\rm CP}^{\rm dir}(B_s\to J/\psi K_{\rm S})$ and 
${\cal A}_{\rm CP}^{\rm mix}(B_s\to J/\psi K_{\rm S})$ or
${\cal A}_{\Delta\Gamma}(B_s\to J/\psi K_{\rm S})$, depending on the 
{\it three} ``unknowns'' $a$, $\theta$ and $\gamma$, as well as on the 
$B^0_s$--$\overline{B^0_s}$ mixing phase $\phi_s$. Consequently, in order 
to determine these ``unknowns'', we need an additional observable, which 
is provided by
\begin{equation}\label{H-def}
H\equiv\left(\frac{1-\lambda^2}{\lambda^2}\right)
\left|\frac{{\cal A}'}{{\cal A}}\right|^2
\frac{\langle\Gamma(B_s\to J/\psi K_{\rm S})\rangle}{\langle\Gamma
(B_d\to J/\psi K_{\rm S})\rangle},
\end{equation}
where the CP-averaged decay rates
$\langle\Gamma(B_s\to J/\psi K_{\rm S})\rangle$ and 
$\langle\Gamma(B_d\to J/\psi K_{\rm S})\rangle$ can be determined from 
the ``untagged'' rates introduced in (\ref{untag-def}) through
\begin{equation}\label{aver-rate}
\langle\Gamma(B_q\to f)\rangle\equiv\frac{\Gamma_q[f(0)]}{2}.
\end{equation}

%
%
%
\boldmath
\subsubsection{Extraction of $\gamma$}
\unboldmath
Since the $U$-spin flavour symmetry of strong interactions implies
\begin{equation}\label{SU3-1}
|{\cal A}'|=|{\cal A}|
\end{equation}
and
\begin{equation}\label{SU3-2}
a'=a,\quad \theta'=\theta,
\end{equation}
we can determine $a$, $\theta$ and $\gamma$ as functions of $\phi_s$ 
by combining $H$ with ${\cal A}_{\rm CP}^{\rm dir}(B_s\to J/\psi K_{\rm S})$ 
and ${\cal A}_{\rm CP}^{\rm mix}(B_s\to J/\psi K_{\rm S})$ or
${\cal A}_{\Delta\Gamma}(B_s\to J/\psi K_{\rm S})$. In contrast to 
isospin relations, EW penguins do not lead to any problems 
in these $U$-spin relations. As we have already noted in 
Section~\ref{sec:Bs}, the $B^0_s$--$\overline{B^0_s}$ mixing phase 
$\phi_s$ is negligibly small in the Standard Model, and 
can be probed nicely through the search for CP violation in 
$B_s\to J/\psi \phi$ modes. 

Interestingly, the strategy to extract $\gamma$ from $B_{s(d)}\to J/\psi
K_{\rm S}$ decays does not require a non-trivial CP-conserving strong 
phase $\theta$. However, its experimental feasibility depends strongly on 
the value of the ``penguin'' parameter $a$, which is very difficult to 
calculate. In the following, we assume, as in Subsection~\ref{subsec:BpsiK}, 
that $a={\cal O}(\overline{\lambda})$, which is a plausible estimate. 
The general formalism to extract $\gamma$ from the 
$B_{s(d)}\to J/\psi K_{\rm S}$ system was developed in \cite{RF-BdsPsiK}.
Although the corresponding formulae are quite complicated, the basic idea 
is very simple: if $\phi_s$ is used as an input, 
${\cal A}_{\rm CP}^{\rm dir}(B_s\to J/\psi K_{\rm S})$ and
${\cal A}_{\rm CP}^{\rm mix} (B_s\to J/\psi K_{\rm S})$ allow us to eliminate
the strong phase $\theta$, thereby fixing a contour in the $\gamma$--$a$ 
plane in a {\it theoretically clean} way. Another contour can be fixed with 
the help of the $U$-spin relations (\ref{SU3-1}) and (\ref{SU3-2}) 
by combining the observable $H$ with 
${\cal A}_{\rm CP}^{\rm mix}(B_s\to J/\psi K_{\rm S})$. Alternatively,
we may combine $H$ with ${\cal A}_{\Delta\Gamma}(B_s\to J/\psi K_{\rm S})$
to fix a third contour in the $\gamma$--$a$ plane. The intersection of these
contours gives then $\gamma$ and $a$, so that also the strong phase 
$\theta$ can be determined. The general formulae simplify considerably, 
if we keep only terms linear in $a$. Within this approximation, 
we obtain 
\begin{equation}\label{gam-approx}
\tan\gamma\approx\frac{\sin\phi_s+
{\cal A}_{\rm CP}^{\rm mix}(B_s\to J/\psi K_{\rm S})}{(1-H)
\cos\phi_s}\,.
\end{equation}

\begin{figure}
\centerline{\rotate[r]{
\epsfysize=9.7truecm
{\epsffile{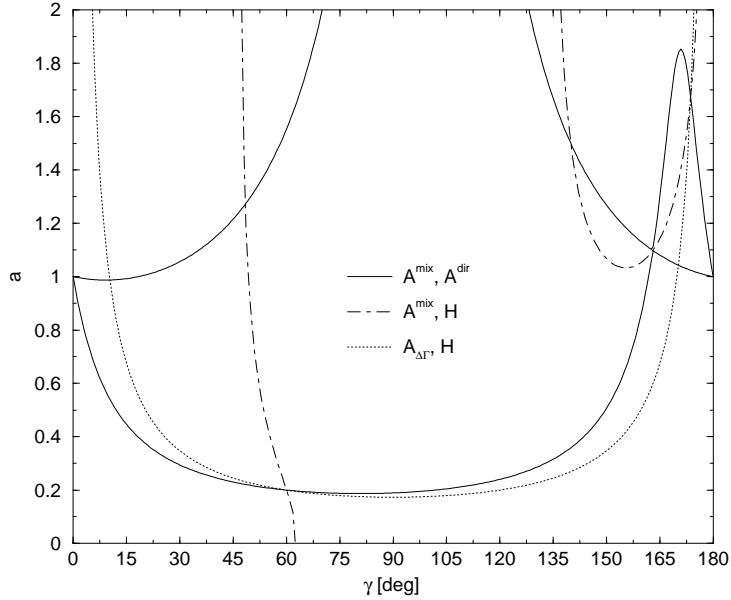}}}}
\caption{Contours in the $\gamma$--$a$ plane fixed through the 
$B_{s(d)}\to J/\psi K_{\rm S}$ observables for a specific 
example discussed in the text.}\label{fig:BsdPsiKScont}
\end{figure}

Let us illustrate this approach by considering an example. Assuming 
$\phi_s=0$, $\gamma=60^\circ$, $a=a'=0.2$ and $\theta=\theta'=30^\circ$, 
we obtain the following $B_{s(d)}\to J/\psi K_{\rm S}$ observables: 
${\cal A}_{\rm CP}^{\rm dir}(B_s\to J/\psi K_{\rm S})=0.20$,
${\cal A}_{\rm CP}^{\rm mix}(B_s\to J/\psi K_{\rm S})=0.31$, 
${\cal A}_{\Delta\Gamma}(B_s\to J/\psi K_{\rm S})=0.93$ and $H=0.86$. 
The corresponding contours in the $\gamma$--$a$ plane are shown 
in Fig.~\ref{fig:BsdPsiKScont}. Interestingly, in the case of these 
contours, we would not have to deal with ``physical'' discrete 
ambiguities for $\gamma$, since values of $a$ larger than 1 would 
simply appear unrealistic. If it should be possible to measure 
${\cal A}_{\Delta\Gamma}(B_s\to J/\psi K_{\rm S})$ with the help of the 
width difference $\Delta\Gamma_s$, the dotted line could be fixed. 
In this example, the approximate expression (\ref{gam-approx}) yields 
$\gamma\approx65^\circ$, which deviates from the ``true'' value of 
$\gamma=60^\circ$ by only 8\%. It is also interesting to note that we 
have ${\cal A}_{\rm CP}^{\rm dir}(B_d\to J/\psi\, K_{\rm S})=-0.89\%$ 
in our example.

An important by-product of the strategy described above is that 
the quantities $a'$ and $\theta'$ allow us to take into account the 
penguin contributions in the determination of $\phi_d$ from 
$B_d\to J/\psi K_{\rm S}$, which are expected to enter at the
$\overline{\lambda}^3$ level and are hence presumably very small.
However, these uncertainties may become an important issue for the 
LHC because of the tremendous experimental accuracy for the 
CP-violating $B_d\to J/\psi K_{\rm S}$ observables that can be achieved 
there \cite{LHC-Report}. 

If we use (\ref{SU3-2}), we obtain an interesting relation, which is the 
counterpart of (\ref{CP-BR-rel}):
\begin{equation}\label{Bspsik-CP-BR-rel}
\frac{{\cal A}_{\rm CP}^{\rm dir}(B_d\to J/\psi
K_{\rm S})}{{\cal A}_{\rm CP}^{\rm dir}(B_s\to J/\psi K_{\rm S})}=
-\left|\frac{{\cal A}'}{{\cal A}}\right|^2
\frac{\langle\Gamma(B_s\to J/\psi K_{\rm S})\rangle}{\langle\Gamma
(B_d\to J/\psi K_{\rm S})\rangle}.
\end{equation}

\boldmath
\subsubsection{Theoretical Uncertainties}
\unboldmath
Let us finally turn to the theoretical uncertainties. In contrast to
the $B\to\pi K$ strategies discussed in Section~\ref{sec:BpiK}, 
flavour-symmetry arguments are sufficient for the extraction of $\gamma$,
i.e.\ we do not have to worry about rescattering effects. The theoretical
accuracy is therefore only limited by $SU(3)$-breaking effects. Whereas 
the solid curve in Fig.\ \ref{fig:BsdPsiKScont} is {\it theoretically 
clean}, the dot-dashed and dotted lines are affected by $U$-spin-breaking 
corrections. Because of the suppression of $a'e^{i\theta'}$ in 
(\ref{Bd-ampl2}) through $\lambda^2$, these contours are essentially 
unaffected by possible corrections to (\ref{SU3-2}), and rely predominantly 
on the $U$-spin relation $|{\cal A'}|=|{\cal A}|$. In the ``na\"\i ve'' 
factorization approximation, we have
\begin{equation}\label{SU3-break}
\left.\left|\frac{{\cal A'}}{{\cal A}}\right|\right|_{\rm fact}=\,
\frac{F_{B_d^0K^0}(M_{J/\psi}^2;1^-)}{F_{B_s^0\overline{K^0}}
(M_{J/\psi}^2;1^-)},
\end{equation}
where the form factors $F_{B_d^0K^0}(M_{J/\psi}^2;1^-)$ and 
$F_{B_s^0\overline{K^0}}(M_{J/\psi}^2;1^-)$ parametrize the quark--current 
matrix elements $\langle K^0|(\overline{b} s)_{\rm V-A}|B^0_d\rangle$ and 
$\langle\overline{K^0}|(\overline{b} d)_{\rm V-A}|B^0_s\rangle$, respectively 
\cite{BSW}. Quantitative studies of (\ref{SU3-break}), which could be 
performed, for instance, with the help of QCD sum-rule or lattice techniques,
are not yet available. In the light-cone sum-rule approach, sizeable 
$SU(3)$-breaking effects were found in the case of the $B_{d,s}\to K^\ast$ 
form factors \cite{BaBr}. It should be emphasized that also non-factorizable 
corrections, which are not included in (\ref{SU3-break}), may play an 
important r\^ole. 
We are optimistic that we will have a better picture of 
$SU(3)$ breaking by the time the $B_s\to J/\psi K_{\rm S}$ measurements 
can be performed in practice. A feasibility study performed by CMS indicates 
that $\gamma$ can be determined at the LHC through the 
$B_{s(d)}\to J/\psi K_{\rm S}$ approach with an experimental accuracy of
$\sim9^\circ$ after three years of taking data \cite{LHC-Report}. A 
variant of the $B_{s(d)}\to J/\psi K_{\rm S}$ strategy employing
$B_{s(d)}\to J/\psi \eta$ decays was recently discussed in \cite{skands}.

\boldmath
\subsection{The $B_{d(s)}\to D_{d(s)}^+D_{d(s)}^-$ and
$B_{d(s)}\to K^0\overline{K^0}$ Systems}\label{subsec:BdsDD}
\unboldmath
\boldmath
\subsubsection{Extracting $\gamma$ from $B_{d(s)}\to D_{d(s)}^+D_{d(s)}^-$
Decays}
\unboldmath
Usually, the decay $B_d\to D_d^+D_d^-$ is considered as a tool to probe 
the $B^0_d$--$\overline{B^0_d}$ mixing phase. In fact, if penguins played 
a negligible r\^ole in this mode, we could determine $\phi_d=2\beta$ 
from the corresponding mixing-induced CP asymmetry. However, the penguin 
topologies, which may in principle also contain important contributions 
from rescattering processes, may well be sizeable.\footnote{Note that
the QCD factorization formula (\ref{QCD-factor}) does not apply to 
$B_{d(s)}\to D_{d(s)}^+D_{d(s)}^-$ decays.} Interestingly, these 
penguin topologies allow us to determine $\gamma$ \cite{RF-BdsPsiK}, if 
the overall normalization of $B_d\to D_d^+D_d^-$ is fixed through the 
CP-averaged, i.e.\ ``untagged'', rate of its $U$-spin partner 
$B_s\to D_s^+D_s^-$, and if $\phi_d$ is determined separately, for 
instance with the help of the ``gold-plated'' mode 
$B_d\to J/\psi K_{\rm S}$. 


The decays $B_{d (s)}^0\to D^{+}_{d(s)} D^{-}_{d(s)}$ are transitions 
into a CP eigenstate with eigenvalue $+1$ and originate from 
$\overline{b}\to\overline{c}c\overline{d}\,(\overline{s})$ quark-level 
processes. We have to deal both with current--current and with penguin 
contributions, corresponding to the topologies shown in 
Fig.~\ref{fig:bpipi}. In analogy to (\ref{Bd-ampl2}) and (\ref{Bs-ampl}), 
the transition amplitudes can be written as 
\begin{eqnarray}
A(B_s^0\to D_s^+D_s^-)&=&\left(1-\frac{\lambda^2}{2}\right)\tilde{\cal A}'
\left[1+\left(\frac{\lambda^2}{1-\lambda^2}\right)\tilde a'e^{i\tilde
\theta'}e^{i\gamma}\right]\label{BDDs-ampl}\\
A(B_d^0\to D_d^+D_d^-)&=&-\lambda\,\tilde{\cal A}\left(1-\tilde a
e^{i\tilde\theta}e^{i\gamma}\right),\label{BDDd-ampl}
\end{eqnarray}
where the quantities $\tilde{\cal A}$, $\tilde{\cal A}'$ and 
$\tilde a e^{i\tilde\theta}$, $\tilde a' e^{i\tilde\theta'}$ take the
same form as in the $B_{s (d)}\to J/\psi K_{\rm S}$ case. The dynamics
underlying these parameters is, of course, rather different, as can be
seen by comparing the relevant Feynman diagrams.

Since the phase structures of the $B_d^0\to D_d^+D_d^-$ and 
$B_s^0\to D_s^+D_s^-$ amplitudes are completely analogous to those 
of $B_s^0\to J/\psi K_{\rm S}$ and $B_d^0\to J/\psi K_{\rm S}$, 
respectively, the $U$-spin approach discussed in the previous subsection 
can be applied by just making appropriate replacements of variables. 
If we neglect tiny phase-space effects, which can be taken into account 
straightforwardly (see \cite{RF-BdsPsiK}), we have
\begin{equation}
\tilde H=\left(\frac{1-\lambda^2}{\lambda^2}\right)
\left|\frac{\tilde{\cal A'}}{\tilde{\cal A}}\right|^2
\frac{\langle\Gamma(B_d\to D_d^+D_d^-)\rangle}{\langle
\Gamma(B_s\to D_s^+D_s^-)\rangle}.
\end{equation}
The $B_{d(s)}\to D_{d(s)}^+ D_{d(s)}^-$ counterpart of (\ref{gam-approx}) 
takes the following form:
\begin{equation}\label{gam-approx2}
\tan\gamma\approx\frac{\sin\phi_d-{\cal A}_{\rm CP}^{\rm mix}(B_d\to 
D_d^+D_d^-)}{(1-\tilde H)\cos\phi_d}\,,
\end{equation}
where the different sign of the mixing-induced CP asymmetry is due to
the different CP eigenvalues of the $B_d\to D_d^+D_d^-$ and $B_s\to 
J/\psi K_{\rm S}$ final states. In analogy to the $B_{s(d)}\to J/\psi 
K_{\rm S}$ system, contours in the $\gamma$--$\tilde a$ plane can be 
determined, allowing a transparent determination of $\gamma$, 
$\tilde a$ and $\tilde\theta$; an example is discussed in \cite{RF-BdsPsiK}.
Note that there are also strategies to extract $\cos\phi_d$, as noted in
\ref{subsubsec:ambig-resol}, complementing $a_{\psi K_{\rm S}}=\sin\phi_d$.

The theoretical accuracy is again only limited by $U$-spin-breaking 
effects, since no dynamical assumptions about rescattering processes 
are required. The relevant corrections affect the relation 
$|\tilde{\cal A}'|=|\tilde{\cal A}|$. In the factorization approximation, 
we have
\begin{equation}\label{SU3-breakBDD}
\left.\left|\frac{\tilde{\cal A'}}{\tilde{\cal A}}
\right|\right|_{\rm fact}=\,
\frac{(M_{B_s}-M_{D_s})\,\sqrt{M_{B_s}M_{D_s}}\,(w_s+1)}{(M_{B_d}-M_{D_d})
\,\sqrt{M_{B_d}M_{D_d}}\,(w_d+1)}\frac{f_{D_s}\,\xi_s(w_s)}{f_{D_d}\,
\xi_d(w_d)},
\end{equation}
where the restrictions form the Heavy-Quark Effective Theory for the 
$B_q\to D_q$ form factors have been taken into account by introducing 
appropriate Isgur--Wise functions $\xi_q(w_q)$ with $w_q=M_{B_q}/(2M_{D_q})$ 
\cite{NS}. Studies of the light-quark dependence of the Isgur--Wise
function were performed within Heavy-Meson Chiral Perturbation Theory, 
indicating an enhancement of $\xi_s(w_s)/\xi_d(w_d)$ at the level of $5\%$ 
\cite{HMChiPT1}. Applying the same formalism to $f_{D_s}/f_D$ gives 
values at the 1.2 level \cite{HMChiPT2}, which is of the same order of 
magnitude as the results of recent lattice calculations \cite{FBlat,lat1}. 
Further studies are needed to get a better picture of the $SU(3)$-breaking 
corrections to the ratio $|\tilde{\cal A'}/\tilde{\cal A}|$. Concerning
the experimental feasibility, preliminary analyses by LHCb have shown that 
an experimental precision on $\gamma$ of a few degrees seems to be 
achievable \cite{LHC-Report}.

\boldmath
\subsubsection{Extracting $\gamma$ from $B_{d(s)}\to K^{0}
\overline{K^{0}}$ Decays}
\unboldmath
The formalism developed for the $B_{d(s)}\to D_{d(s)}^+D_{d(s)}^-$
system applies also to the pair of $U$-spin-related 
decays $B_s^0\to K^0\overline{K^0}$ and $B_d^0\to K^0\overline{K^0}$, 
originating from $\overline{b}\to\overline{s}d\overline{d}$ and 
$\overline{b}\to\overline{d}s\overline{s}$ quark-level transitions,
respectively \cite{RF-ang}. Although these decays do not receive 
contributions from current--current operators at the ``tree'' level, their 
amplitudes can be written, within the Standard Model, in complete analogy 
to (\ref{BDDs-ampl}) and (\ref{BDDd-ampl}). Consequently, as far as the 
extraction of $\gamma$ is concerned, the channels 
$B_s^0\to K^0\overline{K^0}$ and $B_d^0\to K^0\overline{K^0}$ take the 
r\^oles of $B_s^0\to D_s^+D_s^-$ and $B^0_d\to D^+_dD^-_d$, respectively. 
The determination of the corresponding hadronic parameters would be
of particular interest to obtain insights into penguin topologies with
internal up- and charm-quark exchanges, as becomes obvious from the
discussion in Subsection~\ref{subsec:u-c-pens}.

If the decays $B_s^0\to K^{\ast0}\overline{K^{\ast0}}$ and 
$B_d^0\to K^{\ast0}\overline{K^{0\ast}}$ are considered, the corresponding
angular distributions provide many more observables, so that the 
$B^0_d$--$\overline{B^0_d}$ mixing phase $\phi_d$ is no longer required
as an input for the determination of $\gamma$, but can rather be extracted
as well \cite{RF-ang}. From a practical point of view, these strategies 
are, however, more complicated than those involving two pseudoscalar mesons 
in the final states. It should be emphasized that $B_{d,s}\to K^{(\ast)0}
\overline{K^{(\ast)0}}$ decays represent very sensitive probes for new 
physics, since they are penguin-induced modes (see also 
\ref{subsubsec:BphiK-NP}).

\boldmath
\subsection{The $B_d\to\pi^+\pi^-$, $B_s\to K^+K^-$ 
System}\label{subsec:BsKK-Uspin}
\unboldmath
As we have seen in Subsection~\ref{subsec:Bpipi}, the decay 
$B_d\to\pi^+\pi^-$ is usually considered as a tool to determine 
$\alpha=180^\circ-\beta-\gamma$. Unfortunately, the extraction of 
this angle from ${\cal A}_{\rm CP}^{\rm mix}(B_d\to\pi^+\pi^-)$ is 
affected by large penguin uncertainties, and the strategies to 
control them through additional data are challenging. Let us now
discuss a new approach to extract CKM phases from $B_d\to\pi^+\pi^-$
with the help of the decay $B_s\to K^+K^-$ \cite{RF-BsKK}.

\boldmath
\subsubsection{Extraction of $\beta$ and $\gamma$}
\unboldmath
Looking at the Feynman diagrams shown in Fig.\ \ref{fig:bpipi}, we observe 
that $B_s\to K^+K^-$ is obtained from the $B_d\to\pi^+\pi^-$ topologies by 
interchanging all down and strange quarks. The corresponding decay amplitude 
can be parametrized as follows:
\begin{equation}\label{BsKK-ampl0}
A(B_s^0\to K^+K^-)=\left(\frac{\lambda}{1-\lambda^2/2}\right)
{\cal C}'\left[e^{i\gamma}+\left(\frac{1-\lambda^2}{\lambda^2}\right)
d'e^{i\theta'}\right],
\end{equation}
where ${\cal C}'$ and $d'e^{i\theta'}$ take the same form as 
the parameters ${\cal C}$ and $d e^{i\theta}$ in the
\mbox{$B_d\to\pi^+\pi^-$} amplitude (\ref{Bpipi-ampl}).
Applying the formalism discussed in Subsection~\ref{subsec:CPasym}, we arrive
at parametrizations of the following structure:\footnote{For the explicit 
expressions, see \cite{RF-BsKK}.}
\begin{eqnarray}
{\cal A}_{\rm CP}^{\rm dir}(B_d\to\pi^+\pi^-)&=&
\mbox{function}(d,\theta,\gamma)\label{ASYM-1}\\
{\cal A}_{\rm CP}^{\rm mix}(B_d\to\pi^+\pi^-)&=&
\mbox{function}(d,\theta,\gamma,\phi_d=2\beta)
\end{eqnarray}
\vspace*{-0.9truecm}
\begin{eqnarray}
{\cal A}_{\rm CP}^{\rm dir}(B_s\to K^+K^-)&=&
\mbox{function}(d',\theta',\gamma)\\
{\cal A}_{\rm CP}^{\rm mix}(B_s\to K^+K^-)&=&
\mbox{function}(d',\theta',\gamma,\phi_s\approx0),\label{ASYM-4}
\end{eqnarray}
where the Standard-Model expectation $\phi_s\approx0$ can be probed
straightforwardly through $B_s\to J/\psi\phi$. Consequently, we have four 
CP-violating observables at our disposal, depending on six ``unknowns''. 
However, since $B_d\to\pi^+\pi^-$ and $B_s\to K^+K^-$ are related to 
each other by interchanging all down and strange quarks, the $U$-spin 
symmetry implies
\begin{equation}\label{U-spin-rel}
d'e^{i\theta'}=d e^{i\theta}.
\end{equation}
Using this relation, the {\it four} observables (\ref{ASYM-1})--(\ref{ASYM-4}) 
can be expressed in terms of the {\it four} quantities $d$, $\theta$, 
$\phi_d=2\beta$ and $\gamma$, which can hence be determined. 

In comparison with the $U$-spin strategies discussed in 
Subsections~\ref{subsec:BsdPsiKS} and \ref{subsec:BdsDD}, an 
important advantage of the $B_d\to\pi^+\pi^-$, $B_s\to K^+K^-$ 
approach is that the $U$-spin symmetry is only applied
in the form of (\ref{U-spin-rel}), where $d e^{i\theta}$ and
$d'e^{i\theta'}$ are actually given by ratios of amplitudes
(see (\ref{D-DEF})). In particular, no overall normalization
of decay amplitudes has to be fixed through $U$-spin arguments, 
where $U$-spin-breaking effects are expected to be much larger. 
We shall come back to this issue below. Moreover, we again do not 
have to make any dynamical assumptions about rescattering processes, 
which is an important conceptual advantage in comparison with 
the $B\to\pi K$ strategies to determine $\gamma$ reviewed in 
Section~\ref{sec:BpiK}.

\boldmath
\subsubsection{Minimal Use of the $U$-Spin Symmetry}\label{sssec:min-U-spin}
\unboldmath
The $U$-spin-symmetry arguments can be minimized, if we employ 
$\phi_d=2\beta$ as an input, which can be determined straightforwardly 
through $B_d\to J/\psi K_{\rm S}$ (see also \ref{subsubsec:ambig-resol}). 
The CP-violating observables ${\cal A}_{\rm CP}^{\rm dir}(B_d\to\pi^+\pi^-)$ 
and ${\cal A}_{\rm CP}^{\rm mix}(B_d\to\pi^+\pi^-)$ allow us then to 
eliminate the strong phase $\theta$ and to determine $d$ as a function of
$\gamma$. Analogously, we may use ${\cal A}_{\rm CP}^{\rm dir}(B_s\to K^+K^-)$
and ${\cal A}_{\rm CP}^{\rm mix}(B_s\to K^+K^-)$ to eliminate 
$\theta'$, and to determine $d'$ as a function of $\gamma$. The 
corresponding contours in the $\gamma$--$d$ and $\gamma$--$d'$ planes can 
be fixed in a {\it theoretically clean} way. Using now the $U$-spin relation 
$d'=d$, these contours allow the determination both of the CKM angle 
$\gamma$ and of the hadronic quantities $d$, $\theta$, $\theta'$ 
\cite{RF-BsKK}. It should be emphasized that no $U$-spin relation 
involving the strong phases $\theta$ and $\theta'$ has to be employed
in this approach. Alternatively, we could also eliminate $d$ and $d'$,
and could determine $\gamma$ (as well as $d$ and $d'$) with the help of 
theoretically clean contours in the $\gamma$--$\theta$ and 
$\gamma$--$\theta'$ planes through $\theta=\theta'$. In the following,
we shall, however, focus on the former possibility.

\begin{figure}
\centerline{\rotate[r]{
\epsfysize=9.7truecm
{\epsffile{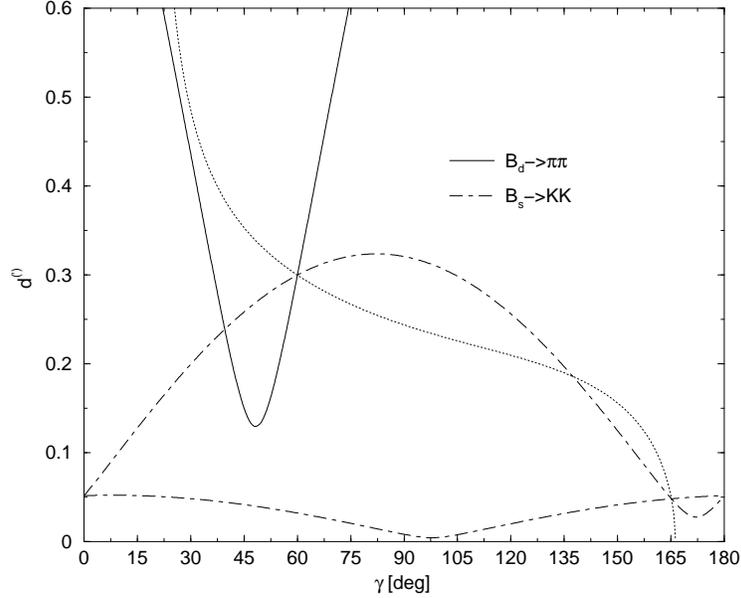}}}}
\caption{The contours in the $\gamma$--$d^{(')}$ planes fixed through 
the CP-violating $B_d\to\pi^+\pi^-$ and $B_s\to K^+K^-$ observables for 
a specific example discussed in the text.}\label{fig:BsKKcont}
\end{figure}

Let us illustrate this approach with a simple example. Assuming
$d=d'=0.3$, $\theta=\theta'=210^\circ$, $\phi_s=0$, $\phi_d=44^\circ$
and $\gamma=60^\circ$, which is in accordance with theoretical estimates
and the fits of the unitarity triangle, we obtain the following observables:
${\cal A}_{\rm CP}^{\rm dir}(B_d\to\pi^+\pi^-)=+19\%$, 
${\cal A}_{\rm CP}^{\rm mix}(B_d\to\pi^+\pi^-)=+62\%$, 
${\cal A}_{\rm CP}^{\rm dir}(B_s\to K^+K^-)=-17\%$ and
${\cal A}_{\rm CP}^{\rm mix}(B_s\to K^+K^-)=-27\%$.
In Fig.~\ref{fig:BsKKcont}, the corresponding contours in the $\gamma$--$d$ 
and $\gamma$--$d'$ planes are represented by the solid and dot-dashed lines, 
respectively. Their intersection yields a twofold solution for $(\gamma,d)$, 
given by $(39^\circ,0.24)$ and our input value of $(60^\circ,0.30)$. It
should be noted that the $B_s\to K^+K^-$ contour would imply in this example
{\it model-independently} that $d'<0.32$. The dotted line is related to the 
quantity
\begin{equation}\label{K-intro}
{\cal K}\equiv-\frac{1}{\epsilon}\left(\frac{d\sin\theta}{d'\sin\theta'}\right)
\left[\frac{{\cal A}_{\rm CP}^{\rm dir}(B_s\to K^+K^-)}{{\cal A}_{\rm 
CP}^{\rm dir}(B_d\to\pi^+\pi^-)}\right]=
\frac{1-2 d\cos\theta\cos\gamma+d^2}{\epsilon^2+
2\epsilon d'\cos\theta'\cos\gamma+d'^2}
\end{equation} 
with 
\begin{equation}\label{eps-CKM-def}
\epsilon\equiv\frac{\lambda^2}{1-\lambda^2},
\end{equation}
which can be combined with ${\cal A}_{\rm CP}^{\rm mix}(B_s\to K^+K^-)$ 
through the $U$-spin relation (\ref{U-spin-rel}) to fix another contour 
in the $\gamma$--$d$ plane. We shall come back to ${\cal K}$ in 
\ref{subsubsec:U-spin-Replace}. Combining all contours in 
Fig.~\ref{fig:BsKKcont} with one another, we obtain a single solution 
for $\gamma$, which is given by the ``true'' value of $60^\circ$. As a
``by-product'', also the hadronic parameters $d$ and $\theta$, 
$\theta'$ can be determined, which would allow an interesting comparison 
with theoretical predictions, thereby providing valuable insights into 
hadronic physics. 

From an experimental point of view, this approach is very promising for 
CDF-II ($\left.\Delta\gamma\right|_{\rm exp}={\cal O}(10^\circ)$) 
\cite{wuerth}, and ideally suited for the LHC experiments, in particular
LHCb ($\left.\Delta\gamma\right|_{\rm exp}={\cal O}(1^\circ)$) 
\cite{LHC-Report}; similar comments apply to BTeV. Moreover, it has 
interesting features concerning the theoretical cleanliness, which 
is our next topic.

\boldmath
\subsubsection{Theoretical Uncertainties}
\unboldmath
The theoretical accuracy of the determination of $\gamma$ and
the hadronic parameters $d$, $\theta$, $\theta'$ discussed in 
\ref{sssec:min-U-spin} is {\it only} limited by $U$-spin-breaking 
corrections to $d'=d$. In particular, it is not affected by final-state 
interaction or rescattering effects. For $d'\not=d$, there would
be a relative shift of the solid and dot-dashed curves in 
Fig.~\ref{fig:BsKKcont}. However, in this example, the extracted value 
of $\gamma$ would be very stable under such effects. 

Let us, in order to put the $U$-spin-breaking corrections to
(\ref{U-spin-rel}) on a more quantitative basis, apply the 
picture of the Bander--Silverman--Soni (BSS) mechanism \cite{BSS} (see 
\ref{subsubsec:Tree-Pen}) to calculate the quantity $d e^{i\theta}$
defined in (\ref{D-DEF}). Following the formalism developed in 
\cite{RF-1,RF-EWP1} and using expression (\ref{Pen-ME}), we obtain
\begin{equation}\label{d-approx}
d e^{i\theta}=\frac{1}{R_b}\left[\frac{{\cal A}_t+
{\cal A}_c}{{\cal A}_{\rm CC}+{\cal A}_t+{\cal A}_u}\right],
\end{equation}
where
\begin{eqnarray}
{\cal A}_{\rm CC}&=&\frac{1}{3}\,\overline{C}_1+\overline{C}_2\label{AT}\\
{\cal A}_t&=&\frac{1}{3}\left[\overline{C}_3+\overline{C}_9+
\chi\left(\overline{C}_5+\overline{C}_7\right)\right]+\overline{C}_4+
\overline{C}_{10}+\chi\left(\overline{C}_6+\overline{C}_8\right)\label{A0}\\
{\cal A}_j&=&\frac{\alpha_s}{9\pi}\left[\frac{10}{9}-G(m_j,k,m_b)\right]
\left[\overline{C}_2+\frac{1}{3}\frac{\alpha}{\alpha_s}\left(3\,\overline{C}_1
+\overline{C}_2\right)\right]\left(1+\chi\right),\label{Aq}
\end{eqnarray}
with $j\in\{u,c\}$. Kinematical considerations at the perturbative quark 
level imply the following ``physical'' range for $k^2$ \cite{BL}:
\begin{equation}
\frac{1}{4}\lsim\frac{k^2}{m_b^2}\lsim\frac{1}{2},
\end{equation}
and 
\begin{equation}
\chi=\frac{2M_\pi^2}{(m_u+m_d)(m_b-m_u)}
\end{equation}
is due to the use of the equations of motion for the quark fields.
Since the ``penguin'' parameter $d e^{i\theta}$ is actually a ratio 
of certain amplitudes, the decay constants and form factors 
arising typically in factorization (for an example, see (\ref{T-fact}))
cancel in (\ref{d-approx}). It should be noted that this expression
gives values for $d$ and $\theta$ of the same order of magnitude as 
those employed in the example leading to the contours shown in 
Fig.~\ref{fig:BsKKcont}. The expression for the $B_s\to K^+K^-$ parameter 
$d'e^{i\theta'}$ takes the same form as (\ref{d-approx}), where $\chi$ 
is replaced in (\ref{A0}) and (\ref{Aq}) by
\begin{equation}
\chi'=\frac{2M_K^2}{(m_u+m_s)(m_b-m_u)}.
\end{equation}
Consequently, in our approach to evaluate $d e^{i\theta}$ and 
$d'e^{i\theta'}$, $U$-spin-breaking corrections enter only through the 
parameters $\chi$ and $\chi'$. However, up to small electromagnetic 
corrections, the chiral structure of strong interactions implies
\begin{equation}\label{mass-rel}
\frac{M_\pi^2}{m_u+m_d}=\frac{M_K^2}{m_u+m_s},
\end{equation}
leading, among other things, to the Gell-Mann--Okubo relation 
(see, for example, \cite{georgi}). In our case, this expression has 
the interesting implication 
\begin{equation}
\chi=\chi', 
\end{equation}
so that the $U$-spin relation (\ref{U-spin-rel}) is not affected by 
$U$-spin-breaking corrections within our formalism. Although (\ref{d-approx}) 
is a simplified expression, as becomes evident from the unphysical $k^2$ 
dependence, it strengthens our confidence into (\ref{U-spin-rel}). 
Further theoretical studies of the $U$-spin-breaking effects in the 
$B_d\to\pi^+\pi^-$, $B_s\to K^+K^-$ system, using, for example, 
QCD factorization \cite{BBNS3}, would be desirable.

Apart from these theoretical considerations, we may also obtain
experimental insights into $U$-spin breaking. A first consistency check 
is provided by $\theta=\theta'$. Moreover, we may determine 
the overall normalizations $|{\cal C}|$ and $|{\cal C}'|$ of the 
$B^0_d\to\pi^+\pi^-$ and $B^0_s\to K^+K^-$ decay amplitudes 
(see (\ref{Bpipi-ampl}) and (\ref{BsKK-ampl0})) with the help of 
the corresponding CP-averaged branching ratios. Comparing them with 
the ``factorized'' result
\begin{equation}\label{U-fact}
\left.\left|\frac{{\cal C}'}{{\cal C}}\right|\right|_{\rm fact}=
\left[\frac{M_{B_s}^2-M_K^2}{M_{B_d}^2-M_\pi^2}\right]
\left[\frac{f_K}{f_\pi}\right]
\left[\frac{F_{B_sK}(M_K^2;0^+)}{F_{B_d\pi}(M_\pi^2;0^+)}\right],
\end{equation}
we have another interesting probe for $U$-spin-breaking effects. Moreover,
the quantity ${\cal K}$ introduced in (\ref{K-intro}) can also be
expressed as
\begin{equation}\label{Kcal-BR}
{\cal K}=\frac{1}{\epsilon}\left|\frac{{\cal C}'}{{\cal C}}\right|^2
\left[\frac{\mbox{BR}(B_d\to\pi^+\pi^-)}{\mbox{BR}(B_s\to K^+K^-)}\right]
\frac{\tau_{B_s}}{\tau_{B_d}},
\end{equation}
so that the $U$-spin relation (\ref{U-spin-rel}) implies
\begin{equation}\label{CP-REL-BsKK}
\frac{{\cal A}_{\rm CP}^{\rm dir}(B_s\to K^+K^-)}{{\cal 
A}_{\rm CP}^{\rm dir}(B_d\to\pi^+\pi^-)}=
-\left|\frac{{\cal C}'}{{\cal C}}\right|^2
\left[\frac{\mbox{BR}(B_d\to\pi^+\pi^-)}{\mbox{BR}(B_s\to K^+K^-)}\right]
\frac{\tau_{B_s}}{\tau_{B_d}}.
\end{equation}
In order to obtain further insights, the $B_d\to\rho^+\rho^-$, 
$B_s\to K^{\ast+} K^{\ast-}$ system would be of particular interest, 
allowing us to determine $\gamma$ together with the mixing phases $\phi_d$ 
and $\phi_s$, and tests of several $U$-spin relations \cite{RF-ang}. 
Here the observables of the corresponding angular distributions have
to be measured.

\subsubsection{Searching for New Physics}
Since penguin processes play a key r\^ole in the decays $B_s\to
K^+K^-$ and $B_d\to\pi^+\pi^-$, they -- and the strategy to determine
$\gamma$, where moreover the unitarity of the CKM matrix is employed --
may well be affected by new physics. Interestingly, the Standard Model 
implies a rather restricted region in the space of the CP-violating 
observables of the $B_s\to K^+K^-$, $B_d\to\pi^+\pi^-$ system, as noted
in \cite{FMat}. A future measurement of observables lying significantly 
outside of this allowed region would be an immediate indication for new 
physics. On the other hand, if the observables should lie within the 
region predicted by the Standard Model, we can extract a value for  
$\gamma$ by following the strategy discussed above. This value for  
$\gamma$ may well be in conflict with those provided by other approaches, 
which would then also indicate the presence of new physics.

\boldmath
\subsubsection{Replacing $B_s\to K^+K^-$ through 
$B_d\to\pi^\mp K^\pm$}\label{subsubsec:U-spin-Replace}
\unboldmath
Although the $B_d\to\pi^+\pi^-$, $B_s\to K^+K^-$ approach is very 
promising for hadron machines, it cannot be implemented at the asymmetric 
$e^+e^-$ $B$-factories operating at the $\Upsilon(4S)$ resonance 
(BaBar and Belle), since $B_s$-mesons are there not accessible. However, 
there is a variant of this strategy, employing $B_d^0\to\pi^- K^+$ instead 
of $B_s^0\to K^+K^-$ \cite{RF-bpipi}. Since these modes differ only in 
their spectator quarks, we expect
\begin{equation}\label{ACP-rep}
{\cal A}_{\rm CP}^{\rm dir}(B_s\to K^+K^-)\approx{\cal A}_{\rm CP}^{\rm dir}
(B_d\to\pi^\mp K^\pm)
\end{equation}
\begin{equation}\label{BR-rep}
\mbox{BR}(B_s\to K^+K^-)
\approx\mbox{BR}(B_d\to\pi^\mp K^\pm)\,\frac{\tau_{B_s}}{\tau_{B_d}},
\end{equation}
and obtain from (\ref{Kcal-BR})
\begin{equation}\label{H-res}
{\cal K}\approx\frac{1}{\epsilon}\left(\frac{f_K}{f_\pi}\right)^2
\left[\frac{\mbox{BR}(B_d\to\pi^+\pi^-)}{\mbox{BR}(B_d\to\pi^\mp K^\pm)}
\right]=
\left\{\begin{array}{ll}
7.3\pm2.9 & \mbox{(CLEO \cite{CLEO-BpiK})}\\
7.2\pm2.3 & \mbox{(BaBar \cite{babar-BpiK})}\\
8.5\pm3.7 & \mbox{(Belle \cite{belle-BpiK}).}
\end{array}\right.
\end{equation}
The advantage of (\ref{H-res}) is that it allows the determination of 
${\cal K}$ without a measurement of the decay $B_s\to K^+K^-$. On the 
other hand, this relation relies not only on $SU(3)$ flavour-symmetry 
arguments, but also on a certain dynamical assumption. The point is that 
$B_s\to K^+K^-$ receives also contributions from ``exchange'' and 
``penguin annihilation'' topologies, which are absent in 
$B_d\to\pi^\mp K^\pm$. It is usually assumed that these contributions 
play a minor r\^ole \cite{ghlr-ewp}. However, they may in principle be 
enhanced through large rescattering effects \cite{FSI}. Although these 
topologies do {\it not} 
lead to any problems in the strategies discussed below if ${\cal K}$ is 
fixed through a measurement of $B_s\to K^+K^-$  -- even if they should 
turn out to be sizeable -- they may affect (\ref{ACP-rep})--(\ref{H-res}). 
Their importance can be probed, in addition to (\ref{ACP-rep}) and 
(\ref{BR-rep}), with the help of the decay $B_s\to\pi^+\pi^-$. The 
na\"\i ve expectation for the corresponding branching ratio is 
${\cal O}(10^{-8})$; a significant enhancement would signal that the 
``exchange'' and ``penguin annihilation'' topologies cannot be neglected. 

The $B$-factory results in (\ref{H-res}) may be used to obtain a 
rather restricted range for the hadronic parameter $d$, 
as well as upper bounds on $|{\cal A}_{\rm CP}^{\rm dir}(B_d\to\pi^+\pi^-)|$ 
and $|{\cal A}_{\rm CP}^{\rm dir}(B_d\to\pi^\mp K^\pm)|$. In order to derive 
these constraints, we make use of the right-hand side of (\ref{K-intro}),
and assume that (\ref{U-spin-rel}) holds also for the $B_d\to\pi^\mp K^\pm$,
$B_d\to\pi^+\pi^-$ system. The quantity ${\cal K}$ allows us then 
to determine
\begin{equation}\label{C-def}
C\equiv\cos\theta \cos\gamma
\end{equation}
as a function of $d$ \cite{RF-bpipi}:
\begin{equation}\label{C-expr}
C=\frac{a-d^2}{2\,b\,d},
\end{equation}
where
\begin{equation}\label{a-b-def}
a=\frac{1-\epsilon^2{\cal K}}{{\cal K}-1}\qquad\mbox{and}\qquad 
b=\frac{1+\epsilon\, {\cal K}}{{\cal K}-1}.
\end{equation}
Since $C$ is the product of two cosines, it has to 
lie between $-1$ and $+1$, thereby implying an allowed range for $d$. 
Taking into account (\ref{C-expr}) and (\ref{a-b-def}), we obtain (for
${\cal K}<1/\epsilon^2$) 
\begin{equation}\label{d-bounds1}
\frac{1-\epsilon\,\sqrt{{\cal K}}}{1+\sqrt{{\cal K}}}\leq 
d\leq\frac{1+\epsilon\,\sqrt{{\cal K}}}{|1-\sqrt{{\cal K}}|}.
\end{equation}
Note that this range takes the same form as (\ref{rho-range}).
In the special case of ${\cal K}=1$, there is only a lower bound on $d$, 
given by $d_{\rm min}=(1-\epsilon)/2$; for ${\cal K}<1$, $C$ takes a 
minimal value that implies the following allowed range for $\gamma$: 
\begin{equation}
|\cos\gamma|\geq C_{\rm min}=\frac{\sqrt{(1-\epsilon^2H)(1-{\cal K})}}{1+
\epsilon\,{\cal K}}
\approx\sqrt{1-{\cal K}}.
\end{equation}
From a conceptual point of view, this bound on $\gamma$ is completely 
analogous to the one derived in \cite{FM} (see \ref{subsec:FM-bound}). 
Unfortunately, it is only of academic interest in the present case, as 
(\ref{H-res}) indicates ${\cal K}>1$, which we shall assume in the following 
discussion. So far, we have treated $\theta$ and $\gamma$ as ``unknown'', 
i.e.\ free parameters. However, for a given value of $\gamma$, we have 
\begin{equation}\label{C-range}
-|\cos\gamma|\leq C\leq+|\cos\gamma|,
\end{equation}
and obtain constraints on $d$ that are stronger than (\ref{d-bounds1}):
\begin{equation}\label{d-range}
d_{\rm min}^{\rm max}=\pm b|\cos\gamma|+\sqrt{a+b^2\cos^2\gamma}.
\end{equation}

\begin{figure}
\centerline{\rotate[r]{
\epsfysize=10.4truecm
{\epsffile{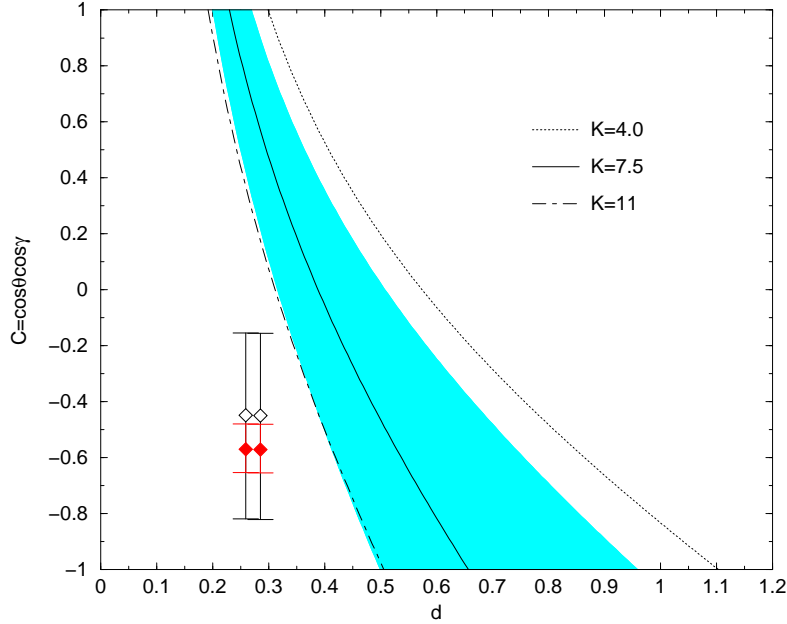}}}}
\caption{$C=\cos\theta\cos\gamma$ as a function of $d$ for various 
values of ${\cal K}$. The ``diamonds'' with error bars represent the 
QCD factorization results \cite{BBNS3} for the ranges for $\gamma$ 
obtained in \cite{Rome-rev} (narrow range) and \cite{AL} (wide range). 
The shaded region corresponds to a variation of $\xi_d$ within the
interval $[0.8,1.2]$ for ${\cal K}=7.5$.}\label{fig:d}
\end{figure}

In Fig.~\ref{fig:d}, we show the dependence of $C$ on $d$ for values of 
${\cal K}$ corresponding to (\ref{H-res}). The ``diamonds''
represent the predictions of the QCD factorization approach 
\cite{BBNS3}: 
\begin{equation}\label{d-QCD-fact}
d=\left\{\begin{array}{l}
\left(28.5\pm5.1\mp5.7\right)\times 10^{-2}\\
\left(25.9\pm4.3\mp5.2\right)\times 10^{-2},
\end{array}\right.
\quad
\theta-180^\circ=\left\{\begin{array}{l}
\left(8.2\pm3.8\right)^\circ\\
\left(9.0\pm4.1\right)^\circ,
\end{array}\right.
\end{equation}
where the notation is as in (\ref{rho-QCD-fact}) and (\ref{qC-QCD-fact}). 
The ``error bars'' in Fig.~\ref{fig:d} correspond to the ranges for $\gamma$ 
obtained in \cite{Rome-rev} and \cite{AL} (see (\ref{Rome-gamma}) and
Table~\ref{tab:CKM-fits}), while the filled and opaque diamonds are 
evaluated with the central values of (\ref{d-QCD-fact}) for the
preferred values of $\gamma=54.8^\circ$ and $63^\circ$,
respectively. The shaded region illustrates the variation of 
\begin{equation}\label{U-spin-viol}
\xi_d\equiv d'/d
\end{equation}
between 0.8 and 1.2 for ${\cal K}=7.5$, where the regions on the left-
and right-hand sides of the solid line correspond to $\xi_d>1$ and 
$\xi_d<1$, respectively. As noted in \cite{RF-bpipi}, the impact of 
a sizeable phase $\Delta\theta$ in $\theta'=\theta+\Delta\theta$, 
representing the second kind of possible corrections to (\ref{U-spin-rel}), 
is very small. 

Looking at Fig.\ \ref{fig:d}, we observe that the experimental values of 
${\cal K}$ imply a rather restricted range for $d$. In particular, 
we get the lower bound of $d\gsim0.2$. Moreover, the curves are
not in favour of an interpretation of the ``QCD factorization'' 
results (\ref{d-QCD-fact}) within the Standard Model, although
the experimental uncertainties are too large to draw definite
conclusions. This discrepancy could be resolved for values of $\gamma$ 
being significantly larger than $90^\circ$ \cite{RF-bpipi}. On the 
other hand, it may also be a manifestation of large $\Lambda_{\rm QCD}/m_b$ 
corrections \cite{charming-pens2}, as we have discussed in 
Subsection~\ref{subsec:BpiK-calc}. In any case, the data tell us that we 
have definitely to care about penguin effects in $B_d\to\pi^+\pi^-$.

\begin{figure}
\centerline{\rotate[r]{
\epsfysize=10.0truecm
{\epsffile{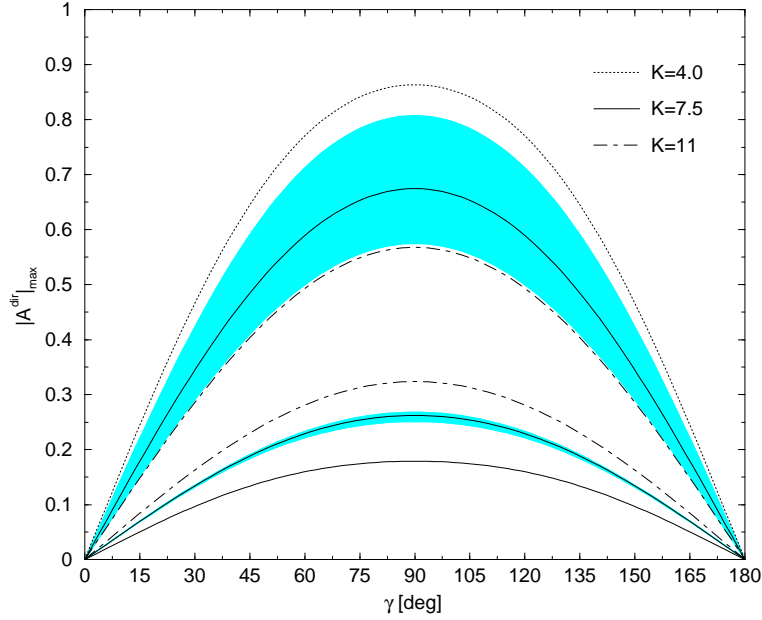}}}}
\caption{The maximal direct CP asymmetries for $B_d\to\pi^+\pi^-$ (upper
curves) and $B_s\to K^+K^-\approx B_d\to\pi^\mp K^\pm$ (lower curves)
as functions of $\gamma$ 
for various values of ${\cal K}$. The shaded regions correspond to a 
variation of $\xi_d$ within $[0.8,1.2]$ for ${\cal K}=7.5$.}\label{fig:Adir}
\end{figure}

The quantity ${\cal K}$ allows us also to derive 
constraints on direct CP asymmetries. The corresponding formulae,
which can be found in \cite{RF-bpipi}, simplify considerably for
$\gamma=90^\circ$:
\begin{equation}\label{ACPBdpipi-approx}
\left|{\cal A}_{\rm CP}^{\rm dir}(B_d\to\pi^+\pi^-)
\right|_{\gamma=90^\circ}^{\rm max}=
2\,\sqrt{\frac{(1-\epsilon^2{\cal K})(\xi_d^2{\cal K}-1)}{\left(\xi_d^2
-\epsilon^2\right)^2{\cal K}^2}}\approx\frac{2}{\xi_d\sqrt{{\cal K}}}
\end{equation}
\begin{equation}\label{ACPBsKK-approx}
\left|{\cal A}_{\rm CP}^{\rm dir}(B_s\to K^+ K^-)
\right|_{\gamma=90^\circ}^{\rm max}=2\,\epsilon\,\xi_d\,\sqrt{\frac{(1-
\epsilon^2{\cal K})(\xi_d^2{\cal K}-1)}{\left(\xi_d^2-
\epsilon^2\right)^2}}\approx2\,\epsilon\,\sqrt{{\cal K}}.
\end{equation}
Interestingly, the latter expression is essentially {\it unaffected} by any 
correction to (\ref{U-spin-rel}) for ${\cal K}={\cal O}(10)$; its 
theoretical accuracy is practically only limited by (\ref{U-fact}), which 
enters in the determination of ${\cal K}$ through (\ref{Kcal-BR}).

In Fig.~\ref{fig:Adir}, we show the dependence of the maximally allowed 
direct CP asymmetries for $B_d\to\pi^+\pi^-$ and 
$B_s\to K^+K^-\approx B_d\to\pi^\mp K^\pm$ on $\gamma$ for 
various values of ${\cal K}$. The shaded regions correspond to a 
variation of the parameter $\xi_d$ within the interval $[0.8,1.2]$ for 
${\cal K}=7.5$. The values for ${\cal K}$ given in (\ref{H-res}) 
disfavour very large direct CP violation in $B_d\to\pi^\mp K^\pm$, 
which is also consistent with the experimental results summarized in 
Table~\ref{tab:BPIK-Asym}, and the most recent BaBar update given
by ${\cal A}_{\rm CP}(B^0_d\to\pi^-K^+)=(7\pm8\pm2)\%$ \cite{BABAR-Bpipi-CP}.
On the other hand, there is a lot of space left for large direct CP violation 
in $B_d\to\pi^+\pi^-$ (see also \cite{charming-pens2}), which is in 
accordance with (\ref{Bpipi-CP-asym-res}).  

As can be seen in Fig.~\ref{fig:Adir}, a measurement of non-vanishing CP 
asymmetries $|{\cal A}_{\rm CP}^{\rm dir}|_{\rm exp}$ would allow us to 
exclude immediately a certain range of $\gamma$ around $0^\circ$ and 
$180^\circ$, since the values corresponding to 
$|{\cal A}_{\rm CP}^{\rm dir}|_{\rm exp}>
|{\cal A}_{\rm CP}^{\rm dir}|_{\rm max}$ would be excluded. However,
in order to probe this angle, the mixing-induced CP asymmetry 
${\cal A}_{\rm CP}^{\rm mix}(B_d\to\pi^+\pi^-)$ is actually more 
powerful. If we assume that $\phi_d$ has been measured through 
$B_d\to J/\psi K_{\rm S}$ and use $\cos\theta=C/\cos\gamma$ to eliminate the
strong phase $\theta$, we obtain 
${\cal A}_{\rm CP}^{\rm mix}(B_d\to\pi^+\pi^-)$ as a monotonic function 
of $d^2$, taking its extremal values for $d=d^{\rm max}_{\rm min}$ 
given in (\ref{d-range}). For a fixed value of $\gamma$, the allowed 
range for ${\cal A}_{\rm CP}^{\rm mix}(B_d\to\pi^+\pi^-)$ is usually 
very large. However, a measured value of this observable may, on the 
other hand, imply a rather restricted range for $\gamma$, as illustrated 
in more detail in \cite{RF-bpipi}. Finally, if in addition to ${\cal K}$ 
and ${\cal A}_{\rm CP}^{\rm mix}(B_d\to\pi^+\pi^-)$ also direct CP violation 
in $B_d\to\pi^+\pi^-$ or $B_d\to\pi^\mp K^\pm$ is observed, both the angle
$\gamma$ and the hadronic parameters $d$ and $\theta$ can be determined. 

At CLEO, BaBar and Belle, significantly improved measurements of the 
$B_d\to\pi^+\pi^-$ and $B_d\to\pi^\mp K^\pm$ branching ratios will be 
obtained, and direct CP violation in these channels will hopefully be 
observed. The latter two experiments should also determine the 
mixing-induced CP asymmetry of the $B_d\to\pi^+\pi^-$ mode. On the 
other hand -- in addition to similar measurements -- run II of the 
Tevatron will also provide first access to the $B_s\to K^+K^-$ channel. 
In the LHC era, the very rich physics potential of these decays can then 
be fully exploited.

\boldmath
\subsection{The $B_{(s)}\to\pi K$ System}\label{subsec:BspiK-Uspin}
\unboldmath
Another interesting pair of decays, which are related to each other by
interchanging all down and strange quarks, is the $B^0_d\to\pi^-K^+$, 
$B^0_s\to\pi^+K^-$ system \cite{BspiK}. Starting from the general 
Standard-Model parametrizations for the corresponding decay amplitudes
and eliminating the CKM factors $\lambda_t^{(s)}$ and $\lambda_t^{(d)}$ 
through the unitarity relation (\ref{CKM-UT-REL}), it is an easy 
exercise to show that we may -- in the strict $U$-spin limit -- write
\begin{equation}\label{PAR1}
A(B^0_d\to\pi^-K^+)=-P\left(1-re^{i\delta}e^{i\gamma}\right)
\end{equation}
\begin{equation}\label{PAR2}
A(B^0_s\to\pi^+K^-)=P\sqrt{\epsilon}
\left(1+\frac{1}{\epsilon}\, r e^{i\delta}e^{i\gamma}\right),
\end{equation}
where $P$ denotes a CP-conserving complex amplitude, $\epsilon$ was 
introduced in (\ref{eps-CKM-def}), $r$ is a real parameter and
$\delta$ a CP-conserving strong phase. At first sight, 
it appears as if $\gamma$, $r$ and $\delta$ could be determined from 
the ratio of the CP-averaged rates and the two CP asymmetries provided 
by these modes.\footnote{Note that these observables are independent of $P$.} 
However, because of the relation 
\begin{displaymath}
|A(B^0_d\to\pi^-K^+)|^2-|A(\overline{B^0_d}\to\pi^+K^-)|^2=4 r \sin\delta
\sin\gamma 
\end{displaymath}
\begin{equation}\label{RU-rel}
=-\left[|A(B^0_s\to\pi^+K^-)|^2-|A(\overline{B^0_s}\to\pi^-K^+)
|^2\right],
\end{equation}
we have actually only two independent observables, so that the three
parameters $\gamma$, $r$ and $\delta$ cannot be determined. To this
end, the overall normalization $P$ has to be fixed, requiring a further
input. Making the same dynamical assumptions as in \ref{subsec:FM-bound},
i.e.\ assuming that the phase factor $e^{i\gamma}$ plays a negligible r\^ole
in $B^+\to\pi^+K^0$ and that colour-suppressed EW penguins can be
neglected as well, the isospin symmetry implies
\begin{equation}\label{PAR3}
P=A(B^+\to\pi^+K^0).
\end{equation}
In order to extract $\gamma$ and the hadronic parameters, it is useful
to introduce observables $R_s$ and $A_s$ by replacing $B_d\to\pi^\mp K^\pm$
through $B_s\to\pi^\pm K^\mp$ in (\ref{R-param}) and (\ref{A0-def}),
respectively. Using (\ref{PAR1}), (\ref{PAR2}) and (\ref{PAR3}), 
we then obtain 
\begin{equation}
R_s=\epsilon+2r\cos\delta\cos\gamma+\frac{r^2}{\epsilon}
\end{equation}
\begin{equation}
A_s=-2r\sin\delta\sin\gamma=-A_0.
\end{equation}
Together with the parametrization for $R$ given in (\ref{R-param}), 
these observables allow the determination of all relevant parameters. 
The extraction of $\gamma$ and $\delta$ is analogous to the 
$B_d\to\pi^\mp K^\pm$, $B^\pm\to\pi^\pm K$ approach
\cite{PAPIII,defan,BpiK-mixed} discussed in \ref{ref-gamma-strat}.
However, now the advantage is that the $U$-spin counterparts 
$B_s\to\pi^\pm K^\mp$ of $B_d\to\pi^\mp K^\pm$ allow us to determine
also the parameter $r$ without using arguments related to factorization
\cite{BspiK}:
\begin{equation}
r=\sqrt{\epsilon\left[\frac{R+R_s-1-\epsilon}{1+\epsilon}\right]}.
\end{equation}
On the other hand, as emphasized above, we still have to make dynamical
assumptions concerning rescattering and colour-suppressed EW penguin 
effects. A variant of this approach using the CKM angle $\beta$ 
as an additional input was proposed in \cite{CW-BspiK}. 

In addition to the dynamical assumptions, the theoretical accuracy is 
further limited by $SU(3)$-breaking effects. A consistency check is 
provided by the relation $A_s=-A_0$, which is due to (\ref{RU-rel}). 
In the factorization approximation, the relevant 
$SU(3)$-breaking effects are governed by the following ratio of decay 
constants and form factors:
\begin{equation}
\frac{f_\pi}{f_K}\left[\frac{F_{B_sK}(M_\pi^2;0^+)}{F_{B\pi}(M_K^2;0^+)}
\right],
\end{equation}
which is estimated to be a few percent smaller than one \cite{BspiK}.

In comparison with the $B_s\to\pi^+\pi^-$, $B_s\to K^+K^-$ approach 
discussed above, the $B_{(s)}\to \pi K$ strategy has the following 
points in its favour:
\begin{itemize}
\item It does not require time-dependent analyses.
\item It is independent of the $B^0_s$--$\overline{B^0_s}$ mixing phase 
$\phi_s$.
\end{itemize}
However, it should be no problem for future ``hadronic'' $B$ experiments 
to measure time-dependent $B_s$ decay rates, and $\phi_s$ can be taken
into account straightforwardly through $B_s\to J/\psi \phi$. The 
$B_s\to\pi^+\pi^-$, $B_s\to K^+K^-$ strategy \cite{RF-BsKK}
has the following advantages:
\begin{itemize}
\item It is a pure $U$-spin strategy, i.e.\ it does not involve any
dynamical assumptions about rescattering or EW penguin effects. 
\item The relevant $U$-spin relation involves only ratios of certain
strong amplitudes that are not affected by $U$-spin-breaking corrections
within the BSS mechanism.
\item It allows the determination of $\phi_d=2\beta$ and $\gamma$. Using 
$\phi_d$ from $B_d\to J/\psi K_{\rm S}$, the $U$-spin input for the 
extraction of $\gamma$ can be minimized, providing also $\theta$
and $\theta'$.
\end{itemize}

\boldmath
\subsection{Other $U$-Spin Approaches}\label{subsec:other-Uspin}
\unboldmath
As we have already noted several times in this review, the observables
of the angular distributions of $B\to V_1V_2$ modes, where
$V_1$ and $V_2$ continue to decay through strong interactions, offer
interesting insights into CP violation and the hadronization dynamics 
of non-leptonic $B$ decays. 

The general formalism to extract CKM phases and hadronic parameters from 
the time-dependent angular distributions of neutral $B_q\to V_1V_2$ decays, 
taking also into account penguin contributions, was developed 
in \cite{RF-ang}. If we fix the $B^0_q$--$\overline{B^0_q}$ mixing 
phase $\phi_q$ separately, it is possible to determine a CP-violating 
weak phase $\omega$, which is usually given by the angles of the unitarity 
triangle, and interesting hadronic quantities in a {\it theoretically clean} 
way as a function of a {\it single} hadronic parameter. If we determine this 
parameter, for instance, by comparing $B_q\to V_1 V_2$ with an 
$SU(3)$-related mode, all remaining parameters, including $\omega$, 
can be extracted. If we are willing to make more extensive use of 
flavour-symmetry arguments, it is possible to determine $\phi_q$ as 
well. This approach can be applied, for example, to the $U$-spin pairs
$B_d\to\rho^+\rho^-$, $B_s\to K^{\ast+} K^{\ast-}$ or 
$B_d\to K^{\ast0} \overline{K^{\ast0}}$, 
$B_s\to K^{\ast0} \overline{K^{\ast0}}$, providing many more cross-checks
of interesting $U$-spin relations than their counterparts involving only
pseudoscalar mesons in the final state. The formalism presented in 
\cite{RF-ang} is very general and can be applied to many other decays.

In our discussion of $U$-spin strategies, we have always encountered 
certain relations between CP asymmetries and CP-averaged branching 
ratios of $U$-spin related decays (see (\ref{CP-BR-rel}), 
(\ref{Bspsik-CP-BR-rel}) and (\ref{CP-REL-BsKK})). They are equivalent 
to 
\begin{equation}\label{general-U-rel}
|A(B\to f)|^2-|A(\overline{B}\to\overline{f})|^2=
-\left[|A(U B\to U f)|^2-|A(U \overline{B}\to U \overline{f})|^2\right],
\end{equation}
where $U$ denotes a $U$-spin transformation. This relation was derived,
within a general framework, in \cite{Gronau-U}, where also further 
strategies to explore $U$-spin-breaking effects were proposed. 

Applications of (\ref{general-U-rel}) to the $U$-spin-related rare 
decays $B^\pm\to K^{\ast\pm}\gamma$ and $B^\pm\to\rho^\pm \gamma$ were
recently discussed in \cite{hurth-mannel}. Introducing
\begin{eqnarray}
\Delta\mbox{BR}(B\to K^\ast\gamma)&\equiv&\mbox{BR}(B^+\to K^{\ast+}\gamma)-
\mbox{BR}(B^-\to K^{\ast-}\gamma)\\
\Delta\mbox{BR}(B\to \rho\gamma)&\equiv&\mbox{BR}(B^+\to \rho^+\gamma)-
\mbox{BR}(B^-\to \rho^-\gamma),
\end{eqnarray}
this relation implies
\begin{equation}\label{Rare-Decay-Rel}
\Delta\mbox{BR}(B\to K^\ast\gamma)+\Delta\mbox{BR}(B\to \rho\gamma)=0.
\end{equation}
The corresponding $U$-spin-breaking corrections were investigated in 
\cite{BoBu}, using a framework that is similar to the QCD factorization 
approach for two-body non-leptonic $B$ decays. These authors find
\begin{equation}
\Delta\mbox{BR}(B\to K^\ast\gamma)=-7\times 10^{-7},\quad
\Delta\mbox{BR}(B\to \rho\gamma)=+4\times 10^{-7},
\end{equation}
so that the sum of these quantities leaves an $U$-spin-breaking
remainder of $-3\times 10^{-7}$. Actually, this result is not too
surprising, if we remember that we also found -- employing, however,
``na\"\i ve'' factorization -- large $U$-spin-breaking corrections to 
analogous relations between non-leptonic $B$ decays, as can be seen, 
for example, in (\ref{CP-BR-rel}). Let us finally note that the $U$-spin 
symmetry has also interesting applications in the $D$ system 
\cite{gro-ro-Uspin}.

\subsection{Summary}
There are several pairs of $U$-spin-related $B$ decays, which allow 
determinations of weak phases and hadronic parameters. In the case of 
the $B_{s(d)}\to J/\psi K_{\rm S}$, $B_{d(s)}\to D^+_{d(s)} D^-_{d(s)}$ 
and $B_{d(s)}\to K^0\overline{K^0}$ systems, $\gamma$ can be extracted 
in a transparent manner, where $\phi_d=2\beta$ is required as an additional 
input from $B_d\to J/\psi K_{\rm S}$ in the case of the latter two $U$-spin 
pairs. It should be emphasized that no dynamical assumptions about 
rescattering or EW penguin effects are required to this end, which is 
an important conceptual advantage in comparison with the $B\to\pi K$
strategies to determine $\gamma$. Consequently, the theoretical accuracy 
is only limited by $U$-spin-breaking corrections, where those affecting
relations between the overall normalizations of decay amplitudes are the 
most serious ones. 

A particularly interesting strategy is provided by the $B_d\to\pi^+\pi^-$, 
$B_s\to K^+K^-$ system, allowing the determination both of $\beta$ and 
$\gamma$ and of hadronic paramaters as functions of the 
$B^0_s$--$\overline{B^0_s}$ mixing phase $\phi_s$, which is negligibly 
small in the Standard Model, and can be probed straightforwardly through 
$B_s\to J/\psi \phi$. In comparison with the $U$-spin strategies listed
in the previous paragraph, here the advantage is that the $U$-spin symmetry 
has to be applied only to certain ratios of strong amplitudes, where 
$U$-spin-breaking effects due to form factors and decay constants cancel. 
Moreover, there are no $U$-spin-breaking corrections within the BSS 
mechanism. Employing $\phi_d=2\beta$ as an input, the use of the $U$-spin 
symmetry can be minimized. From a theoretical point of view, this is the 
cleanest presently known $U$-spin approach to determine $\gamma$. Moreover, 
it is very promising for run II of the Tevatron and ideally suited for the 
LHC era. Eventually, it may allow us to extract $\gamma$ with an uncertainty 
of only a few degrees. The hadronic parameters, which can be determined as a 
``by-product'', are very interesting for comparisons with theoretical 
calculations. Needless to note, new physics may well lead to discrepancies 
with other strategies.

Approximately, we may replace $B_s\to K^+K^-$, which cannot be studied 
at the $e^+e^-$ $B$-factories operating at $\Upsilon(4S)$, through 
$B_d\to\pi^\mp K^\pm$. The corresponding experimental results imply 
already valuable constraints on hadronic parameters -- in particular 
large penguin effects in $B_d\to\pi^+\pi^-$ -- and upper bounds on direct 
CP asymmetries. In the case of $B_d\to\pi^+\pi^-$, 
a lot of space is left for large direct CP violation. 

Further $U$-spin strategies were proposed. For example, 
$B_s\to \pi^\pm K^\mp$ modes can be combined nicely with the 
$B_d\to\pi^\mp K^\pm$, $B^\pm\to\pi^\pm K$ system to determine $\gamma$. 
Moreover, angular distributions of certain $B\to V_1V_2$ decays offer 
important tools to explore CP violation and $U$-spin breaking, and 
various relations between observables can be derived, including also those
of the rare decays $B^\pm\to K^{\ast\pm}\gamma$ and $B^\pm\to\rho^\pm \gamma$.
It will be exciting to fill these strategies with experimental data!

\boldmath
\section{Models with Minimal Flavour Violation}\label{sec:MFV}
\unboldmath
\setcounter{equation}{0}
\subsection{General Remarks}
So far, we have always discussed new-physics effects in a model-independent 
manner. Let us now focus on the simplest class of extensions of the Standard 
Model, which is given by models with ``minimal flavour violation'' (MFV).
In such scenarios for new physics, which are very predictive, as we will
see below, the contributions of any new operators beyond those present in 
the Standard Model are negligible, and all flavour-changing transitions are 
still governed by the CKM matrix, with no new complex phases beyond the 
CKM phase \cite{MFV,UUT}. If one assumes, in addition, that all new-physics 
contributions that are not proportional to $V_{td(s)}$ play a negligible 
r\^ole \cite{UUT}, all Standard-Model expressions for decay amplitudes, 
as well as for particle--antiparticle mixing, can be generalized to the 
MFV models through a straightforward replacement of the initial Wilson 
coefficients for the renormalization-group evolution from 
$\mu={\cal O}(M_W)$ down to appropriate ``low-energy'' scales $\mu$. 
Within the Standard Model, these coefficients are governed by $m_t$-dependent 
Inami--Lim functions \cite{IL}, describing penguin and box diagrams with 
full $W$, $Z$ and top-quark exchanges \cite{PB-expansion} (see 
Subsection~\ref{subsec:ham}). In the MFV models, the 
Inami--Lim functions are replaced by new functions $F_i$, which 
acquire now additional dependences on new-physics parameters. If we 
consider, for instance, expressions (\ref{100a}) and (\ref{M12-expr}),
which are related to $\varepsilon$ and the off-diagonal element of the 
$B^0_{q}$--$\overline{B^0_{q}}$ mass matrix, respectively, we have 
just to replace the Inami--Lim function $S_0(x_t)$ resulting from box 
diagrams with $(t,W^\pm)$ exchanges through an appropriate new function, 
which we denote by $F_{tt}$ 
\cite{UUT,AL}:
\begin{equation}\label{Ftt}
S_0(x_t)\to F_{tt}.
\end{equation}
In the case of the rare kaon decays $K\to\pi\nu\overline{\nu}$, we 
have to deal with a generalized Inami--Lim function $X$, replacing
the Standard-Model function $X_0(x_t)$.

Examples for MFV models of the kind specified above are the Two-Higgs-Doublet 
Model II (THDM) \cite{THDM}, and the constrained MSSM, if 
$\tan\overline{\beta}=v_2/v_1$ is not too large. 
In the analyses of MFV models performed in \cite{Brev01,UUT,BB-Bound,BePe},
it was assumed implicitly that the new functions $F_i$, summarizing 
the Standard-Model and new-physics contributions to $\varepsilon$, 
$\Delta M_{d,s}$ and $K\to\pi\nu\overline{\nu}$ decays, have the 
{\it same} sign as the standard Inami--Lim functions. This assumption is 
certainly correct in the THDM and the MSSM. On the other hand, as
pointed out in \cite{BF-MFV}, it cannot be excluded that there exist 
MFV models in which the relevant functions $F_i$ have a sign 
{\it opposite} to the corresponding Standard-Model Inami--Lim functions. 
In fact, in the case of the decay $B\to X_s\gamma$, such 
a situation is even possible in the MSSM, if particular values 
of the supersymmetric parameters are chosen. Beyond MFV, scenarios 
in which the new-physics contributions to neutral meson mixing and 
rare kaon decays are larger than the Standard-Model contributions 
and have opposite signs were considered in \cite{BITALY}. Due to the 
presence of new complex phases and new sources for flavour violation
in these general scenarios, their predictive power is, however, much 
smaller than that of the MFV models considered here. In SUSY scenarios,
a vast range of possibilities opens up once the restrictive requirement
of MFV is abandonned. Since a detailed discussion of this active research 
field is beyond the scope of this review, we refer the interested reader 
to the papers given in \cite{SUSY} and 
references therein. 

The following interesting features of MFV models were recently pointed out:
\begin{itemize}
\item There exists a universal unitarity triangle (UUT) \cite{UUT}, which
is common to all these models and the Standard Model, and can be constructed 
by using measurable quantities that depend on the CKM parameters, but are 
not affected by the new-physics parameters. Simply speaking, these quantities 
do not depend on the $F_i$.
\item The interplay between $\Delta M_d$ and $\varepsilon$ implies
bounds on $\sin 2\beta$ \cite{BB-Bound,BF-MFV}, depending only on 
$|V_{cb}|$ and $|V_{ub}/V_{cb}|$, as well as on the non-perturbative 
parameters $\hat B_K$, $\sqrt{\hat B_{B_d}}f_{B_d}$ and $\xi$, which 
enter the standard analysis of the unitarity triangle.
\item For given BR$(K^+\to\pi^+\nu\overline{\nu})$ and $a_{\psi K_{\rm S}}$,
only two values for BR$(K_{\rm L}\to\pi^0\nu\overline{\nu})$, corresponding 
to the two signs of $X$, are possible for the full class of MFV models 
\cite{BF-MFV}. Moreover, there are absolute lower and upper bounds on 
BR$(K_{\rm L}\to\pi^0\nu\overline{\nu})$ as functions of 
BR$(K^+\to\pi^+\nu\overline{\nu})$.
\end{itemize}
The last point, where $a_{\psi K_{\rm S}}$ denotes, as usual, the 
mixing-induced
CP asymmetry of the ``gold-plated'' decay $B_d\to J/\psi K_{\rm S}$ (see 
(\ref{e12})), provides a remarkable connection between the $B$- and 
$K$-meson systems. It should be noted that direct CP violation in 
$B_d\to J/\psi K_{\rm S}$ is negligibly small in MFV models, as in the 
Standard Model. Let us now discuss the three points listed above in more 
detail.

\subsection{The Universal Unitarity Triangle}
Since $\lambda$ and $|V_{cb}|=A\lambda^2$ are determined from tree-level
$K$- and $B$-meson decays, these parameters are not affected by 
new-physics effects in MFV models. A similar comment applies to 
the determination of $R_b\propto|V_{ub}/V_{cb}|$. On the other hand, at 
first sight, it appears as if the extraction of $\overline{\rho}$ and 
$\overline{\eta}$  was more complicated in MFV models because of additional 
new-physics parameters. However, as was pointed out in \cite{UUT}, the 
``true'' values of these parameters can still be determined in a transparent 
way with the help of quantities, which are not affected by the new-physics 
contributions within MFV models. In particular, a ``universal unitarity 
triangle'' (UUT) can be constructed for the Standard Model and the whole 
class of MFV models. Other ``reference'' unitarity triangles to search for
new physics were proposed in \cite{reference-UT}. 

If we take into account the substitution given in (\ref{Ftt}) and note 
that the same function $F_{tt}$ enters in (\ref{M12-expr}) for the $B_d$- 
and $B_s$-meson systems, we observe that (\ref{Rt-Ms}) allows us to fix 
$R_t=|V_{td}/(\lambda V_{cb})|$ also in the case of MFV models. On the 
other hand, the determination of $R_t$ through (\ref{RT}) depends on 
$|F_{tt}|$, and hence cannot be used for the construction of the UUT. 

Whereas $\Delta M_d/\Delta M_s$ should be available first, there are 
also other possibilities to determine $R_t$ in the case of MFV models 
through $|V_{td}/V_{ts}|$, which is -- up to corrections of 
${\cal O}(\lambda^2)$ -- equal to $|V_{td}/V_{cb}|$ \cite{UUT}:  
\begin{equation}\label{BXS}
\frac{\mbox{BR}(B\to X_d\nu\overline{\nu})}{\mbox{BR}(B\to
X_s\nu\overline{\nu})}=\left|\frac{V_{td}}{V_{ts}}\right|^2
\end{equation}
\begin{equation}\label{BMUMU}
\frac{\mbox{BR}(B_d\to\mu^+\mu^-)}{\mbox{BR}(B_s\to\mu^+\mu^-)}=
\frac{\tau_{B_d}}{\tau_{B_s}}\frac{M_{B_d}}{M_{B_s}}\left(
\frac{f_{B_d}}{f_{B_s}}\right)^2\left|\frac{V_{td}}{V_{ts}}\right|^2.
\end{equation}
The cleanest relation is (\ref{BXS}), which is essentially free of 
hadronic uncertainties (for detailed discussions, see \cite{BF-rev,Brev01}).  
In (\ref{BMUMU}), the theoretical accuracy is limited by $SU(3)$-breaking 
effects in the $B_d$- and $B_s$-meson decay constants, whereas (\ref{Rt-Ms})
involves, in addition, the ratio $\sqrt{\hat B_{B_s}}/\sqrt{\hat B_{B_d}}$. 
These $SU(3)$-breaking parameters can be determined eventually from lattice 
calculations with high precision. 

As can be seen in Fig.\ \ref{fig:UT} (a), if we measure -- in addition 
to $R_t$ -- the angle $\beta$, we may fix the apex of the UUT, yielding
$\overline{\rho}$ and $\overline{\eta}$. Since no new phases appear in
MFV models, one may think that the determination of $\beta$ from 
$a_{\psi K_{\rm S}}$ is also not affected by new physics in such scenarios. 
However, due to a subtlety, this is actually not the case \cite{BF-MFV}. 
The point is that we have assumed in the formalism developed in 
Section~\ref{sec:neutral} that $S_0(x_t)>0$, as emphasized 
in the paragraph after (\ref{theta-def}). However, since $S_0(x_t)$ is now
replaced by the new parameter $F_{tt}$, which needs no longer be positive, 
the following phase $\phi_d$ is actually probed through mixing-induced
CP violation in $B_d\to J/\psi K_{\rm S}$:
\begin{equation}\label{phid-def}
\phi_d=2\beta+\mbox{arg}(F_{tt}).
\end{equation}
Consequently, expression (\ref{e12}) is generalized as follows:
\begin{equation}\label{CP-mod}
a_{\psi K_{\rm S}}={\rm sgn}(F_{tt})\sin2\beta.
\end{equation}

On the other hand, if we make the replacement (\ref{Ftt}) in (\ref{100a}) 
and (\ref{RT}), and do not assume that $F_{tt}>0$, as in the Standard Model,
we obtain 
\begin{equation}\label{main}
\sin 2\beta={\rm sgn}(F_{tt})\frac{1.65}{ R^2_0\eta_2}
\left[\frac{0.204}{A^2 B_K}
-\overline{\eta} P_c(\varepsilon)\right],
\end{equation}
where the first term in the parenthesis is typically by a factor 2--3 
larger than the second term. Consequently, the sign of $F_{tt}$ 
determines the sign of $\sin 2\beta$. Moreover, as (\ref{100a}) 
implies $\overline{\eta}<0$ for $F_{tt}<0$, also the sign of the second 
term in the parenthesis is changed. This means that, for a given set of 
input parameters, not only the sign of $\sin 2\beta$, but also its 
magnitude is affected by a reversal of the sign of $F_{tt}$. However,
if we use (\ref{main}) to predict $a_{\psi K_{\rm S}}$, we observe that 
the sign of the resulting CP asymmetry is unaffected:
\begin{equation}
a_{\psi K_{\rm S}}=\frac{1.65}{ R^2_0\eta_2}
\left[\frac{0.204}{A^2 B_K}
-\overline{\eta} P_c(\varepsilon)\right],
\end{equation}
i.e.\ it is {\it positive}, which is consistent with the experimental 
results discussed in \ref{ssec:beta-extr}. 

In \cite{UUT,B95}, a construction of the UUT by means of $a_{\psi K_{\rm S}}$ 
and $R_t$ determined through $\Delta M_d/\Delta M_s$ was presented. Generally, 
for given values of $(a_{\psi K_{\rm S}},R_t)$, there are eight solutions for 
$(\overline{\rho},\overline{\eta})$. However, only two solutions -- 
corresponding to the two possible signs of $F_{tt}$ -- are consistent with 
the upper bound on $|\beta|$ related to (\ref{ubound}). For the derivation 
of the explicit expressions for $\overline{\rho}$ and $\overline{\eta}$, it 
is useful to consider the following quantity \cite{BF-MFV}:
\begin{equation}\label{beta-def}
\mbox{sgn}(F_{tt})\,{\rm ctg}\beta=
\frac{1-\overline{\rho}}{|\overline{\eta}|}\equiv f(\beta),
\end{equation}
as (\ref{Rt-def}) implies
\begin{equation}
R_t^2=(1-\overline{\rho})^2+\overline{\eta}^2=\left[f(\beta)^2+1\right]
\overline{\eta}^2.
\end{equation}
Consequently, admitting also negative $F_{tt}$, we obtain
\begin{equation}\label{r-e}
\overline{\eta}=\mbox{sgn}(F_{tt})\left[\frac{R_t}{\sqrt{f(\beta)^2+1}}
\right],\quad \overline{\rho}=1-f(\beta)|\overline{\eta}|.
\end{equation}
If we take into account the constraint $\overline{\rho}\leq R_b<1$, 
we conclude that $f(\beta)$ is always positive. Moreover, because of 
(\ref{CP-mod}), we may write
\begin{equation}\label{rs-beta2}
f(\beta)=\frac{1\pm\sqrt{1-a_{\psi K_{\rm S}}^2}}{a_{\psi K_{\rm S}}}=
{\rm sgn}(F_{tt})\left[\frac{1\pm|\cos2\beta|}{\sin2\beta}\right].
\end{equation}
Since the upper bound $|\beta|\lsim 28^\circ$ corresponding to 
(\ref{ubound}) implies $|{\rm ctg}\beta|=f(\beta)\gsim1.9$, the ``$-$'' 
solution in (\ref{rs-beta2}) is ruled out, so that the measurement of 
$a_{\psi K_{\rm S}}$ determines $f(\beta)$ {\it unambiguously} through
\begin{equation}\label{rs-unambig}
f(\beta)=\frac{1+\sqrt{1-a_{\psi K_{\rm S}}^2}}{a_{\psi K_{\rm S}}}.
\end{equation}
Finally, with the help of (\ref{r-e}), we arrive at
\begin{equation}\label{rho-eta}
\overline{\eta}={\rm sgn}(F_{tt})R_t
\sqrt{\frac{1-\sqrt{1-a_{\psi K_{\rm S}}^2}}{2}},\quad
\overline{\rho}=1-\left[\frac{1+\sqrt{1-
a_{\psi K_{\rm S}}^2}}{a_{\psi K_{\rm S}}}\right]|\overline{\eta}|.
\end{equation}
Neglecting terms of ${\cal O}(a_{\psi K_{\rm S}}^4)$, these expressions
simplify to 
\begin{equation}
\overline{\eta}={\rm sgn}(F_{tt})R_t\left[1+\frac{a_{\psi K_{\rm S}}^2}{8}
\right]\frac{a_{\psi K_{\rm S}}}{2},\quad
\overline{\rho}=1-R_t\left[1-\frac{a_{\psi K_{\rm S}}^2}{8}\right].
\end{equation}
The function $f(\beta)$ plays also a key r\^ole for the analysis of 
the $K\to\pi\nu\overline{\nu}$ system, which will be the topic of 
Subsection~\ref{subsec:Kpinunu-MFV}. 

The expressions given in (\ref{rho-eta}) are the ``master formulae''
for the determination of the apex of the UUT. They also allow us -- 
in combination with $\lambda$ and $A$ -- to calculate $\varepsilon$, 
$\Delta M_d$ and $\Delta M_s$, as well as the branching ratios for the 
rare decays in (\ref{BXS}) and (\ref{BMUMU}). Since these quantities 
depend on the values of the corresponding generalized Inami--Lim functions 
$F_i$, characterizing the various MFV models, a comparison with the data may 
exclude the Standard Model, and may allow us to differentiate between various 
MFV scenarios. In particular, it may well be that the observed pattern 
of observables can only be described by one specific MFV model.

Since $R_t$ cannot yet be determined through (\ref{Rt-Ms}), the UUT
cannot yet be constructed in practice. However, valuable information
can nevertheless be obtained from interesting bounds on $\sin2\beta$
\cite{BB-Bound,BF-MFV}, which hold in the MFV models specified above.

\boldmath
\subsection{Bounds on $\sin2\beta$}
\unboldmath
The starting point for the derivation of the bounds on $\sin2\beta$ 
is (\ref{main}), where the dependence on new-physics enters implicitly 
through $\overline{\eta}$. Let us first assume that ${\rm sgn}(F_{tt})=+1$.
Varying over all positive values of $F_{tt}$ that are consistent
with the experimental values for $\Delta M_{d,s}$, $|V_{ub}/V_{cb}|$ and 
$|V_{cb}|$, and scanning all the relevant input parameters in the ranges 
given in Table~\ref{tab:inputparams} yields the following lower
bound on $\sin2\beta$ \cite{BB-Bound}:
\begin{equation}\label{BB-bound-pos}
\left(\sin2\beta\right)_{\rm min}=0.42,
\end{equation}
which corresponds to $\beta\geq 12^\circ$. This bound could be 
considerably improved when the values of $V_{cb}$, $|V_{ub}/V_{cb}|$, 
$\hat B_K$, $\sqrt{\hat B_{B_d}} f_{B_d}$, $\xi$ and -- in particular of 
$\Delta M_s$ -- will be known better \cite{Brev01,BB-Bound}. A handy 
approximate formula for $\sin 2\beta$ as a function of these parameters 
has recently been given in \cite{BePe}. Using less conservative ranges 
of parameters, these authors find $(\sin 2\beta)_{\rm min}=0.52$.

\begin{table}
\begin{center}
\begin{tabular}{|c|c|c|}
\hline
Quantity & Central & Error  
\\\hline
$\lambda$ & 0.222 &       \\
$|V_{cb}|$ & 0.041 & $\pm 0.002$      \\$|V_{ub}/V_{cb}|$ & $0.085$ & 
$\pm 0.018 $\\
$|V_{ub}|$ & $0.00349$ & $\pm 0.00076$ \\ 
$\hat B_K$ & 0.85 & $\pm 0.15$   \\
$\sqrt{\hat B_{B_d}} f_{B_{d}}$ & $230\,{\rm MeV}$ & $\pm 40\,{\rm MeV}$  \\
$m_t$ & $166\,{\rm GeV}$ & $\pm 5\,{\rm GeV}$   \\
$(\Delta M)_d$ & $0.487\,\mbox{ps}^{-1}$ & $\pm 0.014\,\mbox{ps}^{-1}$ \\
$(\Delta M)_s$  & $>15.0\,\mbox{ps}^{-1}$ & \\$\xi$ & $1.15$ & $\pm 0.06$  
\\
\hline
\end{tabular}
\caption{The ranges of the input parameters for the bounds on
$\sin2\beta$.}\label{tab:inputparams}
\end{center}
\end{table}

Let us now consider the case ${\rm sgn}(F_{tt})=-1$. 
Repeating the analysis that lead to (\ref{BB-bound-pos}) for $F_{tt}<0$ 
yields the following bound \cite{BF-MFV}:
\begin{equation}\label{BB-bound-neg}
\left(-\sin 2\beta\right)_{\rm min}=0.69.
\end{equation}
This result is rather sensitive to the minimal value of 
$\sqrt{\hat B_{B_d}}f_{B_d}$. Taking 
$(\sqrt{\hat B_{B_d}}f_{B_d})_{\rm min}=170\,\mbox{MeV}$ instead 
of 190\,MeV used in (\ref{BB-bound-neg}) gives the bound of 0.51. 
For the same choice, the bound in (\ref{BB-bound-pos}) is decreased to 
0.35. For $(\sqrt{\hat B_{B_d}}f_{B_d})_{\rm min}\ge 195\,\mbox{MeV}$ 
there are no solutions for $\sin 2\beta$ for the ranges of parameters 
given in Table~\ref{tab:inputparams}; only for $\hat B_K\ge 0.96$, 
$|V_{cb}| \ge 0.0414$ and $|V_{ub}/V_{cb}|\ge 0.094$ solutions for 
$\sin 2\beta$ exist. Finally, if we assume a smaller uncertainty for
$\Delta M_d$ of $\pm 0.009\,\mbox{ps}^{-1}$, corresponding to 
(\ref{new-DMd-range}), the numerical value in (\ref{BB-bound-neg}) is
shifted to 0.71, whereas (\ref{BB-bound-pos}) is unaffected.

Since the sign of $a_{\psi K_{\rm S}}$ does -- in contrast to 
$\sin2\beta$ -- not depend on ${\rm sgn}(F_{tt})$, it is actually 
more appropriate to consider bounds on this CP-violating observable 
\cite{BF-MFV}:
\begin{equation}\label{apsiKS-bounds}
\left(a_{\psi K_{\rm S}}\right)_{\rm min}=\left\{\begin{array}{ll}
0.42 & \mbox{($F_{tt}>0$)}\\
0.69 & \mbox{($F_{tt}<0$).}
\end{array}\right.
\end{equation}
The feature that the constraint is substantially stronger for negative 
$F_{tt}$ is not surprising, since in this case the contributions to 
$\varepsilon$ proportional to $V_{ts}^*V_{td}$ interfere destructively 
with the charm contribution. Consequently, $a_{\psi K_{\rm S}}=|\sin2\beta|$ 
has to be larger to fit $\varepsilon$. For $0.42<a_{\psi K_{\rm S}}<0.69$, 
the MFV models with $F_{tt}<0$ would be excluded; for 
$a_{\psi K_{\rm S}}<0.42$, even all MFV models would be ruled out, 
thereby implying that new CP-violating phases and/or new operators are 
required. In anticipation of such a scenario, which was favoured by
the ``old'' $B$-factory data, an extension of the MFV supersymmetric 
models, which could comfortably accommodate lower values of 
$a_{\psi K_{\rm S}}$, was discussed in \cite{Ali-Lunghi}, introducing an 
additional flavour-changing structure beyond the CKM matrix. Generalizations 
of MFV models were also considered in \cite{BCRS}, allowing significant 
contributions of non-standard $\Delta F=2$ operators to the low-energy 
effective Hamiltonian. Their contributions to $\Delta M_d/\Delta M_s$
would in principle allow $\gamma$ to be larger than $90^\circ$ (see
Subsection~\ref{Bs-gen}). For a selection of more general discussions of 
SUSY models, the reader is referred to \cite{SUSY}.

In the spirit of (\ref{apsiKS-bounds}), the two cases $F_{tt}>0$ and
$F_{tt}<0$ can be distinguished through the measurement of 
$a_{\psi K_{\rm S}}$. As pointed out in \cite{BF-MFV}, there are also 
strategies to determine the sign of $F_{tt}$ directly, allowing interesting 
consistency checks. For $a_{\psi K_{\rm S}}=0.79\pm0.10$, corresponding to 
the average given in (\ref{s2b-world}), not even the case of MFV models with 
negative $F_{tt}$ could be excluded. In view of the most recent Belle result 
\cite{belle-CP-obs} (see (\ref{B-factory})), the upper bounds given in 
(\ref{Rb-bounds}) and (\ref{ubound}), which are implied by $R_b$ and hold 
also in MFV models, may play an important r\^ole to probe these 
scenarios for new physics in the future. Let us now turn to more refined
strategies, using in addition rare kaon decays.

\boldmath
\subsection{$K\to\pi\nu\overline{\nu}$ Decays
and Connections with $B$ Physics}\label{subsec:Kpinunu-MFV}
\unboldmath
In MFV models, the short-distance contributions to 
$K^+\to\pi^+\nu\overline{\nu}$ and $K_{\rm L}\to\pi^0\nu\overline{\nu}$
proportional to $V^*_{ts}V_{td}$ are described by a function $X$, 
resulting from $Z^0$-penguin and box diagrams. As pointed out in
\cite{BBSIN}, if $\sin 2\beta$ is expressed in terms of the branching
ratios for these rare kaon decays, the function $X$ drops out. Being 
determined from two branching ratios, there is a four-fold ambiguity 
in the determination of $\sin 2\beta$ that is reduced to a two-fold 
ambiguity for $\overline{\rho}<1$, as required by the size of $R_b$. 
The left over solutions correspond to two signs of $\sin 2\beta$ that 
can be adjusted to agree with the analysis of $\varepsilon$. In 
the Standard Model, the THDM and the MSSM, the functions $F_{tt}$ and 
$X$ are both positive, resulting in $\sin 2\beta$ given by 
(\ref{sin}). However, in general, this needs not be the case, thereby
complicating the $K\to\pi\nu\overline{\nu}$ analysis; it was recently
extended to MFV models with arbitrary signs of $F_{tt}$ and $X$
in \cite{BF-MFV}.

\boldmath
\subsubsection{The Decay $K^+\to\pi^+\nu\overline{\nu}$}
\unboldmath
Within MFV models, the reduced $K^+\to\pi^+\nu\overline{\nu}$ branching 
ratio $B_1$ introduced in (\ref{b1b2}) can be expressed as follows:
\begin{equation}\label{bkpn}
B_1=\left[{{\rm Im}\lambda_t\over\lambda^5}|X|\right]^2+
\left[{{\rm Re}\lambda_c\over\lambda}{\rm sgn}(X) P_c(\nu\overline{\nu}) +
{{\rm Re}\lambda_t\over\lambda^5}|X|\right]^2,
\end{equation}
where $\lambda_c\equiv V^\ast_{cs}V_{cd}=-\lambda (1-\lambda^2/2)$,
and $\lambda_t\equiv V^\ast_{ts}V_{td}$ with
\begin{equation}\label{RE-IM}
{\rm Im}\lambda_t=\eta A^2 \lambda^5,\quad 
{\rm Re}\lambda_t=-\left(1-\frac{\lambda^2}{2}\right)A^2\lambda^5
(1-\overline{\rho}).
\end{equation}
It is now an easy exercise to show that the measured 
$K^+ \to \pi^+ \nu \overline{\nu}$ branching ratio determines 
an ellipse in the $\overline{\rho}$--$\overline{\eta}$ plane,
\begin{equation}\label{ellipse}
\left(\frac{\overline{\rho}-\rho_0}{\overline{\rho}_1}\right)^2+
\left(\frac{\overline{\eta}}{\overline{\eta}_1}\right)^2=1,
\end{equation}
centered at $(\rho_0,0)$ with 
\begin{equation}\label{110}
\rho_0 = 1 + {\rm sgn}(X) \frac{P_c(\nu\overline{\nu})}{A^2 |X|},
\end{equation}
and having the squared axes
\begin{equation}\label{110a}
\overline{\rho}_1^2 = r^2_0, \quad \overline{\eta}_1^2 = 
\left( \frac{r_0}{\sigma}\right)^2 \quad\mbox{with}\quad\,\,
r^2_0 = \frac{\sigma B_1}{A^4 |X|^2}.
\end{equation}
Note that $\sigma$ was already defined in (\ref{sigma-def}). The ellipse 
(\ref{ellipse}) intersects with the circle described by (\ref{Rb-def}), 
thereby allowing us to extract $\overline{\rho}$ and $\overline{\eta}$:
\begin{equation}\label{113}
\overline{\rho} = \frac{1}{1-\sigma^2} \left[\rho_0 \mp \sqrt{\sigma^2
\rho_0^2 +(1-\sigma^2)(r_0^2-\sigma^2 R_b^2)} \right], \quad
\overline{\eta} = {\rm sgn}(F_{tt})\sqrt{R_b^2-\overline{\rho}^2}.
\end{equation}
The deviation of $\rho_0$ from unity measures the relative importance
of the internal charm contribution. For $X>0$, we have, as usual, 
$\rho_0>1$ so that the ``$+$'' solution in (\ref{113}) is excluded 
because of $\overline{\rho}<1$. On the other hand, for $X<0$, the center 
of the ellipse is shifted to $\rho_0<1$, and for 
$|X|\leq P_c(\nu\overline{\nu})/A^2$ can even be at $\rho_0\leq0$.

\boldmath
\subsubsection{Unitarity Triangle from
$K_{\rm L}\to\pi^0\nu\overline{\nu}$ and $K^+\to\pi^+\nu\overline{\nu}$}
\unboldmath
The reduced $K_{\rm L}\to\pi^0\nu\overline{\nu}$ branching ratio $B_2$ 
specified in (\ref{b1b2}) is given by
\begin{equation}\label{bklpn}
B_2=\left[{{\rm Im}\lambda_t\over\lambda^5}|X|\right]^2.
\end{equation}
Following \cite{BBSIN}, but admitting both signs of $X$ and $F_{tt}$, we 
obtain
\begin{equation}\label{rhetb}
\overline{\rho}=1+\left[{\pm\sqrt{\sigma(B_1-B_2)}+
{\rm sgn}(X) P_c(\nu\overline{\nu})\over A^2 |X|}\right],\quad
\overline{\eta}= {\rm sgn}(F_{tt}){\sqrt{B_2}\over\sqrt{\sigma} A^2 |X|}.
\end{equation}
The dependence on $|X|$ cancels in the following quantity \cite{BF-MFV}:
\begin{equation}\label{cbbnew}
r_s\equiv{1-\overline{\rho}\over\overline{\eta}}={\rm ctg}\beta=
{\rm sgn}(F_{tt})\,\sqrt{\sigma}\left[{\mp\sqrt{\sigma(B_1-B_2)}-
{\rm sgn}(X)P_c(\nu\overline{\nu})\over\sqrt{B_2}}\right],
\end{equation}
allowing the determination of $\sin2\beta$ through (\ref{sin}). Note
that (\ref{cbbnew}) reduces to (\ref{cbb}) in the case of positive
values of $F_{tt}$ and $X$. Because of 
$a_{\psi K_{\rm S}}={\rm sgn}(F_{tt})\sin2\beta$, it is actually more 
appropriate to consider this CP-violating observable, where the 
${\rm sgn}(F_{tt})$ factor is cancelled through the one of $\sin2\beta$.

\begin{table}[t]
\begin{center}
\begin{tabular}{|c|c|c|c|}\hline
BR$(K^+\to\pi^+\nu\overline{\nu})$ $[10^{-11}]$ & 
$ a_{\psi K_{\rm S}}=0.42 $ &  $ a_{\psi K_{\rm S}}=0.69 $ &
$ a_{\psi K_{\rm S}}=0.82 $  \\ \hline
5.0 & 0.45~(2.0) & 1.4~(5.8) & 2.2~(8.6) \\
10.0 & 1.2~(3.5) & 3.8~(10.0) & 5.9~(15.0)\\
15.0 & 2.1~(4.8) & 6.3~(14.0) & 9.9~(21.1)\\
20.0 & 3.0~(6.2) & 9.0~(17.9) & 14.1~(27.0)\\
25.0 & 3.9~(7.5) & 11.8~(21.7) & 18.4~(32.8)\\
30.0 & 4.9~(8.7) & 14.6~(25.4) & 22.7~(38.6)\\
\hline
\end{tabular}
\end{center}
\caption{BR$(K_{\rm L}\to\pi^0\nu\overline{\nu})$ in units of $10^{-11}$ 
in MFV models with ${\rm sgn}(X)=+1~(-1)$ for specific values of 
$a_{\psi K_{\rm S}}$ and BR$(K^+\to\pi^+\nu\overline{\nu})$. We set 
$P_c(\nu\overline{\nu})=0.40$.}\label{ANA}
\end{table}

\boldmath
\subsubsection{BR$(\klnn)$ from $a_{\psi K_{\rm S}}$
and BR$(\kpnn)$}\label{sssec:GOLD}
\unboldmath
Since $a_{\psi K_{\rm S}}$ and BR$(K^+\to\pi^+\nu\overline{\nu})$ will 
be known rather accurately prior to the measurement of 
BR$(K_{\rm L}\to\pi^0\nu\overline{\nu})$, it is of particular interest 
to calculate BR$(K_{\rm L}\to\pi^0\nu\overline{\nu})$ as a function of 
$a_{\psi K_{\rm S}}$ and BR$(K^+\to\pi^+\nu\overline{\nu})$. Employing
the quantity $f(\beta)$ introduced in (\ref{beta-def}), we obtain
\cite{BF-MFV}
\begin{equation}\label{B1B2}
B_1=B_2+\left[\frac{f(\beta)\sqrt{B_2}+
{\rm sgn}(X)\sqrt{\sigma}P_c(\nu\overline{\nu})}
{\sigma}\right]^2.
\end{equation}
In comparison with (\ref{cbbnew}), the advantage of (\ref{B1B2}) 
is the absence of the sign ambiguities due to ${\rm sgn}(F_{tt})$ and
the $\mp$ in front of $\sqrt{\sigma(B_1-B_2)}$. As we have seen in 
(\ref{rs-unambig}), the measurement of $a_{\psi K_{\rm S}}$ determines 
$f(\beta)$ unambiguously. This finding, in combination with (\ref{B1B2}), 
implies the following interesting feature of the MFV models \cite{BF-MFV}:
\begin{itemize}
\item For given $a_{\psi K_{\rm S}}$ and BR$(K^+\to\pi^+\nu\overline{\nu})$, 
only two values of BR$(K_{\rm L}\to\pi^0\nu\overline{\nu})$, which
correspond to the two possible signs of $X$, are allowed for the full class 
of MFV models, independently of any new parameters present in these models. 
\end{itemize}
Consequently, measuring BR$(K_{\rm L}\to\pi^0\nu\overline{\nu})$ will 
either select one of these two possible values, or will rule out all MFV 
models. 

In Table~\ref{ANA}, we give BR$(K_{\rm L}\to\pi^0\nu\overline{\nu})$
in MFV models with $\mbox{sgn}(X)=+1$ $(-1)$ for specific values of 
$a_{\psi K_{\rm S}}$ and BR$(K^+\to\pi^+\nu\overline{\nu})$.
Note that the second column gives the absolute lower bound on 
BR$(K_{\rm L}\to\pi^0\nu\overline{\nu})$ in the MFV models as a function
of BR$(K^+\to\pi^+\nu\overline{\nu})$. This bound follows simply from the 
lower bound in (\ref{BB-bound-pos}). On the other hand, the last column 
gives the corresponding absolute upper bound. This bound is the 
consequence of (\ref{ubound}). The third column gives the lower bound on 
BR$(K_{\rm L}\to\pi^0\nu\overline{\nu})$ corresponding to the bound in 
(\ref{BB-bound-neg}) that applies for negative $F_{tt}$. 

\begin{figure}
\centerline{\rotate[r]{
\epsfysize=10.0truecm
{\epsffile{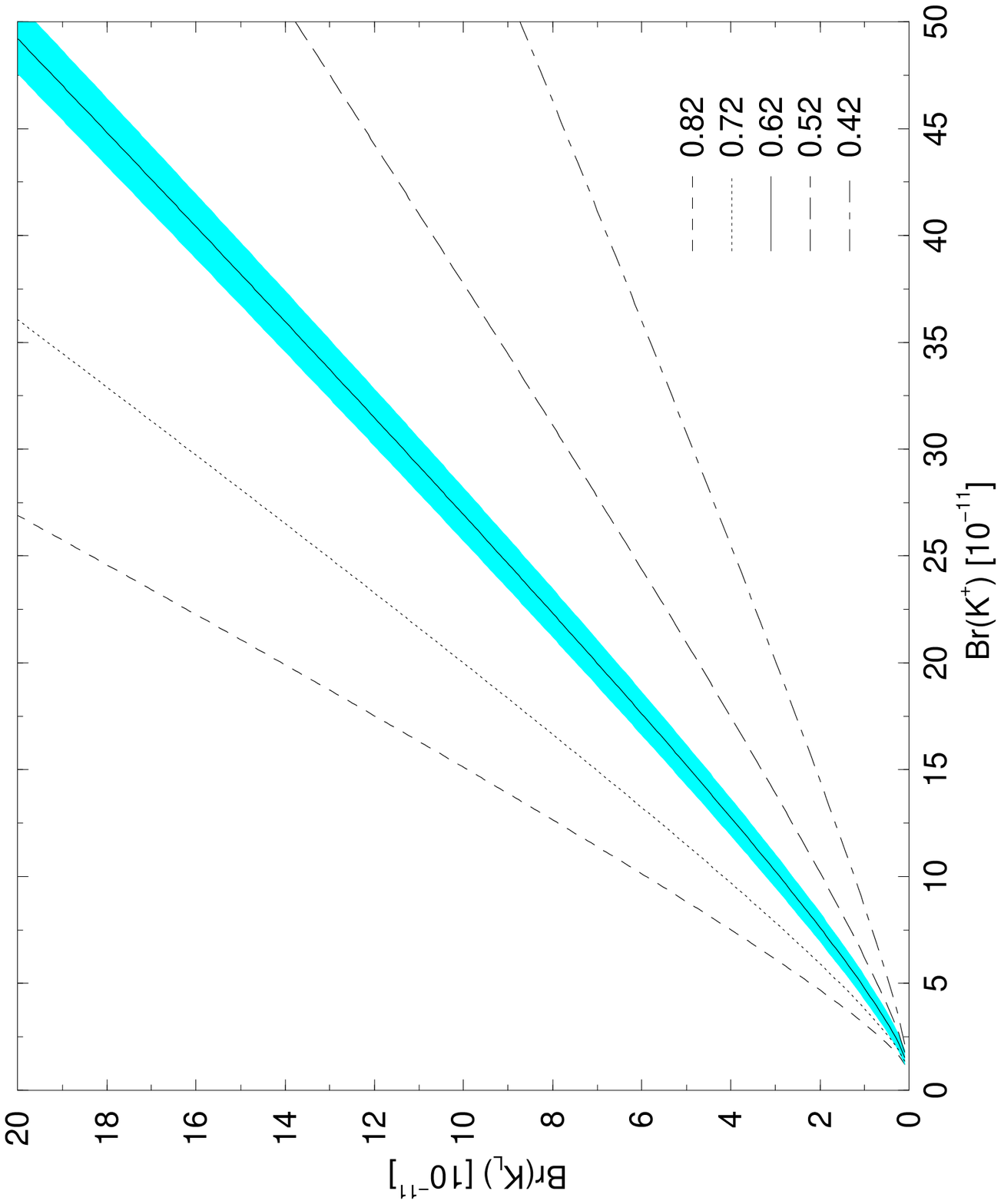}}}}
\caption{BR$(\klnn)$ as a function of BR$(\kpnn)$ for several 
values of the mixing-induced CP asymmetry $a_{\psi K_{\rm S}}$ of
$B_d\to J/\psi K_{\rm S}$ in the case of ${\rm sgn}(X)=+1$. The band 
illustrates the uncertainty due to $P_c(\nu\overline{\nu})=0.40\pm0.06$ for 
$a_{\psi K_{\rm S}}=0.62$.}\label{fig:Xpos}
\end{figure}

\begin{figure}
\centerline{\rotate[r]{
\epsfysize=10.0truecm
{\epsffile{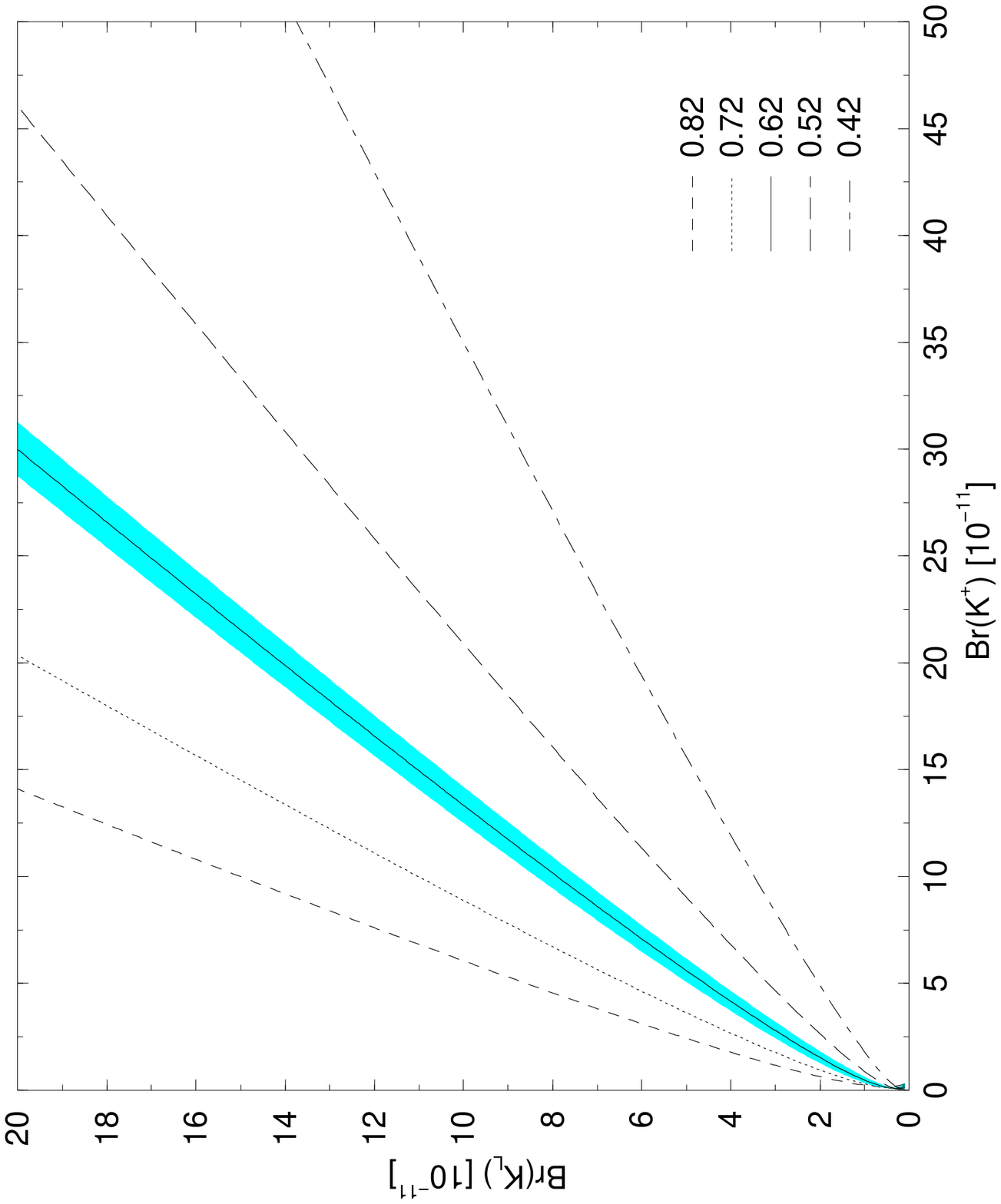}}}}
\caption{BR$(\klnn)$ as a function of BR$(\kpnn)$ for several values 
of the mixing-induced CP asymmetry $a_{\psi K_{\rm S}}$ of
$B_d\to J/\psi K_{\rm S}$ in the case of ${\rm sgn}(X)=-1$. The band 
illustrates the uncertainty due to $P_c(\nu\overline{\nu})=0.40\pm0.06$ 
for $a_{\psi K_{\rm S}}=0.62$.}\label{fig:Xneg}
\end{figure}

The remarkable correlation between the branching ratios of the rare 
$K\to\pi\nu\overline{\nu}$ decays and mixing-induced CP violation in 
$B_d\to J/\psi K_{\rm S}$ within MFV models becomes more transparent in 
Figs.~\ref{fig:Xpos} and \ref{fig:Xneg}. In Fig.~\ref{fig:Xpos}, we show 
BR$(\klnn)$ as a function of BR$(\kpnn)$ for various values of 
$a_{\psi K_{\rm S}}$ in the case of ${\rm sgn}(X)=+1$. The corresponding 
plot for ${\rm sgn}(X)=-1$ is given in Fig.~\ref{fig:Xneg}. It should be 
emphasized that these curves are universal for all MFV models. Looking at 
Table~\ref{ANA} and Figs.~\ref{fig:Xpos} and \ref{fig:Xneg}, we observe
that the measurement of BR$(\klnn)$, BR$(\kpnn)$ and $a_{\psi K_{\rm S}}$ 
will easily allow us to check whether a MFV model is actually realized 
and -- if so -- to distinguish between the two signs of $X$. 
The uncertainty due to $P_c(\nu\overline{\nu})$ is non-negligible, but 
should be decreased through improved knowledge of the charm-quark mass.

It is interesting to note that the upper bound on BR$(\klnn)$ in the 
last column of Table~\ref{ANA} is substantially stronger than the 
model-independent bound following from isospin symmetry \cite{NIR96}:
\begin{equation}
\label{iso}
\mbox{BR}(\klnn) < 4.4 \times \mbox{BR}(\kpnn).
\end{equation}
Indeed, taking the experimental bound 
BR$(\kpnn)\le 5.9\times 10^{-10}~(90\%~\mbox{C.L.})$ from AGS E787 
\cite{Adler00} yields \cite{BF-MFV}
\begin{equation}\label{KL-bound}
\mbox{BR}(\klnn)_{\rm MFV}\le\left\{\begin{array}{ll}
4.9 \times 10^{-10} &  {\rm sgn}(X)=+1\\
7.1 \times 10^{-10}  &  {\rm sgn}(X)=-1.
\end{array}\right.
\end{equation}
This should be compared with BR$(\klnn) < 26 \times 
10^{-10}~(90\%~\mbox{C.L.})$ following from (\ref{iso}), and with the 
present upper bound from the KTeV experiment \cite{KTeV00X}, which is
given by BR$(\klnn)<5.9 \times 10^{-7}$.

The decay $\kpnn$ has already been observed at Brookhaven, with the 
branching ratio given in (\ref{Brook}). In the Standard Model, one expects 
the $K\to\pi\nu\overline{\nu}$ branching ratios listed in (\ref{Kpinunu-SM}). 
As can be seen in Table~\ref{ANA} and Figs.~\ref{fig:Xpos} and 
\ref{fig:Xneg}, the bounds in (\ref{KL-bound}) can be improved considerably 
through better measurements of BR$(\kpnn)$ and $a_{\rm \psi K_{\rm S}}$. 
A new experiment at Brookhaven, AGS E949, is expected to reach a sensitivity 
for $\kpnn$ of $10^{-11}$/event. In the more distant future, the CKM
experiment at Fermilab aims at a sensitivity of $10^{-12}$/event, which
would correspond to a measurement of BR$(K^+\to\pi^+\nu\overline{\nu})$
at the level of $10\%$. The exploration of $\klnn$ is even more
challenging. However, dedicated experiments at Brookhaven, Fermilab and 
KEK aim nevertheless at measurements of BR$(\klnn)$ at the level of
$10\%$, which may be available around 2005. For a much more detailed 
discussion of the experimental prospects for $K\to\pi\nu\overline{\nu}$
analyses, we refer the reader to \cite{littenberg}.

\boldmath
\subsubsection{Upper Bound on BR$(\klnn)$ from
BR$(B\to X_s\nu\overline{\nu})$}
\unboldmath
Within MFV models, the branching ratio for the inclusive rare decay 
$B\to X_s\nu\overline{\nu}$ can be written as follows \cite{Brev01}:
\begin{equation}\label{BXS-expr}
\mbox{BR}(B\to X_s\nu\overline{\nu})=1.57\times 10^{-5}
\left[\frac{\mbox{BR}(B\to X_c e\overline{\nu}_e)}{0.104}\right]
\left|\frac{V_{ts}}{V_{cb}}\right|^2
\left[\frac{0.54}{f(z)}\right]~X^2,
\end{equation}
where $f(z)=0.54\pm0.04$ is the phase-space factor for 
$B\to X_c e\overline{\nu}_e$ with $z=m_c^2/m_b^2$, and 
BR$(B\to X_c e\overline{\nu}_e)=0.104\pm 0.004$. Formulae (\ref{bklpn}) 
and (\ref{BXS-expr}) imply an interesting relation between the decays
$\klnn$ and $B\to X_s\nu\overline{\nu}$, which is given as follows:
\begin{equation}\label{KLBXS}
\mbox{BR}(\klnn)=42.3\times ({\rm Im}\lambda_t)^2
\left[\frac{0.104}{\mbox{BR}(B\to X_c e\overline{\nu}_e)}\right]
\left|\frac{V_{cb}}{V_{ts}}\right|^2
\left[\frac{f(z)}{0.54}\right]\mbox{BR}(B\to X_s\nu\overline{\nu}).
\end{equation}
This expression holds for all MFV models and represents another connection 
between $K$- and $B$-meson decays, in addition to those discussed above 
and in \cite{BB98,BBSIN,NIR,BePe,BePe0}.

In order to obtain another upper bound on BR$(\klnn)$, we 
take into account the following experimental constraint \cite{ALPH}:
\begin{equation}\label{BBOUND}
\mbox{BR}(B\to X_s\nu\overline{\nu})<6.4 \times 10^{-4}\quad 
(90\%~\mbox{C.L.}),
\end{equation}
as well as the bounds on $|{\rm Im}\lambda_t|$, which were 
derived in \cite{BF-MFV} for MFV models:\footnote{Note the relation
$J_{\rm CP}=\lambda\left(1-\lambda^2/2\right)|\mbox{Im}\lambda_t|$ between
the Jarlskog Parameter $J_{\rm CP}$ and $\mbox{Im}\lambda_t$.}
\begin{equation}
|{\rm Im}\lambda_t|_{\rm max}=\left\{\begin{array}{ll}
1.74\times 10^{-4}& \mbox{($F_{tt}>0$)}  \\
 1.70\times 10^{-4}& \mbox{($F_{tt}<0$),} \\
\end{array}\right.
\quad
|{\rm Im}\lambda_t|_{\rm min}=\left\{\begin{array}{ll}
0.55\times 10^{-4}& \mbox{($F_{tt}>0$)}  \\
1.13\times 10^{-4}& \mbox{($F_{tt}<0$).} \\
\end{array}\right.
\end{equation}
Setting, in addition to (\ref{BBOUND}), 
${\rm Im}\lambda_t=1.74\times 10^{-4}$, $|V_{ts}|=|V_{cb}|$, 
$f(z)=0.58$ and BR$(B\to X_c e\overline{\nu}_e)=0.10$, (\ref{KLBXS}) 
implies the upper bound \cite{BF-MFV}
\begin{equation}\label{L-Bound}
\mbox{BR}(\klnn)<9.2\times 10^{-10}\quad (90\%~\mbox{C.L.}),
\end{equation}
which is not much weaker than the bound in (\ref{KL-bound}). 
Since the experimental bound in (\ref{BBOUND}) should be improved 
through the $B$-factories, also (\ref{L-Bound}) should be improved 
in the next couple of years.

\boldmath
\subsubsection{Towards a Determination of $X$}
\unboldmath
The strategy discussed in \ref{sssec:GOLD}, which is based 
on Figs.~\ref{fig:Xpos} and \ref{fig:Xneg} and involves only
$a_{\psi K_{\rm S}}$, BR$(\kpnn)$ and BR$(\klnn)$, allows an
elegant check whether a MFV model is actually realized in nature 
and -- if so -- to determine the sign of $X$. In order to determine 
also $|X|$, which would be a very important information, providing 
valuable constraints on the MFV models, $\Delta M_d/\Delta M_s$ is needed 
as an additional input, as illustrated in more detail in \cite{BF-MFV}. 
Within the Standard Model, we have $X\approx 1.5$. Constraints on this
parameter were recently derived in \cite{BF-MFV,BePe}, yielding 
$|X|< 6.8$.

\subsection{Summary}
The simplest class of extensions of the Standard Model is given by 
new-physics scenarios with minimal flavour violation. For such 
models, a universal unitarity triangle can be constructed
through quantities, which are not affected by the new-physics
parameters, for instance through the ratio $\Delta M_d/\Delta M_s$
and the mixing-induced CP asymmetry $a_{\psi K_{\rm S}}$ of
$B_d\to J/\psi K_{\rm S}$. Although no new complex phases
are present in such MFV models, we may encounter discrepancies
between the UUT and the contours in the $\overline{\rho}$--$\overline{\eta}$
plane implied by $\Delta M_d$ and $\varepsilon$, thereby indicating 
physics beyond the Standard Model. 

At present, the UUT cannot yet be constructed in practice. However, 
interesting bounds on $a_{\psi K_{\rm S}}$ can be derived, allowing 
already a comparison with the $B$-factory results. Within MFV models, 
both $\varepsilon$ and $B^0_{d,s}$--$\overline{B^0_{d,s}}$ mixing are 
governed by a single generalized Inami--Lim function $F_{tt}$. 
If $a_{\psi K_{\rm S}}$ is found to be smaller than 0.69, all models with 
$F_{tt}<0$ would be excluded. In the case of $a_{\psi K_{\rm S}}<0.42$, 
also the MFV models with $F_{tt}>0$ would be ruled out, which would 
imply new CP-violating weak phases and/or new operators. The most recent
$B$-factory data are no longer in favour of small values of 
$a_{\psi K_{\rm S}}$, and the present world average does not even allow
us to exclude MFV models with negative $F_{tt}$. In fact, in view of the 
most recent Belle result, the upper bound 
$\left(a_{\psi K_{\rm S}}\right)_{\rm max}=0.82$ that is due to 
$R_b^{\rm max}=0.46$ may play an important r\^ole in the future. 

Further insights are provided by remarkable connections between $B$ physics 
and the rare kaon decays $\kpnn$ and $\klnn$, which are characterized in 
MFV models through a generalized Inami--Lim function $X$. In particular, 
BR$(\klnn)$ can be predicted, for given $a_{\psi K_{\rm S}}$, as a function 
of BR$(\kpnn)$. These correlations depend only on the sign of $X$, i.e.\
for given BR$(\kpnn)$ and $a_{\psi K_{\rm S}}$, only two values for
BR$(\klnn)$, corresponding to the two signs of $X$, are possible in 
the full class of MFV models, thereby offering a simple check whether 
such a model is actually realized in nature. In this context, it should 
be noted that BR$(\kpnn)$ and $a_{\psi K_{\rm S}}$ will be known rather 
accurately prior to the measurement of BR$(\klnn)$. In addition, there 
are various bounds on BR$(\klnn)$, where also the inclusive rare decay 
$B\to X_s\nu\overline{\nu}$ would be very useful. 


%
%
%
\boldmath
\section{Further Interesting Aspects of $B$ Physics}\label{sec:Other}
\unboldmath
\setcounter{equation}{0}
In addition to the decays to explore CP violation considered in this 
review, there are also other interesting aspects of $B$ physics,
which we could not cover here. They are mainly related to rare
decays of the following kind: $B\to K^\ast\gamma$ and $B\to \rho\gamma$,
which appeared briefly in Subsection~\ref{subsec:other-Uspin}, 
$B\to K^\ast\mu^+\mu^-$ and $B_{s,d}\to \mu^+\mu^-$. The corresponding
inclusive decays, for example $B\to X_s\gamma$, are also of particular
interest, suffering from smaller theoretical uncertainties. Within the 
Standard Model, these transitions occur only at the one-loop level, 
exhibit small branching ratios at the $10^{-4}$--$10^{-10}$ level, 
do not -- apart from $B\to \rho\gamma$ -- show sizeable CP-violating 
effects, and depend on $|V_{ts}|$ or $|V_{td}|$. A measurement of these 
CKM factors through such decays would be complementary to the one from 
$B^0_{s,d}$--$\overline{B^0_{s,d}}$ mixing. Since rare $B$ decays are 
absent at the tree level in the Standard Model, they represent interesting 
probes to search for new physics. Concerning the status of 
$B\to X_s\gamma$, the present situation is as follows \cite{misiak}:
\begin{eqnarray}
\mbox{BR}(B\to X_s\gamma)_{\rm exp}&=&\left(3.11\pm0.39\right)\times10^{-4}\\
\mbox{BR}(B\to X_s\gamma)_{\rm th}&=&\left(3.73\pm0.30\right)\times10^{-4},
\end{eqnarray}
corresponding to a difference between experiment and theory at the 
$1.4 \sigma$ level. It will be interesting to see whether a serious
discrepancy will arise in the future, which could complement the one
between a recent measurement of the anomalous magnetic moment of the myon
and the Standard-Model expectation at the $2.6 \sigma$ level 
\cite{myon-moment}.

For detailed discussions of the many interesting aspects of rare $B$ decays, 
and the hard theoretical work that went into these transitions, the reader 
is referred to the overview articles listed in \cite{Brev01,misiak,Rare}.

\boldmath
\section{Conclusions and Outlook}\label{sec:concl}
\unboldmath
\setcounter{equation}{0}
I hope that this review could convince the reader that the phenomenology 
of CP violation in the $B$ system is a very interesting topic, and provides 
a fertile testing ground for the Standard-Model description of CP 
violation. In this respect, non-leptonic $B$ decays play a key r\^ole,
allowing various direct measurements of the angles of the unitarity 
triangle of the CKM matrix. Here the goal is to overconstrain this 
triangle as much as possible, not only through determinations of its 
angles, but also through measurements of its sides. Concerning the latter
aspect, the observation of $B^0_s$--$\overline{B^0_s}$ mixing, which 
is expected at run II of the Tevatron, will be an important ingredient. 
There is a strong hope that discrepancies will show up in this rich
research programme, which may eventually open a window to the physics beyond 
the Standard Model. 

Let us summarize in the following the most promising channels:
\begin{itemize}
\item The ``gold-plated'' mode $B_d\to J/\psi K_{\rm S}$ allows us
to determine $\sin2\beta$ through its mixing-induced CP asymmetry 
$a_{\psi K_{\rm S}}$. Making use of this and similar channels, the
BaBar and Belle collaborations could recently observe CP violation
in the $B$ system, thereby establishing this phenomenon for the first 
time outside the kaon system. Since the corresponding results for 
$\sin2\beta$ are not fully consistent with each other, the measurement 
of this quantity will continue to be a very exciting topic. Taking into 
account also previous results from the CDF and ALEPH collaborations, the 
resulting average for $\sin2\beta$ is now in good agreement with the range 
implied by the ``standard analysis'' of the unitarity triangle. However, 
new physics may even hide in such a situation, since only $\sin2\beta$ and 
not $\beta$ itself is measured. An important further step would be a 
determination of $\cos2\beta$. In this context, a time-dependent analysis 
of the $B_d\to J/\psi[\to \ell^+\ell^-] K^\ast[\to \pi^0 K_{\rm S}]$ angular
distribution would be useful. The preferred mechanism for new physics to 
affect these measurements is through contributions to 
$B^0_d$--$\overline{B^0_d}$ mixing. 

\item In the Standard Model, the penguin mode $B_d\to \phi K_{\rm S}$ 
allows also a determination of $\sin2\beta$ through its mixing-induced CP 
asymmetry. A comparison with $a_{\psi K_{\rm S}}$ may indicate new-physics 
contributions to decay amplitudes. Since the $B_d\to \phi K_{\rm S}$ decay
is due to $\overline{b}\to\overline{s}$ flavour-changing neutral-current 
processes, it is very sensitive 
to such effects. In order to get the full picture, $B^\pm \to \phi K^\pm$ 
and $B^\pm \to J/\psi K^\pm$ modes should be analysed as well.
\item The CP-violating observables of $B_d\to\pi^+\pi^-$ should be measured, 
although an interpretation in terms of $\alpha$ is problematic because of 
penguin topologies, requiring further inputs (see below). In addition to 
certain $B\to\pi K$ modes, $B_d\to\pi^+\pi^-$ may establish direct CP 
violation in the $B$ system. The measurement of $B^\pm\to\pi^\pm\pi^0$ 
should be refined, and efforts be made to constrain $B_d\to\pi^0\pi^0$.
\item The physics potential of $B\to\pi K$ decays is very promising
for the $B$-factories to determine $\gamma$. Since these modes are 
governed by penguin topologies, they are sensitive probes for new physics. 
Consequently, discrepancies of the extracted values of $\gamma$ with other 
approaches may well show up. In order to probe the importance of rescattering 
effects, $B^\pm\to K^\pm K$ modes should be measured as well.
\end{itemize}
All these decays are accessible at the $e^+e^-$ $B$-factories operating
at the $\Upsilon(4S)$ resonance. Further exciting perspectives open up 
at hadron colliders, where also decays of $B_s$-mesons can be studied,
involving the following major aspects:
\begin{itemize}
\item The present lower bound on $\Delta M_s$ has already an important
impact on the allowed range in the $\overline{\rho}$--$\overline{\eta}$ 
plane, implying $\gamma<90^\circ$. Much more stringent constraints can 
be obtained through a measurement of $\Delta M_s$, allowing in particular 
the construction of the ``universal unitarity triangle'' for MFV models.
\item The ``gold-plated'' mode for $B$-physics studies at hadron machines 
is $B_s\to J/\psi\phi$, which is the $B_s$-meson counterpart of 
$B_d\to J/\psi K_{\rm S}$. Its angular distribution allows us to determine 
the $B^0_s$--$\overline{B^0_s}$ mixing parameters $\Delta M_s$ and
$\Delta\Gamma_s$, and to probe the tiny mixing phase $\phi_s$, which is 
related to another unitarity triangle. Since CP violation is very small 
in $B_s\to J/\psi\phi$ within the Standard Model, it is an important tool 
to search for new physics.
\item The decay $B_s\to K^+K^-$ is related to $B_d\to \pi^+\pi^-$ 
through the $U$-spin symmetry of strong interactions and allows an 
interesting determination of $\beta$ and $\gamma$ that is -- apart from
other advantages -- not affected by penguin or rescattering processes. 
Further strategies to extract $\gamma$ are provided by other $U$-spin
related modes.
\end{itemize}
It is expected that already run II of the Tevatron will make important
contributions to these topics. In the LHC era, the physics potential of the
$B_s$ system can then be fully exploited. As we have seen in this review, 
there are also many other interesting strategies to explore CP violation, 
involving, for example, $B\to \rho\pi$, $B_d\to D^{\ast\pm}\pi^\mp$, 
$B^\pm_u\to K^\pm D$, $B^\pm_c\to D_s^\pm D$ or $B_s\to D_s^\pm K^\mp$ 
decays. However, in the practical implementation of these approaches, 
we have to deal with more or less serious challenges.

Although there is usually a strong emphasis on the extraction of
weak phases from studies of CP violation, it should not be forgotten
that the corresponding strategies provide, in several cases, also 
valuable insights into hadron dynamics. For example, also CP-conserving 
strong phases or certain penguin parameters can be determined, allowing
a comparison with theoretical predictions. Concerning the theoretical 
description of non-leptonic $B$ decays, interesting progress has recently 
been made for a large class of decays in the heavy-quark limit, including
$B\to\pi K, \pi\pi$ modes. As important applications, $SU(3)$-breaking
effects in the $B\to\pi K$ strategies to probe $\gamma$ can be controlled
more reliably, and rescattering processes are found to play a minor
r\^ole. Further progress, concerning mainly the importance of 
$\Lambda_{\rm QCD}/m_b$ corrections, will hopefully be made. 

In the Standard Model and its extensions with minimal flavour 
violation, there are interesting connections between $B$ physics 
and the rare kaon decays $K^+\to\pi^+\nu\overline{\nu}$ and  
$K_{\rm L}\to\pi^0\nu\overline{\nu}$. In such models, these decays
and $(\varepsilon,\Delta M_{d,s})$ are characterized by two 
generalized Inami--Lim functions $X$ and $F_{tt}$, respectively. 
For given values of $a_{\psi K_{\rm S}}$ and
BR$(K^+\to\pi^+\nu\overline{\nu})$, BR$(K_{\rm L}\to\pi^0\nu\overline{\nu})$
can be predicted, where only two values are allowed for the whole class of 
MFV models, corresponding to the two possible signs of $X$. Following
these lines, MFV scenarios can be confirmed or excluded in an elegant
manner. There are dedicated experiments at Brookhaven, Fermilab and KEK 
to explore $K\to\pi\nu\overline{\nu}$ decays, aiming at measurements
of their branching ratios at the $10\%$ level. 

Interestingly, already the CP asymmetry $a_{\psi K_{\rm S}}$ may provide 
insights into MFV models, since bounds on this observable can be derived. 
If $a_{\psi K_{\rm S}}$ is found to be smaller than 0.69, the MFV models 
with $F_{tt}<0$ would be excluded, whereas for $a_{\psi K_{\rm S}}<0.42$ 
also those with $F_{tt}>0$, i.e.\ all MFV models would be ruled out. In 
this exciting case, new CP-violating phases and/or new operators would be 
required. Due to reduced uncertainties of the relevant input parameters,
these bounds can be improved in the future. The most recent $B$-factory 
data are no longer in favour of small values of $a_{\psi K_{\rm S}}$, 
and the present world average does not even allow us to exclude MFV models 
with negative $F_{tt}$. Consequently, an important r\^ole may be played 
in the future by the upper bound on $a_{\psi K_{\rm S}}$ that is implied 
by the CKM factor $R_b$, yielding 
$\left(a_{\psi K_{\rm S}}\right)_{\rm max}=0.82$ for
$R_b^{\rm max}=0.46$. 

In view of the rich experimental programmes of this decade and the
strong interaction between theory and experiment, I have no doubt
that an exciting future is ahead of us!

\vspace*{0.9truecm}

\noindent
{\it Acknowledgements}

\vspace*{0.3truecm}

\noindent
First of all, I would like to thank Patricia Ball, Andrzej Buras, Amol Dighe,
Isard \mbox{Dunietz}, Thomas Mannel, Joaquim Matias, Ulrich Nierste and 
Daniel Wyler for the collaborations on several of the topics covered in 
this review. I am also grateful to Ahmed Ali, Peter Buchholz, 
Wilfried Buchm\"uller, Andreas Ringwald, Fridger Schrempp, Alan Schwartz, 
Guy Wilkinson and Frank W\"urthwein for interesting discussions.


\vspace*{0.9truecm}


\newpage


\begin{thebibliography}{999}  

\bibitem{CP-discovery}
J.H. Christenson, J.W. Cronin, V.L. Fitch and R. Turlay, 
{\it Phys.\ Rev.\ Lett.}~{\bf 13} (1964) 138.

\bibitem{babar-CP-obs}BaBar Collaboration (B. Aubert {\it et al.}),
{\it Phys.\ Rev.\ Lett.}~{\bf 87} (2001) 091801.

\bibitem{belle-CP-obs}Belle Collaboration (K. Abe {\it et al.}),
{\it Phys.\ Rev.\ Lett.}~{\bf 87} (2001) 091802.

\bibitem{basics}G. Branco, L. Lavoura and J. Silva, {\it
CP Violation} (Oxford Science Publications, Clarendon Press, Oxford, 1999);\\
I.I. Bigi and A.I. Sanda, {\it CP Violation} (Cambridge
Monographs on Particle Physics, Nuclear Physics and Cosmology,
Cambridge University Press, Cambridge, 2000).

\bibitem{B-revs}J.L. Rosner, lectures given at TASI 2000,
June 4--30, 2000, Boulder, Colorado, EFI-2000-47 [hep-ph/0011355];\\ 
R. Fleischer, lectures given at NATO ASI 2000, June 26 -- July 7, 2000, 
Cascais, Portugal, DESY 00-170 [hep-ph/0011323];\\ 
Y. Nir, lectures given at 27th SLAC Summer Institute on Particle 
Physics, July 7--16, 1999, Stanford, California, 
IASSNS-HEP-99-96 [hep-ph/9911321].

\bibitem{Confs}M. Gronau, talk given at BCP4,  
February 19--23, 2001, Ise-Shima, Japan [hep-ph/0104050];\\
Proceedings of the UK Phenomenology Workshop on Heavy Flavour and 
CP Violation, Durham, England, September 17--22, 2000, T. Hurth 
{\it et al.}, {\it J. Phys.}~{\bf G27} (2001) 1277;\\
D. Wyler, talk given at RADCOR 2000, September 11--15, 2000,
Carmel, California, ZU-01-01 [hep-ph/0101259];\\
J. Ellis, {\it Nucl.\ Phys.\ Proc.\ Suppl.}~{\bf 99} (2001) 331;\\
P. Ball, talk given at 4th Rencontres du Vietnam, July 19--25, 2000,
Hanoi, Vietnam, CERN-TH-2000-299 [hep-ph/0010024]. 

\bibitem{BF-rev}A.J. Buras and R. Fleischer, in 
{\it Heavy Flavours II}, eds.\ A.J. Buras and M. Lindner, 
World Scientific, Singapore (1998) p.\ 65 [hep-ph/9704376].

\bibitem{Brev01}A.J. Buras, lectures given at the Erice International School 
of Subnuclear Physics: Theory and Experiment Heading for New Physics, 
August 27 -- September 5, 2000, Erice, Italy, 
TUM-HEP-402/01 [hep-ph/0101336].

\bibitem{Studies}{\it The BaBar Physics Book}, eds.\ P. Harrison and 
H.R. Quinn, SLAC report 504 (1998);\\ 
Workshop on $B$ Physics at the Tevatron Run II and Beyond, see\\
{\tt http://www-theory.fnal.gov/people/ligeti/Brun2/};\\
Report of the $b$-decay Working Group of the Workshop {\it Standard Model 
Physics (and More) at the LHC}, P. Ball {\it et al.}, 
CERN-TH/2000-101 [hep-ph/0003238].

\bibitem{littenberg}L. Littenberg, BNL-67772 [hep-ex/0010048].

\bibitem{new-phys}For reviews, see Y. Grossman, Y. Nir and R. Rattazzi,  
in {\it Heavy Flavours II}, eds.\ A.J. Buras and M. 
Lindner, World Scientific, Singapore (1998) p.\ 755 [hep-ph/9701231];\\
A. Masiero and O. Vives, 
{\it Ann.\ Rev.\ Nucl.\ Part.\ Sci.}~{\bf 51} (2001) 161;\\
M. Gronau and D. London {\it Phys.\ Rev.}~{\bf D55} (1997) 2845;\\ 
Y. Nir and H.R. Quinn, {\it Annu.\ Rev.\ Nucl.\ Part.\ 
Sci.}~{\bf 42} (1992) 211;\\ 
R. Fleischer, in the proceedings of the 7th International Symposium 
on Heavy Flavor Physics, July 7--11, 1997, Santa Barbara, California, ed.\ 
C. Campagnari, World Scientific, Singapore (1999) p.\ 155 [hep-ph/9709291];\\ 
L. Wolfenstein, {\it Phys.\ Rev.}~{\bf D57} (1998) 6857.

\bibitem{MFV}M. Ciuchini, G. Degrassi, P. Gambino and G.F. Giudice,
{\it Nucl.\ Phys.}~{\bf B534} (1998) 3.

\bibitem{UUT}A.J. Buras, P. Gambino, M. Gorbahn, S. J\"ager and 
L. Silvestrini, {\it Phys.\ Lett.}~{\bf B500} (2001) 161.

\bibitem{akhmedov}For overviews, see 
E. Akhmedov, lectures given at NATO ASI 2000, June 26 -- July 7, 2000, 
Cascais, Portugal, hep-ph/0011353;\\
B. Kayser, lectures given at TASI 2000,
June 4--30, 2000, Boulder, Colorado, hep-ph/0104147;\\
V. Barger, talk given at PASCOS 99, Granlibakken, Tahoe City, California, 
December 10-16, 1999, MADPH-00-1168 [hep-ph/0005011];\\
S.M. Bilenkii, C. Giunti and W. Grimus,
{\it Prog.\ Part.\ Nucl.\ Phys.}~{\bf 43} (1999) 1.

\bibitem{lindner}A. De Rujula, M.B. Gavela and P. Hernandez,
{\it Nucl.\ Phys.}~{\bf B547} (1999) 21;\\
K. Dick, M. Freund, M. Lindner and A. Romanino, 
{\it Nucl.\ Phys.}~{\bf B562} (1999) 29.

\bibitem{sakharov}A.D. Sakharov, {\it JETP Lett.}~{\bf 5} (1967) 24.

\bibitem{buchmueller}For an overview, see W. Buchm\"uller, lectures 
given at NATO ASI 2000, June 26 -- July 7, 2000, Cascais, Portugal, 
DESY-00-194 [hep-ph/0101102]. 

\bibitem{shapos}V.A. Rubakov and M.E. Shaposhnikov, 
{\it Usp.\ Fiz.\ Nauk} {\bf 166} (1996) 493; 
{\it Phys.\ Usp.}~{\bf 39} (1996) 461;\\ 
A. Riotto and M. Trodden, {\it Annu.\ Rev.\ Nucl.\ Part.\ 
Sci.}~{\bf 49} (1999) 35.

\bibitem{bigi-D}I.I. Bigi, talk given at BCP4,  
February 19--23, 2001, Ise-Shima, Japan, UND-HEP-01-BIG 02 
[hep-ph/0104008].

\bibitem{Ddecay}S. Bergmann and Y. Nir, 
{\it J.\ High Energy Phys.}~{\bf 9909} (1999) 031;\\
Y.~Nir, {\it Nuovo Cim.}~{\bf 109A} (1996) 991;\\
G. Blaylock, A. Seiden and Y. Nir,
{\it Phys.\ Lett.}~{\bf B355} (1995) 555;\\
L. Wolfenstein, {\it Phys.\ Rev.\ Lett.}~{\bf 75} (1995) 2460.

\bibitem{dipole}X.-G. He, B.H. McKellar and S. Pakvasa,
{\it Int.\ J.\ Mod.\ Phys.}~{\bf A4} (1989) 5011;\\
W. Bernreuther and M. Suzuki, {\it Rev.\ Mod.\ Phys.}~{\bf 63} (1991) 313;\\
S.M. Barr, {\it Int.\ J.\ Mod.\ Phys.}~{\bf A8} (1993) 209;\\
W. Bernreuther, PITHA-98-25 [hep-ph/9808453];\\
V. Barger, T. Falk, T. Han, J. Jiang, T. Li and T. Plehn,
{\it Phys.\ Rev.}~{\bf D64} (2001) 056007.
 
\bibitem{hyperon}G. Valencia, {\it AIP Conf.\ Proc.}~{\bf 539} (2000) 80;\\
S. Pakvasa, UH-511-955-00, hep-ph/0002210;\\
X.-G. He, H. Murayama, S. Pakvasa and G. Valencia,
{\it Phys.\ Rev.}~{\bf D61} (2000) 071701.

\bibitem{SM}S.L. Glashow, {\it Nucl.\ Phys.}~{\bf 22} (1961) 579;\\ 
S. Weinberg, {\it Phys.\ Rev.\ Lett.}~{\bf 19} (1967) 1264;\\ 
A. Salam, in {\it Elementary Particle Theory}, ed.\ N. Svartholm,  
Almqvist and Wiksell, Stockholm (1968).

\bibitem{cab}N. Cabibbo, {\it Phys.\ Rev.\ Lett.}~{\bf 10} (1963) 531.

\bibitem{km}M. Kobayashi and T. Maskawa, {\it Progr.\ Theor.\ 
Phys.}~{\bf 49} (1973) 652.

\bibitem{peccei}R.D. Peccei, UCLA-98-TEP-21 [hep-ph/9807514]. 

\bibitem{SMpar}L.-L. Chau and W.-Y. Keung, {\it Phys.\ Rev.\ Lett.}~{\bf 53} 
(1984) 1802;\\
H. Harari and M. Leurer, {\it Phys.\ Lett.}~{\bf B181} (1986) 123;\\
H. Fritzsch and J. Plankl, {\it Phys.\ Rev.}~{\bf D35} (1987) 1732;\\
F.J. Botella and L.-L. Chao, {\it Phys.\ Lett.}~{\bf B168} (1986) 97.

\bibitem{FX}H. Fritzsch and Z.-Z. Xing, {\it Phys.\ 
Lett.}~{\bf B413} (1997) 396.

\bibitem{schu}M. Schmidtler and K.R. Schubert, {\it Z. Phys.}~{\bf C53}
(1992) 347.

\bibitem{blo}A.J. Buras, M.E. Lautenbacher and G. Ostermaier, 
{\it Phys.\ Rev.}~{\bf D50} (1994) 3433. 

\bibitem{wolf}L. Wolfenstein, {\it Phys.\ Rev.\ Lett.}~{\bf 51} (1983)
1945.

\bibitem{jarlskog}C. Jarlskog, {\it Phys.\ Rev.\ Lett.}~{\bf 55}
(1985) 1039; {\it Z. Phys.}~{\bf C29} (1985) 491.

\bibitem{BBG}J. Bernabeu, G. Branco and M. Gronau, {\it Phys.\ 
Lett.}~{\bf B169} (1986) 243.

\bibitem{AKL}R. Aleksan, B. Kayser and D. London, {\it Phys.\ Rev.\ 
Lett.}~{\bf 73} (1994) 18.

\bibitem{JS}C. Jarlskog and R. Stora, {\it Phys.\ Lett.}~{\bf B208} (1988) 268.

\bibitem{ut}L.L. Chau and W.-Y. Keung, {\it Phys.\ Rev.\ 
Lett.}~{\bf 53} (1984) 1802.

\bibitem{Neubert-rev}For a review, see M. Neubert, in 
{\it Heavy Flavours II}, eds.\ A.J. Buras and M. Lindner, 
World Scientific, Singapore (1998) p.\ 239 [hep-ph/9702375].

\bibitem{Rome-rev}M. Ciuchini {\it et al.}, 
{\it J.\ High Energy Phys.}~{\bf 0107} (2001) 013.

\bibitem{CLEO-Vub}CLEO Collaboration (B.H. Behrens {\it et al.}),
{\it Phys.\ Rev.}~{\bf D61} (2000) 052001. 

\bibitem{LEP-Vub}LEP Working Group on $|V_{ub}|$:\\ 
{\tt http://battagl.home.cern.ch/battagl/vub/vub.html.}

\bibitem{PDG}Particle Data Group, D.E. Groom {\it et al.}, {\it Eur. 
Phys. J.}~{\bf C15} (2000) 1. 

\bibitem{AL}A. Ali and D. London, {\it Eur.\ Phys.\ J.}~{\bf C18} (2001) 
665;\\
see also {\it Eur.\ Phys.\ J.}~{\bf C9} (1999) 687;
{\it Phys.\ Rep.}~{\bf 320} (1999) 79; hep-ph/0002167.

\bibitem{PS}S. Plaszczynski and M.-H. Schune, LAL-99-67 [hep-ph/9911280].

\bibitem{GNPS}Y. Grossman, Y. Nir, S. Plaszczynski and M.-H. Schune,
{\it Nucl.\ Phys.}~{\bf B511} (1998) 69.

\bibitem{hoecker}A. H\"ocker, H. Lacker, S. Laplace, F. Le Diberder, 
{\it Eur.\ Phys.\ J.}~{\bf C21} (2001) 225.

\bibitem{BB-Bound}A.J. Buras and R. Buras, {\it Phys.\ Lett.}~{\bf B501} 
(2001) 223.

\bibitem{BF-MFV}A.J. Buras and R. Fleischer, 
{\it Phys.\ Rev.}~{\bf D64} (2001) 115010.

\bibitem{fulvia}F. De Fazio, DPT-00-46 [hep-ph/0010007]. 

\bibitem{OPE}K.G. Wilson, {\it Phys.\ Rev.}~{\bf 179} (1969) 1499;\\
K.G. Wilson and W. Zimmermann, {\it Comm.\ Math.\ Phys.}~{\bf 24} (1972) 87;\\
W. Zimmerman, in the proceedings of the 1970 Brandeis Summer Institute
in Theoretical Physics, eds.\ S. Deser, M. Grisaru and H. Pendleton,
MIT Press (1971) p.\ 369; {\it Ann.\ Phys.}~{\bf 77} (1973) 570.

\bibitem{IL}T. Inami and C.S. Lim, 
{\it Progr.\ Theor.\ Phys.}~{\bf 65} (1981) 297.

\bibitem{PB-expansion}G. Buchalla, A.J. Buras and M.K. Harlander,
{\it Nucl.\ Phys.}~{\bf B349} (1991) 1.

\bibitem{CC-REFS}F.J. Gilman and M.B. Wise, {\it Phys.\ Rev.}~{\bf D20}
(1979) 2392;\\
G. Altarelli, G. Curci, G. Martinelli and S. Petrarca,
{\it Phys.\ Lett.}~{\bf B99} (1981) 141;\\
A.J. Buras and P.H. Weisz, {\it Nucl.\ Phys.}~{\bf B333} (1990) 66.

\bibitem{MS-bar}W.A. Bardeen, A.J. Buras, D.W. Duke and T. Muta,
{\it Phys.\ Rev.}~{\bf D18} (1978) 3998.

\bibitem{Buras-a1-a2}A.J. Buras, {\it Nucl.\ Phys.}~{\bf B434} (1995) 606.

\bibitem{BBL-rev}G. Buchalla, A.J. Buras and M.E. Lautenbacher, 
{\it Rev.\ Mod.\ Phys.}~{\bf 68} (1996) 1125.

\bibitem{Buras-lect}A.J. Buras, in the proceedings of the Les Houches 
1997 Summer School on Theoretical Physics: Probing the Standard Model of 
Particle Interactions, July 28 -- September 5, 1997, Les Houches, France, 
eds.\ R. Gupta, A. Morel, E. de Rafael and F. David,  
North-Holland, Amsterdam (1998) [hep-ph/9806471].

\bibitem{Munich}A.J. Buras, M. Jamin, M.E. Lautenbacher and P.H. Weisz,
{\it Nucl.\ Phys.}~{\bf B400} (1993) 37;\\
A.J. Buras, M. Jamin and M.E. Lautenbacher,
{\it Nucl.\ Phys.}~{\bf B400} (1993) 75 and {\bf B408} (1993) 209.

\bibitem{Rome}M. Ciuchini, E. Franco, G. Martinelli and L. Reina,
{\it Phys.\ Lett.}~{\bf B301} (1993) 263;
{\it Nucl.\ Phys.}~{\bf B415} (1994) 403.

\bibitem{NDR-HV}G. 't Hooft and M. Veltman, {\it Nucl.\ Phys.}~{\bf B44}
(1972) 189;\\
P. Breitenlohner and D. Maison, {\it Comm.\ Math.\ 
Phys.}~{\bf 52} (1977) 11, 39, 55.

\bibitem{BMU}A.J. Buras, M. Misiak and J. Urban, 
{\it Nucl.\ Phys.}~{\bf B586} (2000) 397.

\bibitem{DF2-NP}M. Ciuchini, E. Franco, V. Lubicz, G. Martinelli, 
I. Scimemi and L. Silvestrini, {\it Nucl.\ Phys.}~{\bf B523} (1998) 501;\\
A.J. Buras, S. J\"ager and J. Urban, {\it Nucl.\ Phys.}~{\bf B605} 
(2001) 600.

\bibitem{RF-1}R. Fleischer, {\it Z.\ Phys.}~{\bf C58} (1993) 483.

\bibitem{RF-EWP1}R. Fleischer, {\it Z. Phys.}~{\bf C62} (1994) 81.

\bibitem{BSS}M. Bander, D. Silverman and A. Soni, 
{\it Phys.\ Rev.\ Lett.}~{\bf 43} (1979) 242.

\bibitem{BphiK-old}D. London and R.D. Peccei,
{\it Phys.\ Lett.}~{\bf B223} (1989) 257;\\
N.G. Deshpande and J. Trampetic,  
{\it Phys.\ Rev.}~{\bf D41} (1990) 895 and 2926;\\ 
J.-M. G\'erard and W.-S. Hou, {\it Phys.\ Rev.}~{\bf D43} (1991) 2909; 
{\it Phys.\ Lett.}~{\bf B253} (1991) 478. 

\bibitem{kramer}G. Kramer, W.F. Palmer and H. Simma,
{\it Nucl.\ Phys.}~{\bf B428} (1994) 77; {\it Z.\ Phys.}~{\bf C66} 
(1995) 429.

\bibitem{AKL-calc1}A. Ali, G. Kramer and C. L\"u,
{\it Phys.\ Rev.}~{\bf D58} (1998) 094009.

\bibitem{AKL-calc2}A. Ali, G. Kramer and C. L\"u,
{\it Phys.\ Rev.}~{\bf D59} (1999) 014005.

\bibitem{BL}H. Simma and D. Wyler, {\it Phys.\ Lett.}~{\bf B272} (1991)
395.

\bibitem{ward}B.F. Ward, {\it Phys.\ Rev.}~{\bf D51} (1995) 6253.

\bibitem{GIM}S.L. Glashow, J. Iliopoulos and L. Maiani, {\it
Phys.\ Rev.}~{\bf D2} (1970) 1285.

\bibitem{BF-u-c-pens}A.J. Buras and R. Fleischer,  
{\it Phys.\ Lett.}~{\bf B341} (1995) 379. 

\bibitem{RF-BdKK}R. Fleischer, {\it Phys.\ Lett.}~{\bf B341} (1994) 205.

\bibitem{RF-ang}R. Fleischer, {\it Phys.\ Rev.}~{\bf D60} (1999) 073008.

\bibitem{charming-pens1}M. Ciuchini, E. Franco, G. Martinelli and 
L. Silvestrini, {\it Nucl.\ Phys.}~{\bf B501} (1997) 271. 

\bibitem{ILNPS}C. Isola, M. Ladisa, G. Nardulli, T.N. Pham and P. Santorelli,
{\it Phys.\ Rev.}~{\bf D64} (2001) 014029.

\bibitem{charming-pens2}M. Ciuchini, E. Franco, G. Martinelli, M. Pierini
and L. Silvestrini, 
{\it Phys.\ Lett.}~{\bf B515} (2001) 33.

\bibitem{FSI}L. Wolfenstein, {\it Phys.\ Rev.}~{\bf D52} (1995) 537;\\
J.F. Donoghue, E. Golowich, A.A. Petrov and J.M. Soares, 
{\it Phys.\ Rev.\ Lett.}~{\bf 77} (1996) 2178;\\
B. Blok, M. Gronau and J.L. Rosner, {\it Phys.\ Rev.\ Lett.}~{\bf 78}
(1997) 3999 and {\bf 79} (1997) 1167;\\
M. Neubert, {\it Phys.\ Lett.}~{\bf B424} (1998) 152;\\
J.-M. G\'erard and J. Weyers, {\it Eur.\ Phys.\ J.}~{\bf C7} (1999) 1;\\
A. Falk, A. Kagan, Y. Nir and A. Petrov,  
{\it Phys.\ Rev.}~{\bf D57} (1998) 4290;\\
D. Atwood and A. Soni, {\it Phys.\ Rev.}~{\bf D58} (1998) 036005.

\bibitem{BFM}A.J. Buras, R. Fleischer and T. Mannel,  
{\it Nucl.\ Phys.}~{\bf B533} (1998) 3.

\bibitem{RF-rev}R. Fleischer, 
{\it Int.\ J. Mod.\ Phys.}~{\bf A12} (1997) 2459.

\bibitem{DH-PhiK}N.G. Deshpande and X.-G. He, {\it Phys.\ Lett.}~{\bf B336} 
(1994) 471.

\bibitem{RF-EWP2}R. Fleischer, {\it Phys.\ Lett.}~{\bf B321} (1994) 259.

\bibitem{RF-EWP3}R. Fleischer, {\it Phys.\ Lett.}~{\bf B332} (1994) 419.

\bibitem{EWP-BpiK}N.G. Deshpande and X.-G. He,
{\it Phys.\ Rev.\ Lett.}~{\bf 74} (1995) 26 [E: ibid., p.\ 4099].

\bibitem{ghlr-ewp}M. Gronau, O. Hern\'andez, D. London and J. Rosner,  
{\it Phys.\ Rev.}~{\bf D52} (1995) 6374.

\bibitem{PAPI}A.J. Buras and R. Fleischer, {\it Phys.\ Lett.}~{\bf B365} 
(1996) 390.

\bibitem{BF-BpiK1}A.J. Buras and R. Fleischer,
{\it Eur.\ Phys.\ J.}~{\bf C11} (1999) 93.

\bibitem{PAPIII}R. Fleischer, {\it Phys.\ Lett.}~{\bf B365} (1996) 399.

\bibitem{BSW}M. Bauer, B. Stech and M. Wirbel,
{\it Z.\ Phys.}~{\bf C34} (1987) 103;\\
M. Wirbel, B. Stech and M. Bauer,
{\it Z.\ Phys.}~{\bf C29} (1985) 637.

\bibitem{NS}M. Neubert and B. Stech, in {\it Heavy Flavours II}, 
eds.\ A.J. Buras and M. Lindner, World Scientific, Singapore (1998)
p.\ 294 [hep-ph/9705292].

\bibitem{facto}J. Schwinger, {\it Phys.\ Rev.\ Lett.}~{\bf 12} (1964) 630;\\ 
D. Farikov and B. Stech, {\it Nucl.\ Phys.}~{\bf B133} (1978) 315;\\ 
N. Cabibbo and L. Maiani, {\it Phys.\ Lett.}~{\bf B73} (1978) 418
[E: ibid.~{\bf B76} (1978) 663].

\bibitem{fact-idea}J.D. Bjorken, {\it Nucl.\ Phys.\ (Proc.\ 
Suppl.)}~{\bf B11} (1989) 325;\\ 
M. Dugan and B. Grinstein, {\it Phys.\ Lett.}~{\bf B255} (1991) 583;\\ 
H.D. Politzer and M.B. Wise, {\it Phys.\ Lett.}~{\bf B257} (1991) 399.

\bibitem{BuSi}A.J. Buras and L. Silvestrini, {\it Nucl.\ Phys.}~{\bf B548} 
(1999) 293.

\bibitem{Luo-Ros}Z. Luo and J.L. Rosner, 
{\it Phys.\ Rev.}~{\bf D64} (2001) 094001.

\bibitem{Fact-new}M. Diehl and G. Hiller, 
{\it J.\ High Energy Phys.}~{\bf 0106} (2001) 067.

\bibitem{largeN}A.J. Buras and J.-M. G\'erard, {\it Nucl.\ Phys.}~{\bf B264} 
(1986) 371;\\
A.J. Buras, J.-M. G\'erard and R. R\"uckl, 
{\it Nucl.\ Phys.}~{\bf B268} (1986) 16.

\bibitem{RF-Nc}R. Fleischer, {\it Nucl.\ Phys.}~{\bf B412} (1994) 201.

\bibitem{BBNS1}M. Beneke, G. Buchalla, M. Neubert and C.T. Sachrajda, 
{\it Phys.\ Rev.\ Lett.}~{\bf 83} (1999) 1914.

\bibitem{BBNS2}M. Beneke, G. Buchalla, M. Neubert and C.T. Sachrajda, 
{\it Nucl.\ Phys.}~{\bf B591} (2000) 313.

\bibitem{BBNS3}M. Beneke, G. Buchalla, M. Neubert and C.T. Sachrajda,
{\it Nucl.\ Phys.}~{\bf B606} (2001) 245.

\bibitem{PQCD}H.-n.\ Li and H.L. Yu, {\it Phys.\ Rev.\ Lett.}~{\bf 74} 
(1995) 4388 and {\it Phys.\ Rev.}~{\bf D53} (1996) 2480;\\
C.H. Chang and H.-n.\ Li, {\it Phys.\ Rev.}~{\bf D55} (1997)
5577;\\ 
T.W. Yeh and H.-n.\ Li, {\it Phys.\ Rev.}~{\bf D56} (1997) 1615;\\ 
H.Y. Cheng, H.-n.\ Li and K.C. Yang, {\it Phys.\ Rev.}~{\bf D60} (1999) 
094005.

\bibitem{PQCD-anni}Y.Y. Keum, H.-n.\ Li and A.I. Sanda, 
{\it Phys.\ Lett.}~{\bf B504} (2001) 6; 
{\it Phys.\ Rev.}~{\bf D63} (2001) 054008.

\bibitem{PQCD-comp}Y.Y. Keum and H.-n.\ Li, 
{\it Phys.\ Rev.}~{\bf D63} (2001) 074006.

\bibitem{buchalla}G. Buchalla, lectures given at TASI 2000,
June 4--30, 2000, Boulder, Colorado, CERN-TH-2001-041 [hep-ph/0103166].

\bibitem{BJW90}A.J. Buras, M. Jamin, and P.H. Weisz,
{\it Nucl.\ Phys.}~{\bf B347} (1990) 491.

\bibitem{HN}S. Herrlich and U. Nierste, {\it Nucl.\ Phys.}~{\bf B419} 
(1994) 292; {\it Phys.\ Rev.}~{\bf D52} (1995) 6505; 
{\it Nucl.\ Phys.}~{\bf B476} (1996) 27.

\bibitem{lellouch}L. Lellouch, {\it Nucl.\ Phys.\ (Proc.\ Suppl.)}
{\bf B94} (2001) 142.

\bibitem{large-N-est}W.A. Bardeen, A.J. Buras and J.-M. G\'erard,
{\it Phys.\ Lett.}~{\bf B211} (1988) 343;\\
J. Bijnens and J. Prades, {\it Nucl.\ Phys.}~{\bf B444} (1995) 523;
{\it J.\ High Energy Phys.} {\bf 0001} (2000) 002;\\
T. Hambye, G.O. K\"ohler and P.H. Soldan, {\it Eur.\ Phys.\ J.}~{\bf C10}
(1999) 271.

\bibitem{chiral-est}S. Bertolini, J.O. Eeg, M. Fabbrichesi and E.I. Lashin,
{\it Nucl.\ Phys.}~{\bf B514} (1998) 63.

\bibitem{eps-prime-anat}J.M. Flynn and L. Randall,
{\it Phys.\ Lett.}~{\bf B224} (1989) 221 [E: {\bf B235} (1989) 412];\\
G. Buchalla, A.J. Buras and M.K. Harlander,
{\it Nucl.\ Phys.}~{\bf B337} (1990) 313.

\bibitem{superweak}L. Wolfenstein,  
{\it Phys.\ Rev.\ Lett.}~{\bf 13} (1964) 562.

\bibitem{KTeV}KTeV Collaboration (A. Alavi-Harati {\it et al.}), 
{\it Phys.\ Rev.\ Lett.}~{\bf 83} (1999) 22;\\ 
for the most recent update given in (\ref{epsprime-res}), see\\
{\tt http://kpasa.fnal.gov:8080/public/ktev.html}.

\bibitem{NA48}NA48 Collaboration (V. Fanti {\it et al.}), {\it Phys.\
Lett.}~{\bf B465} (1999) 335;\\ 
for the most recent update given in (\ref{epsprime-res}), see
{\tt http://na48.web.cern.ch/NA48/}.

\bibitem{epsprime}S. Bertolini, M. Fabbrichesi and J.O. Eeg,
{\it Rev.\ Mod.\ Phys.}~{\bf 72} (2000) 65;\\
M. Jamin, HD-THEP-99-51 [hep-ph/9911390];\\
M. Ciuchini and G. Martinelli, {\it Nucl.\ Phys.\ (Proc.\ Suppl.)}~{\bf B99}
(2001) 27;\\
T. Hambye and P.H. Soldan,
{\it Nucl.\ Phys.\ (Proc.\ Suppl.)}~{\bf B96} (2001) 323;\\
E. Pallante, A. Pich and I. Scimemi, 
{\it Nucl.\ Phys.}~{\bf B617} (2001) 441.

\bibitem{Bu-Ge}A.J. Buras and J.-M. G\'erard, 
{\it Phys.\ Lett.}~{\bf B517} (2001) 129.

\bibitem{kpinunu-LD}D. Rein and L.M. Segal, {\it Phys.\ Rev.}~{\bf D39} 
(1989) 3325;\\
J.S. Hagelin and L.S. Littenberg, {\it Prog.\ Part.\ Nucl.\ Phys.}~{\bf 23}
(1989) 1;\\
M. Lu and M.B. Wise, {\it Phys.\ Lett.}~{\bf B324} (1994) 461;\\
S. Fajfer, {\it Nuovo Cim.}~{\bf A110} (1997) 397;\\
C.Q. Geng, I.J. Hsu and Y.C. Lin, {\it Phys.\ Rev.}~{\bf D54} (1996) 877. 

\bibitem{BI}G. Buchalla and G. Isidori, {\it Phys.\ Lett.}~{\bf B440} 
(1998) 170.

\bibitem{FLP}A.F. Falk, A. Lewandowski and A.A. Petrov, {\it Phys.\
Lett.}~{\bf B505} (2001) 107.

\bibitem{Kpi-NLO}G. Buchalla and A.J. Buras,
{\it Nucl.\ Phys.}~{\bf B398} (1993) 285 and {\bf B400} (1993) 225;\\
M. Misiak and J. Urban, {\it Phys.\ Lett.}~{\bf B451} (1999) 161.
 
\bibitem{BB98}G. Buchalla and A.J. Buras,
{\it Nucl.\ Phys.}~{\bf B548} (1999) 309 and {\bf B412} (1994) 106.

\bibitem{Adler00}S. Adler {\it et al.}, 
{\it Phys.\ Rev.\ Lett.}~{\bf 84} (2000) 3768. 

\bibitem{BBSIN}G. Buchalla and A.J. Buras, 
{\it Phys.\ Lett.}~{\bf B333} (1994) 221;
{\it Phys.\ Rev.}~{\bf D54} (1996) 6782.

\bibitem{NIR}Y. Grossman and Y. Nir, {\it Phys.\ Lett.}~{\bf B398} (1997) 163.

\bibitem{WW}V.F. Weisskopf and E.P. Wigner, {\it Z. Phys.}~{\bf 63} (1930)
54 and {\bf 65} (1930) 18.

\bibitem{BuSlSt}A.J. Buras, W. S{\l}ominski and H. Steger,
{\it Nucl.\ Phys.}~{\bf B245} (1984) 369.

\bibitem{UKJS}J. Urban, F. Krauss, U. Jentschura and G. Soff, 
{\it Nucl.\ Phys.}~{\bf B523} (1998) 40. 

\bibitem{FBlat}L. Lellouch and C.D. Lin (UKQCD Collaboration), 
{\it Phys.\ Rev.}~{\bf D64} (2001) 094501;\\
J. Flynn and C.D. Lin, {\it J.\ Phys.}~{\bf G27} (2001) 1245;\\
C.T. Sachrajda, {\it Nucl.\ Instrum.\ Meth.}~{\bf A462} (2001) 23.

\bibitem{QCDSR}E. Bagan, P. Ball, V.M. Braun and H.G. Dosch,
{\it Phys.\ Lett.}~{\bf B278} (1992) 457;\\
M. Neubert, {\it Phys.\ Rev.}~{\bf D45} (1992) 2451;\\
S. Narison, {\it Phys.\ Lett.}~{\bf B322} (1994) 247.

\bibitem{cleo-eps-B}CLEO Collaboration (D.E. Jaffe {\it et al.}), 
{\it Phys.\ Rev.\ Lett.}~{\bf 86} (2001) 5000.

\bibitem{GKN}Y. Grossman, B. Kayser and Y. Nir, 
{\it Phys.\ Lett.}~{\bf B415} (1997) 90.

\bibitem{LEPBOSC}LEP $B$ Oscillation Working Group, see\\ {\tt 
http://lepbosc.web.cern.ch/LEPBOSC/combined$_{-}$results/}.

\bibitem{FM-BpsiK}R. Fleischer and T. Mannel, 
{\it Phys.\ Lett.}~{\bf B506} (2001) 311.

\bibitem{dim-6}W. Buchm\"uller and D. Wyler, {\it Nucl.\ Phys.}~{\bf B268}
(1986) 621.

\bibitem{bisa}A.B. Carter and A.I. Sanda,  
{\it Phys.\ Rev.\ Lett.}~{\bf 45} (1980) 952; {\it Phys.\ Rev.}~{\bf D23}
(1981) 1567;\\ 
I.I. Bigi and A.I. Sanda, {\it Nucl.\ Phys.}~{\bf B193} (1981) 85.

\bibitem{RF-BdsPsiK}R. Fleischer, {\it Eur.\ Phys.\ J.}~{\bf C10} (1999) 
299.

\bibitem{NiSi}Y. Nir and D. Silverman, {\it Nucl.\ Phys.}~{\bf B345} (1990)
301.

\bibitem{opal}OPAL Collaboration (K. Ackerstaff {\it et al.}), 
{\it Eur.\ Phys.\ J.}~{\bf C5} (1998) 379.

\bibitem{cdf}CDF Collaboration (T. Affolder {\it et al.}),
{\it Phys.\ Rev.}~{\bf D61} (2000) 072005.

\bibitem{aleph}ALEPH Collaboration (R. Barate {\it et al.}),
{\it Phys.\ Lett.}~{\bf B492} (2000) 259.

\bibitem{babar}BaBar Collaboration (B. Aubert {\it et al.}), 
{\it Phys.\ Rev.\ Lett.}~{\bf 86} (2001) 2515.

\bibitem{belle}Belle Collaboration (A. Abashian {\it et al.}), 
{\it Phys.\ Rev.\ Lett.}~{\bf 86} (2001) 2509.

\bibitem{BABAR1}D. Hitlin (BaBar Collaboration), plenary talk given at
ICHEP 2000, July 27 -- August 2, 2000, Osaka, Japan, BABAR-PROC-00-14
[hep-ex/0011024].

\bibitem{BELLE1}H. Aihara (Belle Collaboration), plenary talk given at
ICHEP 2000, July 27 -- August 2, 2000, Osaka, Japan, hep-ex/0010008. 

\bibitem{low-sin}A.L. Kagan and M. Neubert, 
{\it Phys.\ Lett.}~{\bf B492} (2000) 115;\\ 
J.P. Silva and L. Wolfenstein, {\it Phys.\ Rev.}~{\bf D63} (2001) 056001;\\
Z.-Z. Xing, hep-ph/0008018;\\ 
G. Eyal, Y. Nir and G. Perez, 
{\it J. High Energy Phys.}~{\bf 0008} (2000) 028;\\
G. Barenboim, F.J. Botella and O. Vives, 
{\it Phys.\ Rev.}~{\bf D64} (2001) 015007;\\
H. Fritzsch and Z.Z. Xing, {\it Phys.\ Lett.}~{\bf B506} (2001) 109;\\
Y.-L. Wu and Y.-F. Zhou, 
{\it Int.\ J.\ Mod.\ Phys.}~{\bf A16} (2001) 4547;\\
G. Bhattacharyya, A. Datta and A. Kundu, 
{\it Phys.\ Lett.}~{\bf B514} (2001) 47.

\bibitem{BePe}S. Bergmann and G. Perez, 
{\it Phys.\ Rev.}~{\bf D64} (2001) 115009.

\bibitem{Ali-Lunghi}A. Ali and E. Lunghi, 
{\it Eur.\ Phys.\ J.}~{\bf C21} (2001) 683.

\bibitem{BCRS}A.J. Buras, P.H. Chankowski, J. Rosiek and 
{\L}. S{\l}awianowska, {\it Nucl.\ Phys.}~{\bf B619} (2001) 434.

\bibitem{LHC-Report}P. Ball {\it et al.}, Report of the $b$-decay Working 
Group of the Workshop {\it Standard Model Physics (and More) at the LHC}, 
CERN-TH/2000-101 [hep-ph/0003238].

\bibitem{hierarchy}M. Gronau, O.F. Hern\'andez, D. London and J.L. Rosner,
{\it Phys.\ Rev.}~{\bf D52} (1995) 6356 and 6374.

\bibitem{growo}Y. Grossman and M.P. Worah, {\it Phys.\ Lett.}~{\bf B395}
(1997) 241.

\bibitem{cleo-dir}CLEO Collaboration (G. Bonvicini {\it et al.}), 
{\it Phys.\ Rev.\ Lett.}~{\bf 84} (2000) 5940.

\bibitem{babar-dir}J. Nash (BaBar Collaboration), talk given at 
Lepton Photon 01, July 23--28, 2001, Rome, Italy.

\bibitem{BABAR-BR-B}BaBar Collaboration (B. Aubert {\it et al.}), 
{\it Phys.\ Rev.}~{\bf D65} (2002) 032001.

\bibitem{BELLE-BR-B}Belle Collaboration (K. Abe {\it et al.}), 
BELLE-CONF-0101.

\bibitem{BABAR-LIFE}BaBar Collaboration (B. Aubert {\it et al.}), 
{\it Phys.\ Rev.\ Lett.}~{\bf 87} (2001) 201803.

\bibitem{BELLE-LIFE}H. Tajima (Belle Collaboration), Belle Preprint
2001-8, contributed to the proceedings of the XXXVIth Rencontres de
Moriond session devoted to QCD and High Energy Hadronic Interactions,
March 17--24, 2001, Bourg-Saint-Maurice, France [hep-ex/0105024].

\bibitem{DQSTL}I. Dunietz, H.R. Quinn, A. Snyder, W. Toki and H.J.
Lipkin, {\it Phys.\ Rev.}~{\bf D43} (1991) 2193.

\bibitem{DDF1}A.S. Dighe, I. Dunietz and R. Fleischer, 
{\it Eur.\ Phys.\ J.}~{\bf C6} (1999) 647.

\bibitem{DDF2}A.S. Dighe, I. Dunietz and R. Fleischer, 
{\it Phys.\ Lett.}~{\bf B433} (1998) 147.

\bibitem{DFN}I. Dunietz, R. Fleischer and U. Nierste, 
{\it Phys.\ Rev.}~{\bf D63} (2001) 114015.

\bibitem{ambig}Ya.I. Azimov, V.L. Rappoport and V.V. Sarantsev,
{\it Z. Phys.}~{\bf A356} (1997) 437;\\
Y. Grossman and H.R. Quinn, {\it Phys.\ Rev.}~{\bf D56} (1997) 7259;\\
J. Charles, A. Le Yaouanc, L. Oliver, O. P\`ene
and J.-C. Raynal, {\it Phys.\ Lett.}~{\bf B425} (1998) 375;\\
B. Kayser and D. London, {\it Phys.\ Rev.}~{\bf D61} (2000) 116012;\\
H.R. Quinn, T. Schietinger, J.P. Silva, A.E. Snyder,
{\it Phys.\ Rev.\ Lett.}~{\bf 85} (2000) 5284.

\bibitem{LoSo}D. London and A. Soni, 
{\it Phys.\ Lett.}~{\bf B407} (1997) 61. 

\bibitem{FM-PhiK}R. Fleischer and T. Mannel, 
{\it Phys.\ Lett.}~{\bf B511} (2001) 240.

\bibitem{belle-BphiK}Belle Collaboration (K. Abe {\it et al.}),
BELLE-CONF-0113.

 
\bibitem{cleo-BphiK}CLEO Collaboration (R.A. Briere {\it et al.}),  
{\it Phys.\ Rev.\ Lett.}~{\bf 86} (2001) 3718.

\bibitem{babar-BphiK}BaBar Collaboration (B. Aubert {\it et al.}),
{\it Phys.\ Rev.\ Lett.}~{\bf 87} (2001) 151801.

\bibitem{BphiK-recent-calc}X.-G. He, J.P. Ma and C.-Y. Wu,
{\it Phys.\ Rev.}~{\bf D63} (2001) 094004;\\
H.-Y. Cheng and K.-C. Yang, 
{\it Phys.\ Rev.}~{\bf D64} (2001) 074004;\\
S. Mishima, hep-ph/0107163;\\
C.-H. Chen, Y.-Y. Keum and H.-n.\ Li, 
{\it Phys.\ Rev.}~{\bf D64} (2001) 112002.

\bibitem{GIW}Y. Grossman, G. Isidori and M.P. Worah, {\it Phys.\  
Rev.}~{\bf D58} (1998) 057504. 

\bibitem{RF-BsKK}R. Fleischer, {\it Phys.\ Lett.}~{\bf B459} (1999) 306.

\bibitem{alpha-uncert}M. Gronau, {\it Phys.\ Lett.}~{\bf B300} (1993)
163;\\ 
R. Aleksan {\it et al.}, {\it Phys.\ Lett.}~{\bf B356} (1995) 95;\\
F. DeJongh and P. Sphicas, {\it Phys.\ Rev.}~{\bf D53} (1996) 4930;\\
P.S. Marrocchesi and N. Paver, {\it Int.\ J. Mod.\ Phys.}~{\bf A13}
(1998) 251;\\ 
A. Ali, G. Kramer and C.-D. L\"u,  in \cite{AKL-calc2};\\
B.F. Ward, in \cite{ward}.

\bibitem{SiWo}J.P. Silva and L. Wolfenstein, {\it Phys.\ Rev.}~{\bf D49} 
(1994) R1151;
{\it Phys.\ Rev.}~{\bf D62} (2000) 014018.

\bibitem{CLEO-BpiK}CLEO Collaboration (D. Cronin--Hennessy {\it et al.}),
{\it Phys.\ Rev.\ Lett.}~{\bf 85} (2000) 515.

\bibitem{babar-BpiK}BaBar Collaboration (B. Aubert {\it et al.}),
{\it Phys.\ Rev.\ Lett.}~{\bf 87} (2001) 151802.

\bibitem{belle-BpiK}Belle Collaboration (K. Abe {\it et al.}),
{\it Phys.\ Rev.\ Lett.}~{\bf 87} (2001) 101801.

\bibitem{RF-bpipi}R. Fleischer, {\it Eur.\ Phys.\ J.}~{\bf C16} (2000) 87.

\bibitem{BABAR-Bpipi-CP}J. Dorfan (BaBar Collaboration), talk given at 
Lepton Photon 01, July 23--28, 2001, Rome, Italy;\\
BaBar Collaboration (B. Aubert {\it et al.}), BABAR-CONF-01/05 
[hep-ex/0107074].

\bibitem{GL}M. Gronau and D. London, {\it Phys.\ Rev.\ 
Lett.}~{\bf 65} (1990) 3381.

\bibitem{GPY}M. Gronau, D. Pirjol and T.-M. Yan, {\it Phys.\
Rev.}~{\bf D60} (1999) 034021.

\bibitem{gardner}S. Gardner, {\it Phys.\ Rev.}~{\bf D59} (1999) 077502.

\bibitem{cleo-pi0pi0}CLEO Collaboration (D.M. Asner {\it et al.}),
{\it Phys.\ Rev.}~{\bf D65} (2002) 031103.

\bibitem{GrQu}Y. Grossman and H.R. Quinn, {\it Phys.\ Rev.}~{\bf D58}
(1998) 017504.

\bibitem{charles}J. Charles, {\it Phys.\ Rev.}~{\bf D59} (1999) 054007.

\bibitem{GLSS}M. Gronau, D. London, N. Sinha and R. Sinha,
{\it Phys.\ Lett.}~{\bf B514} (2001) 315.

\bibitem{Brhopi}H. Lipkin, Y. Nir, H. Quinn and A. Snyder, 
{\it Phys.\ Rev.}~{\bf D44} (1991) 1454.

\bibitem{qs}A. Snyder and H. Quinn, {\it Phys.\ Rev.}~{\bf D48}
(1993) 2139.

\bibitem{qusi}H. Quinn and J. Silva, {\it Phys.\ Rev.}~{\bf D62}
(2000) 054002. 

\bibitem{dea}A. Deandrea, R. Gatto, M. Ladisa, G. Nardulli and
P. Santorelli, {\it Phys.\ Rev.}~{\bf D62} (2000) 036001.

\bibitem{et-got}R. Enomoto and M. Tanabashi, {\it Phys.\ Lett.}~{\bf B386}
(1996) 413;\\ 
S. Gardner, H.B. O'Connell and A.W. Thomas, {\it Phys.\ Rev.\
Lett.}~{\bf 80} (1998) 1834.

\bibitem{eos}R. Enomoto, Y. Okada and Y. Shimizu, {\it Phys.\ Lett.}~{\bf 
B433} (1998) 109.

\bibitem{bbgm}I. Bediaga, R.E. Blanco, C. G\"obel and R. Mendez--Galain,
{\it Phys.\ Rev.\ Lett.}~{\bf 81} (1998) 4067.

\bibitem{BF-alpha}A.J. Buras and R. Fleischer, {\it Phys.\
Lett.}~{\bf B360} (1995) 138.

\bibitem{FM-alpha}R. Fleischer and T. Mannel, {\it Phys.\
Lett.}~{\bf B397} (1997) 269.

\bibitem{wuerth}F. W\"urthwein, private communication;\\
M. Tanaka  (CDF Collaboration),
{\it Nucl.\ Instrum.\ Meth.}~{\bf A462} (2001) 165.

\bibitem{BDpi}R.G. Sachs, EFI-85-22 (unpublished);\\
I. Dunietz and R.G. Sachs, {\it Phys.\ Rev.}~{\bf D37} (1988) 3186 [E:
ibid.~{\bf D39} (1988) 3515];\\ 
I. Dunietz, {\it Phys.\ Lett.}~{\bf B427} (1998) 179.

\bibitem{Babar-book}{\it The BaBar Physics Book}, eds.\ P. Harrison and 
H. Quinn, SLAC report 504 (1998).

\bibitem{diehl}M. Diehl and G. Hiller, 
{\it Phys.\ Lett.}~{\bf B517} (2001) 125.

\bibitem{FD2}R. Fleischer and I. Dunietz, {\it Phys.\ 
Lett.}~{\bf B387} (1996) 361.

\bibitem{petrak}For an analysis of the physics potential at the
$\Upsilon(5S)$ resonance, see S. Petrak, BABAR NOTE 507 (1999).

\bibitem{DG-hist}J.S. Hagelin, {\it Nucl.\ Phys.}~{\bf B193} (1981) 123;\\
E. Franco, M. Lusignoli and A. Pugliese, {\it Nucl.\ Phys.}~{\bf B194} 
(1982) 403;\\
L.L. Chau, {\it Phys. Rep.}~{\bf 95} (1983) 1;\\
M.B. Voloshin, N.G. Uraltsev, V.A. Khoze and M.A. Shifman, {\it Sov.\ J. 
Nucl.\ Phys.}~{\bf 46} (1987) 112;\\
A. Datta, E.A. Paschos and U. T{\"u}rke, 
{\it Phys.\ Lett.}~{\bf B196} (1987) 382;\\
A. Datta, E.A. Paschos and Y.L. Wu,
{\it Nucl.\ Phys.}~{\bf B311} (1988) 35;\\
R. Aleksan, A. Le Yaouanc, L. Oliver, O. Pene and J.C. Raynal,
{\it Phys.\ Lett.}~{\bf B316} (1993) 567.

\bibitem{stocchi}A. Stocchi, talk given at ICHEP 2000, July  27 -- August 2, 
2000, Osaka, Japan, LAL-00-55 [hep-ph/0010222]. 

\bibitem{Buras-Rt}A.J. Buras, in the proceedings of ICHEP 1996,
July 25--31, 1996, Warsaw, Poland, eds.\ Z. Ajduk and A.K. Wroblewski, 
World Scientific, Singapore (1997) p.\ 243 [hep-ph/9610461].

\bibitem{BBD}M. Beneke, G. Buchalla and I. Dunietz, 
{\it Phys.\ Rev.}~{\bf D54} (1996) 4419.

\bibitem{DGamma-cal}M. Beneke, G. Buchalla, C. Greub, A. Lenz and 
U. Nierste, {\it Phys.\ Lett.}~{\bf B459} (1999) 631.

\bibitem{hashimoto}S. Hashimoto and N. Yamada (JLQCD Collaboration),
talk given at BCP4, February 19--23, 2001, Ise-Shima, Japan, KEK-CP-108
[hep-ph/0104080];\\
S. Hashimoto, K. Ishikawa, T. Onogi, M. Sakamoto, 
N. Tsutsui and N. Yamada, {\it Phys.\ Rev.}~{\bf D62} (2000) 114502.

\bibitem{rome}D. Becirevic, D. Meloni, A. Retico, V. Gimenez, 
V. Lubicz and G. Martinelli, {\it Eur.\ Phys.\ J.}~{\bf C18} (2000) 157.

\bibitem{BeLe}M. Beneke and A. Lenz, {\it J.\ Phys.}~{\bf G27} (2001) 1219.

\bibitem{dunietz}I. Dunietz, {\it Phys.\ Rev.}~{\bf D52} (1995) 3048.

\bibitem{FD1}R. Fleischer and I. Dunietz, {\it Phys.\ 
Rev.}~{\bf D55} (1997) 259.

\bibitem{RF-Bs}R. Fleischer, {\it Phys.\ Rev.}~{\bf D58} (1998) 093001.

\bibitem{Hawaii}R. Fleischer, in the proceedings of the 2nd Conference 
on B Physics and CP Violation, March 24--28, 1997, Honolulu, Hawaii,
eds.\ T.E. Browder, F.A. Harris and S. Pakvasa, World Scientific, 
Singapore (1998) p.\ 469 [hep-ph/9705404].

\bibitem{ChiWo}C.-W. Chiang and L. Wolfenstein, 
{\it Phys.\ Rev.}~{\bf D61} (2000) 074031.

\bibitem{gro-DG}Y. Grossman, {\it Phys.\ Lett.}~{\bf B380} (1996) 99.

\bibitem{ADK}R. Aleksan, I. Dunietz and B. Kayser, 
{\it Z. Phys.}~{\bf C54} (1992) 653.

\bibitem{GL0}M. Gronau and D. London, {\it Phys.\ Lett.}~{\bf B253} (1991) 
483.

\bibitem{LSS}D. London, N. Sinha and R. Sinha, 
{\it Phys.\ Rev.\ Lett.}~{\bf 85} (2000) 1807.

\bibitem{FP}A. Falk and A. Petrov,  
{\it Phys.\ Rev.\ Lett.}~{\bf 85} (2000) 252.

\bibitem{DDLR}A. Dighe, I. Dunietz, H. Lipkin and J. Rosner, 
{\it Phys.\ Lett.}~{\bf B369} (1996) 144.

\bibitem{pol}J. Rosner, {\it Phys.\ Rev.}~{\bf D42} (1990) 3732.

\bibitem{CDF-angular}CDF Collaboration (T. Affolder {\it et al.}),
{\it Phys.\ Rev.\ Lett.}~{\bf 85} (2000) 4668.

\bibitem{LR-model1}See, for instance, D. Chang,  
{\it Nucl.\ Phys.}~{\bf B214} (1983) 435;\\ 
G. Ecker and W. Grimus, {\it Nucl.\ Phys.} {\bf B258} (1985) 328; 
{\it Z.\ Phys.} {\bf C30} (1986) 293.

\bibitem{BBMR}G. Barenboim, J. Bernabeu, J. Matias and M. Raidal,
{\it Phys.\ Rev.}~{\bf D60} (1999) 016003.

\bibitem{LR-model2}P. Ball, J.-M. Fr\`ere and J. Matias,  
{\it Nucl.\ Phys.}~{\bf B572} (2000) 3. 

\bibitem{BF-LR}P. Ball and R. Fleischer, {\it Phys.\ Lett.}~{\bf
B475} (2000) 111.

\bibitem{sil}D. Silverman, {\it Phys.\ Rev.}~{\bf D58} (1998) 095006.

\bibitem{gw}M. Gronau and D. Wyler, {\it Phys.\ Lett.}~{\bf B265} (1991)
172.

\bibitem{cleo-BDK}CLEO Collaboration (M. Athanas {\it et.\ al.}),
{\it Phys.\ Rev.\ Lett.}~{\bf 80} (1998) 5493.

\bibitem{belle-BDK}Belle Collaboration (K. Abe {\it et.\ al.}), 
{\it Phys.\ Rev.\ Lett.}~{\bf 87} (2001) 111801.

\bibitem{ads}D. Atwood, I. Dunietz and A. Soni, {\it Phys.\ Rev.\
Lett.}~{\bf 78} (1997) 3257.

\bibitem{ADS2}D. Atwood, I. Dunietz and A. Soni,
{\it Phys.\ Rev.}~{\bf D63} (2001) 036005;\\
D. Atwood, talk given at BCP4, February 19--23, 2001, Ise-Shima, Japan,
AMES-HET-01-03 [hep-ph/0103345].

\bibitem{gro-BDK}M. Gronau, {\it Phys.\ Rev.}~{\bf D58} (1998) 037301.

\bibitem{xing-BDK}Z.-Z. Xing, {\it Phys.\ Rev.}~{\bf D58} (1998) 093005.

\bibitem{FM}R. Fleischer and T. Mannel, {\it Phys.\ Rev.}~{\bf D57} (1998) 
2752.

\bibitem{GroRo-BDK}M. Gronau and J.L. Rosner,
{\it Phys.\ Lett.}~{\bf B439} (1998) 171.

\bibitem{JaKo}J. Jang and P. Ko, {\it Phys.\ Rev.}~{\bf D58} (1998) 111302.

\bibitem{dun}I. Dunietz, {\it Phys.\ Lett.}~{\bf B270} (1991) 75.

\bibitem{masetti}M. Masetti, {\it Phys.\ Lett.}~{\bf B286} (1992) 160.

\bibitem{fw}R. Fleischer and D. Wyler, {\it Phys.\ 
Rev.}~{\bf D62} (2000) 057503.

\bibitem{LHC-Bc}M. Gald\'on, R. P\'erez Ochoa, M.A. Sanchis-Lozano, 
J.A. Valls, {\it Nucl.\ Phys.\ (Proc.\ Suppl.)}~{\bf B50} (1996) 311.

\bibitem{Bc-BRs}J.-F. Liu and K.-T. Chao, {\it Phys.\ Rev.}~{\bf D56}
(1997) 4133;\\
P. Colangelo and F. De Fazio, {\it Phys.\ Rev.}~{\bf D61} (2000)
034012.

\bibitem{FOCUS}FOCUS Collaboration (J.M. Link {\it et al.}), 
{\it Phys.\ Lett.}~{\bf B485} (2000) 62.

\bibitem{BGLNP}S. Bergmann, Y. Grossman, Z. Ligeti, Y. Nir and A.A. Petrov,
{\it Phys.\ Lett.}~{\bf B486} (2000) 418.

\bibitem{E791}E791 Collaboration (E.M. Aitala {\it et al.}), 
{\it Phys.\ Lett.}~{\bf B445} (1999) 449; 
{\it Phys.\ Rev.\ Lett.}~{\bf 83} (1999) 32.

\bibitem{CLEO-yD}CLEO Collaboration (D. Cronin-Hennessy {\it et al.}),
hep-ex/0102006. 

\bibitem{CLEO-D}CLEO Collaboration (R. Godang {\it et al.})
{\it Phys.\ Rev.\ Lett.}~{\bf 84} (2000) 5038.

\bibitem{tanaka}J. Tanaka (Belle Collaboration), talk given at BCP4, 
February 19--23, 2001, Ise-Shima, Japan, hep-ex/0104053.

\bibitem{MeSi}C.C. Meca and J.P. Silva,
{\it Phys.\ Rev.\ Lett.}~{\bf 81} (1998) 1377.

\bibitem{Silva-Soffer}J.P. Silva and A. Soffer,
{\it Phys.\ Rev.}~{\bf D61} (2000) 112001.

\bibitem{GRL}M. Gronau, J. Rosner and D. London, {\it Phys.\ Rev.\
Lett.}~{\bf 73} (1994) 21.

\bibitem{GHLR-SU3}O. Hern\'andez, D. London, M. Gronau and J. Rosner,
{\it Phys.\ Lett.}~{\bf B333} (1994) 500; {\it Phys.\ Rev.}~{\bf D50} (1994) 
4529;\\
M. Gronau, O. Hernandez, D. London and J. Rosner, 
{\it Phys.\ Rev.}~{\bf D52} (1995) 6356.

\bibitem{CLEO97}CLEO Collaboration (R. Godang {\it et al.}),
{\it Phys.\ Rev.\ Lett.}~{\bf 80} (1998) 3456.

\bibitem{CLEO-BpiK-asym}CLEO Collaboration (S. Chen {\it et al.}),
{\it Phys.\ Rev.\ Lett.}~{\bf 85} (2000) 525.

\bibitem{Belle-BpiK-asym}Belle Collaboration (K. Abe {\it et al.}),
{\it Phys.\ Rev.}~{\bf D64} (2001) 071101.

\bibitem{defan}R. Fleischer, {\it Eur.\ Phys.\ J.}~{\bf C6} (1999) 451.

\bibitem{BpiK-mixed}M. Gronau and J. Rosner, 
{\it Phys.\ Rev.}~{\bf D57} (1998) 6843.

\bibitem{NR}M. Neubert and J. Rosner, {\it Phys.\ Lett.}~{\bf B441} (1998)
403;
{\it Phys.\ Rev.\ Lett.}~{\bf 81} (1998) 5076.

\bibitem{neubert-BpiK}M. Neubert, {\it J. High Energy Phys.}~{\bf 9902} 
(1999) 014.

\bibitem{BF-neut}A.J. Buras and R. Fleischer, 
{\it Eur.\ Phys.\ J.}~{\bf C16} (2000) 97.

\bibitem{FKNP}A. Falk, A. Kagan, Y. Nir and A. Petrov, in \cite{FSI}.

\bibitem{RF-FSI}R. Fleischer, {\it Phys.\ Lett.}~{\bf B435} (1998) 221.

\bibitem{GR-FSI}M. Gronau and J. Rosner, {\it Phys.\ Rev.}~{\bf D58} (1998) 
113005.

\bibitem{FSI-strat}M. Gronau and D. Pirjol, 
{\it Phys.\ Lett.}~{\bf B449} (1999) 321;\\
K. Agashe and N. Deshpande, {\it Phys.\ Lett.}~{\bf B451} (1999) 215
and {\bf B454} (1999) 359.

\bibitem{HSW}W.-S. Hou, J.G. Smith and Frank W\"urthwein, NTU-HEP-99-25
[hep-ex/9910014].

\bibitem{FM-BpiK-NP}R. Fleischer and T. Mannel, TTP-97-22
[hep-ph/9706261].

\bibitem{BPIK-NP}D. Choudhury, B. Dutta and A. Kundu, 
{\it Phys.\ Lett.}~{\bf B456} (1999) 185;\\ 
X.-G. He, C.-L.Hsueh and J.-Q. Shi, 
{\it Phys.\ Rev.\ Lett.}~{\bf 84} (2000) 18.

\bibitem{GNK}Y. Grossman, M. Neubert and A. Kagan, {\it J. High Energy 
Phys.}~{\bf 9910} (1999) 029.

\bibitem{FMat}R. Fleischer and J. Matias, {\it Phys.\ 
Rev.}~{\bf D61} (2000) 074004.

\bibitem{matias}J. Matias, {\it Phys.\ Lett.}~{\bf B520} (2001) 131.

\bibitem{QCDSrules}A. Khodjamirian, R. R\"uckl and C.W. Winhart,
{\it Phys.\ Rev.}~{\bf D58} (1998) 054013;\\
E. Bagan, P. Ball and V.M. Braun,
{\it Phys.\ Lett.}~{\bf B417} (1998) 154;\\
P. Ball, {\it J.\ High Energy Phys.} {\bf 9809} (1998) 005.

\bibitem{HY}W.-S. Hou and K.-C. Yang, {\it Phys.\ Rev.}~{\bf D61} (2000)
073014.

\bibitem{neubert-proc}M. Neubert, CLNS-00-1660 [hep-ph/0001334], presented 
at the ICTP Summer School in Particle Physics, June 7 -- July 9, 1999,
Trieste, Italy.

\bibitem{BrLe}S.J. Brodsky and G. Lepage, {\it Phys.\ Lett.}~{\bf B87}
(1979) 359 and {\it Phys.\ Rev.}~{\bf D22} (1980) 2157;\\
R. Field {\it et al.}, {\it Nucl.\ Phys.}~{\bf B186} (1981) 429.

\bibitem{Rosner-charm}J.L. Rosner, 
{\it Phys.\ Rev.}~{\bf D64} (2001) 094002.

\bibitem{khod}A. Khodjamirian, {\it Nucl.\ Phys.}~{\bf B605} (2001) 558.

\bibitem{kagan}A.L. Kagan, {\it Phys.\ Rev.}~{\bf D51} (1995) 6196.

\bibitem{pirjol}D. Pirjol, {\it Phys.\ Rev.}~{\bf D60} (1999) 054020.

\bibitem{Lipkin}H. Lipkin, {\it Phys.\ Lett.}~{\bf B415} (1997) 186.

\bibitem{snowmass}I. Dunietz, in the proceedings of the Workshop on
$B$ Physics at Hadron Accelerators, Snowmass, Colorado, eds.\ P. McBride
and C. Shekhar Mishra, p.\ 83 (1993).

\bibitem{BspiK}M. Gronau and J. Rosner, {\it Phys.\ Lett.}~{\bf
B482} (2000) 71.

\bibitem{BaBr}P. Ball and V.M. Braun, {\it Phys.\ Rev.}~{\bf D58} (1998) 
094016.

\bibitem{skands}P.Z. Skands, {\it J.\ High Energy Phys.}~{\bf 0101} (2001) 008.

\bibitem{HMChiPT1}E. Jenkins and M.J. Savage, {\it Phys.\ Lett.}~{\bf B281}
(1992) 331.

\bibitem{HMChiPT2}B. Grinstein, E. Jenkins, A.V. Manohar, M.J. Savage and
M.B. Wise, {\it Nucl.\ Phys.}~{\bf B380} (1992) 369.

\bibitem{lat1}UKQCD Collaboration (L. Lellouch {\it et al.}), 
{\it Nucl.\ Phys.\ (Proc.\ Suppl.)}~{\bf B73} (1999) 357.

\bibitem{georgi}H. Georgi, {\it Weak Interactions and Modern
Particle Theory} (Addison--Wesley Publishing Company, Redwood City,
California, 1984).

\bibitem{CW-BspiK}C.-W. Chiang and L. Wolfenstein, 
{\it Phys.\ Lett.}~{\bf B493} (2000) 73.

\bibitem{Gronau-U}M. Gronau, {\it Phys.\ Lett.}~{\bf B492} (2000) 297.

\bibitem{hurth-mannel}T. Hurth and T. Mannel, 
{\it Phys.\ Lett.}~{\bf B511} (2001) 196.

\bibitem{BoBu}S.W. Bosch and G. Buchalla, 
{\it Nucl.\ Phys.}~{\bf B621} (2002) 459;\\
see also  M. Beneke, T. Feldmann and D. Seidel, 
{\it Nucl.\ Phys.}~{\bf B612} (2001) 25.

\bibitem{gro-ro-Uspin}M. Gronau and J.L. Rosner,
{\it Phys.\ Lett.}~{\bf B500} (2001) 247.

\bibitem{THDM}L.F. Abbott, P. Sikivie and M.B. Wise,
{\it Phys.\ Rev.}~{\bf D21} (1980) 1393.

\bibitem{BITALY}A.J. Buras and L. Silvestrini,
{\it Nucl.\ Phys.}~{\bf B546} (1999) 299;\\
A.J. Buras, G. Colangelo, G. Isidori, A. Romanino and
L. Silvestrini,  {\it Nucl.\ Phys.}~{\bf B566} (2000) 3. 

\bibitem{SUSY}A. Masiero and O. Vives,
{\it Nucl.\ Phys.\ Proc.\ Suppl.}~{\bf B99} (2001) 228;\\
S.A. Abel and J.M. Fr\`ere, {\it Phys.\ Rev.}~{\bf D55} (1997) 1623;\\
A.L. Kagan and M. Neubert, {\it Phys.\ Rev.\ Lett.}~{\bf 83} (1999) 4929;\\
M. Brhlik, L.L. Everett, G.L. Kane, S.F. King and O. Lebedev,
{\it Phys.\ Rev.\ Lett.}~{\bf 84} (2000) 3041;\\
R. Barbieri, R. Contino and A. Strumia,
{\it Nucl.\ Phys.}~{\bf B578} (2000) 153;\\
K.S. Babu, B. Dutta and R.N. Mohapatra,
{\it Phys.\ Rev.}~{\bf D61} (2000) 091701;\\
A. Masiero, M. Piai, A. Romanino and L. Silvestrini,
{\it Phys.\ Rev.}~{\bf D64} (2001) 075005;\\
M. Dine, E. Kramer, Y. Nir and Y. Shadmi,
{\it Phys.\ Rev.}~{\bf D63} (2001) 116005.

\bibitem{reference-UT}T. Goto, N. Kitazawa, Y. Okada and M. Tanaka,
{\it Phys.\ Rev.}~{\bf D53} (1996) 6662;\\
Y. Grossman, Y. Nir and M.P. Worah,
{\it Phys.\ Lett.}~{\bf B407} (1997) 307;\\
A.L. Kagan and M. Neubert, in \cite{low-sin};\\
M. Randhawa and M. Gupta, {\it Phys.\ Lett.}~{\bf B516} (2001) 446.

\bibitem{B95}A.J. Buras, {\it Phys.\ Lett.}~{\bf B333} (1994) 476;
{\it Nucl.\ Instr.\ Meth.}~{\bf A368} (1995) 1.

\bibitem{NIR96}Y. Grossman, Y. Nir and R. Rattazzi, in \cite{new-phys}.

\bibitem{KTeV00X}A. Alavi-Harati {\it et al.}, {\it Phys.\ Rev.}~{\bf D61} 
(2000) 072006.

\bibitem{BePe0}S. Bergmann and G. Perez, {\it J.\ High Energy 
Phys.}~{\bf 0008} (2000) 034.

\bibitem{ALPH}ALEPH Collaboration (R. Barate {\it et al.}),
{\it Eur.\ Phys.\ J.}~{\bf C19} (2001) 213.

\bibitem{misiak}M. Misiak, talk given at the XXXVIth Rencontres de Moriond, 
March 10--17, 2001, Les Arcs, France [hep-ph/0105312].

\bibitem{myon-moment}Muon $g-2$ Collaboration (H.N. Brown {\it et al.}),
{\it Phys.\ Rev.\ Lett.}~{\bf 86} (2001) 2227.

\bibitem{Rare}T. Hurth, CERN-TH-2001-146 [hep-ph/0106050];\\
T. Mannel, talk given at BCP4, February 19--23, 2001, 
Ise-Shima, Japan, TTP01-09 [hep-ph/0103310];\\
C. Greub, talk given at the 8th International Symposium on Heavy Flavor 
Physics, July 25 -- 29, 1999, Southampton, England, BUTP-99-22
[hep-ph/9911348];\\ 
A. Ali, in the proceedings of the 7th International Symposium on Heavy 
Flavor Physics, July 7--11, 1997, Santa Barbara, California, ed.\ 
C. Campagnari, World Scientific, 
Singapore (1999) p.\ 196 [hep-ph/9709507].



\end{thebibliography}
\end{document}